\documentclass[
    a4paper, 
    fontsize=10pt, 
    twoside=true, 
	numbers=noenddot, 
	fontmethod=tex, 
]{kaobook}

\ifxetexorluatex
	\usepackage{polyglossia}
	\setmainlanguage{english}
\else
	\usepackage[english]{babel} 
\fi
\usepackage[english=british]{csquotes}	

\usepackage{blindtext} 

\usepackage{kaobiblio}
\usepackage[framed=true]{kaotheorems}

\usepackage{kaorefs}

\graphicspath{{figures/}{figures/side}} 

\makeindex[columns=3, title=Alphabetical Index, intoc] 

\makeglossaries 
\newglossaryentry{darwinism}
{
    name=Darwinism,
    description={Charles Darwin's theory explaining the mechanism of evolution by natural selection. According to this theory, evolution results from the interaction of three principles: heredity, variation and survival (natural selection). It describes how species adapt to their environment over time}
}

\newglossaryentry{evolution}
{
    name=evolution,
    description={The gradual change in inherited characteristics of biological populations over successive generations. It results from several different processes: (i) mutations, changes in the genetic sequence (ii) recombinations, exchanges of genetic material between individuals (iii) genetic drift, random changes in gene frequency and (iv) natural selection (see entry: \emph{selection})}
}

\newglossaryentry{genotype}
{
    name=genotype,
    description={The genetic makeup of an organism, representing the specific combination of genes present in its DNA. It serves as the blueprint for the organism's traits and characteristics}
}

\newglossaryentry{trait}
{
    name=trait,
    description={A specific characteristic or feature of an organism that can be inherited or influenced by environmental factors. Traits contribute to the overall phenotype and are subject to evolutionary pressure}
}

\newglossaryentry{phenotype}
{
    name=phenotype,
    description={The observable characteristics and traits of an organism. In general, the phenotype is determined by both the genomic makeup (genotype) and environmental factors. It includes features like appearance, behavior, and physiological functions}
}

\newglossaryentry{statistical genetics}
{
    name=statistical genetics,
    description={In the sense of Neher-Shraiman, a statistical multilocus theory that explains how the laws of quantitative genetics -- i.e., the study of phenotypic variation among individuals -- emerge from the stochastic evolutionary dynamics in the space of genotypes}
}

\newglossaryentry{inheritance}
{
    name=inheritance,
    description={The process by which genetic information is passed from one generation to the next. This transfer of genetic material occurs during reproduction, ensuring the continuity and maintenance of traits within a species}
}

\newglossaryentry{variation}
{
    name=variation,
    description={Diversity observed in the traits and characteristics among individuals within a population, arising from genetic mutations, recombination, and other sources. It introduces differences in physical and behavioral attributes, providing the raw material upon which natural selection acts. It is crucial for a population's adaptability to changing environments}
}

\newglossaryentry{selection}
{
    name=selection,
    description={The mechanism through which certain heritable traits confer advantages to individuals, increasing their likelihood of survival and reproduction. It is driven by the interplay between organisms and their environment, favoring traits that enhance an organism's fitness for its ecological niche}
}

\newglossaryentry{fitness}
{
    name=fitness,
    description={Organism's ability to survive and reproduce in a given environment. It is a measure of the relative reproductive success of individuals with specific traits. Often defined as proportional to the average number of offspring of an individual}
}

\newglossaryentry{fitness landscape}
{
    name=fitness landscape,
    description={A metaphorical representation used in evolutionary biology to illustrate the relationship between genotypes and their associated fitness in a given environment. In this landscape, each point represents a unique genotype, and the elevation at that point represents the corresponding fitness. Peaks on the landscape represent optimal genotypes with high fitness. The structure of the fitness landscape influences the paths evolution may take}
}

\newglossaryentry{function}
{
    name=function,
    description={The specific role or task performed by a component (e.g., molecule, cell, organ) within a living organism. They are essential for the organism's survival, growth, and reproduction, therefore subject to evolutionary pressure}
}

\newglossaryentry{self-referential}
{
    name=self-referential,
    description={Said of biological dynamics, where the update rules change during the time evolution of the system, in a manner that depends on the state and thus on the history of the system.}
}

\newglossaryentry{caenorhabditis elegans}
{
    name=caenorhabditis elegans,
    description={Abbreviated as \ce. A free-living transparent nematode (roundworm) about 1 mm in length. It is often used as a model organism in biological research, due to its simplicity, well-defined anatomy, and short life cycle. Despite this, it possesses a wide behavioral array. Beyond the basics of locomotion, foraging, and feeding, the worm can discern and navigate towards or away from various chemicals, odors, temperature gradients, and food sources. Furthermore, it demonstrates social awareness, detecting the presence, density, and even sex of neighboring nematodes}
}

\newglossaryentry{nervous system}
{
    name=nervous system,
    description={A network of specialized cells (neurons) that coordinate and regulate the activities of an organism. In \ce, the nervous system of an adult hermaphrodite consists of 302 neurons, uniquely identifiable. Most of them are found in clusters, called \emph{ganglia}. Neuronal processes extend from the ganglia and travel in longitudinal nerve bundles to different regions of the nervous system. The most prominent are the nerve ring, ventral nerve cord and dorsal nerve cord}
}

\newglossaryentry{neuron}
{
    name=neuron,
    description={Also, or nerve cell. It is the basic structural and functional unit of the nervous system. Neurons transmit information using electrical and chemical signals. They consist of a cell body, dendrites (receiving inputs), and an axon (transmitting outputs). Neurons play a crucial role in processing and transmitting information in the nervous system. The morphology of a neuron can vary substantially. In \ce, they are mostly unipolar or bipolar.}
}

\newglossaryentry{synapse}
{
    name=synapse,
    description={A specialized junction between two neurons, where information is transferred from one cell to another. There can be electrical or chemical. The former, also called \emph{gap junctions}, are specialized channels that directly connect the cytoplasm of adjacent cells, allowing various molecules, ions, and electrical impulses to pass between the cells. The latter, \emph{chemical synapses}, function as specialised junctions that facilitate the one-way relay of chemical signals, or neurotransmitters, from a presynaptic to one or more postsynaptic cells}
} 

\usepackage{nomencl}
\makenomenclature 

\usepackage{CJKutf8}

\usepackage{bm}
\usepackage{algorithm2e}  
\usepackage{tikz}
\usepackage{pgfplots}
\usepackage{siunitx}

\usepackage{fontawesome5}
\usepackage{booktabs,tabularx}
\usepackage[table]{xcolor}
\usepackage{caption}
\usepackage{subcaption}
\usepackage{xargs}
\usepackage{multicol}

\usepackage{listings}
\newcommand{\av}[1]{\langle #1 \rangle}
\newcommand{\de}{\partial}  
\DeclareMathOperator*{\argmax}{arg\,max}
\newcommand{\ce}{\emph{C. elegans}}         
\newcommand{\h}{(\faBell[regular])}  

\definecolor{darkbrown}{HTML}{DFD7BF}
\definecolor{lightbrown}{HTML}{F2EAD3}

\definecolor{sensorybase}{HTML}{fda0fd}
\colorlet{sensory}{sensorybase!40}
\definecolor{interneuronbase}{HTML}{ff442f}
\colorlet{interneuron}{interneuronbase!40}
\definecolor{motorbase}{HTML}{5cafff}
\colorlet{motor}{motorbase!40}
\definecolor{modulatorybase}{HTML}{ffc000}
\colorlet{modulatory}{modulatorybase!40}

\begin{document}

\pagestyle{plain}

\frontmatter









\newcommand{\sectionlinetwo}[2]{%
  \nointerlineskip \vspace{.5\baselineskip}\hspace{\fill}
  {\color{#1}
    \resizebox{0.7\linewidth}{1.5ex}
    {{%
    {\begin{tikzpicture}
    \node (C) at (0,0) {};
    \node (D) at (9,0) {};
    \path (C) to [ornament=#2] (D);
    \end{tikzpicture}}}}}%
    \hspace{\fill}
    \par\nointerlineskip \vspace{.5\baselineskip}
  }

  
\begin{titlepage}
\pdfbookmark{Titlepage}{Titlepage}

    \begin{center}
        {
        \large
        Doctoral Thesis\\
        \vspace{2cm}

        \par\noindent\rule{\textwidth}{4pt}
        
        \vspace{.5cm}
        {\Huge \textbf{The exploration-exploitation paradigm}}\bigskip\\
        \LARGE A biophysical approach
        }
        \vspace{.5cm}
        \par\noindent\rule{.5\textwidth}{1.5pt}
        \vspace{2cm}
        \\
        \includegraphics[width=0.2\textwidth]{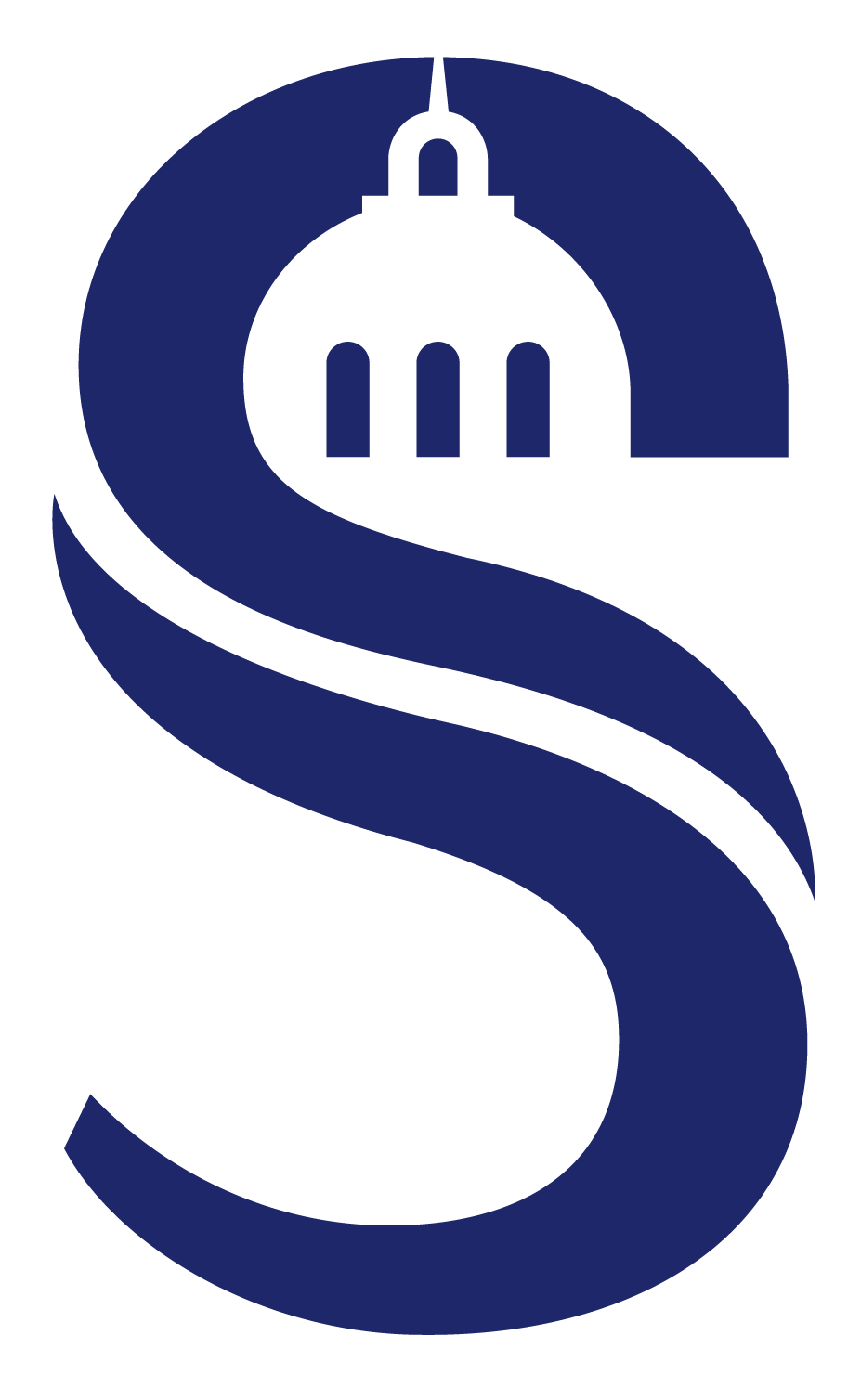}\\
        \vspace{2cm}

        \huge \textbf{Vito Dichio} 
        
        \bigskip
        
        \vfill  
        \large    
        A thesis submitted in partial fulfillment of the requirements for the degree of \\
        \Large \emph{Doctor of Philosophy} \\
        \vspace{1cm}
        
        \large October 2023
    \end{center}        
\end{titlepage}

\thispagestyle{empty}
\pdfbookmark{Titleback}{Titleback}


\hfill

\vspace{\stretch{1}}

\vspace{\stretch{1}}
\includegraphics[width=0.3\textwidth]{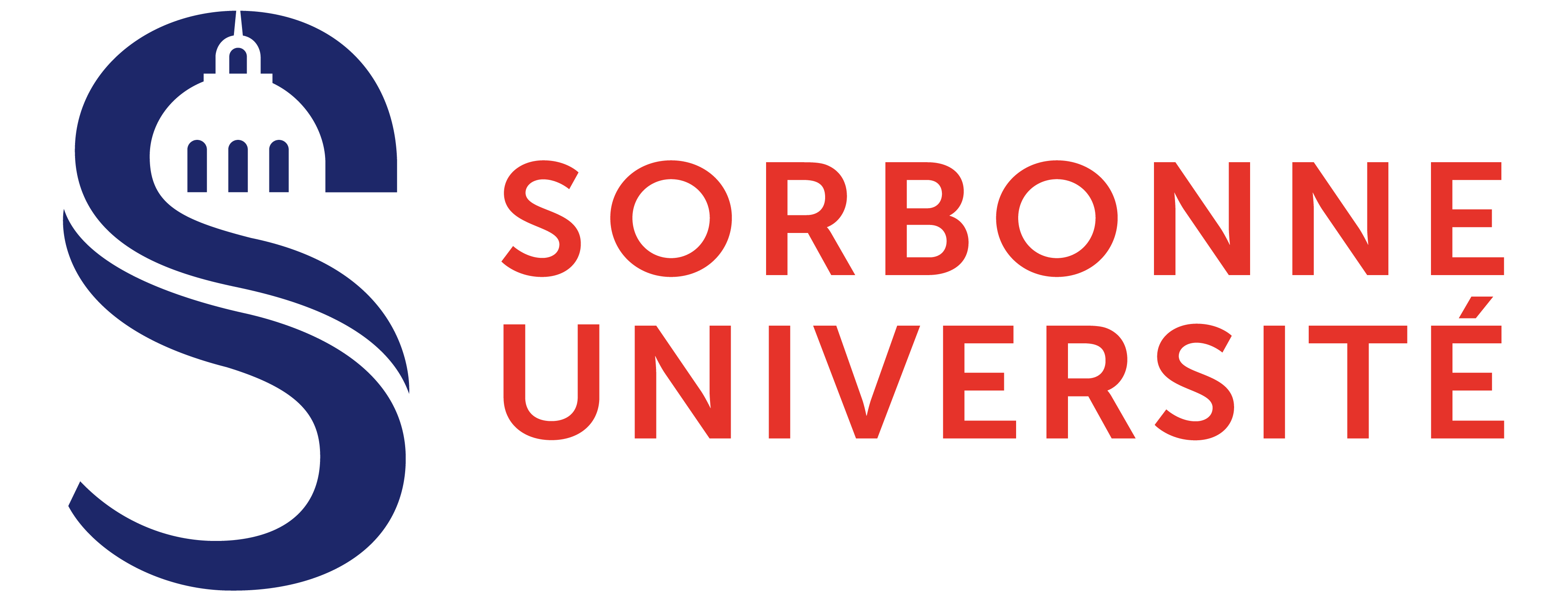} 
\hfill
\includegraphics[width=0.3\textwidth]{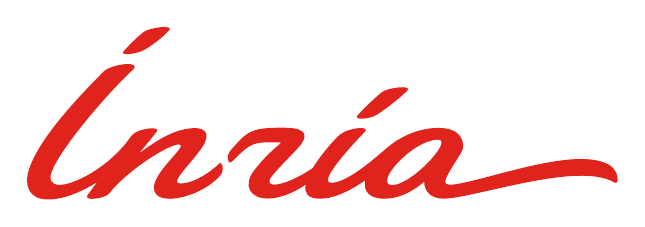}
\hfill
\includegraphics[width=0.3\textwidth]{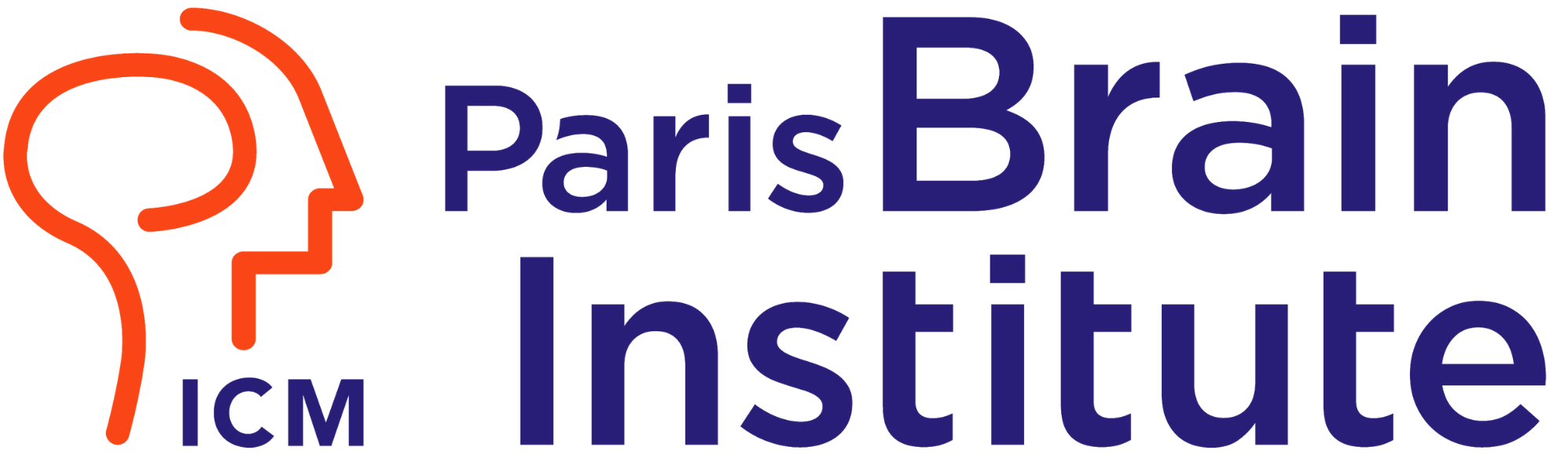}

\begin{center}
    \normalsize \textbf{Sorbonne université}\\
    \small 21, Rue de l'école de médecine, 75006 Paris, France\medskip\\
    \normalsize \textbf{Institut national de recherche en informatique et en automatique}\\
    \small 2, Rue Simone IFF, 75012 Paris, France\medskip\\
    \normalsize \textbf{Institut du cerveau}\\ 
    \small 47, bd de l'hôpital 75013 Paris, France\medskip\\
\end{center}

\bigskip

\vspace{1.5cm}

\begin{center}
\large
Members of the Jury: \\
\bigskip

\normalsize \textbf{Fabrizio De Vico Fallani} (Directeur de thèse)\\
\small \emph{Institut national de recherche en informatique et en automatique, Insitut du cerveau} \medskip\\

\normalsize\textbf{Guido Caldarelli} (Président, Rapporteur)\\
\small \emph{Università Ca' Foscari} \medskip\\

\normalsize\textbf{Marta Sales-Pardo} (Rapporteure)\\
\small \emph{Universitat Rovira i Virgili} \medskip\\

\normalsize\textbf{Demian Battaglia} (Examinateur)\\
\small \emph{Université d'Aix-Marseille, Université de Strasbourg} \medskip\\

\normalsize\textbf{Ga\v{s}per Tka\v{c}ik} (Examinateur)\\
\small \emph{Institute of Science and Technology Austria} \medskip\\
\end{center}

\vspace{2cm}

\begin{center}
    \includegraphics[width=0.4\textwidth]{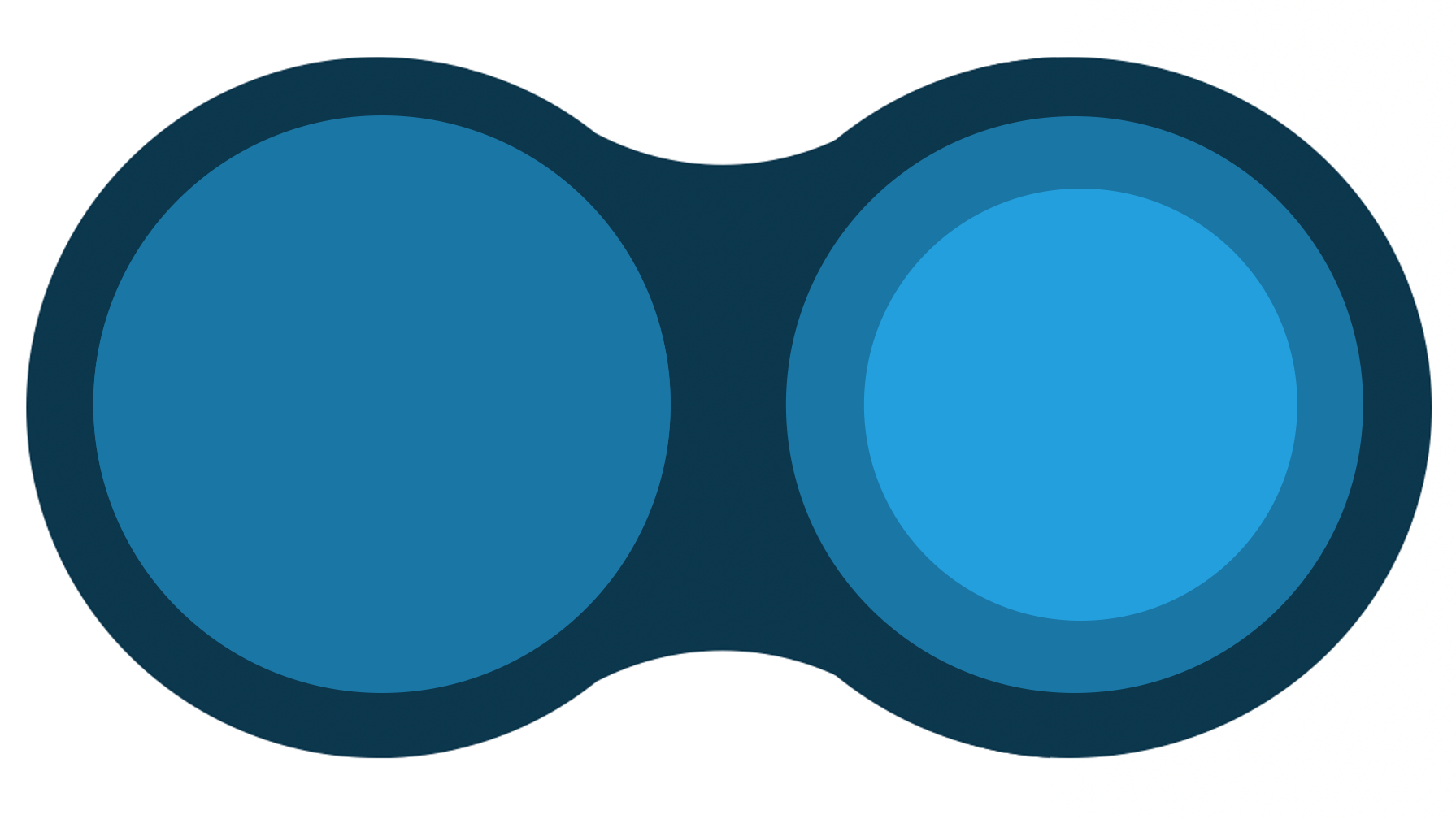}
\end{center}
        
\vspace{1cm}

\newcommand{\mail}[1]{\href{mailto:#1}{\texttt{#1}}}
\noindent
\faEnvelope[regular]\quad \mail{dichio.vito@gmail.com} 


\newpage

\vspace*{\fill}

\normalsize
    \begin{center}
	Among Chuang-tzu's many skills, he was an expert draftsman. The king asked him to draw a crab. Chuang-tzu replied that he needed five years, a country house, and twelve servants. Five years later the drawing was still not begun. "I need another five years," said Chuang-tzu. The king granted them. At the end of these ten years, Chuang-tzu took up his brush and, in an instant, with a single stroke, he drew a crab, the most perfect crab ever seen.
    \end{center}
	\begin{flushright} -- Italo Calvino, Six memos for the next millennium \end{flushright}
 
    \vspace{3cm}

    \begin{center}
    \begin{CJK}{UTF8}{min}
    富士山に一度も登らぬバカ、二度登るバカ\smallskip\\
    (Fools those who never climb the Mount Fuji, fools those who do it twice)
    \end{CJK}
    \end{center}
    \begin{flushright} -- Japanese proverb \end{flushright}
    \vspace*{\fill}
    
\newpage
    
    \begin{figure}[p]
        \centering
        \includegraphics[width=\columnwidth]{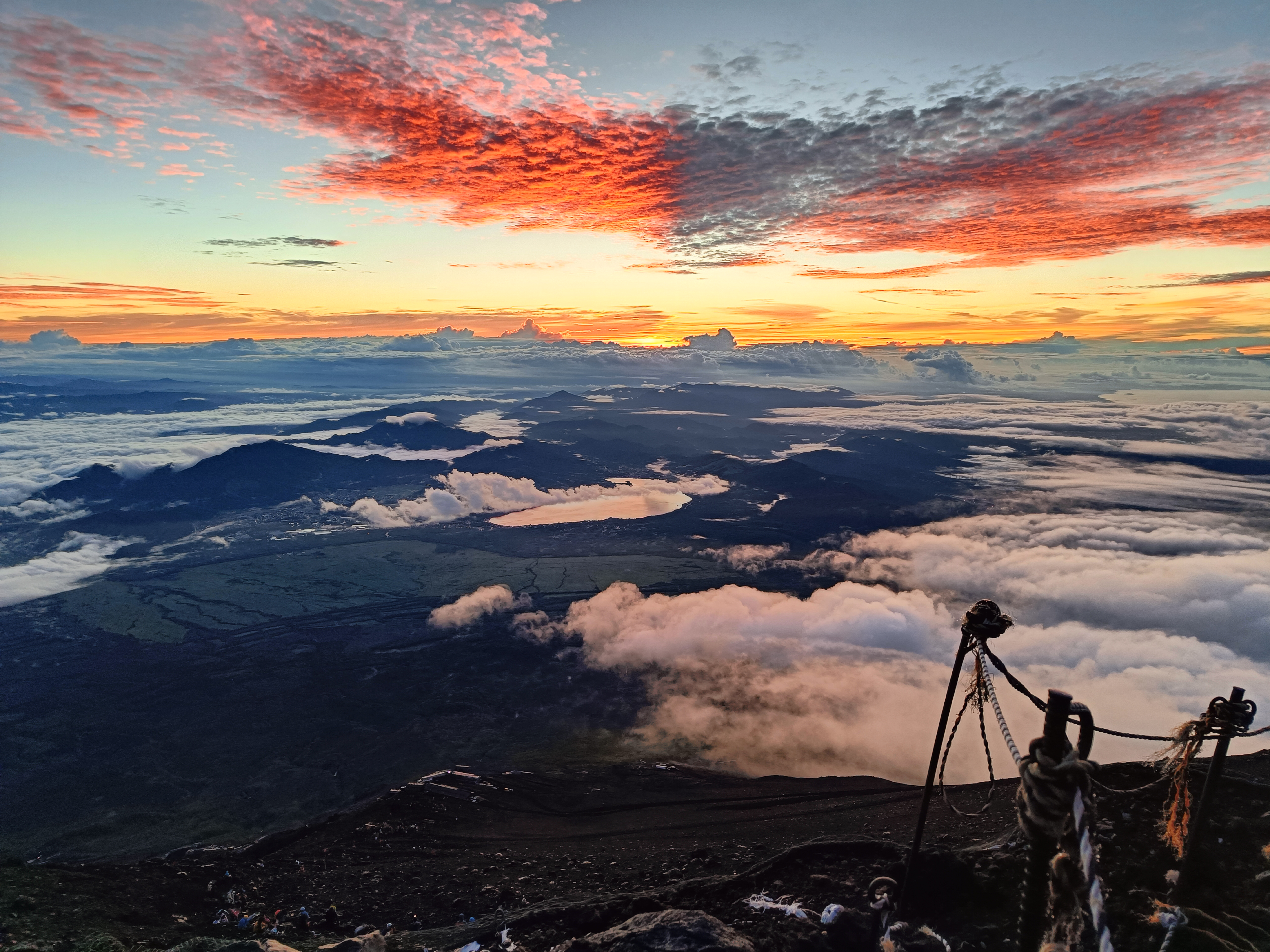}
    \begin{center}
    \small
        \emph{Peak of the Mount Fuji (Japan), 3376m, 4.40 am (sunrise), 8 August 2023.}
    \end{center}
    \end{figure}





\begingroup 

\setlength{\textheight}{230\hscale} 

\etocstandarddisplaystyle 
\etocstandardlines 

\tableofcontents 


\let\cleardoublepage\bigskip
\let\clearpage\bigskip



\endgroup

\chapter*{Acknowledgments}
\addcontentsline{toc}{chapter}{Acknowledgments} 

\begin{center}
    \emph{Below are the words I read after being awarded the title of Doctor of Philosophy, 23 October 2023, Auditorium of the Institut du Cerveau, Paris, end of the ceremony. Surrounded by people I love, I tried in vain to hold back a some tears.}
\end{center}

As you may know, if you are here, in recent times I have climbed a mountain. At the foot of the mountain, it was foggy, it was late evening, we could not see far. As we started, step by step, the path seemed smooth. I had never climbed a mountain; I am not familiar with such surroundings. Still, I though, it is easy enough, the path is smooth. Then, it started to rain. A rainstorm, for hours, it was night time. We could not stop, we had to go. The stones were slippery, the rain paralysing. It is time to stop, I thought, one has to accept it, at some point, it’s not a big deal, it is just a mountain. However, I had friends around me. Friends walking alongside me, probably with similar thoughts to the mines. Midway up a fucking mountain, night-time, under a heavy rain, I still thought that, among those people, that was to one of the most amazing paths I had taken in my life. At some point, climbing the mountain, we were so high that the clouds were left below, the rain with them. We were literally above the clouds. At that point I realised it was only a matter of time before we made it to the end. At 4 in the morning, on the top of mountain, I gazed out over the horizon, it was clear, it was immense. It was the sunrise, the most spectacular sunrise I have ever seen in my life.
\bigskip

In the first place, once more, I would like to express my most heartfelt thanks to the members of my Jury, for having accepted the invitation. At the cost of sounding rhetoric, it was an honour for me to stand on this stage and defend and discuss my research with you. 

Then, what to say about Fabrizio. Fabrizio has been more than a supervisor, a friend – and (by the way) an excellent central defender on the football pitch. Thanks in particular for constant encouragement over the last three years. Also, for having granted me the possibility of spending some fruitless time in exploring the many wrong ideas I have had. More than the plots I have shown you here, more than the paper we have written, it is there that I have grown the most as a scientist.

There are so many people I should say thanks to, a PhD is a joint effort between a host of people. Before switching language, let me collectively thank all those people I love that have crowded my days, in my recent life. Those that have been fundamental for my scientific path, Mario, Erik, Hong-Li, Fulvio. Those that, this afternoon, decided to dedicate to my dissertation a few hours, here in this room – I know it’s not just for the Italian cookies outside – and those that are listening to these words remotely, on Zoom, your presence means the world to me, in few selected moments of one’s life, it is important to just be there. Those that have come by Rue Bonaparte, for a dinner, a drink, one night or few days. Those that I have met in some school or conference, in some corner of the world. And to this corner of the world, Paris, that I have come to call home, where I have most likely spent the most incredible years of my life. To all of you, thanks from the bottom of my heart, thanks for the times we had.

Voilà. Ça c’est le moment de partager quelques mots en français, j’y tiens beaucoup. Pour commencer, je voudrais remercier l’énorme famille de cet Institut, et d’ARAMIS. Ça fait trois ans que chaque fois qu’il y a une soutenance chez nous, il/elle dit que vous êtes incroyables, ambiance de ouf. Et vous savez quoi, c’est vrai, chaque fois c’est vrai. Donc, c’est à moi de le dire maintenant. Et je suis heureux que c’est à moi de le dire maintenant, mais j’en suis un peu triste aussi, je ne le cache pas. Je n'ai pas l'illusion de pouvoir avoir deux fois la chance que j’ai eu quand je vous ai rencontrés. Tellement de choses de cette thèse vont me manquer, et vous, bien sûr, plus que tout.
Un énorme merci à Juliana, Charley, Remy, pour le parcours qu’on a fait ensemble. À Ravi et la team du foot du Dimanche. À ceux qui sont passés par la salle des stagiaires, les meilleurs postes de travail d’ARAMIS, by far. À Elisa, Domitille, Sophie et la team du quatrième. 

Camille, Elise. Après le cinquième Spritz Campari, dans une place de Bologna, je vous ai dit un truc, et je voudrais que tout le monde sache. Vous êtes la joie de vivre en personne, j’adore ce que vous êtes, et ce que je suis avec vous, grâce à vous. Merci aussi pour les slangs que vous m’avez appris. Aujourd’hui, je me nachav. No, je déconne, je déconne. 

Et enfin, Tristan, Tristano. On a partagé bien plus qu’un bureau dans les dernières trois années, on a partagé un parcours de vie. Tu étais déjà là pour m’aider quand je ne comprenais rien de la procédure d’inscription au doctorat. Mais tu étais là aussi la dernière fois que je n’arrivais pas à parler, trop de larmes, je m’en souviens très bien. Au milieu, plein de choses. Il y a des stéréotypes sur les Parisiens et chaque fois que quelqu’un me demande je dis que non, ce n’est pas vrai, au contraire. Dans des cas comme ça, je pense surtout à toi, voilà. 

Due ringraziamenti in Italiano. Ai tanti amici italiani qui a Parigi, a giudicare dalla mia personale e completamente unbiased statistica, una persona su due a Parigi è italiana. Grazie per tutte le chiacchiere, le serate, gli sprazzi di casa. Se non mi sono mai sentito solo in questi anni è perché casa era anche qui, casa eravate voi, \#teamMattarella.

Per concludere, la casa con la C maiuscola. Nella vita un po’ sconclusionata che mi trovo a vivere ci sono tuttavia alcuni punti fissi. Ad esempio, tornare giù nel mio paese e passare a prendere un caffè dalle zie. Oppure, chiamare in serata quasi ogni giorno il mio fratellino, che racconta i cazzi suoi e non fa domande. Ecco, ci tengo a ringraziare, per una volta in modo esplicito, i miei fratelli e sorelle, Cì, Roz e Aaaaangela. Non so come sia successo di preciso, ma siamo tutti arrivati adulti a volerci tutti un sacco bene. Forse non l’ho mai detto, ma sono davvero felice di essere vostro fratello, ancor di più di essere il fratello scherziero -- eh sì sì è scherziero, lo è. 

Gli ultimi due ringraziamenti li ho lasciati in fondo. A mia mamma, sora Anna. Mi dispiace per tutte le telefonate a cui non ho risposto negli anni, ma anche quando non l’ho fatto, mi è sempre stato caro il sapere che qualcuno, in qualunque situazione, in qualunque condizione, anche dall’altra parte del mondo, stesse pensando a me. Infine, a mio papà Raffaele, a l’ingegnere. A un certo punto tutto è sembrato andare un po’ a rotoli, però stiamo ancora qua, io ho di nuovo una cravatta al collo e prendo un altro titolo di studio. Stasera andiamo a cena, e la prossima volta mi racconti se ha piovuto, se si può seminare, cosa dicevano gli antichi il giorno di San Vito.

Ah dimenticavo, un grazie anche a \emph{cur' scem' d' Pelo}, il cagnolino più bello del mondo, che parla in dialetto e mi segue da Lemenzano.

\begin{center}
    \rule{.1\textwidth}{1pt}
\end{center}
\emph{Something I did not say that day but which was the only paragraph of a first version of these acknowledgments and which I would like to leave on the records. /} My heartfelt gratitude to the many libraries around the world that have graciously hosted me during the countless hours spent in writing these pages.

\begin{itemize}
    \item[$\circ$] \emph{Bibliothèque Sainte-Geneviève}, 10 place du Panthéon, Paris, France
    \item[$\circ$] \emph{Bibliothèque Mazarine}, 23 quai de Conti, Paris France
    \item[$\circ$] \emph{Bibliothèque Richelieu (BnF)}, 5 Rue Vivienne, Paris, France
    \item[$\circ$] \emph{Bibliothèque François-Mitterrand (BnF)}, Quai François Mauriac, Paris, France
    \item[$\circ$] \emph{Wienbibliothek im Rathaus}, Felderstraße 1, Wien, Austria
    \item[$\circ$] \emph{Universitätsbibliothek Wien}, Universitaetsring 1, Wien, Austria
    \item[$\circ$] \emph{Tokyo Metropolitan Central Library}, 5 Chome-7-13 Minamiazabu, Minato City, Tokyo, Japan
    \item[$\circ$] \emph{Katsushika City Library}, 6 Chome-7-13, Kanamachi, Katsushika City, Tokyo, Japan
    \item[$\circ$] \emph{Biblioteca Provinciale "T. Stigliani"}, Piazza Vittorio Veneto, Matera, Italia
\end{itemize}

\chapter*{Abstract}
\addcontentsline{toc}{chapter}{Abstract} 

The study of living systems is notoriously challenging. The often-quoted daunting complexity of biological systems is primarily due to the intricacies of their interactions, their multiple organisation levels and their dynamic nature. In the quest to understand this complexity, parallels drawn with standard physics – in particular, statistical physics -- are both useful and of limited use. On the one hand, they provide a rich set of theoretical and methodological building blocks for constructing theories and designing experiments. On the other hand, life also unfolds according to principles that are unparalleled in the physics of conventional matter. 

A crucial difference lies in the notion of function: biological systems are shaped by the need to perform specific tasks. A general problem for living systems is to find and promote those configurations that yield improved or optimal functions, we call this the exploration-exploitation (EE) problem. One specific instance of the above is found in evolutionary biology. There, random genetic mutations sustain the exploration of the configuration space, with those leading to higher reproductive success being favoured by natural selection.

Inspired by the latter, we develop a novel formalism that encodes a general exploration-exploitation dynamics for biological networks. In particular, our EE dynamics is represented as an exploration of a functional landscape and consists of stochastic configuration changes combined with the state-dependent optimisation of an objective function ($F$ metric). We begin by investigating its main features through the study of simple, analytically tractable functional landscapes. We deploy simulations for more general and complex applications. 

We then turn to the brain wiring problem, i.e., the development of an individual's nervous system during its early life. We argue that this is another specific instance of the EE problem and therefore can be addressed by using our theoretical framework. In particular, we focus on brain maturation in the nematode \ce, the only organism for which a complete network of neurons and neuronal connections has been reconstructed, at multiple developmental time points (seven). We fix the network at birth and use the adult stage to infer (i) a parsimonious maxent (ERG) description of the $F$ metric for the worm brain and (ii) the two parameters of our EE dynamics.
According to the topography of its functional landscape, the adult brain is characterised by a tendency to form both triads and high degree nodes. We demonstrate that our EE dynamics in such landscape is capable of tracking down the entire developmental history. In particular, we show that the trajectory we obtain closely reproduces the other experimental time points that we did not use for inference. This is true both in the space of model statistics and for a number of other network properties. Additionally, we discuss a micro-level interpretation of the EE dynamics in terms of the underlying synapse formation process.

Our study is a first step towards the system-level understanding of the development of a natural brain and can be extended (i) to encompass more complex functional landscapes, (ii) to different organisms than the \ce\ and (iii) to several different problems than the brain wiring. Indeed, we posit that the exploration-exploitation paradigm is among those life-specific principles that we are just beginning to uncover.

\chapter*{Résumé en français}
\addcontentsline{toc}{chapter}{Résumé en français} 

L'étude des systèmes vivants est notoirement difficile. La complexité déconcertante des systèmes biologiques, souvent citée, est principalement due à la complexité de leurs interactions, à leurs multiples niveaux d'organisation et à leur nature dynamique. Dans la quête de compréhension de cette complexité, les parallèles établis avec la physique standard - en particulier la physique statistique - sont à la fois utiles et d'une utilité limitée. D'une part, ils fournissent un riche ensemble d'éléments théoriques et méthodologiques pour construire des théories et concevoir des expériences. D'autre part, la vie biologique se déroule aussi selon des principes qui sont sans équivalent dans la physique de la matière conventionnelle. 

Une différence cruciale réside dans la notion de fonction : les systèmes biologiques sont façonnés par la nécessité d'accomplir des tâches spécifiques. Un problème général pour les systèmes vivants est de trouver et de promouvoir les configurations qui produisent des fonctions améliorées ou optimales, ce que nous appelons le problème de l'exploration-exploitation (EE). Un exemple spécifique de ce problème se trouve dans la biologie évolutive. Dans ce cas, des mutations génétiques aléatoires soutiennent l'exploration de l'espace de configuration, celles qui correspondent à un succès reproductif plus élevé étant favorisées par la sélection naturelle.

Inspirés par ce dernier cas, nous développons un nouveau formalisme qui encode une dynamique générale d'exploration-exploitation pour les réseaux biologiques, représentée comme une exploration d'un paysage fonctionnel. En particulier, notre dynamique d'EE consiste en des changements de configuration stochastiques combinés à l'optimisation dépendante de l'état d'une fonction objective (métrique $F$). Nous commençons par étudier ses principales caractéristiques à travers l'étude de paysages fonctionnels simples et analytiquement traitables. Nous déployons des simulations pour des applications plus générales et plus complexes.
 
Nous nous penchons ensuite sur le problème du câblage du cerveau, c'est-à-dire le développement du système nerveux d'un individu tout au long de sa vie. Nous soutenons que ce dernier est un autre exemple spécifique du problème de l'EE et qu'il peut donc être traité à l'aide de notre cadre théorique. En particulier, nous nous concentrons sur la maturation du cerveau chez le nématode \ce, le seul organisme pour lequel un réseau complet de neurones et de connexions neuronales a été reconstruit, à plusieurs moments du développement. Nous fixons le réseau à la naissance et utilisons le stade adulte pour déduire (i) une description max.ent. parcimonieuse (ERG) de la métrique $F$ pour le cerveau du ver et (ii) les deux paramètres de notre dynamique EE.

Selon la topographie de son paysage fonctionnel, le cerveau adulte est caractérisé par une tendance à former des triades et des nœuds de degré supérieur. Nous montrons que notre dynamique d'EE dans un tel paysage est capable de retracer toute l'histoire du développement. En particulier, nous montrons que la trajectoire que nous obtenons reproduit étroitement les autres points temporels expérimentaux que nous n'avons pas utilisés pour l'inférence. Ceci est vrai à la fois dans l'espace des statistiques du modèle et pour un certain nombre d'autres propriétés du réseau. En outre, nous discutons d'une interprétation micro-niveau de la dynamique de l'EE en termes de processus sous-jacent de formation des synapses.

Notre étude est un premier pas vers la compréhension au niveau du système du développement d'un cerveau naturel et peut être étendue (i) à des paysages fonctionnels plus complexes, (ii) à d'autres organismes que le \ce\ et (iii) à d'autres problèmes que le câblage du cerveau. En effet, nous pensons que le paradigme de l'exploration-exploitation fait partie de ces principes spécifiques à la vie que nous commençons à peine à découvrir.

\chapter*{A foreword}
\addcontentsline{toc}{chapter}{A foreword} 

\emph{I have to say, I was somewhat reminded of Middle Age theological debates about how many angels
can dance on the head of a needle} -- wrote an anonymous reviewer. He/she was reporting on the manuscript of a review I had submitted some time ago with my collaborators. A delightfully unfavourable assessment and, frankly, the reviewer was right. That first draft of our manuscript was indeed far too long, too mathsy and, ultimately, too obscure -- a reader unfamiliar with those matters would never have made it to the end. The editor did not reject our paper, the report said 'major revision', it felt like a last chance. A famous quote from Blaise Pascal goes: \emph{I have made this letter longer because I have not had time to make it shorter.} In the weeks that followed, I came to understand its meaning first hand. Months later, the review was accepted for publication.

There is a truth that everyone in academia knows, the students know it, the seniors know it, the blackboards know it, the orchids in the office know it and everyone repeats the same curse in chorus: \emph{no one will ever read your thesis.} After all, why should anyone? Everything publishable has already been published or is at most under revision. With a few rare exceptions, PhD reports end up being more of a \emph{souvenir} of your roaring 20s than a real scientific contribution. Your parents will keep a copy in plain sight on the brightest bookshelf, and you will pick it up from time to time, to read the acknowledgements page again and remember with nostalgia who was there, who not anymore, who not yet.

Done this way, I believe, it is a missed opportunity, besides a waste of time. As students, we spend a considerable amount of our PhD time crafting such a document, in the midst of (probably) the peak of our scientific creativity. How to rescue these pages from their doomed fate? This question has haunted me throughout the writing. I obviously have no illusions of success but I still believe it was worth trying. Concretely, this meant making some stylistic and editorial choices that, in my subjective and absolutely questionable judgement, resulted in a clearer and more useful manuscript.

First and foremost, the content had to be original. By this I do not mean the inclusion of new topics or results -- all of those discussed here have already been made available to the scientific community by standard means. Rather, it meant minimising the overlap with the content of the papers: often adding to, sometimes subtracting from, in any case striving to offer a broader or complementary discussion. 

Second, the content had to be coherent. This starkly contrasts with the typical developmental trajectory of a PhD project, which is often winding, riddled with branching points, wrong ideas, dead ends. Therefore, I deliberately omitted from this manuscript a minor but non trivial fraction of the work from these years. Instead, I preferred to direct the reader's attention on a streamlined narrative of the core scientific idea of this project, in its formal, press release version.

Finally, the content had to be compact. Less is more, they say, and I tend to agree. To paraphrase a famous quote, everything should be as short as possible, but no shorter. My ambition was to write a self-contained document in which every single discussion had a purpose later on in the text. At the same time, I insisted in keeping the discussion focused on the essentials, avoiding long and tedious expositions of well-known subjects. Indeed, I suspect that the scientific community is not dying to learn about my view on the Markov processes, or my hot takes on the Bayes theorem.

An original, coherent and compact narrative: this was my North Star. Since much of the concrete meaning of these adjectives depends of their endpoint consumer, an essential question was: who am I talking to? The reader I had in mind was primarily a physicist, not necessarily conversant with the wonders of biology but with a basic knowledge of the fundamentals of statistical mechanics. However, I hope that a biologist who happens upon this work will find it digestible. As a physicist myself, I cannot be overly confident. Still, I think it is a reasonable hope. 

A great deal of help in the writing came from the \LaTeX\ class \href{https://github.com/fmarotta/kaobook}{kaobook v 0.9.8} by Federico Marotta -- which I have come to love and recommend \emph{urbi et orbi}. Thanks to its wide margins, I was able to establish a hierarchy of the information to be conveyed, and to separate the essentials from the frills. Indeed, I have employed sidenotes for various purposes: to offer historical context, to expand on discussions, and to share more personal views. All of them can be ignored, and the narrative in the main body of the text should stand. 

A final note, which I would like to label as the \emph{AI statement} for this manuscript. There is not a single sentence on these pages whose content has been written alone by AI tools such as  \href{https://openai.com/blog/chatgpt}{Chat GPT}. I am far too jealous of my thoughts to blindly entrust them to algorithms that I neither understand (at present, nobody does) nor trust. I firmly claim the full intellectual authorship of these pages. On the other hand, there is not a single sentence on these pages that has not been grammatically reviewed and often improved by using both \href{https://openai.com/blog/chatgpt}{Chat GPT} and \href{https://www.deepl.com/write}{DeepL}. Ignoring the existence and usefulness of these tools or, worse, opposing them, would be irrational, useless and detrimental.

\begin{flushleft}
Tokyo, Japan\\
August 2023
\end{flushleft}
\begin{flushright}
\emph{Vito Dichio}
\end{flushright}

\chapter*{Overview of research contributions}
\addcontentsline{toc}{chapter}{Overview of research contributions} 

The source material for this manuscript has been shared with the community through the following scientific papers -- in inverse chronological order.
\begin{itemize}
    \item[\cite{dichio2023b}] \textbf{Vito Dichio} and Fabrizio De Vico Fallani. \href{https://doi.org/10.48550/arXiv.2306.17300}{\emph{The exploration-exploitation paradigm for networked biological systems}}. In: arXiv e-prints 2306.17300 (2023)
        \begin{itemize}
            \item[] \emph{Pocket abstract}: The stochastic exploration of the configuration space and the exploitation of optimal functional states underlie many biological processes. The evolutionary dynamics stands out as a remarkable example. Here, we introduce a novel formalism that mimics evolution and encodes a general exploration-exploitation dynamics for biological networks. We apply it to the brain wiring problem, focusing on the maturation of the \ce\ connectome. We demonstrate that a parsimonious maxent description of the adult brain combined with our framework is able to track down the entire developmental trajectory. 
        \end{itemize}
    \item[\cite{dichio2023c}]\textbf{Vito Dichio} and Fabrizio De Vico Fallani. \href{https://iopscience.iop.org/article/10.1088/1361-6633/ace6bc/meta}{\emph{Statistical models of complex brain networks}}. In: Reports on Progress in Physics 86.10 (2023), p. 102601
        \begin{itemize}
            \item[] \emph{Pocket abstract}: Exponential random graph (ERG) models are a class of maximum entropy inference models for network data. Here, we provide a pedagogical introduction to the theory and methods of ERG models, including a critical discussion of the limitations of the approach. We review recent investigations based on their application to the study of functional, large-scale brain networks -- fMRI, EEG data. ERG models can be used to characterise networks, discriminate between different states and predict the outcome of neurological diseases such as stroke. 
        \end{itemize}
    \item[\cite{dichio2023}] \textbf{Vito Dichio}, Hong-Li Zeng, and Erik Aurell. \href{https://dx.doi.org/10.1088/1361-6633/acc5fa}{\emph{Statistical genetics in and out of quasi-linkage equilibrium}}. In: Reports on Progress in Physics 86.5 (2023), p. 052601 
        \begin{itemize}
            \item[] \emph{Pocket abstract}:  Statistical genetics applies the principles of statistical physics to the study of evolution. In this review, we focus on quasi-linkage equilibrium (QLE) regime, which exists when mutations and recombinations happen at a sufficiently fast rate with respect to the strength of natural selection. Under QLE conditions, it is possible to infer epistatic interactions between genes. We also review recent investigations of the loss of a QLE state, e.g., the transition to a clonal competition state. We introduce and characterize a new transition from QLE to a non-random coexistence (NRC) regime.
        \end{itemize}
\end{itemize}
\bigskip

The results of this thesis have been further disseminated as contributed talks at the following international scientific conferences -- in inverse chronological order.

\begin{multicols}{2}
\begin{itemize}[left=0pt]
    \item[$\circ$] \emph{Statphys28}, Tokyo, Japan, August 2023 
    \item[$\circ$] \emph{NetSci 2023}, Vienna, Austria, July 2023
\end{itemize}
\columnbreak
\begin{itemize}[left=0pt]
    \item[$\circ$] \emph{APS March Meeting}, Las Vegas, USA, March 2023
    \item[$\circ$] \emph{CCS 2022}, Palma de Mallorca, Spain, October 2022
\end{itemize}
\end{multicols}

\begin{figure}[h!]
    \centering
    \includegraphics[width=12cm]{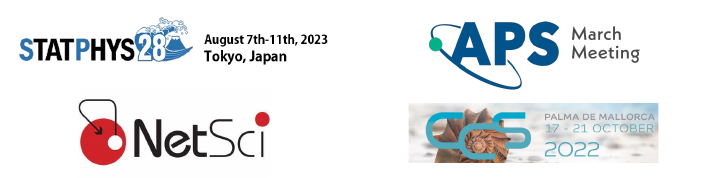}
\end{figure}
\newpage
In addition to the principal subject of this thesis, I have been involved in the following two projects during my PhD. As they are outside the interest and scope of this manuscript, they will not be discussed further. 

\begin{itemize}
    \item[\cite{zeng2022}] Hong-Li Zeng, Yue Liu, \textbf{Vito Dichio}, and Erik Aurell \href{https://doi.org/10.1103/PhysRevE.106.044409}{\emph{Temporal epistasis inference from more than 3 500 000 SARS-CoV-2 genomic sequences}}. In: Physical Review E 106.4 (2022), p. 044409
        \begin{itemize}
            \item[] \emph{Pocket abstract}: Building on our previous work, we use direct coupling analysis (DCA) to determine epistatic interactions between loci of the SARS-CoV-2 virus. Genomes are grouped by month of sampling (up to October 2021). We find that DCA terms are more stable over time than correlations, but still change over time as mutations disappear from the global population or reach fixation. We identify putative epistatic interaction mutations involving loci in the genomic region encoding the spike protein.
        \end{itemize}
    \item[\cite{zeng2021}] Hong-Li Zeng, Eugenio Mauri, \textbf{Vito Dichio}, Rémi Monasson, Simona Cocco and Erik Aurell \href{https://doi.org/10.1088/1742-5468/ac0f64}{\emph{Inferring epistasis from genomic data with comparable mutation and outcrossing rate}}. In: Journal of Statistical Mechanics: Theory and Experiment 2021.8 (2021), p. 08350
        \begin{itemize}
            \item[] \emph{Pocket abstract}: We address the problem of inferring the epistatic fitness from the evolutionary dynamics of a population, under quasi-linkage equilibrium conditions.
            We extend current state-of-the-art methods and build on a recently proposed technique that uses a Gaussian approximation for the genome probability distribution. We validate the results with \emph{in silico} experiments. 
        \end{itemize}
\end{itemize}
\index{preface}


\mainmatter 
\setchapterstyle{kao} 

\chapter{Introduction}
\begin{quote}
\begin{flushright}
\emph{Homo liber nulla de re minus quam de morte cogitat, et eius sapientia non mortis sed vitae meditatio est}\sidenote{Tr.: There is nothing over which a free man ponders less than death; his wisdom is to meditate not on death but on life.}.

--- Baruch Spinoza
\end{flushright}
\end{quote}
\bigskip

At a typical physics conference, the parallel sessions dedicated to biological physics are often a sight to behold, as the most spectacular phenomena appear on the blackboard, \emph{pardon}, on the screen. On one side of the room, the presenter discusses the collective behaviour of insect swarms, or the evolutionary dynamics in experimental populations of \emph{E. coli}, or the electrical activity of individual neurons in the mouse hippocampus, or the layer formation in bacterial colonies  \cite{cavagna2023,good2017,meshulam2019,copenhagen2021} (...) 

If we now turn our gaze to the other side of the room, we observe a diverse crowd of scientists. Some just passing by, perhaps taking a classical break from their quantum session, or taking a breath from a deadly series of talks on the latest theory of the universe in a few dozens of dimensions. However, most of the people in the room would probably describe their research field as biophysics, or physics of life, or biological physics\sidenote{Identikit: their interest in biological problems arose very late in their university education, or even later. Their natural home is a physics department, but they are often found elsewhere. They cannot resist throwing in a reference to E. Schrodinger's \emph{What is life?} whenever it is remotely possible, as I will do at the end of this chapter.
}. The questions are: what exactly is biological physics? What are these physicists looking for in biology?

Despite the long history of the subject \cite{loofbourow1940,frauenfelder1999,bialek2017}, it is only recently that the community has organised itself, and only recently that biological physics has been recognised as a genuine, distinct sub-discipline of physics\sidenote{In 2022, the first decadal survey of biological physics -- \emph{Physics of Life} \cite{nas2022} -- was published by the National Academies of the United States, a kind of historic moment. The survey is an extraordinarily rich and vivid portrait of the state of the art, including the (many) open challenges for the near future. If I may suggest: highly recommended.
}  \cite{nas2022}. Perhaps because of this, it remains a somewhat nebulous or quirky field for many, and there are some misconceptions and misunderstandings about the subject that I will try to clear up briefly before delving into the contents of this manuscript\sidenote{What follows is a personal but of course not entirely original perspective on the subject. Therefore -- full disclosure -- let me acknowledge my main intellectual debts, the works that have molded the most my own views, as found in these first pages: \cite{bialek2012,bialek2017,nas2022,goldenfeld2011,nelson2008,burnham1998}.The responsibility for any possible inaccuracy or fallacy is, of course, mine alone.}.

\subsection*{Physics, biology, biological physics}

There exist two broad ways in which academic disciplines define themselves: either by the object or by the style of their investigation. 

As physicists, we undoubtedly belong to the second class. Physics spans the entire range of natural scales, from quarks to clusters of galaxies, and the frontiers of its exploration have been and are ever broadening. The \emph{leitmotif} of our inquiry is the nature of the questions being formulated, and the nature of answers being pursued. In particular, we seek a parsimonious mathematical understanding of the phenomena, distilled into few general principles\sidenote{This very statement, as I understand it, is at the heart of what \emph{thinking like a physicist} is supposed to mean.}.
\begin{quote}
    (...) the physics community clings to the romantic notion that Physics is one subject. Not only is the book of Nature written in the language of mathematics, but also there is only one book, and we expect that if we really grasped its content, it could be summarized in very few pages.
    \begin{flushright}
        \emph{W. Bialek, Biophysics, 2012}
    \end{flushright}
\end{quote}
This, we strive to achieve by a tight dialogue between experiment and theory. Through the former, we question Nature, yearning for clues or for verdicts. Through the latter, we draw an understanding from what has been seen, and prescribe what ought to be seen\sidenote{See footnote 4, chapter 1 in \cite{bialek2012}.}. If the approximate reasoning is granted -- and often deemed necessary --, we still insist on the quantitative agreement between theory and experiment, between predictions and numerical facts about the world.

The character of the scientific enterprise is quite different in biology, which definitely belongs to the first class. In fact, biology is defined -- Greek vocabulary in hands -- as the study of living systems. Not only biology, but the also its many branches are strongly tied to the specific piece of the natural (biological) world they study. So that an ecologist, a geneticist and a cognitive neuroscientist may have very little in common, not only in terms of the system of interest, but also in terms of the nature of the questions that are formulated and the answers that are reasonably within reach\sidenote{I am aware of the dangers of such a vague statement, as here I may have awakened the Cerberus of biology. Consider, for example, the level of experimental precision that geneticists can achieve with CRISPR gene editing versus the noisy and aggregated measurements of large-scale brain activity (e.g. fMRI), common in cognitive neuroscience \cite{barrangou2016,logothetis2008}. These techniques allow for equally interesting but fundamentally different questions. Indeed, the kind of theoretical questions one can meaningfully ask is constrained by the kind of experimental answers one can get.}. Moreover, the vast majority of questions in biology are still investigated almost exclusively experimentally, so that theory is a much more unequal partner to experiment than in physics. 

Where does biological physics fits within this picture? By now, the answer should be evident. Biological physics is the investigation \emph{à la} physicist of the biological phenomena. The agenda is (i) identify the general principles that govern the phenomena of life, (ii) articulate them in a mathematical language and (iii) make quantitatively accurate predictions in agreement with experimental data.

By their very nature, the principles we seek should transcend the details of this or that particular system. Even more, they are expected to intersect with and manifest in several of the standard sub-disciplines of biology, and to cast a variety of seemingly disparate biological problems into a single, more fundamental physics problem\sidenote{There are many possible starting points. For example, staying alive involves solving a number of highly non-trivial physics problems (sensing the environment, navigating in physical space, converting energy...). Therefore, one possible line of investigation is to ask what are the physical problems that living systems have to solve. Others: how do living systems represent information? how do functions emerge from microscopic components? how do systems navigate parameter space? what are the physical limits of biological processes? how did life begin from a soup of molecules? \cite{nas2022}}. In articulating principles, we borrow the formal and conceptual tools of statistical physics and information theory, but also mechanics and thermodynamics. Finally, for a genuine biological physics to deliver on its promises, we shall pursue in biology the same level of quantitative agreement with data that is standard in other physics domains.

It is important to emphasise that in biological physics, the symbiosis between physics and biology leads to an enrichment in both directions. However, one of them has only recently been fully appreciated\sidenote{An early view of biological physics regarded it as an \emph{application} of the tools of physics to the problems of biology \cite{loofbourow1940}. Today we find this view limiting, as it overlooks what is perhaps the most intellectually stimulating direction.}. As Stan Ulam said once, \emph{ask not what physics can do for biology -- ask what biology can do for physics} \cite{frauenfelder2014}. 

There is clearly something unique about the state of matter we call life, that has no equivalent in the physics of conventional matter. It is not a new force of nature that we are missing, the very carbon atoms and interactions that constitute the pencil I write with also form the neurons that guide my hand. What we are instead missing is a precise understanding of how evolution, adaptation and learning have shaped my own brain over very different time scales, so that I can now write about them. The enterprise of explaining these three processes exclusively in terms of standard condensed matter physics is doomed to failure. This because they are all related to the notion of \emph{biological function}, which is essential for life but foreign to standard physics. Its centrality in biology cannot be overstated.

This to say, biology is not merely a playground for our physics tools. There is a new physics to be learnt from living systems, and this is the enterprise that we, as theorists, as biological physicists, are committed to.

\subsection*{From the principles to models}

To carry out the programme of biological physics, we need to project general and abstract statements about the physics of life (principles) into models of real biological systems. In other words, we need to \emph{make them work}. The path is anything but straightforward.

While a principle manifests across diverse problems, it is essential to tackle each of them individually, zeroing in on a specific problem, or \emph{context}\sidenote{For instance, think about \emph{homeostasis}, i.e., any self-regulating process by which a biological system maintains stable properties despite perturbations. In animal physiology, it may refer to the ability of maintaining a stable internal temperature (thermal homeostasis). In ecology, it may refer to the need of keeping a stable quantity of essential nutrients for the existing species -- i.e., conserving the ecological stoichiometry. These are two different contexts for the same general principle \cite{cannon1939,ivanov2006,hessen2013}.\label{sn-env_hom}}. When we do so, the semantics of our statements translate into the language of mathematics and are rendered as a set of equations. This requires us to specify a formal representation for the biological systems under study\sidenote{As an example, throughout this manuscript we will represent systems as strings of zeros and ones, a binary, discrete representation. See below.} -- which is in general context dependent. The equations have a number of parameters, whose biological interpretation is again context-bound. At this juncture, the details of the equations -- including the values of the parameters -- are unspecified\sidenote{Here is where we can tread the well-trodden path of analysis of a typical problem in theoretical physics. This involves fixing the details in a convenient way -- so to make calculations simpler, or at least possible -- and starting to understand the resulting \emph{toy models} (exact solutions, approximate solutions, asymptotic behaviours...). The hope is that we can learn from them about more complex cases. I am tempted to call this stage of analysis 'preliminary', but there are cases where people have been stuck in it for decades (and still are). Naturally, one can always resort to simulations. This is what we do after all in theoretical physics, we solve what we can, as much as we can, and we simulate the rest. 
}.

To set them, the analysis must be further narrowed down to a particular system\sidenote{Consider again the environmental homeostasis in s.n. \ref{sn-env_hom}. A particular system could be a freshwater pond. A nutrient runoff from adjacent land causes an algae bloom in a pond. In turn, this causes a decrease of oxygen levels, affecting aquatic life. Yet, certain bacteria and plants can absorb these excess nutrients, curbing algae growth and restoring the pond's balance \cite{smith2003}.}. Only at this granular level is a model defined. Each and every (biological) physicist repeats the same mantra over and over again\sidenote{Beware of the opposites: particular principles, general models.} 
\begin{quote}
    \emph{General principles, particular models. General principles, particular models. General principles, particular models (...)}
\end{quote}
Modelling may involve the formulation of additional theoretical assumptions or the setting of parameter values. Both of these endeavours benefit directly from the data we have on hand. They help us not only to trim our theoretical picture of the system, but also to fix (infer) the values of the parameters of the theory. 

The task of modelling is full of nuances, especially when it comes to biological systems. Therefore, let me comment briefly on a few aspects.

\subsubsection{Simple systems}
The exploration of a physical principle usually begins from its simplest instances. For example, quite understandably the vast majority of physics students first encounter the principle of least action in classical -- rather than quantum -- mechanics. The first Lagrangian written on the blackboard is likely that of the simple -- rather than Kapitza -- pendulum. 

Starting with simple systems is not just a pragmatic approach, it is a philosophical stance on the nature of understanding. By peeling back the layers of complexity, we do find an easier access to the underlying principles, that may remain otherwise obscured. The situation is no different when it comes to the biological matter, except that \emph{the simplest biological systems already are of jaw-dropping complexity.}

This is true no matter where we look in biology. For instance, consider cyanobacteria: among the earliest life forms, they perform oxygenic photosynthesis via an intricate molecular apparatus\sidenote{Which we are very much grateful for, given that it is thought to have been responsible for the rise of atmospheric oxygen $O_2$ some 2.5 billion years ago.}. The roundworm \ce, despite the misfortune of having one of the smallest nervous systems known (302 neurons), is capable of locomotion, mating, chemosensation and more. \emph{Archaea}, single-celled microorganisms, possess specialized membranes, enabling them to flourish in the most hostile environments \cite{sanchez2020,girard2007,koga2012}. My census of "simple" biological systems could continue: complexity is a ubiquitous and perhaps necessary feature of living systems.

This indubitably makes our job as physicists more challenging, and we should be more vigilant than ever before about the pitfalls that lie in wait. Yet, I do not intend to dishearten the reader, on the contrary. It is precisely the intrinsic complexity\sidenote{A matter of semantics.
I am aware of my loose use of the terms of \emph{simplicity} — and \emph{complexity} in this section. Indeed, it is tricky to define them formally and a significant debate exists around them -- which I honestly find somewhat futile. My point here is to highlight that while it is fairly easy to tell what is \emph{the simplest} or \emph{the most complex}, it is much harder to say what is \emph{simple} or \emph{complex}. For instance, everyone agrees that the \ce\ has one of the simplest nervous systems, yet it is much more problematic (and probably pointless) to declare the \ce\ brain to be simple.} of living systems that makes the whole enterprise of biological physics so magnetic and, ultimately, rewarding.

\subsubsection{Parameters}
Our models have parameters. It is common sense that the more realistic we want our model to be, the more effects, therefore parameters, we shall include. Pushed to its limit, this reasoning would suggest that a \emph{biological truth} is attainable only in the limit of infinite parameters. So says a influential book on the topic \cite{burnham1998}:
\begin{quote}
    We believe that “truth” (full reality) in the biological sciences has essentially infinite dimension (...) It is generally a mistake to believe that there is a simple “true model” in the biological sciences and that during data analysis this model can be uncovered and its parameters estimated.
    \begin{flushright}
        \emph{K. P. Burnham and D. R. Anderson,\\ Model selection and inference, 1998}
    \end{flushright}
\end{quote}
I do think that this point of view misses the focus of what we are trying to do. If by truth is meant the account of every possible fine-scale detail of a biological system then truth is unattainable, therefore uninteresting. On the contrary, it is very much interesting to ask: are all details \emph{really} necessary?

Two contrasting pictures are the following\sidenote{The landscape of possible answers is actually more multifaceted, for a more in-depth discussion of this topic, see \cite{bialek2017}.}. One possibility is that (almost) all details are indeed necessary, and the multiplicity, or irreducibility, of parameters is hence an intrinsic characteristic of biological systems. This would spell doom for the physicists' dream of an understanding of the life phenomena in terms of a handful of principles (and parameters). An opposing view goes something like this. The vast majority of the microscopic details of a biological system are irrelevant, since the system functions are "robust" properties of the model, independent of the configuration of those details\sidenote[][-0.5cm]{This independence of large-scale properties from microscopic details should not sound new to those familiar with the theoretical apparatus of statistical mechanics.}. Instead, the relevant features of the system are controlled by (a few) parameters, that are fine-tuned on evolutionary scales by natural selection. It is from this tension between robustness of the functional outcomes and optimal tuning, that life unfolds.

There is no need for me to say where I stand between the two, the reader has already guessed\sidenote[][-1.5cm]{Does this matter? After all, a model is what it is, regardless of my philosophical stance of the underlying biology. Well, in my opinion, it does matter, as it is tough to navigate the ocean without a star to steer by.}. 

\subsubsection{Data} 

In recent decades, most scientific fields have experienced an exponential surge in data volume\sidenote{The sense of Wordsworthian sublime and dread at the same time of many scholars is conveyed by the apocalyptic vocabulary often employed: the data \emph{deluge}, or \emph{flood}, or \emph{avalanche}, or \emph{explosion} (...)}, and particularly so in biology. Genomic sequencing now enables researchers to determine millions of DNA sequence reads in a single run, spanning from viruses to the entirety of the human genome. High-throughput mass-spectrometry churns out extensive datasets about protein composition and structure. Serial-section electron microscopy offers detailed three-dimensional reconstructions of an ever-growing number of natural brains  \cite{collins2003,mardis2008,aebersold2016,briggman2012} (...) 

This has sparked a widespread data-centric enthusiasm. Some went so far as to say that data are all you need, the end of theory has come \cite{anderson2008}\sidenote{This is wrong in so many ways that it would be difficult to account for them all here. See \cite{leonelli2014,hosni2018,gitelman2013}.}. 
Then, there is machine-learning. By summer 2023, there is no need to extol the impressive effectiveness of black-box artificial neural networks \cite{halevy2009,lecun2015}, we are all astonished, all amazed. The paradigm seems to be: take the largest amount of data out there, feed them to your machine-learning architecture, \emph{et voilà}, get the most accurate predictions. There seems to be no theory here, do we really need a theory?

I think so. Making predictions about the facts of nature is an essential part of what we do as (biological) physicists. However, this is the means by which we achieve the goal of our scientific enterprise, not the goal itself. What we do is formulate hypotheses (theories) using transparent and interpretable mathematical models, based on empirical observations. Our theories make quantitative predictions, and if they accurately describe the data, then we claim to have achieved some form of understanding of the natural phenomenon. Understanding is the goal, prediction the means. Any finite amount of data will not suffice \emph{alone} in this, in order for data to speak meaningfully, it must be meaningfully questioned\sidenote{
Let me go further. There is a feeling that by piling on layers of artificial neurons we are not getting an inch closer to understanding what is going on. That is not necessarily good or bad, it depends on the question. I think that machine learning has finally freed theoretical physics from the anxiety of providing fast answers to (very hard) quantitative questions about biological systems. The process of our science may be slow, it may take time to disprove our wrong assumptions and identify the right ones. If you want a prompt prediction, machine-learn it. If you want to understand what is going on, ask the theorist -- and be patient!}. A fundamental part of our job is and will remain to take a piece of paper, a pencil, sit down under a tree and wait for an apple to fall on our head.

This said, as theorists, we do share the data-enthusiasm. Even if data are not enough, data are definitely good. Data and theory should coexist and enrich each other: data informing theory building\sidenote{There is room here for all sorts of inference methods, white or at least grey boxes.} and theory guiding data mining. It is a safe bet to predict that such a symbiotic relationship will become increasingly essential in the future of biological physics.

\subsection*{On the representation}

There is one final point of importance that deserves further attention, the \emph{representation} \cite{cilliers2002}. As the subject is vast and to make the discussion concrete, let me start by defining the mathematical representation of interest for most part of this manuscript. 

A graph, or network\sidenote{Here and everywhere in this text, I will use these two terms interchangeably.}, in its simplest form, is a collection of points (nodes) connected by lines (edges), fig. \reffigshort{f-graph} \cite{newman2018}. A simple\sidenote{A graph is said to be simple if its edges are undirected, unweighted (binary) and has no self-loops -- the number of nodes is finite.} graph $G$ can be identified with a symmetric, binary matrix, with zero diagonal, i.e.,

\begin{equation}\label{e-graph_def}
    G = 
    \begin{bmatrix}
        0 &  &   &  \\
        a_{21} & 0 &  &  \\
        \vdots & \vdots & \ddots &  \\
        a_{N1} & a_{N2} & \dots & 0
    \end{bmatrix} = G^T\ ,
\end{equation}
where $a_{ij}\in \{0,1\}$ indicates the absence or presence of an edge\sidenote{If $N$ is the number of nodes, there are $L=N(N-1)/2$ possible edges, therefore an equivalent representation is that of a string of $L$ binary values $00110\dots$. There are $2^L$ possible graphs.} within the dyad $(ij)$ -- i.e., between the nodes $i,j$. 
\begin{marginfigure}
    \centering
	\includegraphics[width = 4 cm]{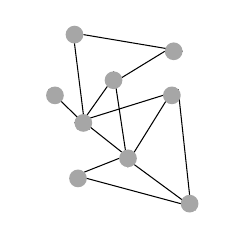}
    \caption{A simple graph is a collection of points (nodes) and lines (edges).}
    \labfig{f-graph}
\end{marginfigure}

Representing a system as a graph has proved to be a valuable theoretical tool for the analysis of complex systems -- including biological systems \cite{newman2018,latora2017}. Despite (and perhaps because of) its widespread adoption, there are periodic cries of alarm from the community warning of its potential misuse \cite{butts2009,peel2022}.

The first (maybe trivial) point is to fully acknowledge the fundamental difference between the system, the data and their representation. The data are what we measure or observe about a system, an empirical fact. The representation is the way in which we represent the system, an abstraction. The construction or choice of a representation is therefore a genuinely theoretical act, involving assumptions about the class of systems that we are studying and the data we have on hand. A crucial one regards the choice of relevant variables, which is a foundational issue of any scientific approach\sidenote{Why should (ir)relevant variables even exist? It is very fundamental question, not exclusive of the studies of complex/biological systems. A classic and superb piece of literature on the topic is \emph{The unreasonable effectiveness of mathematics in natural sciences} by E. Wigner \cite{wigner1990}.}. It is also important to say that there exist no intrinsically \emph{correct} representation for any data or systems, as it depends on the answers being sought.

When it comes to networks, this translates to two fundamental questions: \emph{what are the nodes? what are the edges?} Consider the human brain \cite{sporns2016}. Down to the cellular scale, $\sim10^{11}$ individual neurons (nodes) and their synapses (edges) form a intricate web of connections, which naturally lend themselves to a network representation\sidenote{A similar system of neurons and synapses -- though not in the case of the human brain -- will be a main focus of this manuscript. For this, a graph representation is certainly the simplest and most natural.}. At a much larger scale, one of the most common experimental techniques, an electroencephalogram (EEG) allows to measure electrical activity in the brain using around a hundred of sensors (nodes) attached to the scalp. These signals can be then correlated, yielding a measure of functional relatedness (edges) between different brain regions. These two networks describe the same biological systems, but they differ substantially in terms of \emph{what} they represent -- neuron vs brain regions, direct and physical vs indirect and functional interactions\sidenote{There is a number of unimportant details I am overlooking here. Of course, since the more fundamental representation of a brain is in terms of neurons and synapses, there should be a way of deducing the large-scale behaviours from the the firing patterns of individual neurons. To my knowledge, however, we have no clue of how this could be done. Worse, a graph of the physical neural connections (which support the transfer of information) has never been drawn and is currently experimentally out of reach. For the human brain, the road is long.}.

The more artificial our definitions of \emph{nodes} and \emph{edges}, the greater the chance of introducing spurious effects over which we have no control. The approach we have taken in this work has been to focus on systems for which the path from the system to data and their representation is as short as possible and under theoretical control. For a genuinely theoretical approach, the latter is a \emph{conditio sine qua non}\sidenote{Let me emphasise this point. In choosing a representation, we may of course make approximations that take us away from the "true" system. But when we do, we (hopefully) know what we are leaving out, what we are including, and we have strong theoretical control over what is going on in our model.}. It is a practical matter, too. If the model works, we declare victory and rejoice of it. If it does not, whose fault is it? Is it a fallacy of the model? Or of the data we are using? Or of uncontrolled spurious biases in the representation? 

As for me, if I fail, whenever I fail, every time I fail, I would like to take full responsibility for my own failure.

\section*{Overview of the manuscript}

The body of this manuscript is structured into four chapters. My aim has been to prioritise a coherent and continuous flow of ideas from one to another and to reflect the historical development of the project. At the heart of this manuscript, the principle formulated in ch. \ref{c-EE} and implemented in ch. \ref{c-Celegans}. Yet, scientific ideas seldom emerge from nowhere, the one at the heart of this thesis was definitely not an \emph{Eureka!} moment, but rather a gradual sedimentation of intuitions\sidenote{Many incorrect (not shown), but some, or at least one, worth pursuing.}. In ch. \ref{c-ergm} and ch. \ref{c-Darwin} I therefore briefly lay out the methodological and theoretical ground from which later chapters have sprung. More specifically:

\begin{itemize}
    \item[$\circ$] \textbf{Chapter \ref{c-ergm}} introduces the task of inference from network data in the context of exponential random graph (ERG) models. The discussion is organised in two parts. The first deals with general theoretical aspects by placing the ERG approach in the larger context of maximum entropy inference. The second covers a range of methodological issues that arise in practical applications. The inference method introduced here find application in ch. \ref{c-Celegans}.
    \item[$\circ$] \textbf{Chapter \ref{c-Darwin}} delves into evolutionary dynamics, starting with a concise overview of key concepts in evolutionary biology. Later, a recently proposed model of multilocus evolution is presented and its salient features discussed. The chapter concludes by examining extensions to genetic algorithms. The concepts introduced here constitute a theoretical background for ch. \ref{c-EE}.
    \item[$\circ$] \textbf{Chapter \ref{c-EE}} introduces the exploration-exploitation (EE) paradigm. First, the rationale is thoroughly discussed and the basic formalism is established. As a first step, the theory is solved for a set of toy models, from which general conclusions can be drawn. Finally, a simulation framework for the theory is discussed and tested against the analytical solutions.
    \item[$\circ$] \textbf{Chapter \ref{c-Celegans}} deals with modelling the growth of the  \ce\ brain using the EE framework. This system is naturally represented as a graph. The chapter begins with an illustration of the brain wiring problem and an overview of the \ce\ nervous system. A model of its development (from birth to adulthood) is formulated and made to work. The chapter concludes with a discussion of a biological mechanistic interpretation of the model, and an detailed outline of potential extensions.
    \item[$\circ$] \textbf{Chapter \ref{c-conclusions}} concludes this manuscript by summarising the lines of investigation followed in this project and and the main findings\sidenote{An appendix is attached at the end of the manuscript. Appendix \ref{a-EE_math} presents mathematical details of the toy models in ch. \ref{c-EE}. Appendix \ref{a-measures} discusses the set of network measures used in ch. \ref{c-Celegans}. Finally, a Glossary provides brief definitions of key biological concepts discussed in the manuscript.}.
\end{itemize}
\bigskip 

This manuscript is essentially the story of a scientific idea, from what it has blossomed, how it has grown, what it might become. As E. Schrodinger wrote once, \emph{I do not know whether my way of approach is really the best and simplest. But, in short, it was mine (...) And I could not find any better or clearer way towards the goal than my own crooked one} \cite{schroedinger1944}.
\setchapterpreamble[u]{\margintoc}
\chapter{Maxent inference for graphs}\label{c-ergm}

\begin{quote}
\begin{flushright}
Those who ignore statistics are condemned to reinvent it. 

---Bradley Efron
\end{flushright}
\end{quote}
\bigskip

The core problem in statistical inference is to recover the parameters $\bm{\chi}\in\mathbb{R}^r$ of a statistical model 
\begin{equation}\label{e-model}
    P(D;\bm{\chi}):\mathcal{D}\times\mathbb{R}^r\mapsto[0,1]
\end{equation}
from the data $D^*\in\mathcal{D}$, assuming that $D^*\sim P(D;\bm{\chi})$. A perfect reconstruction of $\bm{\chi}$ from finite data is neither possible in theory nor in practice. Therefore, we rather seek a good approximation $\bm{\chi}^*$, close to the \emph{true} values. The two fundamental tasks of statistical modelling thus consist in specifying (i) the model $P(D;\bm{\chi})$ and (ii) how to infer $\bm{\chi}^*$ from $D^*$ \cite{cox2006}. 

There is more than one interpretation for the notion of probability in \eqref{e-model} and, by consequence, for the statistical uncertainty that follows from it. One, phenomenological, interprets the uncertainty as the empirical variation in the data-generating process. Another, epistemological, interprets it as our uncertainty about the outcome, arising from the limited information at our disposal. In some fortunate cases, as for statistical mechanics, the two overlap. Regardless of the interpretation, however, we distinguish the notion of statistical uncertainty from that of measurement error: here, the data are assumed to be noiseless representations of the underlying system.

The present chapter focuses on the statistical inference based on \emph{exponential random graph} (ERG) models. The data consist of unweighted, undirected graphs $G$ \eqref{e-graph_def}. The starting point is an exponential, maxent distribution (sec. \ref{s-ERG}) \cite{cimini2019,park2004,hunter2008}. The role of the modeller is to specify a model within the ERG framework and draw conclusions about the data from the inferred parameters (sec. \ref{s-ERG_handbook}).

A minimal version of the ERG modelling is considered here, for the sake of clarity. A number of generalisations of the present framework have been proposed, a survey of which is beyond scope of this manuscript, we refer to \cite{krivitsky2022} for an entry point.

\begin{kaopaper}[title=Main reference]
\faFile*[regular] \textbf{Vito Dichio}, Fabrizio De Vico Fallani (2022). \emph{Statistical models of complex brain networks: a maximum entropy approach}. In: Reports on
Progress in Physics 86.10 (2023), p. 102601 \cite{dichio2023c}.
\end{kaopaper}

\section{Exponential random graph models}[ERG models]\label{s-ERG}

\emph{Where a streamlined theoretical and methodological minimum of ERG models is provided. Some emphasis is placed on the philosophy of the approach -- often overlooked in the literature and the source of a number of misconceptions.}\bigskip  

The interest in statistical exponential families dates back to the dawn of modern statistics\sidenote[][-0.6cm]{A very large family, indeed. The Bernoulli, Poisson, Gaussian, binomial, multinomial, Boltzmann, Rayleigh (...) distributions all belong to it. See \cite{sundberg2019,nielsen2009} for a taxonomy.} \cite{darmois1935,koopman1936,pitman1936}. A number of mathematical properties makes them particularly apt for purposes of statistical inference \cite{barndorff2014,brown1986}. In the context of graph they burst onto the scene in the 1980s, mainly motivated by the study of social network interactions \cite{holland1981,ove1986,strauss1986}. Later on, they attracted the attention of the physics community, encouraged by their formal similarity to the well-developed theory of classical statistical mechanics \cite{park2004,cimini2019}. 

When tailored to graphs, exponential distributions are referred to as \emph{exponential random graph} (ERG) models and in this section we provide a minimalist theoretical introduction to them\sidenote{The discussion here will be general, focusing on methodological aspects. We refer to \cite{lusher2013,ghafouri2020,dichio2023c} for recent reviews of the applications of ERG models, including those to social sciences, economics, neuroscience (...)}.

\subsection{A maximum entropy approach}\label{ss-max_ent_appr}
Let $G^*\in\mathcal{G}$ be an observed graph (data). Let us assume that all relevant information about the data can be reduced to a vector of \emph{statistics} $\bm{x}(G^*)\in\mathbb{R}^r$. We postpone the discussion on the choice of the $\bm{x}:\mathcal{G}\mapsto\mathbb{R}^r$ (model selection problem) to sec. \ref{ss-mod_sel} and, until then, we consider it as given.

According to the \emph{maximum entropy (maxent) principle}\sidenote{It was formulated for the first time in the 1957 by Edwin Thompson Jaynes (1922-1998) \cite{jaynes1957,jaynes1957b}. Jaynes dedicated much of his career to advocating for the principle of maximum entropy as a fundamental tool for statistical inference. He held a strong belief in Bayesian probability and often defended its interpretation as an extension of logic \cite{jaynes2003}. 
} 
\cite{jaynes1957,jaynes1957b,jaynes1982} the most unbiased model of the data, consistent with the current state of knowledge, is found by maximising the Shannon entropy
\begin{equation}\label{e-shannon_ent}
    H(P) = -\sum_{G\in\mathcal{G}}P(G)\log P(G)
\end{equation}
subject to the normalization $\sum_{G\in\mathscr{G}}P(G) = 1 $ and to the (soft) constraints:
\begin{equation}
    \sum_{G\in\mathscr{G}}\bm{x}(G)P(G) = \bm{x}(G^*)\ .\label{e-maxent_constr}
\end{equation}
This constrained maximization problem is easily solved with the method of Lagrange multipliers \cite{park2004,dichio2023c} and yields:
\begin{equation}\label{e-ergm}
    P(G|\bm{\theta}^*)= \frac{e^{\bm{\theta}^*\cdot\bm{x}(G)}}{\sum_{\Tilde{G}\in\mathscr{G}}e^{\bm{\theta}^*\cdot\bm{x}(\Tilde{G})}}\ ,
\end{equation}
where the parameters $\bm{\theta}^*\in\mathbb{R}^r$ are set so to satisfy \eqref{e-maxent_constr}. We refer to \eqref{e-ergm} as the ERG model of the data\sidenote{The maxent derivation here illustrated is nowadays standard. However, it is not the way the ERG models were originally introduced. Instead, the original formulation was based on the Hammersley-Clifford theorem for Markov graphs \cite{ove1986}, and built on a previous work of J. Besag in the context of spatial models of lattice systems \cite{besag1974}.}. Before proceeding further, let us elucidate some fundamental underpinnings of the maxent approach.

\subsubsection{On the rationale}
Intuitively, the Shannon entropy is associated to the uncertainty of a random variable \cite{shannon1948,cover1999}.

An unconstrained maximisation of \eqref{e-shannon_ent}, subject only to the normalisation, would yield a flat distribution where each possible graph has probability $2^{-L}$. This corresponds to the case where no information is encoded in the probability distribution. On the contrary, maximising \eqref{e-shannon_ent} subject to the constraints \eqref{e-maxent_constr} yields the distribution in which no information \emph{other} then that contained in the constrains is taken into account. In this sense, the maxent distribution \eqref{e-ergm} is the most \emph{unbiased}.\sidenote[][-1cm]{The argument here is deliberately qualitative, to avoid slowing down the discussion. It can be made more quantitative, though. See for instance \cite{jaynes1982,jaynes2003}.} 

The maxent simply prescribes the optimal approach to integrate any prior knowledge about the system -- here, observed statistics -- into the probability distribution. In the words of E. T. Jaynes \cite{jaynes2003}: 
\begin{quote}
    The information available defines constraints fixing some properties of the initial probability distribution, but not all of them. The ambiguity remaining is to be resolved by the policy of honesty; frankly acknowledging the full extent of its ignorance by taking into account all possibilities allowed by its knowledge.
    \begin{flushright}
        \emph{E.T. Jaynes, Probability Theory - The Logic of Science, 2003}
    \end{flushright}
\end{quote}

An important point is the following. In presenting the maxent method we have carefully avoided referring to $\bm{x}(G^*)$ as \emph{sufficient statistics}. They are, but the matter is more subtle.

By definition, a statistic $\bm{x}(G)$ is said to be \emph{sufficient} for the model $P(G|\bm{\theta})$ with parameters $\bm{\theta}$, if and only if the data reduction $x:\mathcal{G}\mapsto\mathbb{R}^r$ implies no information loss\sidenote{More precisely, let $P(D|\chi)$ be a parametric distribution for the data $D$, with parameters $\chi$. Let $t(D)$ be any statistic of the data. According to the data processing inequality, 
\begin{equation*}
    I(\chi;t(D))\le I(\chi;D)\ ,
\end{equation*}
where $I$ is the mutual information \cite{cover1999}. In words, any manipulation of the data $D$ can either reduce or preserve the information about the parameters $\chi$. In this latter case, we call $t(D)$ sufficient statistics.
}. This is true in the case of the statistics $\bm{x}(G)$ and the ERG model \eqref{e-ergm}, since the data $G$ appear in the distribution only through the statistics $\bm{x}(G)$. But this is so by design, as a result the maxent construction: any choice of $\bm{x}(G)$ would be sufficient for the resulting ERG model. In other words, the notion of sufficient statistics is determined a priori, as an hypothesis, rather assessed a posteriori, as a property of the distribution. We are therefore led to concede that different modellers with different amounts of information about the physical system will come up with different ERG distributions, leading to different predictions. The vast majority of these models, presumably, will be wrong. 

The above is somehow bewildering if we embrace an \emph{orthodox} school of thought, for which probabilities are long run frequencies of repeated experiments \cite{dienes2011}. In this latter case in fact, we clearly would not want probabilities to depend on the state of knowledge of the modeller. This is indeed the major source of criticism to the maxent modelling approach \cite{aurell2016,auletta2017}: the nature out there remains indifferent to our knowledge or lack thereof. The argument is evidently true, but it misses the point. Again in the words of E.T. Jaynes \cite{jaynes2003}:
\begin{quote}
    The principle of maximum entropy is not an oracle telling which predictions must be right; it is a rule for inductive reasoning that tells us which predictions \emph{are most strongly indicated by our present information}.
    \begin{flushright}
        \emph{E.T. Jaynes, Probability Theory - The Logic of Science, 2003}
    \end{flushright}
\end{quote}
Here, differently from the orthodox view, probabilities are considered as epistemic statements, informed guesses on a phenomenon. It can be proved that if the information included in the maxent formulation encompasses all \emph{relevant} constraints operating in a system, then the maxent distribution is the overwhelmingly most likely to be observed experimentally. 
What if, instead, the observations disagree with the predictions of the maxent model? For the maxent modeller, this is not a cause for embarrassment. It simply hints at the presence of additional or different constraints in the systems that have not yet been accounted for. 

\subsubsection{On the similarity with statistical mechanics}
The reader conversant with statistical mechanics (SM) will readily recognise \eqref{e-ergm} as analogous to a Gibbs-Boltzmann distribution, i.e.,
\begin{equation}\label{e-gibbs_boltz}
    P(G|\bm{\theta})=\frac{e^{-\mathscr{H}(G,\bm{\theta})}}{\mathscr{Z(\bm{\theta})}}\ , 
\end{equation}
where $\mathscr{H}(G,\bm{\theta}) = -\bm{\theta}\cdot\bm{x}(G)$ is the (graph) Hamiltonian and 
\begin{equation}\label{e-part_funct}
    \mathscr{Z(\bm{\theta})} = \sum_{\Tilde{G}\in\mathcal{G}}e^{-\mathscr{H}(\Tilde{G},\bm{\theta})}
\end{equation}
is the partition function \cite{park2004,cimini2019}. A Gibbs-Boltzmann distribution (canonical ensamble distribution) describes the statistical behavior of particles in a thermodynamic system at equilibrium. It is known from SM that all sorts of observables can be computed by differentiation from the free energy \cite{huang1987,peliti1997}
\begin{equation}\label{e-free_en}
    \mathscr{F}(\bm{\theta}) = -\log\mathscr{Z}(\bm{\theta}) \ .
\end{equation}
For instance, the expected value of the $\alpha$-th statistic\sidenote{By construction, the ensamble averages of the statistics match the experimental values $\av{x_{\alpha}}= x_{\alpha}^*$, cf. \eqref{e-maxent_constr}.}:
\begin{equation}\label{e-aver}
    -\frac{\de \mathscr{F}(\bm{\theta})}{\de \theta_{\alpha}}  = \frac{1}{\mathscr{Z}(\bm{\theta})}\sum_{\Tilde{G}}x_{\alpha}(\Tilde{G})\ e^{-\mathscr{H}(\Tilde{G},\bm{\theta})} = \av{x_{\alpha}}\ ,
\end{equation}
where we have used \eqref{e-ergm} and introduced the shorthand 
\begin{equation}
    \av{O} = \sum_{\Tilde{G}\in\mathcal{G}} O(\Tilde{G})P(\Tilde{G})
\end{equation}
for a graph observable $O:\mathcal{G}\mapsto\mathbb{R}$. The formal analogy with \eqref{e-gibbs_boltz} is powerful because it allows a number of results and methods from over a hundred years of SM to be translated directly into the ERG context \cite{park2004,bianconi2009}. 

It is certainly not a stroke of luck. In two groundbreaking papers published in 1957, E. T. Jaynes demonstrated that, considering SM as a form of statistical inference, the Gibbs-Boltzmann distribution can be derived directly from the maxent principle \cite{jaynes1957,jaynes1957b}. Indeed, when spoiled from its physical interpretation, the mathematical structure of the computations of SM turns out to be a general property the maxent formalism \cite{jaynes2003}.\sidenote{This is the ultimate reason for the existence of so many journal articles in the literature with titles of the form "\emph{Statistical mechanics of (something else).}"} There is, however, a crucial \emph{caveat}.

SM is more than a statistical theory, it is a physical theory, it agrees with experiments, it \emph{works}. As discussed in the previous section, the maxent argument is independent of any experimental verification. The very reason for the experimental success of SM, viewed as a maxent model, is that its choice of sufficient statistics --  notably, the energy of a microstate -- is \emph{correct} for a thermodynamic system at equilibrium. The latter result is peculiar to the case of SM and does not generalise. In summary, when formulating a maxent model, we are allowed to borrow the formal structure of SM, but not (in general) its interpretation as a physical process.

\subsubsection{On the purposes}
We have recently argued that ERG models are certainly descriptive, might be predictive and they are not explicative \cite{dichio2023c}. 

The latter is a straightforward consequence of the discussion above. ERG models are agnostic about the data-generating process therefore, in general, they cannot answer to the question why does a phenomenon happen\sidenote{Here we adopt a classical (strict) definition of scientific explanation, articulated in a highly influential paper (1948) by Carl Hempel, and particularly close to the spirit of physics -- \emph{(...) the question "Why does the phenomenon happen?" is construed as meaning "according to what general laws, and by virtue of what antecedent conditions does the phenomenon occur?"} \cite{hempel1948}. Nevertheless, this is not at all the only conceivable definition, as the subject has been extensively debated in the philosophy of science, see \cite{salmon2006}.} \cite{hempel1948}.

Prediction is here intended as feature generalisation, i.e., as the ability of a ERG model to predict the values of different statistics of the data $y_{\alpha}(G),\ \alpha=1,\dots,s$ than those used in the specification of the model. This is in line with similar maxent approaches -- e.g., \cite{mora2010,chen2019}. Clearly, the potential predictive power of the ERG model is inherently linked to the selection of statistics, as they are the only means by which an ERG model is informed about the system. In the case where our hypothesis $\bm{x}$ was accurate, the resulting model would be capable of predicting \emph{any other} test statistics $\bm{y}$. In practice, this is very seldom the case. Nonetheless, ERG models constructed with incomplete information can still demonstrate strong predictive performance on specific subsets of test statistics. 

Regardless of their predictive power, ERG models remain inherently descriptive. They provide a framework to characterise a system (read, compute observables), based on any hypothesis about the sufficient statistics. This makes them ideal for constructing null models \cite{cimini2019}. In line with the view of the maxent principle as a \emph{rule for inductive reasoning} (see above), ERG-based null models can always be used to lower bound the complexity of the \emph{true} model. Furthermore, experimental deviations from the null predictions may contain useful information about the system, and suggest possible theoretical refinements.

\subsection{Model inference}\label{ss-ERG_inference}
In deriving the ERG distribution \eqref{e-ergm} we have implicitly stated an inference (or inverse\sidenote{Given a model with known parameters, the \emph{forward} or \emph{direct} problem is to compute the values of the observables (data). This is the case, e.g., in statistical mechanics \cite{peliti1997}. Conversely, given a set of observables (data), the \emph{inverse} problem is to infer the unknown values of the model parameters. This is the case, e.g., for the inverse Ising problem \cite{nguyen2017}.}) problem \cite{mackay2003}. Let us highlight it.

\begin{definition}
    \textbf{(ERG inference)}
    Given a set of observations $\bm{x}(G^*)\in\mathbb{R}^r$, the ERG inference consists in finding the parameters $\bm{\theta}^*\in\mathbb{R}^r$ such that the constraints \eqref{e-maxent_constr} are met. 
\end{definition}
It is instructive to start by considering a simple solvable case.

\subsubsection{Bernoulli random graphs}
Let us consider an ERG model whose only statistic is the number of edges in the graph, i.e., $x(G)=\sum_{i<j}a_{ij}$ \cite{park2004}. Given a graph $G^*$, with $x(G^*)$ edges, the goal of the ERG inference is to find the parameter $\theta^*$ such that $\av{x}=x(G^*)$. The inference problem can be solved in two steps.

First, we solve the forward problem, i.e., we express $\av{x}$ as a function of $\theta$. We start by evaluating \eqref{e-part_funct}:
\begin{equation}\label{e-Bern_part_fun}
    \mathscr{Z}(\theta) = \sum_{G\in\mathcal{G}}e^{\theta \sum_{i<j}a_{ij}} = \prod_{i<j}\sum_{a_{ij}=0,1}e^{\theta a_{ij}}= (1+e^{\theta})^L\ ,
\end{equation}
where $L=N(N-1)/2$ is the number of possible edges. We can obtain a simple analytical expression for \eqref{e-free_en} as well, which reads 
\begin{equation}\label{e-Bern_free_en}
    \mathscr{F}(\theta) = -L\log(1+e^{\theta}), 
\end{equation} therefore, using \eqref{e-aver}:
\begin{equation}\label{e-Bern_edges}
    \av{x} = -\frac{\de\mathscr{F}(\theta)}{\de\theta} = \frac{L}{1+e^{-\theta}}\ ,
\end{equation}
which solves the forward problem.\sidenote{By defining $p=1/(1+e^{-\theta})$, we can rewrite the ERG probability distribution \eqref{e-ergm} as 
\begin{equation}
\begin{split}
            P(G|\theta) &= \frac{e^{\theta x(G)}}{(1+e^{\theta})^L} \label{e-Bern_Bern}\\
        &= p^{x(G)}(1-p)^{L-x(G)}\ ,
\end{split}
\end{equation}
which is the probability of a Bernoulli graph where each of the $L$ possible edges appears independently with probability $p$, hence the name. Note also from \eqref{e-Bern_edges} that $\av{x} = Lp$, as it should be \cite{erdhos1960,albert2002}. 
}

By a simple inversion of the latter formula, we can find the value $\theta^*$ for which the constraint \eqref{e-maxent_constr} is met. Imposing $x(G^*) = \av{x}$, we get
\begin{equation}\label{e-Bern_param}
    \theta^* = \log\Bigg[ \frac{d(G^*)}{1-d(G^*)} \Bigg] \ ,
\end{equation}
where $d(G)=x(G)/L$ is the density of a graph. The latter expression solves the inverse problem. The ERG probability distribution can be finally written as
\begin{equation}\label{e-Bern_pdf}
    P(G|\theta^*) = \frac{e^{\theta^*x(G)}}{(1+e^{\theta^*})^L} = \prod_{i<j} \frac{e^{\theta^* a_{ij}}}{(1+e^{\theta^*})}\ .
\end{equation}

\subsubsection{A general framework: MLE}
The ERG inference defined above can be placed in the broader context of the maximum likelihood estimation (MLE). Given an observed graph $G^*$ and a model $P(G|\bm{\theta})$, according to the maximum likelihood principle, the best choice of the unknown parameters is given by
\begin{equation}\label{e-MLE}
    \bm{\theta}^* = \argmax_{\bm{\theta}}\ \log P(G^*|\bm{\theta}) ,
\end{equation}
where $P(G^*|\bm{\theta})$ is the \emph{likelihood} of the data, given the parameters $\bm{\theta}$\sidenote[][-2cm]{
The MLE estimator \eqref{e-MLE} can be derived from the Bayes theorem \cite{mackay2003}. Accordingly, the posterior distribution $P(\bm{\theta}|G^*)$ -- which represents our knowledge after taking into account the information in the data -- can be expressed as 
\begin{equation}
    P(\bm{\theta}|G^*) = \frac{P(G^*|\bm{\theta})P(\bm{\theta})}{P(G^*)}\ ,
\end{equation}
where $P(\bm{\theta})$ is our prior information on the parameters. Our best choice for the parameters is the one that maximises the posterior distribution above. If $P(\bm{\theta})$ is a uniform distribution in the parameters space (no prior information available) this is the same as maximising the likelihood $P(G^*|\bm{\theta})$. The estimator \eqref{e-MLE} has a number of appealing properties, in particular it converges in probability to the true values (consistency) and reaches the Cramér-Rao lower bound (efficiency) \cite{rossi2018}. 
}. 

It is easy to show that the ERG inference problem can be derived from the maximum likelihood principle. Introducing the notation $\mathscr{L}(\bm{\theta}) = \log P(G^*|\bm{\theta})$, the $r$ equations \eqref{e-maxent_constr} are found by setting to zero the derivatives with respect to the parameters:
\begin{equation}
        0 = \frac{\de \mathscr{L}(\bm{\theta})}{\de\theta_{\alpha}} = \frac{\de}{\de\theta_{\alpha}} \Big[\bm{\theta}\cdot\bm{x}(G^*) - \log \mathscr{Z}(\bm{\theta})\Big]\stackrel{\eqref{e-aver}}{=} x_{\alpha}(G^*) - \av{x_{\alpha}}\ .
\end{equation}

There is of course an elephant in the room of this discussion, hidden in \eqref{e-MLE}. In order to evaluate $P(G^*|\bm{\theta}^*)$, we need to compute $\log \mathscr{Z}(\bm{\theta})$, or $\mathscr{F}(\bm{\theta})$. In the case of the Bernoulli graphs, this could be done analytically \eqref{e-Bern_free_en}, by virtue of the utmost simplicity of the model. In general, the computation of $\mathscr{F}(\bm{\theta})$ is an \emph{extremely} difficult problem, well-known in statistical mechanics\sidenote{One might say -- without fear of contradiction -- that this is \emph{the} problem of statistical mechanics. Just to give a flavor, a Nobel Prize for Physics has recently been assigned (G. Parisi, 2021) for the science that has blossomed from a \emph{trick} in evaluating $\log\mathscr{Z}$, under particular conditions \cite{mezard1987,castellani2005}.}. 

In the vast majority of the cases, there is little choice but to resort to numerical approximations for \eqref{e-MLE}, which we will now briefly describe.\sidenote{In some simple cases \cite{park2004b,park2005}, it is possible to work out the mean field theory of the model (exact in the limit of large network sizes) and possibly perform a diagrammatic expansion around the mean-field solution. However, the calculations, close to those of statistical field theory \cite{parisi1998}, soon become cumbersome and for all practical purposes, they are unworkable.}

\subsubsection{MCMC-MLE}
The fundamental idea to circumvent the explicit evaluation of $\log\mathscr{Z}(\bm{\theta})$ was introduced in the early 1990s \cite{geyer1991}. Let us consider an arbitrary vector of parameters $\bm{\theta}_0\in\mathbb{R}^r$, we can formally rewrite $\mathscr{Z}(\bm{\theta})$ as 
\begin{equation}\label{e-MCMCMLE_trick}
\begin{split}
    \mathscr{Z}(\bm{\theta}) &= \mathscr{Z}(\bm{\theta}_0) \sum_{\Tilde{G}\in\mathscr{G}}e^{(\bm{\theta}-\bm{\theta}_0)\cdot\bm{x}(\Tilde{G})}\ \frac{1}{\mathscr{Z}(\bm{\theta}_0)}e^{\bm{\theta}_0\cdot \bm{x(\Tilde{G})}}\\
    &= \mathscr{Z}(\bm{\theta}_0) \langle e^{(\bm{\theta}-\bm{\theta}_0)\cdot\bm{x}}\rangle_{\bm{\theta}_0}
\end{split}
\end{equation}
where the subscript $\langle\cdot\rangle_{\bm{\theta}_0}$ indicates the expectation value over the distribution $P(G|\bm{\theta}_0)$. The trick is now is to use the a Markov Chain Monte Carlo sampling\sidenote{See also s.n. \ref{sn-MCMC}.} to evaluate approximately the right hand side of \eqref{e-MCMCMLE_trick}. In particular, given a sample of $m$ graphs $G_1,\dots,G_m$ whose stationary distribution is $P(G|\bm{\theta}_0)$, we can approximate
\begin{equation}\label{e-MCMCMLE_approx}
    \frac{\mathscr{Z}(\bm{\theta})}{\mathscr{Z}(\bm{\theta}_0)} = \langle e^{(\bm{\theta}-\bm{\theta}_0)\cdot\bm{x}}\rangle_{\bm{\theta}_0} \simeq \frac{1}{m} \sum_{i=1}^{m} e^{(\bm{\theta}-\bm{\theta}_0)\cdot\bm{x}(G_i)}\ . 
\end{equation}
We know consider the log-likelihood and note that the argument that maximises $\mathscr{L}(\bm{\theta})$, maximises the shifted log-likelihood $\bar{\mathscr{L}}(\bm{\theta}) = \mathscr{L}(\bm{\theta}) - \mathscr{L}(\bm{\theta}_0)$, too. The latter, however, can be numerically approximated: 
\begin{equation}\label{e-MCMCMLE_newlik}
\begin{split}
    \bar{\mathscr{L}}(\bm{\theta}) & = (\bm{\theta} - \bm{\theta}_0 ) \cdot \bm{x}(G^*) - \log\Bigg\{\frac{\mathscr{Z}(\bm{\theta})}{\mathscr{Z}(\bm{\theta}_0)} \Bigg\}\\
    & \stackrel{\eqref{e-MCMCMLE_approx}}{\simeq} (\bm{\theta} - \bm{\theta}_0 ) \cdot \bm{x}(G^*) - \log\Bigg\{ \frac{1}{m} \sum_{i=1}^{m} e^{(\bm{\theta}-\bm{\theta}_0)\cdot\bm{x}(G_i)}\Bigg\}\ . 
\end{split}
\end{equation}
A parameter estimation based on the maximisation of \eqref{e-MCMCMLE_newlik} converges to the same result as \eqref{e-MLE}, in the limit $m\rightarrow\infty$ and it is used in practice as an approximation of the MLE.

{\LinesNumberedHidden
\begin{algorithm}[H]
    \DontPrintSemicolon
    
    \TitleOfAlgo{MCMC-MLE (pseudocode)\sidenote{The algorithm starts from an initial guess of the parameters. Albeit arbitrary, an appropriate choice can aid in achieving rapid convergence. A commonly adopted approach is to use the parameters obtained through pseudo-likelihood (pl) maximization \cite{besag1974,hunter2008}. The parameter space is explored iteratively by solving at each step a maximisation problem based on \eqref{e-MCMCMLE_newlik}. The previous set of parameters serves each time as the starting point for the optimization process. Iterations continue until convergence is reached.
    }.}
    
    $\tau \leftarrow 0$\;
    $\bm{\theta}_{\tau} \leftarrow \bm{\theta}_{\text{pl}}$\;
    \While{conv $=$ \texttt{F}}{
        $\tau \mathrel{+}=  1$\;
        generate $m$ graphs $G_k\sim P(G|\bm{\theta}_{\tau-1})$ by MCMC\;
        $\bm{\theta}_{\tau} = \argmax_{\bm{\theta}} \Big[ (\bm{\theta}-\bm{\theta}_{\tau-1})\cdot \bm{x}(G^*) - \log\big[ \frac{1}{m}\sum_{i=1}^m e^{(\bm{\theta}-\bm{\theta}_{\tau-1})\cdot\bm{x}(G_i)}\big] \Big]$\;
        \If{(convergence criterion)}{
            conv $\leftarrow$ \texttt{T}\;
        }
    } 
\end{algorithm}
}

\subsubsection{Software tools}
In the last two decades, a number of open-source libraries have been developed that can be used to perform a ERG inference, as illustrated above. By far, the most popular implementation is the \texttt{R}-based \texttt{ergm} package, published in the 2008 by \emph{Hunter et al.} \cite{hunter2008} and recently updated \cite{krivitsky2023}.
It stands as a comprehensive tool, providing extensive functionality for fitting ERG models, including model specification, inference and diagnostics. It has served as a foundational library for numerous generalisations, an overview can be found in \cite{krivitsky2022}. Unfortunately, no ERG implementation has attained an equivalent level of maturity outside the \texttt{R} programming language\sidenote{There are historical reasons for this. In fact, since the beginning \cite{holland1981}, ERG models have been particularly popular within the domain of social sciences which, in turn, are particularly fond of \texttt{R}. The level of technical detail that packages such as \texttt{ergm} have reached is such that, over a typical 3-years PhD project, one would rather learn a new programming language than re-write everything from scratch. As the adage goes: \emph{good programmers write good code; great programmers steal great code.}}.

The \texttt{ergm} library is the reference implementation of the ERG inference in this manuscript. In the GitHub folder \href{https://github.com/dichio/ergm_minimal}{\texttt{ergm\_minimal}} -- originally presented in \cite{dichio2023c} -- we have published the scripts for a minimal implementation of an ERG inference workflow\sidenote{Additional resources and examples can be found on the \href{https://statnet.org/workshops/}{statnet} website \cite{krivitsky2022}.}.

\section{User handbook}\label{s-ERG_handbook}

\emph{Where two important and subtle aspects of the ERG method (at the beginning, at the end) are discussed in detail. The limitations of the ERG inference are highlighted, what they can(not) say, where they can(not) work.}\bigskip 

\begin{figure}
    \centering
    \includegraphics{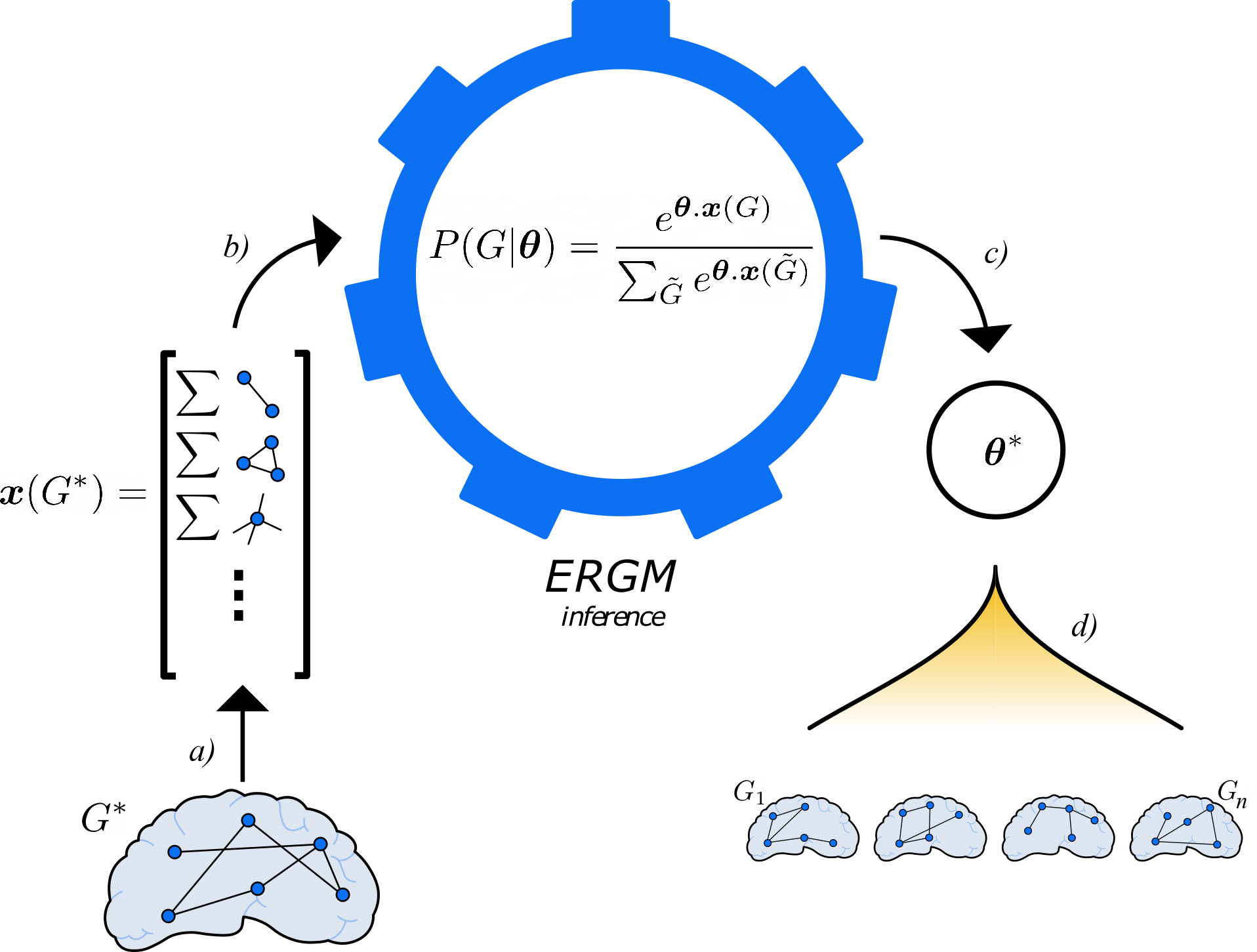}
    \caption{ERG models workflow. a) A graph $G^*$ is given, here, for instance, a brain network. b) The information of the graph $G^*$ is condensed in a set of statistics $\bm{x}(G^*)\in\mathbb{R}^r$. Their choice is up to the modeller (model selection problem) and they represent the input of the ERG model inference. c) The output of an ERG inference 
    is a vector of parameters $\bm{\theta}^*\in\mathbb{R}^r$. d) The inferred parameters can be used to simulate $n$ new networks $G_1,\dots,G_n$ that statistically reproduce the observed one i.e. $\frac{1}{n}\sum_{i=1}^n\bm{x}(G_i)\sim\bm{x}(G^*).$}
    \label{f-ergm}
\end{figure}

Stripped down to the essentials, an ERG inference is a computational device that takes as input a real vector (graph statistics) and returns as output a real vector (estimated parameters), fig. \ref{f-ergm}. 

Two key matters for practitioners are therefore (i) how to select the model statistics and (ii) how to interpret the estimated parameters. In this section, we tackle these two questions. For reasons that will be clear later, we do it in reverse order. 

\subsection{Interpretation of parameters}\label{ss-ERG_int}
In sec. \ref{ss-max_ent_appr} we have argued that the ERG inference can be used for the purpose to characterise a system. Here, we quantify this qualitative statement.

\subsubsection{Bernoulli strikes back}
Let us start again by considering the ERG inference for the Bernoulli graphs \eqref{e-Bern_pdf}, sec. \ref{ss-ERG_inference}.

As discussed above, we do have a mathematical understanding of how the data, through the ERG machinery, determine the inferred parameter, namely \eqref{e-Bern_param}. In particular, if the original graph is maximally random -- i.e., $d(G^*)=1/2$ \sidenote[][-5cm]{This is intuitive, and it is also elementary to show. First, since all dyads are independent, in the large graph limit $d(G)\sim p$, where $p$ is the probability of having an edge between each possible dyads (connection probability). Let us focus on a single dyad. The \emph{randomness} of the connection can be quantified by computing the Shannon entropy of the Bernoulli trial: $P_B(a_{ij}=1)=1$ and $P_B(a_{ij}=0)=1-p$, i.e.,
\begin{equation*}
    H(P_B)=-p\log p -(1-p)\log(1-p), 
\end{equation*}
which is maximum when $p=1/2$ \cite{cover1999}. By consequence, we conclude that a Bernoulli graph with $d(G)=1/2$ corresponds is the maximally random graph - or, simply, random graph.\label{sn-ran_graph}
} -- we obtain $\theta^*=0$. If instead we start from a denser graph $d(G^*)>1/2$, we get $\theta^*>0$. Finally, $d^*<1/2$ implies $\theta^*<0$. There are two general lessons to be learnt from this: (i) the ERG model \eqref{e-Bern_pdf} automatically rules out model effects for which there is no evidence in the data \cite{jaynes2003} and (ii) nonzero parameters are evidence of structure in the data, captured qualitatively by the sign of the parameter and quantitatively by its value.

However, the logic of an interpretation of the parameter $\theta$ is essentially the inverse of the one we have just illustrated. Namely, \emph{given the result of the inference, we want to make a statement about the structure of the original data.} Two possible ways to interpret the ERG parameter $\theta^*$ are the following.
\begin{enumerate}
    \item A \emph{direct interpretation}, using \eqref{e-Bern_edges} and \eqref{e-maxent_constr}:
    \begin{equation}\label{e-Bern_int1}
        d(G^*) = \frac{1}{1+e^{-\theta^*}}\ .
    \end{equation}
    \item A \emph{dyadic (indirect) interpretation}, by computing the effect of the parameter $\theta^*$ on the probability of the existence of an edge between two nodes. Specifically, let us consider the dyad $(ij)$ of the generic graph $G\in\mathcal{G}$, and let $P(a_{ij}=1|\theta^*)$ be the probability of an edge between the nodes $i,j$\sidenote{
    In the case of a Bernoulli graph, we can treat each dyad independently. In fact, using \eqref{e-Bern_pdf} we have 
    \begin{equation*}
    \begin{split}
        P(a_{ij}|G_{\backslash ij},\theta^*) &= \frac{P(G|\theta^*)}{P(G_{\backslash ij}|\theta^*)}\\
        &=\frac{e^{\theta^*a_{ij}}}{1+e^{\theta^*}} =P(a_{ij}|\theta^*)\ ,
    \end{split}
    \end{equation*}
    where $G_{\backslash ij}$ stands for all other dyads in the graph than $ij$.
    }.
    Using \eqref{e-Bern_pdf} we have
    \begin{equation}\label{e-Bern_int2}
        \log \frac{P(a_{ij}=1|\theta^*)}{P(a_{ij}=0|\theta^*) } = \theta^*\ ,
    \end{equation}
    Suppose for instance $\theta^*>0$. The probability of having an edge within the dyad $ij$ is then higher than its complementary (absence of an edge) by a factor $\exp{\theta^*}$. This is a property of the ERG graph distribution. Nevertheless, we can leverage \eqref{e-maxent_constr} to argue that, by ERG construction, it is also a property of the original graph, which therefore is denser-than-random.
\end{enumerate}
Whenever a direct relation \eqref{e-Bern_int1} is available, a dyadic interpretation \eqref{e-Bern_int2} is clearly an unnecessary complication. When models become wilder than meek Bernoulli graphs, however, we soon loose the former, and the latter is all we have left.

\subsubsection{The one and only general interpretation}
In applications of ERG models of any interest, the vector of statistics includes several effects, that are in general mutually dependent. In such cases, as discussed in sec. \ref{ss-ERG_inference}, the inference problem cannot be solved analytically, no such equations as \eqref{e-Bern_param}, \eqref{e-Bern_int1} are available and no direct interpretation is possible.

The parameters $\bm{\theta}^*$ are obtained by numerical approximation, are however still amenable to a dyadic interpretation. Let $P(a_{ij}=1|G_{\backslash ij},\bm{\theta}^*)$ be the probability of an edge within the dyad $(i,j)$\sidenote{Note that dyads are in general dependent, by consequence the probabilities are to be conditioned on the rest of the graph $G_{\backslash ij}$.}. Using \eqref{e-Bern_pdf} we have
\begin{equation}\label{e-erg_int}
    \log\ \frac{P(a_{ij}=1|G_{\backslash ij},\bm{\theta}^*)}{P(a_{ij}=0|G_{\backslash ij},\bm{\theta}^*)} = \bm{\theta}^*\cdot \bm{\Delta x}(G_{ij})\ ,
\end{equation}
where $\bm{\Delta x}(G_{ij})$ is the vector of \emph{change statistics}. Introducing the shorthand $G_{+ij} = \{a_{ij}=1,G_{\backslash ij}\}$ and $G_{-ij} = \{a_{ij}=0,G_{\backslash ij}\}$, the $\alpha$-th element of the change statistics is defined as:
\begin{equation}
    \Delta x_{\alpha}(G_{ij}) = x_{\alpha}(G_{+ij}) - x_{\alpha}(G_{-ij})\ .
\end{equation}
Whether or not the presence of an edge between the nodes $ij$ is favoured is depends on the overall sign of the the right hand side of \eqref{e-erg_int} and, therefore, on the combination of change statistics, weighted by the corresponding parameters\sidenote{The scenario here considered is also at the hearth of Markov chain Monte Carlo (MCMC) methods for ERG models. The goal is to construct a Markov chain that has \eqref{e-ergm} as its equilibrium (stationary) distribution. To this end, at each step of the Markov chain a random change to the current graph is proposed, its effect on the (log) probability of the graph is evaluated by \eqref{e-erg_int}, and accepted or rejected based on the Metropolis-Hastings rule \cite{hunter2008}.\label{sn-MCMC}
}.
We are in the position to state the following:

\begin{definition}\label{d-ERG_int}
    \textbf{(ERG interpretation)}
    Given an ERG model \eqref{e-ergm}, the dyadic interpretation \eqref{e-erg_int} of the parameter $\theta_{\alpha}$ is the change in the log probability of a graph, resulting from switching from $G_{-ij}$ to $G_{+ij}$ 
    (i) per unit increase of the corresponding statistic $\Delta x_{\alpha}(G_{ij})=1$, and (ii) holding fixed the cumulative effect of the other statistics $\sum_{\beta\ne\alpha} \theta_{\beta}\Delta x_{\beta}(G_{ij})$.
\end{definition}
Once again, the ERG interpretation is based on a characterisation of the ensamble distribution $P(G|\bm{\theta}^*)$, \eqref{e-ergm}. This because, by construction \eqref{e-maxent_constr}, the average properties of the latter reflect those of the original graph.

Large positive (negative) values of the parameter $\theta_{\alpha}$ indicate an over- (under-) representation in the original graph of the corresponding $x_{\alpha}$, with respect to the null expectation -- i.e., $\av{x_{\alpha}}$ of an ERG model with $\theta_{\alpha}=0$ and unaltered $\bm{\theta}_{\backslash \alpha}$\sidenote{Against sloppiness. A nasty habit in the ERG literature is to interpret the parameters exclusively in qualitative terms. When it comes to something as subtle as the interpretation of ERG parameters, def. \ref{e-erg_int}, qualitative statements alone run the risk of being too vague, or even misleading, if not false.\\ For instance, let us consider the following common statement: "\emph{$\theta>0$ implies that the corresponding metric is higher than expected by chance}". For ERG models with multiple statistics, this is only true in precise sense of def. \ref{d-ERG_int}. \emph{Chance} here is \emph{the rest of the ERG model} and not a maximally random graph. A common misinterpretation of the ERG parameters indeed arises from forgetting (ii) in def. \ref{d-ERG_int}. 
}. 
This has an important consequence: parameters associated with the same statistic within different ERG models cannot be directly compared. A comparison is possible only if the cumulative effect of the rest of the statistics of the models are held fixed.

\subsection{The model specification problem}\label{ss-mod_sel}
The core question of the \emph{model specification problem} (or, \emph{feature selection problem}) \cite{fisher1922,kira1992,guyon2003} in the context of ERG models is fairly simple to state: \emph{for a given networked system, what is the \emph{best} choice of the statistics $\bm{x}(G)$?} 

Generally speaking, the optimal choice of statistics is one that most accurately embodies our hypotheses regarding the \emph{relevant} characteristics of the system. When using an ERG as a null model, this choice represents the null hypothesis. When using an ERG as a model of the data, this choice represents (at least) our best theoretical approximation to the \emph{true} state space of the system \cite{burnham1998}.

All sort of graph statistics $x:\mathcal{G}\rightarrow\mathbb{R}$ can be designed. The first and simplest class of ERG statistics is the one of \emph{edge covariates}, i.e., 
\begin{equation}\label{e-erg_ec}
    x_{ec}(G) = \sum_{i<j} a_{ij} \gamma_{ij}
\end{equation}
where $\gamma \in \mathbb{R}^{N\times N}$ is a real matrix, with the same algebraic properties as $G$. The latter assigns an attribute to all dyads in a graph, by performing the sum of the values $\gamma_{ij}$ over all existing edges. A simple subcase of edge covariate is obviously the number of edges in a graph -- i.e., \eqref{e-erg_ec} for $\gamma_{ij}=1\ \forall i,j$ --, which we have encountered when defining the Bernoulli random graphs, sec. \ref{ss-ERG_inference}. In the case of spatially embedded graphs, the matrix $\gamma$ can be used to encode the physical distance between each any nodes. Alternatively, it can be used to quantify homophily or heterophily effects\sidenote{\emph{Homophily} (\emph{heterophily}) refers to the tendency of nodes to form connections with others that have similar (different) attributes or characteristics.} on the edge formation based on a nodal (categorical) attribute $\eta$, i.e., $\gamma_{ij} = \delta_{\eta_i,\eta_j}$, where $\eta_i,\eta_j$ are the nodal attributes of $i,j$ and $\delta$ is the Kroenecker delta. ERG models based on statistics of the form \eqref{e-erg_ec} are still amenable to analytical treatment\sidenote{It is a straightforward generalisation of the discussion in sec. \ref{ss-ERG_inference} for Bernoulli random graphs. See the section III.B, \emph{Generalised random graphs}, in \cite{park2004b}}. However, they are of limited interest, since the interest is often in modelling the complex interactions between dyads. 

Thus, we shall turn to consider graph statistics that encode dyadic dependencies. More specifically, we restrict our attention to \emph{Markov graphs} \cite{besag1974,strauss1986}, i.e., we assume that any two dyads that do not share an endpoint are independent, conditional on the rest of the graph\sidenote{In other words,
only when they share a node can any two dyads be statistically dependent, when fixing the rest of the graph. Note that the Markov dependence is a property of the graph probability distribution, and not of the individual graph.
}. A large family of graph statistics for modelling dyadic dependencies falls under the umbrella of \emph{motifs counts}, i.e., the number of times a particular connection pattern occurs in the graph. However, early numerical investigations brought to light a significant hurdle to ERG inference based on simple motif counts: degeneracy.

\subsubsection{Degeneracy, the trap of simplicity}
Let us consider the simplest ERG model of dyadic dependence, which is based on the graph Hamiltonian:
\begin{equation}\label{e-H_2st}
    -\mathscr{H}_{2\star}(G,\bm{\theta}) = \theta_{-} \sum_{i<j}a_{ij} + \theta_{\wedge} \sum_{i<j,k}a_{ik}a_{kj}\ .
\end{equation}
This is referred to as the \emph{two star} model, and includes two terms. The first, of the kind \eqref{e-erg_ec}, counts the number of edges ($-$). The second counts the number of two stars, i.e., a pair of edges that share a common node ($\wedge$)\sidenote{For instance, a two star model could be used to characterise a sparse graph ($\theta_{-}<0$) with many high degree nodes ($\theta_{\wedge}>0$).}. In practice, it has soon be realised that ERG estimations based on \eqref{e-H_2st} are not possible \cite{strauss1986}.

The point at which the approximate estimation described in sec. \ref{ss-ERG_inference} fails is the MCMC sampling of \eqref{e-ergm}. Almost everywhere in the ERG parameter space, graph Markov chains based on \eqref{e-erg_int} get trapped in graph configurations that are either almost empty, or almost fully connected. This means that for the vast majority of the parameter combinations, \eqref{e-erg_int} assigns a negligible probability to all but the two extreme, unrealistic graph configurations (\emph{degeneracy}) \cite{strauss1986}. The subset of the parameter space that yields non-degenerate probability distribution turns out to be negligible and in practice unattainable \cite{snijders2002,rinaldo2008}. The estimation task is therefore impracticable\sidenote[][]{The phenomenology of the two star model has been analytically understood by \emph{Park \& Newman} \cite{park2004,park2004b}. The degenerate behaviour is nothing but a symmetry breaking between high- and low-density phases. A continuous (second-order) phase transition exists is found for this system. An analogous degeneracy problem affects other simple ERG models, e.g., the \emph{Strauss clustering model} \cite{strauss1986,park2005}.
}.

The root of the the degeneracy issue is therefore the geometry of the ERG distribution. Unstable graph statistics lead to degenerate model behaviours, this is the case for the two star model\sidenote{According to \cite{schweinberger2011}, $x(G)$ is stable if there exist $C>0$ and $L_c>0$ such that $U_x\equiv \max_{G\in\mathcal{G}}x(G)\le CL\ \forall L>L_c$, where $L$ is the number of dyads in the network. For instance, since $U_{-} = L$,  the edges term is stable while $U_{\wedge} \sim NL$ implies that the two star term is unstable.}. In such cases, the nearest-neighbor log probabilities \eqref{e-erg_int} are unbounded, and MCMC simulations are a waste of time and resources -- see \cite{schweinberger2011} for a detailed, general discussion. These theoretical investigation had a fairly simple conclusion: inference for ill-posed ERG models is not possible.

Therefore, let us state the following:

\begin{definition}\label{d-ERG_mod_spe}
    \textbf{(ERG model specification)}
    The task of selecting a set of graph statistics that (i) optimally represents a given hypothesis about the relevant features of data and (ii) yields a non-degenerate ERG model.
\end{definition}

A general strategy to cure the degeneracy issue is to "add structure" to the ERG specification. For instance, this includes adding constraints on the block, multilayer, spatial structure of the input graph \cite{schweinberger2020}. Another solution is to stabilise the ERG model by using more sophisticated and more realistic graph statistics, as opposed to simple motifs counts as the two stars in \eqref{e-H_2st}. As this is the case of interest for this report, we describe it in some detail.

\subsubsection{Curving ERG models}
\begin{marginfigure}[-3cm]
	\includegraphics{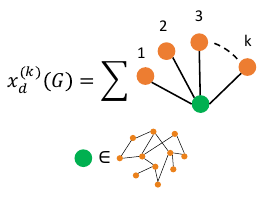}
\end{marginfigure}
A widely adopted choice to design non degenerate ERG models consists in using the so-called \emph{curved statistics} \cite{snijders2006,hunter2006}. Let us start by considering the following \emph{geometrically weighted degree} ($gwd$):
\begin{equation}\label{e-CS_GWD}
    x_{gwd}(G|\lambda) = \sum_k w_{\lambda}^{(k)} x_d^{(k)}(G)\ ,
\end{equation}
where $x_d^{(k)}(G)$ is the number of nodes in the graph $G$ with degree $k$ and
\begin{equation}\label{e-CS_w}
    w_{\lambda}^{(k)} = e^{\lambda}\Big\{1-\big(1-e^{-\lambda}\big)^k\Big\}\ ,
\end{equation}
$\lambda>0$. The $gwd$ statistic is a linear combination of the graph degree distribution. An ERG model containing \eqref{e-CS_GWD} is curved\sidenote[][-4.5cm]{An exponential distribution is \emph{curved} -- in the sense of Efron \cite{efron1975,efron1978} -- when its natural parameters satisfy constraints. Here, for instance, in order to model the information of the graph probability distribution, one should generally include in the graph Hamiltonian one statistic for each of the $N-1$ possible degrees, each associated to an independent parameter $\theta^{(k)}$. In \eqref{e-CS_GWD}, we are imposing the following non-linear constraint on the parameter space:
\begin{equation*}
    \theta^{(k)} = \theta w_{\lambda}^{(k)} \ ,
\end{equation*}
where $w_{\lambda}^{(k)}$ are defined in \eqref{e-CS_w}. Thus, in this view, we are constraining a $N-1$ dimensional parameter space associated to the degree distribution to a two-dimensional subspace, hence the name of "curved" model. However, we take a slightly different view on the roles of $\theta$ and $w_{\lambda}^{(k)}$, see later in the text. 
} and stable\sidenote[][4cm]{More specifically, an ERG model with a curved statistic such as \eqref{e-CS_GWD}, \eqref{e-CS_w} is stable as long as $\lambda>-\log2$ \cite{schweinberger2011}. Here this is always the case, since $\lambda>0$.} \cite{schweinberger2011,schweinberger2020}. To interpret the role of \eqref{e-CS_GWD}, we can reason in an analogous way to sec. \ref{ss-ERG_int} \cite{hunter2007,dichio2023c}. 

\begin{figure}
    \centering
    \includegraphics[width=.9\linewidth]{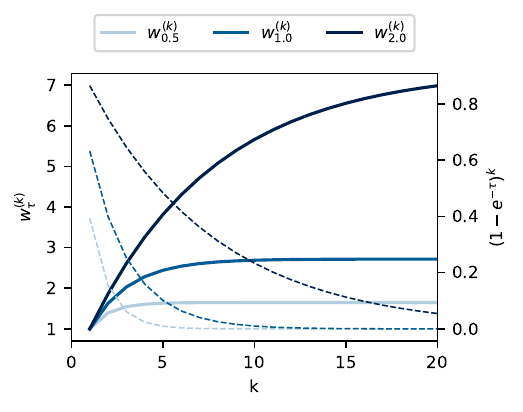}
    \caption{Parametric weights for curved ERG statistics \cite{snijders2006,hunter2006}. Solid lines plot \eqref{e-CS_w} for three sample values of $\lambda$, as a function of $k$. Dashed lines plot the geometric decreasing factor as it appears in \eqref{e-erg_int_gwd}, \eqref{e-erg_int_gwesp}. The parameters $\lambda$ determine the geometry of the curved statistics, therefore they pertain to the model specification problem. In practice, they can be inferred from data together with the maxent parameters.}
    \label{f-cs_w}
\end{figure}

As a result of adding an edge to the dyad $(ij)$, the degrees of both the extremal nodes increase by one unit. Let us call $G_{-ij}, G_{+ij}$ the graphs before and after the edge addition. If $k_i$ is the original degree of the node $i$, then
\begin{equation}\label{e-CS_gwd_cs}
x_d^{(k_i)}(G_{+ij}) = x_d^{(k_i)}(G_{-ij}) - 1\ ,\quad x_d^{(k_i+1)}(G_{+ij}) = x_d^{(k_i+1)}(G_{-ij}) + 1\ ,
\end{equation}
and analogously for the node $j$. For simplicity, let us assume that the edge addition does not produce any other change in the graph statistics than \eqref{e-CS_gwd_cs}. From \eqref{e-erg_int}, one finds:
\begin{equation}\label{e-erg_int_gwd}
\begin{split}
    \log\ \frac{P(a_{ij}=1|G_{\backslash ij},\bm{\theta}^*)}{P(a_{ij}=0|G_{\backslash ij},\bm{\theta}^*)} &= \theta_{gwd} \Big[ x_{gwd}(G_{+ij}|\lambda) - x_{gwd}(G_{-ij}|\lambda)\Big]\\
    & = \theta_{gwd} (1-e^{-\lambda})^{k_i} + \theta_{gwd}(1-e^{-\lambda})^{k_j}\ .
\end{split}
\end{equation}
Note that $(1-e^{-\lambda})^k\rightarrow0$ for $k\rightarrow\infty$. This means that effect of the $gwd$ statistic decreases geometrically with the degree, with the rapidity of the decay controlled by $\lambda$ -- large values of $\lambda$ correspond to slow decays, fig. \ref{f-cs_w}. Let us consider for instance $\theta_{gwd}>0$. In this case, \eqref{e-CS_gwd_cs} always implies an advantage for the edge addition. However, this effect wanes and eventually vanishes for increasing the original degrees of the nodes $i,j$\sidenote{In practice, the presence of nodes with high enough degrees is indistinguishable from the random expectation, i.e., the one we would get in the absence of the $gwd$ term.}. Intuitively, this is what prevents a statistic such as \eqref{e-CS_GWD} from driving the system into a fully connected state. 

Apart from yielding non-degenerate ERG models, the $gwd$ term is also a more plausible statistic than the two stars in \eqref{e-H_2st} for modelling, e.g., the presence of hubs in a graph. On the one hand, hubs would be reflected by a degree distribution skewed towards higher degrees. On the other hand, there must be a cut-off for this effect: due to the presence of physical and/or functional constraints, it is highly unrealistic in biological networks for nodes to be connected to large fractions of the graph. The $gwd$ statistic -- and in particular the $\lambda$ parameter -- is able to tune this trade-off.

The $gwd$ is only one of the possible curved statistics. They are all defined as linear combinations of a distribution of graph counts, with weights \eqref{e-CS_w}. Another example of interest for this work is the \emph{geometrically weighted edgewise shared partner} ($gwesp$):
\begin{marginfigure}[-.75cm]
	\includegraphics{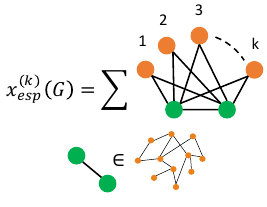}
\end{marginfigure}
\begin{equation}\label{e-CS_GWESP}
    x_{gwesp}(G|\lambda) = \sum_k w_{\lambda}^{(k)} x_{esp}^{(k)}(G)\ ,
\end{equation}
where $w_{\lambda}^{(k)}$ are defined in \eqref{e-CS_w} and $x_{esp}^{(k)}(G)$ is the number of dyads that are connected by an edge and that have exactly $k$ neighbors in common. It can be used as a more nuanced proxy for a tendency towards triadic closure \cite{bianconi2014}, rather than a simple (and unstable) count of triangles. Similarly as above, a single edge addition to the dyad $(il)$ that increases the number of common neighbors between the connected nodes $i,j$ by one unit results in
\begin{equation}\label{e-erg_int_gwesp}
    \log\ \frac{P(a_{il}=1|G_{\backslash il},\bm{\theta}^*)}{P(a_{il}=0|G_{\backslash il},\bm{\theta}^*)}  = \theta_{gwesp} (1-e^{-\lambda})^{k_{ij}}\ ,
\end{equation}
where $k_{ij}$ is the initial number of common partners of the nodes $i,j$. Once again, in the case $\theta_{gwesp}>0$, the tendency to add shared partners is damped for increasing $k_{ij}$, thus sidestepping the pitfall of degeneracy.

ERG inference for curved models builds upon and extends the general framework discussed in sec. \ref{ss-ERG_inference}\sidenote{Parameters estimation of curved ERG models was first discussed in \cite{hunter2006} and implemented in the first release of the \texttt{ergm} package \cite{hunter2008}. See \cite{krivitsky2023} for an overview of the recent developments.}. In practice, the parameters $\bm{\theta}\in\mathbb{R}^r$ which weight the terms in the graph Hamiltonian and those $\bm{\lambda}\in\mathbb{R}_+^{q}$ which govern the geometry of the curved statistics ($q \le r$) are estimated simultaneously \cite{hunter2006}. Nevertheless, we regard them as logically distinct. The former are the Lagrange multipliers derived from the maxent principle, sec. \ref{ss-max_ent_appr}. The latter, pertain to the issue of model specification, in the sense of def. \ref{d-ERG_mod_spe}.

We set aside both graphs and inference methods for the moment. The former will resurface in chapter \ref{c-EE}, the latter one chapter later. Instead, over the next few pages, our discussion will take a sharp turn in both style and subject.
\setchapterpreamble[u]{\margintoc}
\chapter{Darwin and the others}\label{c-Darwin}


\begin{quote}
\begin{flushright}
The alternative to thinking in evolutionary terms\\ is not to think at all.

---Peter Medawar
\end{flushright}
\end{quote}
\bigskip

24 November 1859. The publication of \emph{On the Origin of Species} \cite{darwin1859} by the English naturalist Charles R. Darwin (1809-1882) constitutes a pivotal juncture in the annals of the history of science. The core set of concepts and principles there articulated have survived almost unaltered to the present day, eliciting deep philosophical ramifications\sidenote{\emph{A spectre is haunting the modern world, Darwin’s spectre,
Darwinism.} -- begins a recent book by Michael R. Rose \cite{rose2000}, paraphrasing the opening line of a well-known revolutionary book, a \emph{Manifesto}. Too much emphasis? Apparently, not. Ernst W. Mayr (1904-2005), leading evolutionary biologist, mentions \emph{On the Origin of Species} by Charles Darwin among the three most influential books ever written \cite{mayr2001}, together with \emph{Das Kapital}, by Karl Marx, and the \emph{Bible}, by many authors (or just one).
} \cite{mayr2001}. The theory there presented -- combined with Medel's genetics \cite{huxley1942} -- represents our current conceptual understanding of the emergence of life's complexity \cite{adami2002,adami2012}.

According to Darwin's theory it is the natural selection, or the \emph{survival of the fittest}, that drives the emergence of the biological complexity, which has no equivalent in the inanimate world \cite{wolf2018}. It is arguably a remarkable achievement, for a single theory, to put forth a unifying explanation for the process that leads to intricately structured organisms, starting from a primordial soup of molecules. This process, which is the subject of the present chapter, is referred to as \emph{\gls{evolution}}.

In the eyes of a physicist, evolution is puzzling. If the general picture of Darwin's evolution is accepted, the details are poorly understood. The evolutionary problem is a formidable testing ground for our style of scientific enquiry, trained on spin lattices and pairwise interactions, now facing genetic codes and mostly unknown interactions \cite{goldenfeld2011}. 

First, the quest for a theoretical understanding of the evolutionary process entails learning from the fellow biologists \emph{what} happens and \emph{how} it happens, i.e., the subject matter (sec. \ref{s-subj-matt}). A theoretical framework can be then be established, which in turn allows for hypotheses to be formulated and models to be mathematically stated (sec. \ref{s-mod_ev}). In so doing, connections with related problems -- by analogy and/or by generalization -- are sometimes unveiled, and fruitfully exploited (sec. \ref{s-GA}). 

\begin{kaopaper}[title=Main reference ]
\faFile*[regular] \textbf{Vito Dichio}, Hong-Li Zeng, and Erik Aurell. \emph{Statistical genetics in and out of quasi-linkage equilibrium}. In: Reports on Progress in Physics 86.5 (2023), p. 052601 \cite{dichio2023}.
\end{kaopaper}

\section{Subject matter}\label{s-subj-matt}
\emph{Where an attempt is made to summarise in a few pages the salient features of evolution, in the narrative initiated by Darwin and perpetuated by modern evolutionary biology.}\bigskip  

The Darwinian theory, or \emph{\Gls{darwinism}}, is the foundational theory upon which the entire field of evolutionary biology is built\sidenote{Two classical, comprehensive references that extensively cover the biology outlined in this section are \cite{klug2019,hartl1997}.}.
The subject matter of Darwinism is the evolutionary dynamics of a population - the latter defined as a group of individuals (organisms) of the same species that live in a specific geographical area and reproduce across successive generations. 

With respect to an individual, a distinction is made between the \gls{genotype} and \gls{phenotype}. The genotype refers to the specific set of genes that an organism carries, which encode the instructions for building and operating an organism. These genes can exist in different versions, called alleles. The phenotype, on the other hand, represents the actual observable \glspl{trait} or characteristics of the organism. Phenotypes result from the expression of an organism's genes as well as the influence of environmental factors and the interactions between the two. The genotype-phenotype map (or GP map) is the term used to describe the relationship between an organism's genetic makeup (its genotype) and its observable traits (its phenotype). It translates the information stored in the genes into the physical, macroscopic traits. This translation process is complex and not fully understood, as it involves multiple steps of gene expression, regulation, and interaction, and it is influenced by environmental factors as well \cite{manrubia2021}.

The key ingredient of evolution is \emph{\gls{inheritance}}: offspring inherit traits from their parents through genetic information passed down from generation to generation. At the population level, two opposing drivers define the evolutionary dynamics:
\begin{itemize}
    \item[(a)] Genetic \emph{\gls{variation}}. It broadly refers to the increase of genetic heterogeneity within a population. It primarily stems from by stochastic events that introduce variability in the genetic makeup of individuals. The two most common sources of variability are:
    \begin{itemize}
        \item[i.] Mutations, which involve random alterations of an individual's genotype. Beneficial, deleterious, and neutral mutations respectively enhance, impair, or do not noticeably affect an organism's ability to survive and reproduce in its environment.
        \item[ii.] Recombinations, which involve the exchange and rearrangement of genetic material between genotypes. Importantly, they require a physical transfer of genetic material between two individuals during reproduction in sexual populations. 
    \end{itemize}
    \item[(b)] Natural \emph{\gls{selection}}. It acts upon the genetic heterogeneity within a population, by favoring individuals with traits that enhance their survival and reproductive success, while disadvantaging those with less favorable traits. It is ultimately due to the selective pressure exerted by the environment, which include a variety of factors such as resource availability, predation, and competition. Through natural selection, advantageous traits become more common in a population over time, leading to the adaptation of species to their ecological niches.
\end{itemize}

Inheritance, variation, and selection form the conceptual core of Darwinism, constituting the fundamental principles that underpin the theory. It is worth emphasizing that while variation happens at the level of the genotypes, selection operates on observable traits, hence on phenotypes. 

As stated in the beginning, the subject of the evolutionary process is not the genotype nor the phenotype of isolated individuals, but rather the population as a whole. Evolution unfolds over multiple generations, by gradually changing the genetic composition of populations or, more precisely, the statistical properties of the genotype distribution. One can then take a step forward and frame the problem in terms of the dynamics of the distribution of allele frequencies. This latter change of perspective resonates with classical statistical mechanics, as asserted already in the early 1950s by the R. A. Fisher \cite{fisher1953}\sidenote{Ronald Aylmer Fisher (1890-1962) was a renowned British statistician and biologists. He is regarded as the \emph{founder of modern statistics} \cite{rao1992}. He was also among the founders, together with J. B. S. Haldane (1892-1964) and S Wright (1889-1988), of population genetics -- the subfield of biology that studies the distributions and changes of allele frequencies in a population (bottom-up approach) \cite{hamilton2021}.}:

\begin{quote}
    Now, the frequencies with which the different genotypes occur define the gene ratios characteristic of the population, so that it is often convenient to consider natural population not so much as an aggregate of living individuals as an aggregate of gene ratios. Such a change of viewpoint is similar to that familiar in the theory of gases, where the specification of the population of velocities is often more useful than that of a population of particles.
    \begin{flushright}
        \emph{R. A. Fisher, Croonian Lecture, 1953}
    \end{flushright}
\end{quote}

\section{Modelling evolution} \label{s-mod_ev}
\emph{Where one theoretical approach to evolutionary dynamics is presented and stated in the mathematical language. Where the discussion also hinges on the contemplation of a popular metaphor.}\bigskip 

The diversity observed in natural outcomes of the evolution implies the need for a probabilistic description of the evolutionary dynamics. Indeed, the notion of probability pops up everywhere in formulating the building blocks of Darwinism: genetic mutations happen by chance, recombination events randomly reshuffle the parental genetic material, natural selection enhances - but does not guarantee - the reproductive success of apt individuals (...) The Darwin's theory is inherently a statistical theory.
A model of Darwin's evolution is not expected to predict what \emph{must} happen, but to inform about what \emph{could} happen \sidenote[][*1]{The prominent role of chance in the Darwin's theory caused a sensation in the scientific community at the time, including some scorn reactions. One of his scientific mentors, John Herschel (1792-1871), privately dubbed his theory as \emph{the Law of higgledy-piggledy} \cite{pence2018}.}.

A wide array of theoretical models have been proposed, encompassing various aspects of genetic variation and evolutionary processes. A review of these approaches is far beyond the scope of this section, the interested reader is referred to \cite{charlesworth2017,crow2017,hartl1997}. Here, our attention is rather directed towards a particular such approach, which is instructive, as it comprises the subject matter discussed earlier and formulates the essential theoretical tools employed in this work. It has been originally formulated by \emph{Neher and Shraiman} \cite{neher2011}, and recently reviewed in \cite{dichio2023}.

\subsection{Statistical genetics}\label{ss-NS_theory} 

The interest of physicists to the problem of the evolutionary dynamics stems primarily from a conceptual similarity between thermodynamics and quantitative genetics\sidenote[][-1.5cm]{The study of continuous-varying phenotypic traits (height, weight...), and specifically on population-wide averages over several genetically diverse individuals (top-down approach).}, as first noted by R. A. Fisher in the 1930 \cite{fisher1930}. The lack of an energy-like conserved quantity for the evolutionary dynamics hampers a straightforward translation of the equilibrium thermodynamics laws, while a non-equilibrium picture appear more appropriate \cite{sella2005,rao2022}. The comparison can still be pursued, by seeking a theory for quantitative genetics that parallels the role of statistical mechanics for thermodynamics. Specifically, a phenomenological theory that provides a link between the population-averaged phenotypic traits and the underlying genotype dynamics. Such a theory is referred to as \gls{statistical genetics}.

In the framework proposed by \emph{Neher and Shraiman} \cite{neher2011,dichio2023}, a genotype is represented as a string of $L$ binary variables $g=\{\sigma_1,\dots,\sigma_L\}$, where $\sigma_i=\pm1$. Each locus $\sigma_i$ represents a spin-like biallelic gene\sidenote{Here, for simplicity, one gene = one locus (spin variable).}, the number of genes $L$ in each genotype is fixed. Consequently, the genotype space $\mathcal{G}$ is an $L$-dimensional hypercube. Each possible genotype is found in a population with probability $P(g,t)$, which depends on time. In particular, it changes under the effect of mutations, recombinations and selection.
\begin{itemize}
    \item[(a)] \emph{Mutations}. In a time interval $\Delta t$, mutations change the genotype distribution as follows:
    \begin{equation}\label{e-NS-mut}
        P(g,t+\Delta t) = P(g,t) + \Delta t \mu \sum_{i=1}^L[ P(M_ig,t)-P(g,t) ]\ ,        
    \end{equation}
    where $\mu>0$ is the constant mutation rate, uniform across the genotype, and the operator $M_i$ swaps the $i$-th locus i.e., it replaces $\sigma_i\rightarrow-\sigma_i$.
    \item[(b)] \emph{Recombinations}. One such event consists in the exchange of genetic material between two individuals $g^{*}, g^{**}$, to form an offspring. The result is a novel genotype $g$, which randomly inherits parts of the parental genotypes (crossover). Formally, this can be described by a set of Boolean variables $\xi_i=\{0,1\}$, defining $\sigma_i = \xi_i\sigma_i^{*}+(1-\xi_i)\sigma_i^{**}$. In words, the locus $\sigma_i$ of the offspring $g$ is inherited from $g^{*}$ if $\xi_i=1$, from $g^{**}$ if $\xi_i=0$. Each crossover pattern $\bm{\xi}=\{\xi_i\}$ comes with probability $C(\bm{\xi})$. The change of the genotype distribution induced by recombinations is:
    \begin{equation}\label{e-NS_rec}
    \begin{split}
            P(g,t+\Delta t) = (1&-\Delta t r) P(g,t) +\ \\
            &+\Delta t r \sum_{\bm{\xi},\bar{g}} C(\bm{\xi}) P(g^*,t)P(g^{**},t)\ ,
    \end{split}
    \end{equation}
    where $r$ is the recombination rate and the sum in the second terms runs over all possible crossover patters $\bm{\xi}$ and the genetic material $\bar{g}$ that is discarded during the recombination event, i.e., $\bar{\sigma}_i = (1-\xi_i)\sigma_i^{*}+\xi_i\sigma_i^{**}$\sidenote{Recombination acts as collision process in the theory of gases. An unspoken assumption in \eqref{e-NS_rec} is that all the two-genome distributions factorize $P_2(g^{\alpha},g^{\beta},t)\sim P(g^{\alpha},t)P(g^{\beta},t)$, which is equivalent to the assumption of molecular chaos (\emph{Stosszahlansatz}). A critical discussion of this hypothesis can be found in \cite{dichio2023}.}.
    \item[(c)] \emph{Natural selection}. A fitness-based scheme is formulated. The \gls{fitness} of a genotype is proportional to the average number of offspring of an individual with genotype $g$ \cite{peliti1997}. Since it is a non-negative function, we choose an exponential representation $\text{fitness}(g)\sim\exp{[\Delta t\varphi F(g)]}$ where $\Delta t$ is a time interval, $\varphi>0$ is the selection rate and $F(g)\in\mathbb{R}$ is the fitness function. We postpone to sec. \ref{ss-fit_fun} a thorough discussion of the latter. Natural selection favors individuals with higher-than-the-average reproductive success as follows:
    \begin{equation}\label{e-NS_sel}
        P(g,t+\Delta t) = \frac{e^{\Delta t \varphi F(g)}}{\av{e^{\Delta t \varphi F(g)}}_t}P(g,t)\ ,
    \end{equation}
    where we have used the notation $\av{e^{\Delta t \varphi F(g)}}_t = \sum_g e^{\Delta t \varphi F(g)}P(g,t)$ for the population-average. It is important to stress that, according to \eqref{e-NS_sel}, the notion of fitness is inherently comparative. It is not the intrinsic value of fitness that is important, but rather its value in comparison to average in the population, at a given time.
\end{itemize}
In the limit $\Delta t\rightarrow0$, the combined action of mutations, recombinations and selection can be expressed in a unified phenomenological master equation \cite{gardiner1985}:
\begin{equation}\label{e-NS_me}
\begin{split}
    \frac{d}{dt}P(g,t) &= \big[F(g)-\av{F}_t\big]P(g,t)+\mu \sum_{i=1}^L[ P(M_ig,t)-P(g,t) ] \ +\\
    &+ r \sum_{\bm{\xi},\bar{g}} C(\bm{\xi}) \big[P(g^*,t)P(g^{**},t)-P(g,t)P(\bar{g},t)\big]\ .
\end{split}
\end{equation}
The master equation \eqref{e-NS_me} describes the genotype dynamics in the limit of an infinite population i.e. $M\rightarrow\infty$, where $M$ is the number of individuals in a population. This corresponds to the limit of a perfect sampling of the genotype distribution, which allows to neglect the sampling noise -- referred to as genetic drift. 

The dynamics of an asexual population can be studied by setting $r=0$ in \eqref{e-NS_me}. In sexually reproducing populations, it is frequently observed that recombinations happen at much faster rate than mutations. In this case, the dynamics on the time scale $r^{-1}$ is investigated by using \eqref{e-NS_me} with $\mu\sim0$ \cite{neher2011}. Note however that recombinations do not create nor eliminate alleles in the population. Therefore, for the long-term dynamics of the allele frequencies in a population must be governed by the influx of new mutations. Indeed, mutations play a more fundamental role in the evolutionary process compared to recombinations. Unlike the latter, which act on pre-existing variability, mutations actively spawn new variations\sidenote{A simple example is that of a population of $M$ identical individuals. Recombinations have no effect whatsoever in the genetic composition of the populations. For an evolutionary process to begin, the influx of new alleles (read, mutations) is needed. \label{sn-pop_diversity}}.

With \eqref{e-NS_me} one can in principle to compute the dynamics of the average of any quantitative trait $O(g)$, i.e., of any function in the genotype space. In fact, 
\begin{equation}\label{e-NS_expectation}
    \frac{d}{dt}\av{O}_t = \frac{d}{dt}\sum_gO(g)P(g,t)=\sum_gO(g)\ \boxed{\frac{d}{dt}P(g,t)}\ ,
\end{equation}
using \eqref{e-NS_me} to evaluate the boxed quantity. Moreover, it can be used to evaluate the dynamics of the distribution of any trait, which can be obtained from \eqref{e-NS_me} by projection:
\begin{equation}\label{e-NS_dyn_obs}
    P(O,t) = \sum_g \delta[O-O(g)]P(g,t)\ ,
\end{equation}
where $\delta(O)$ is the Dirac delta function. The equations \eqref{e-NS_expectation}, \eqref{e-NS_dyn_obs} fulfill the purpose of the theoretical framework, that is, to furnish a formal scaffolding that enables (i) hypotheses on the parameter space to be formulated and (ii) computations to be carried out. The results of such computations, in turn, are to be tested against experimental data\sidenote[][-4.5cm]{A review of the results recently obtained starting from \eqref{e-NS_expectation} falls outside the scope of this manuscript. A considerable interest has been devoted to the region of the parameter space corresponding to the \emph{quasi-linkage equilibrium} phase, investigated both in theory \cite{neher2009,neher2011,mauri2021} and numerically \cite{zeng2020b,zeng2021,dichio2023} using efficient simulation tools \cite{zanini2012}. Notably, the framework outlined here has enabled compelling connections with experimental data, as explored and exploited in the case of the SARS-CoV-2 viral genomes \cite{zeng2020,zeng2022}.
}.

\subsection{On the fitness landscape}\label{ss-fit_fun}
A key conceptual aspect of the framework illustrated above pivots around the question: what is the fitness function? It is a coarse-grained metaphor, at once useful and of limited use\sidenote[][-0.5cm]{Moreover, as old as population genetics itself. In fact, it was introduced by S. Wright in the 1932: "\emph{The problem of evolution as I see it is that of a mechanism by which the species may continually find its way from lower to higher peaks in such a field.} (fitness landscape, ed.)" \cite{wright1932}}. 

In the previous section, the fitness function $F$ has been assigned the role of providing a map from the genotype space to the reproductive success. 
It is the environment that exerts a selective pressure of a population (competition for limited resources), hence establishing the fitness of any individual. Selection, however, acts on phenotypic traits, hence it impacts genotypes in an indirect manner. This implies that the mapping from genotypes to  fitness aggregates two distinct types of information: (i) the genotype-phenotype map and (ii) the phenotype-to-reproductive-success map \cite{manrubia2021}. 

\begin{figure}
    \centering
    \includegraphics{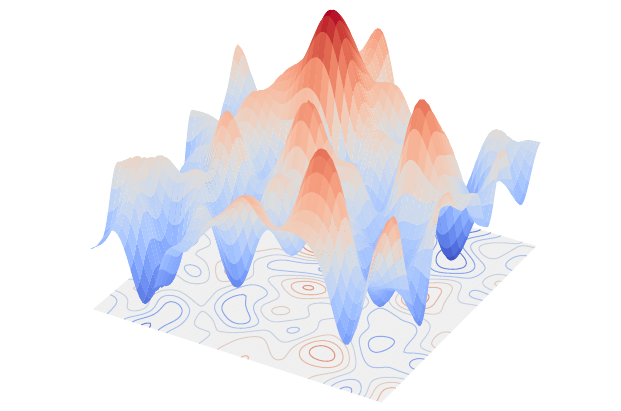}
    \caption{A beautiful, deceptive picture. A long-standing tradition, initiated by Wright \cite{wright1932}, is to depict the notion of a fitness function a surface in an aesthetically pleasing 2D or 3D plot. Here, the $xy$ plane is the state space (genotypes or phenotypes) and the $z$ direction, or "height", stands for fitness. This representation makes intuitive the notion of evolution as an attempt to "climb the fitness hills". However, the true state space is multi-dimensional and intuitions developed in a lower dimensional representation can be highly misleading, both quantitatively and qualitatively (e.g., the notion of "neighbor"). For these reasons, it has been proposed to abandon at all the landscape picture \cite{kaplan2008,mccandlish2011}. On our end, we recognize the efficacy of the abstract landscape illustration and semantics, we advise prudence when using it as a tool for intuitive understanding.
    }
    \label{f-fit_land}
\end{figure}

If and when the theoretical picture of a fitness function holds valid, the evolutionary process can be thought as a dynamic on the \emph{\gls{fitness landscape}}, with the population probability distribution that, driven by natural selection, progressively climbs up the fitness hills, fig. \ref{f-fit_land}. The shape of the fitness landscape is fixed once the parameters of $F$ are specified. In particular, as a function on a $L$-dimensional hypercube, the fitness function can generally be decomposed as follows:
\begin{equation}\label{e-NS_fit_fun}
    F(g) = \bar{F}+\sum_iF_i\sigma_i + \sum_{i<j}F_{ij}\sigma_i\sigma_j+\sum_{i<j<k}F_{ijk}\sigma_i\sigma_j\sigma_{k} + \dots \ ,
\end{equation}
where $\bar{F}$ is a constant and the $F_{i_1,\dots,i_k}$ are the parameters - there are $2^L$ of them, as expected for an exact representation. The higher the order of interactions that play a significant role in \eqref{e-NS_fit_fun}, the more rugged and multi-peaked will be resulting landscape. At the very least, \eqref{e-NS_fit_fun} is useful, as it enables the explicit evaluation of the dynamics of phenotypic traits \eqref{e-NS_expectation}, \eqref{e-NS_dyn_obs}. 

The $2^L$ parameters of \eqref{e-NS_fit_fun} are arguably a huge number\sidenote[][-1cm]{What is the minimal genome size needed to support life? The record-holding organism, to date, is the bacterium \emph{Candidatus Nasuia deltocephalinicola}, with 137 protein-coding genes \cite{moran2014}. A fitness function of the form in \eqref{e-NS_fit_fun} would have $2^{137}\sim10^{41}$ parameters (!).}. We therefore seek a theoretical argument to scale down the dimensionality of $F$, i.e., the number of degrees of freedom and, by consequence, that of parameters. Two possible approaches are the following.
\begin{itemize}
    \item[i.] Landscapes of the form \eqref{e-NS_fit_fun} have been thoroughly investigated within the framework of spin glass theory \cite{mezard1987}. Notably, it is understood from these studies that the essential consequences of the complexity of such a landscape are already manifest in a model with only pairwise interactions \cite{sherrington1975}. Therefore, for the sake of simplicity, one starts exploring a simplified fitness function:
    \begin{equation}\label{e-NS_SK_fit_fun}
        F(g) = \bar{F}+\sum_iF_i\sigma_i + \sum_{i<j}F_{ij}\sigma_i\sigma_j\ ,
    \end{equation}
    which has $\sim L^2$ parameters. As a matter of fact, this is the only (non-trivial) choice that has allowed to solve the evolutionary dynamics in a closed form \cite{mauri2021,dichio2023}.
    \item[ii.] In order to mimic more closely the mechanism of natural selection, it is possible to define a set of phenotypic variables $f_{GP}(g) \in \mathbb{R}^r$, where $f_{GP}:\mathcal{G}\mapsto\mathbb{R}^r$ is the genotype-phenotype map. The cardinality of the phenotype space is necessarily lower than or equal to that of the genotype space, i.e., $r\le 2^L$. A phenotype-fitness map $F:\mathbb{R}^r\mapsto\mathbb{R}$ can be defined on the phenotype space. The fitness function is then written as $F(g) = F(f_{GP}(g))$,
    \begin{equation}
        \mathcal{G}\xrightarrow[]{f_{GP}}\mathbb{R}^r\xrightarrow[]{F}\mathbb{R} \ .
    \end{equation}
    For instance, if $F$ is a simple linear combination of phenotypic traits, the resulting fitness function has a number $r$ of parameters.
\end{itemize}

Historically, skepticism about the concept of a fitness landscape has largely stemmed from the lack of empirical data to outline its actual topography. In recent years, this critique has been somewhat offset with the advent of methods that allow the construction of \emph{empirical fitness landscapes}\sidenote{These approaches consist in creating artificial mutants, each carrying one or more mutations with respect to the wild-type genotype, then measuring their fitness using a fitness proxy (e.g., antibiotic resistance) \cite{devisser2014}. For instance, deep mutational scans \cite{fowler2014} are able to test and assess the phenotypes of all single mutants, and several double- and triple-mutants of a wild-type genotype.} \cite{devisser2014,manrubia2021}.
Though still in their early stages, these experiments carry potential to drive a more comprehensive understanding of the shape and significance of the fitness function.

A different source of criticism has emerged in recent years, against the assumption of a constant environment, implicit in \eqref{e-NS_fit_fun} \cite{mustonen2009,lassig2017}. A time-dependent selection would rather be mediated by a \emph{fitness seascape} $F(g,t)$. In fact, the hypothesis of a fixed environment is never exactly true, as even in the simplest lab environments a number of factors induce a temporal dynamic of the selective pressure, including modifications of the physical environment, frequency-dependent selection, co-evolution effects, interaction between species (ecology) \cite{neher2018}. The assumption of a fixed landscape holds approximately true when the timescale of the environmental changes, broadly speaking, is sufficiently long compared to that of the process under investigation.
Unquestionably, it results in a remarkable simplification of the complexity of the phenomenon.

The picture of a fitness landscape, therefore, should be regarded as approximate in several senses. Yet, with the \emph{caveat} above, it offers an instructive way of thinking about evolution.

\section{Genetic algorithms}\label{s-GA}
\emph{Where evolution is recognised as solving a more abstract optimisation problem, the evolutionary algorithm is therefore isolated and used elsewhere.}\bigskip 

Devoid of all details, the evolutionary problem described in sec. \ref{s-mod_ev} is easily recognized as a particular instance of a general discrete optimization problem. The latter is defined as:
\begin{equation}\label{e-opt_prob}
    \max_{y} f(y)\quad \text{subject to}\quad y\in\mathcal{Y} ,
\end{equation}
where $f:\mathcal{Y}\mapsto\mathbb{R}$ is the \emph{objective function}, $\mathcal{Y}$ is the \emph{feasible set} and the $y_i, \ i=1,\dots, l$ are \emph{binary decision variables} \cite{bierlaire2018}. In sub-case of evolution, the binary decision variables are the biallelic genes $\sigma_i$, the feasible set is the genotype space $\mathcal{G}$, the objective function is the biological fitness $F$. In this (simplified) view, natural selection acts as an optimization process. Over time, evolution leads to the "optimization" of a population for survival and reproduction in their specific environment.

For all problems of the form \eqref{e-opt_prob}, a straightforward algorithm exists. It involves generating a complete list of all possible $y\in\mathcal{Y}$, evaluating the objective function value for each solution, and identifying those $y$ that yield the maximum value of $f$. Practically however, this becomes soon unfeasible, since it involves computing $\sim 2^l$ computations of $f$. The exponential explosion of the running time as a function of the dimension of the problem is often referred to -- with a hint of desperation -- as the \emph{curse of dimensionality} \cite{bellman1961}. In computer science, several approximate yet efficient methods have been devised to explore the feasible solution space of an optimization problem and discover the optimal solutions. These approaches, commonly known as \emph{metaheuristics}\sidenote[][-4cm]{Metaheuristics are problem-solving strategies that provide a general framework for solving optimization problems. They are not tailored to a specific problem (as it is the case of heuristics) but offer a set of guiding principles and strategies that can be applied to a wide range of problem domains. An overview of the existing approaches can be found in \cite{gendreau2010}. A substantial subset of such techniques is inspired by biological systems, as it is the case for the particle swarm optimization \cite{kennedy1995}, ant colony optimization \cite{dorigo2006}, and genetic algorithms (see below).
}, encompass a range of techniques \cite{gendreau2010}, often inspired by natural processes -- simulated annealing being a celebrated example \cite{kirkpatrick1983}.

Genetic algorithms (GAs), in particular, leverage the parallelism demonstrated above with the evolutionary problem, to construct a metaheuristic for \eqref{e-opt_prob}\sidenote[][-1cm]{John H. Holland (1929-2015), father of GAs, opens an influential article in \emph{Scientific American} by proclaiming: \emph{Living organisms are consummate problem solvers. (...) Pragmatic researchers see evolution's remarkable power as something to be emulated rather than envied.} \cite{holland1992}. However, GAs were introduced by Holland and his students in the 1970s not with a specific application as \eqref{e-opt_prob} in mind, but with the general purpose of simulating and studying artificial adaptive systems in a computational environment \cite{holland1975,dejong1993}.} \cite{goldberg1989}. In GAs, candidate solutions to a problem undergo successive generations of selection, recombination, and mutation to converge towards optimal or near-optimal solutions. GAs are population-based, i.e., they operate on a population of potential solutions, each represented as a binary strings. The following example presents a minimal version of the algorithm.

\subsubsection{An illustrative case}
Suppose an optimization problem 
\begin{equation}\label{e-opt_prob_ill}
    \max_{y}\ y_{[10]}|\sin{y_{[10]}}| \quad \text{subject to}\quad y\in\mathbb{Z}_2^5\ ,
\end{equation}
where the feasible set $\mathcal{Y}=\mathbb{Z}_2^5$ is the space of all possible binary bit-like strings\sidenote[][-2.5cm]{In the previous section, we used spin-like binary variables $\sigma_i\in\{-1,1\}$, as it is common in statistical physics. Here, we rather use bit-like binary variables $y_i\in\{0,1\}$, as it is common in computer science. The two sets are obviously equivalent.} of length $5$, $f(y) = y_{[10]}|\sin{y_{[10]}}|\in\mathbb{R}^+ $ is the objective function and $y_{[10]}$ is the decimal representation of the binary string $y$\sidenote{In the language of sec. \ref{s-mod_ev}, we would say that $y$ is a genotype, $y_{[10]}$ its phenotype and $f(y)$ its fitness.}. Clearly, for \eqref{e-opt_prob_ill} an exhaustive list of all possible $2^5=32$ strings in $\mathcal{Y}$ could be written, their $f$ evaluated and the maximum determined. Here, for illustrative purposes, we ask \eqref{e-opt_prob_ill} to be solved using a GA.

The algorithm starts by randomly instantiating an initial population of $M$ strings, say $M=4$. Three evolutionary operators are then iteratively applied at each generation:
\begin{itemize}
    \item[1.] \emph{Selection}. The objective function $f$ is evaluated for all $y^{(i)}, \ i=1,\dots,M$ in the last generation. Each string $y^{(i)}$ has a probability to reproduce proportional to its value of the objective function. In the simplest scheme (biased roulette wheel), the probability corresponds to 
    \begin{equation}\label{e-roulette}
        f(y^{(i)})/\sum_jf(y^{(j)})\ .
    \end{equation}
    See tab. \ref{t-ex_roulette} for an example. As a result of selection, the population average of the objective function $\av{f} = \sum_{j=1}^M f(y^{(j)})$ increases.
    \begin{table}[htbp]
    \centering
    \caption{Example of selection. Each existing string $y^{(i)}$ as a probability \eqref{e-roulette} to be selected proportional to its value of the objective function $f$. In the last column, the result of a sampling of $M=4$ individuals. The population average $\av{f}$ increases from $13.15$ to $16.74$.
    }\label{t-ex_roulette}
    \begin{tabular}{cccccc}
    \toprule
    $i$ & $y^{(i)}$ & $y^{(i)}_{[10]}$ & $f(y^{(i)})$ & \eqref{e-roulette} & new count \\
    \midrule
    1 & $01101$ & 13 & 5.46 & 0.10 & $0$ \\
    2 & $10101$ & 21 & 17.57 & 0.33 & $1$ \\
    3 & $01111$ & 15 & 5.44 & 0.19 & $1$ \\
    4 & $11010$ & 26 & 3.30 & 0.38 & $2$ \\
    \bottomrule
    \end{tabular}
    \end{table}
    \item[2.] \emph{Recombinations}. Pairs of strings undergo recombination with rate $r$. A popular such scheme (single-point crossover) entails swapping a portion of the parental strings, e.g.
    \begin{equation}
        \begin{array}{c}
            \textcolor[HTML]{4169E1}{01111}  \\
            \textcolor[HTML]{990011}{11010}  
        \end{array}
        \rightarrow
        \begin{array}{c}
            \textcolor[HTML]{4169E1}{01}\ |\ \textcolor[HTML]{4169E1}{111}  \\
            \textcolor[HTML]{990011}{11} \ |\ \textcolor[HTML]{990011}{010}  
        \end{array}
        \rightarrow
        \begin{array}{c}
            \textcolor[HTML]{4169E1}{01} \textcolor[HTML]{990011}{010}  \\
            \textcolor[HTML]{990011}{11} \textcolor[HTML]{4169E1}{111}  
        \end{array}
        \ ,
    \end{equation}
    where the position of the cut $|$ is randomly chosen.
    \item[3.] \emph{Mutations}. Each existing bit in the population is flipped with uniform rate $\mu$, e.g.
    \begin{equation}
        01010 \rightarrow 01\textcolor[HTML]{990011}{1}1\textcolor[HTML]{990011}{1}\ .
    \end{equation}
\end{itemize}
By repeatedly applying the three aforementioned steps, the population gradually converges towards the optimal solution \eqref{e-opt_prob_ill}. However, the parameters governing the strength of evolutionary operators must be tuned to achieve a balance between the exploitation of those strings that better approximate the optimal solution (selection) and the exploration of the solution set (recombinations, mutations)\sidenote{In the GAs literature, however, mutations have a secondary role with respect to recombinations. They are mostly regarded as a mere insurance policy against the loss of diversity (s.n. \ref{sn-pop_diversity}) and a premature convergence of the search algorithm \cite{goldberg1989,holland1992}. By consequence, mutation rates are typically set to low values.}.

\subsubsection{Remarks}
A number of successful applications of GAs to real-world optimization problems demonstrate the interest of the scientific community in the approach -- see \cite{reeves2010} and references therein.\sidenote{Curiously, there is no consensus on the exact reasons \emph{why} GAs work. An overview of the debate can be found in \cite{reeves2010,reeves2002}.} Indeed, genetic algorithms exhibit two compelling features that make them highly attractive.
\begin{itemize}
    \item[i.] \emph{GAs are population-based}. Realistic objective function are largely multimodal, meaning that they exhibit multiple peaks and local optima. Any search algorithm in such a landscape should avoid getting ensnared within a local optima and rather pursuit the global optimum. GAs rise to the occasion by unleashing multiple walkers that venture through the solution space simultaneously. In the case where a walker became trapped, the others would aid in circumventing the trap in the subsequent generation. This mitigates -- even though, does not eliminate -- the peril of stagnation around sub-optimal solutions. 
    \item[ii.] \emph{GAs have minimal assumptions about $f$}. As it should be clear from the example above, GAs solely rely on the payoff (objective function) values assigned to individual strings. For instance, they do not require the computation of derivatives, as gradient-based methods do. Even more, GAs do not even necessitate a mathematical expression for $f$. Consistently with the logic of GAs, an objective function can be considered a black box that takes an input (string) and produces a real number as output. This implies the formulation of GAs is \emph{problem independent}, since it does not rely of problem-specific information about the topography of the search landscape.
\end{itemize}
Clearly, enhancing robustness and generality does not come without drawbacks, and GAs are not an exception, notably:
\begin{itemize}
    \item[iii.] \emph{GAs are computationally demanding.} The major computational bottleneck of GAs is the large number of objective function evaluations, which scales linearly with the population size. Therefore, GAs simulations are highly sensitive to the complexity of the search landscape. The latter, in turn, depends on the problem at hand, therefore there is no universal approach to guarantee timely convergence. Approximate methods and/or additional assumptions should be tailored to the specific phenomenon under investigation.
\end{itemize}

\setchapterpreamble[u]{\margintoc}
\chapter{The exploration-exploitation dynamics}\label{c-EE}
\begin{quote}
\begin{flushright}
No one trusts a model except the man who wrote it; everyone trusts an observation, except the man who made it.

---Harlow Shapley 
\end{flushright}
\end{quote}
\bigskip

\begin{marginfigure}[1cm]
	\includegraphics[height=2.5cm]{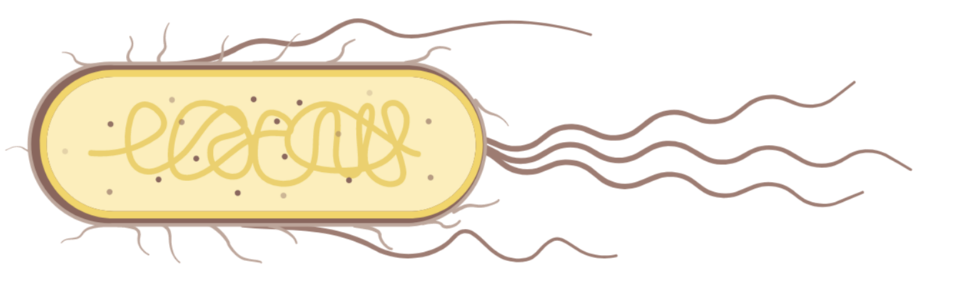}
\end{marginfigure}
To navigate its microscopic world, the bacterium \emph{Escherichia coli} uses chemotaxis to find nutrients and avoid harm. When a desirable sugar molecule is in the vicinity, its receptors alert the bacterium's tiny flagella, which spin counterclockwise to gently glide the \emph{E. coli} towards the source. Conversely, if a harmful substance is detected, the flagella whirl clockwise, causing the \emph{E. coli} to tumble and change direction, scurrying away from the danger \cite{webre2003,sourjik2012}.

\begin{marginfigure}
	\includegraphics[height=3cm]{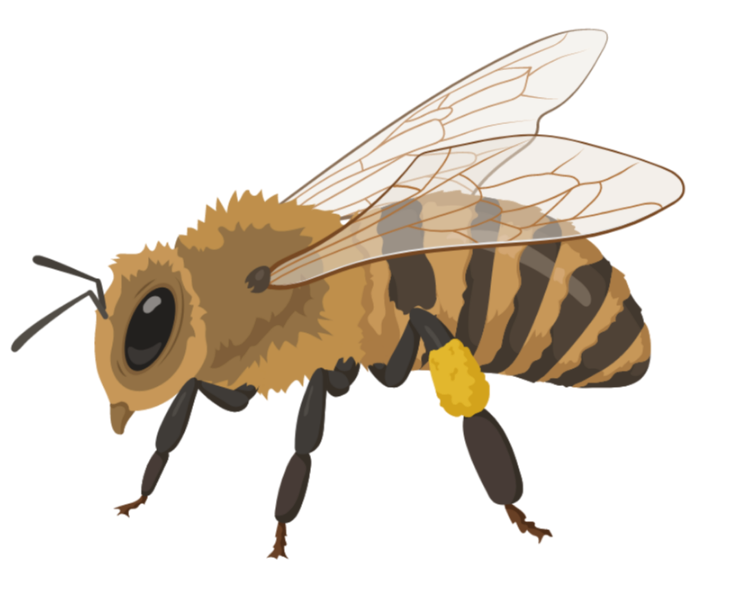}
\end{marginfigure}
When a foraging honeybee (\emph{Apis mellifera}) comes across a rich source of nectar, it makes its way back to the hive to perform the iconic 'waggle dance'. This intricate dance communicates the location of the food source to her hive mates, inviting them to take advantage of the newfound resources. Some, however, deliberately ignore the information in the dance and choose to continue exploring the environment for new nectar sources. This provides the colony with a safety net against the whims of flower nectar production \cite{seeley1991,dyer2002,gruter2009}. 

\begin{marginfigure}[.5cm]
	\includegraphics[height=3cm]{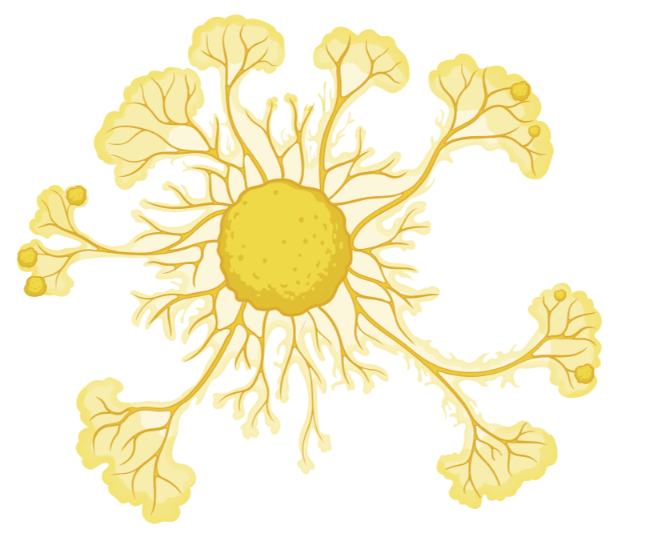}
\end{marginfigure}
The slime mold \emph{Physarum polycephalum} utilizes a form of reactive navigation to explore its environment. It is composed of several oscillating units, with the oscillation frequency varying according to environmental cues. In the presence of attractants like food, the oscillation frequency increases, inducing cytoplasmic flow towards the food source, while repellents like light or salts reduce the frequency. Moreover, as the slime mold forages, it leaves behind a trail of nonliving extracellular slime, which it later avoids. This simple, noneuronal organism is able to solve complex navigation tasks, such as the U-shaped trap problem \cite{reid2012}.

\begin{marginfigure}[.5cm]    
	\includegraphics[height=3cm]{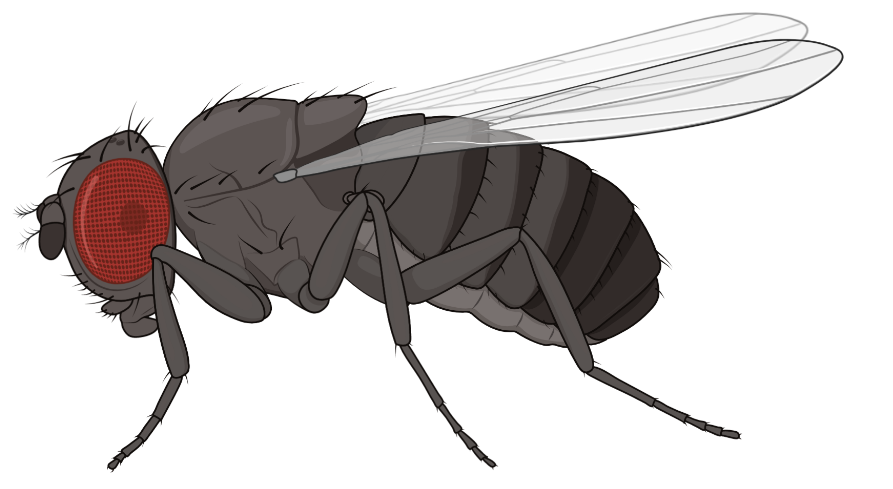}
\end{marginfigure}
When trying to attract a mate, male fruit flies (\emph{Drosophila melanogaster}) serenade females with a courtship song, produced by wing vibration \cite{dickson2008,dukas2015}. Fruit flies display two major modes of song (sine and pulse). Mode transitions, and variable mode durations result in individual courtship sessions that are highly diverse. Recent evidence shows that simple flies use sensory information -- in particular, the female's position and velocity -- to pattern his song sequences over short timescales \cite{coen2014}.
\begin{center}
    (...)
\end{center}

Flicking through the catalogue of biological dynamics, the most extraordinary behaviours stand out. The examples provided above represent a concise selection of noteworthy cases, chosen among numerous other possible instances\sidenote{See also the ant foraging behaviour \cite{czaczkes2015}, the fish predatory strategies \cite{sims2008}, the root growth dynamics \cite{bengough2006}, the immune system response to pathogens \cite{altan2020}.}. There is a clear leitmotif that connects them all, and with the evolutionary dynamics if ch. \ref{c-Darwin}, and with another case that we will discuss in ch. \ref{c-Celegans}. Indeed, they can all be brought under the same conceptual umbrella: \emph{seeking a goal under uncertainty}. Indeed, this seems to be a ubiquitous requirement of life, from the simplest organisms to complex human social behaviours \cite{hills2015,wilson2021}. There are two possible reasons for a concept to be ubiquitous in a scientific domain: either it is trivial, or it is fundamental. Here, we will argue for the latter. 

We start by formalising what we call the \emph{exploration-exploitation} dynamics and providing a mathematical formulation for the case where a biological system can be represented as a graph (sec. \ref{s-EE_fun}). We start by studying simple, solvable models, which are instructive to showcase the essential features of the EE dynamics (sec. \ref{s-EE_scenarios}). Finally, we briefly describe and test the simulations we have designed to cope with complex scenarios (sec. \ref{s-EE_sims}).

\begin{kaopaper}[title=Main reference]
\faFile*[regular] \textbf{Vito Dichio} \& Fabrizio De Vico Fallani. \emph{The exploration-exploitation paradigm for networked biological systems}. In: arXiv e-prints 2306.17300 (2023) \cite{dichio2023b}.
\end{kaopaper}

\section{Fundamentals}\label{s-EE_fun}
\emph{Where the core idea of this dissertation is presented in its final, press-ready version. In particular, where the exploration-exploitation problem for (networked) biological systems is defined and formalised.}\bigskip 

The line of thought is fairly straightforward. Biological systems inherently and universally exhibit randomness. Nonetheless, their dynamics are shaped by functional constraints, therefore randomness and \gls{function} must coexist. Even more, biological systems achieve high-level functions not only \emph{in spite of} randomness but also \emph{through} randomness\sidenote{In the last two decades, with the advent of quantitative biology, there has been a paradigm shift in how we look at the role of chance in living systems. There is growing evidence that randomness is not always a hindrance to biological function, and it can be more than just \emph{noise}, it can be a potential asset in the workings of life. Long established in the context of evolution, it is now recognised, e.g., within the domains of molecular biology \cite{raj2008,kaern2005}, cell biology \cite{balazsi2011}, neuroscience \cite{faisal2008,guo2018}, to name a few.} \cite{tsimring2014,kaneko2006}.

The space in which the randomness unfolds -- typically, the systems' configuration space -- is heavily constrained. In fact, in biology, systems have functions and we can reasonably expect the overwhelming majority of possible configurations to be \emph{non functional} or \emph{poorly functional} \cite{garson2019,bialek2012}. Broadly speaking, therefore, the relation between randomness and biological function has a dual nature. On one side, randomness frequently triggers changes that are detrimental. On the other hand, it serves as an essential mechanism for exploring the range of possible configurations, and identifying those that enhance the system function\sidenote{
In this paragraph, we have used the words \emph{randomness} and \emph{biological function} rather loosely. On the one hand, it was necessary to keep the discussion general. On the other hand, the exact definition of these two concepts is more problematic than it may seem at first sight. The semantic discussion about the meaning of both is instructive, since goes deep into the foundations of biology. Excellent starting points are, e.g, \cite{heams2014,garson2019}. 
}. If the details of this interplay are context-dependent, it is meaningful to look for general principles. 

In the following, we develop a formalism that embeds the discussion above into two fundamental concepts: \emph{exploration, exploitation}. Exploration indicates the stochastic search of the configuration space. Exploitation refers to the use of the configurations that have been found to optimise the system function. The optimisation problem implicit in these definitions, in turn, is formalised by stating (i) how the optimal states are encoded and (ii) how the system approaches them.

To proceed further, we need to define the characteristics of the system configuration space, i.e., we need to select a representation. Henceforth, we will narrow our discussion to consider a graph representation and develop an exploration-exploitation (EE) graph dynamics.

\subsection{A graph dynamic}
Consider a biological system that can be represented as an unweighted, undirected graph $G\in\mathcal{G}$ over $N$ nodes, as defined in \eqref{e-graph_def}. Let $P(G,t)$ be the probability of observing the graph $G$ at time $t$. Exploration and exploitation affect the graph probability distribution as follows.

\subsubsection{Exploration}
The most elementary configuration change involves the reversal of one dyadic state, which means that if there was no edge present, one is created $0\rightarrow1$, or if an edge was already in place, it gets removed $1\rightarrow0$. Let $M_{ij}:\mathcal{G}\mapsto\mathcal{G}$ be the operator which, acting on the graph $G$, inverts the state of its dyad $(ij)$. Again, by simplicity, we assume that dyadic reversals occur throughout the graph at a uniform, time-independent rate $\mu$ (\emph{exploration rate})\sidenote{This is the simplest, non-trivial, exploration scheme. In fact, (i) it is defined for each microscopic degree of freedom (dyad), (ii) it happens at a constant rate and (iii) it does not depend on the dyads (iv) nor on their history. Each of these four assumptions can be relaxed to design more complex exploration algorithms.
}.

In the time interval $\Delta t$, the graph probability distribution changes as a result of the exploration as follows
\begin{equation}\label{e-EE_mut}
    P(G,t+\Delta t) = P(G,t) + \Delta t\mu\sum_{i<j} [P(M_{ij}G,t)-P(G,t)]  \ .
\end{equation}

\subsubsection{Exploitation}
Let us assume that the notion of \emph{biological function} for a given system can be mathematically represented as a function on the graph configuration space $F:\mathcal{G}\mapsto\mathbb{R}$, which we refer to as \emph{functional metric}, or simply \emph{F metric}. As the argument goes, biological systems attempt to maximise their function, hence to increase the value of their $F$ metric\sidenote[][-1cm]{In the following, we will mostly use "$F$ metric" instead of "function", to avoid confusion with the homonymous general mathematical notion.}. Optimal states correspond to the maxima of $F$ or, pictorially, to the peaks in the \emph{F landscape}.

\begin{marginfigure}   
	\includegraphics{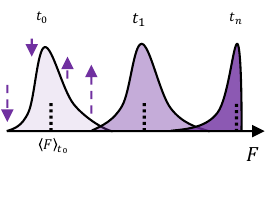}
    \caption{Dynamic of the onedimensional $F$ distribution (illustration). At each point in time, the likelihood of those graphs that have higher $F$ values than the ensamble average (dotted lines) increases (upwards arrows), the likelihood of those that have lower $F$ values decreases (downwards arrows). As a result, the $F$ distribution as a whole shifts towards higher $F$ values.}
	\labfig{f-F_dyn}
\end{marginfigure}

In a time interval $\Delta t$, exploitation changes the graph probability distribution according to: 
\begin{equation}\label{e-EE_sel}
    P(G,t+\Delta t) = \frac{e^{\Delta t \varphi F(G)}}{\av{e^{\Delta t \varphi F}}_t}P(G,t)\ ,
\end{equation}
where $\av{\cdot}_t$ stands for the ensemble average at time $t$, i.e., $\av{e^{\Delta t \varphi F}}_t = \sum_{G} e^{\Delta t \varphi F(G)}P(G,t)$ and $\varphi>0$ (\emph{exploitation rate}) is an overall scaling. In words, the way in which exploitation approaches the most functional configurations is by amplifying (reducing) the likelihood of those graphs that exhibit higher (lower) $F$ values than the ensemble average at time $t$, fig. \reffigshort{f-F_dyn}. The topography of the $F$ landscape plays the delicate role of simultaneously encoding (i) the mechanisms that yield high functional states and (ii) the constraints that restrict the dynamics in the graph configuration space. It generalises the notion of \emph{environment} of evolutionary dynamics to what we might designate as the \emph{functional context}. Clearly, $F$ is problem-dependent. 

In the following, we further assume the existence of a lower dimensional \emph{sufficient} representation of a graph $G$, in terms of $r<2^L$ statistics $\bm{x}(G)\in\mathbb{R}^r$. Therefore, in this latter space the $F$ metric is mathematically defined $F(G)=F(\bm{x}(G))$,
\begin{equation}\label{e-EE_F_def}
    \mathcal{G}\xrightarrow[]{\bm{x}}\mathbb{R}^r\xrightarrow[]{F}\mathbb{R}\ .
\end{equation}

We can collect \eqref{e-EE_mut} and \eqref{e-EE_sel} in a single expression, which defines our EE graph dynamics\sidenote{This dynamics is correctly normalised over the graph ensamble $\av{\cdot}_t$ at each time $t$. Note that the two terms in $\Delta P_{\Delta t}(G,t)$ are normalised independently, as it should be. }:

\begin{equation}\label{e-EE}
\begin{split}
    P(G,&t+\Delta t) = P(G,t)\ + \\
    &+ \underbrace{\Delta t\mu\sum_{i<j} [P(M_{ij}G,t)-P(G,t)] + 
    \Bigg[\frac{e^{\Delta t \varphi F(G)}}{\av{e^{\Delta t \varphi F}}_t}-1\Bigg]P(G,t)}_{\Delta P_{\Delta t}(G,t)} \ .
\end{split}
\end{equation}

It is convenient to define an adimensional parameter to weight the relative strengths of exploration and exploitation. We call it \emph{functional pressure}, 
\begin{equation}\label{e-EE_fp}
    \rho = \varphi/\mu \ .
\end{equation}
For mild functional pressures, $\rho\sim0$ the  dynamic is dominated by random dyadic mutations, and it is similar to a random walk in the graph space $\mathcal{G}$. On the contrary, $\rho\to\infty$ corresponds to the limit of a perfectly exploitative dynamics, where only the most functional graph configurations carry significant probability. 

\subsection{Beyond Darwin, an interpretable GA}
It will not have escaped the attention of the reader that the theoretical structure assembled in the previous section bears resemblance to the evolutionary dynamics we formulated in ch. \ref{c-Darwin}. Of course, this is no mere coincidence.

The EE graph dynamics \emph{is} an evolutionary dynamics -- in the sense of sec. \ref{ss-NS_theory} -- for genotypes of length $L$, based on mutations and natural selection. It is worth stressing the semantic: the EE dynamics defined in \eqref{e-EE_mut} - \eqref{e-EE_sel} is not \emph{analogous} to the mutation-selection dynamics \eqref{e-NS-mut} - \eqref{e-NS_sel}: mathematically speaking, they are the \emph{identical}\sidenote{This is possible in the first place because we have used the same representation for genotypes and graphs, both of which live in the space of $L$-dimensional binary strings -- though, technically, the first $g$ is a vector of length $L$, the second $G$ is a matrix with $L$ degrees of freedom. This is also the reason why we have used the same notation $\mathcal{G}$ for the space of genotypes and graphs, the same $F$ for the fitness function and biological function and so on.}. Naturally, what changes from one case to the other is the interpretation we give to the same mathematical objects. 

In the previous section, we have formulated the EE dynamics on a very general basis. The key logical step is to recognise that the evolutionary dynamics (without recombinations) is a particular instance of the more general EE dynamics where (i) exploration is interpreted as genetic mutations and (ii) exploitation is interpreted as natural selection -- the "biological function" to be maximised is the fitness, or reproductive success. See tab. \ref{t-EE-evol} for a complete list of correspondences. In this sense, the EE dynamics is a generalisation of an evolutionary process without recombinations.

\begin{table}[htbp]
  \centering
  \caption{Translation rules from the vocabulary of EE graph dynamics to that of evolutionary dynamics. We group by colour those terms that refer to the structure of the configuration and state spaces (top), to the the structure of the dynamics (middle) and to the dynamic parameters (bottom). $^*$See sec. \ref{s-EE_sims}.}
  \begin{tabularx}{\textwidth}{>{\raggedright\arraybackslash\hsize=1.3\hsize}X>{\centering\arraybackslash\hsize=0.4\hsize}X>{\raggedleft\arraybackslash\hsize=1.3\hsize}X}
    \textbf{EE graph dyn.} & \textbf{Notation} & \textbf{Evolutionary dyn.} \\
    \toprule
    \rowcolor[HTML]{DFD7BF}
    graph space  & $\mathcal{G}$ & genotype space \\
    \rowcolor[HTML]{DFD7BF}
    number of dyads  & $L$ & genome length \\
    \rowcolor[HTML]{DFD7BF}
    graph (unw., und.) & $G/g$ & genotype \\
    \rowcolor[HTML]{DFD7BF}
    graph statistics & $\bm{x}/f_{GP}$ & phenotype (traits)\\
    \rowcolor[HTML]{DFD7BF}
    dyad (bit-like)  & $a_{ij}/\sigma_{ij}$ & locus (spin-like) \\ \addlinespace[3pt]
    \rowcolor[HTML]{F2EAD3}
    time window  & $T$ & \# generation \\
    \rowcolor[HTML]{F2EAD3}
    \# samples$^*$  & $M$ & population size \\
    \rowcolor[HTML]{F2EAD3}
    bio. function ($F$ metric)  & $F$ & fitness function  \\
    \addlinespace[3pt]
    \rowcolor[HTML]{F5F5F5}
    exploitation r.  & $\varphi$ & natural selection r. \\
    \rowcolor[HTML]{F5F5F5}
    exploration r.  & $\mu$ & mutation r. \\
    \rowcolor[HTML]{F5F5F5}
    $\times$  & $r$ & recombination r.\\ 
    \bottomrule
  \end{tabularx}
  \label{t-EE-evol}
\end{table}

In other words, in writing \eqref{e-EE} we 
borrow the \emph{algorithm} coded by Nature itself to solve the exploration-exploitation problem in the specific context of evolution. The crucial motivation for this is that evolutionary dynamics exhibits a ubiquitous, general and emergent property of the dynamics of living systems: it is \emph{\gls{self-referential}} \cite{goldenfeld2011}. This means that the time-evolution operator that governs the dynamic depends on the state of the system, therefore on its history, as it is manifest in the term $\av{\exp(\Delta t \varphi F)}_t$ in \eqref{e-NS_me} or \eqref{e-EE}. This is \emph{the key, defining feature of biological dynamics}\sidenote[][-3cm]{Self-referentiality has no equivalent in the dynamics of conventional matter. The bewilderment and  and despair of a physicist -- accustomed to the educated, gentle phenomenology of the dynamics of condensed matter -- is well summarised in a recent review on the subject: 
\emph{To a physicist, this sounds strange and mysterious: What is the origin of this feature that sets biological systems apart from physical ones? Aren’t biological systems ultimately physical ones anyway; thus, why is self-reference an exclusive feature of biological systems (whatever they are!)?} \cite{goldenfeld2011}
}.

In sec. \ref{s-GA} we have already discussed a very similar generalisation of the evolutionary dynamics, that of genetic algorithms (GAs). Indeed, our EE dynamics can be regarded as a GA where mutations are the crucial exploration mechanism and, once again, recombinations are left behind. In fact, unlike mutation and selection, it is not clear what recombination should correspond to, outside the context of the evolutionary process\sidenote[][-1cm]{In evolution, in fact, the probability distribution $P(g,t)$ is interpreted as resulting from individuals \emph{existing at the same time in a physical space}, in the infinite population limit. Because of this, individuals can physically exchange genetic material. In general, we will not require such a strict physical interpretation of the EE probability distribution, and will rather use it as an abstract probabilistic tool, see later in ch. \ref{c-Celegans}.}. This is not an embarrassment for GAs, since their generalisation of the evolutionary process is purely \emph{algorithmic}, i.e., they treat evolution as a computational strategy for solving optimization problems, no less, no more.

\begin{figure}[h]   
	\includegraphics[width=6cm]{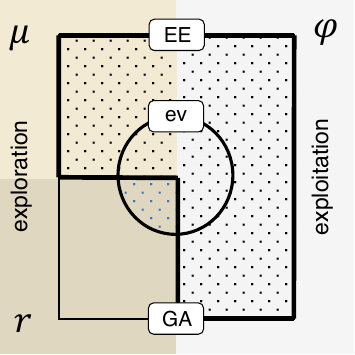}
    \caption{Relationship between exploration - exploitation (EE) dynamics, evolutionary dynamics (evol) and genetic algorithms (GAs). On the left, exploration mechanisms (mutation $\mu$, light brown and recombinations $r$, brown); on the right, exploitation (selection $\varphi$, gray). The EE dynamics (area within thick black line) generalise the evolutionary process (circle) without recombinations. Alternatively, they can be regarded as GAs (rectangle) without recombinations. Dotted area indicates the "interpretability" as a real-world process. Recombinations are only interpretable within the evolutionary context. 
    }
    \label{f-EE_Venn}
\end{figure}

Here, however, we have a more ambitious plan. We want to leave open the possibility of interpreting the EE dynamic as a real-world process, when used outside in the evolutionary context. We do want our EE dynamics -- lifted from evolution -- to be meaningfully instantiated whenever the exploration-exploitation mechanisms cooperate in shaping a biological process. For this to be possible, we must handle the lesson we learn from evolution with caution, keeping what is essential and leaving back the rest, fig. \ref{f-EE_Venn}.

\section{The math of simple scenarios}[Simple scenarios]\label{s-EE_scenarios}
\emph{Where from the investigation of simple case studies, general characteristics of the EE dynamics are unveiled. Pen and paper, let the theorist theorying. 
}\bigskip 

The graph dynamic \eqref{e-EE} is not a model for anything in particular, it is a framework, or a theory for a class of phenomena \cite{bialek2017}. In order to specify a model we must specify the context-specific representation of the notion of biological function. Having set the theoretical stage, much of our role as modellers boils down to one simple, crucial question: \emph{what is $F$}?

As a first step, it is important to disentangle the intrinsic characteristics of the EE dynamic from those originating from the complexity of $F$. To do this, we start by studying \eqref{e-EE} with models of $F$ of minimal complexity, amenable to analytical investigation\sidenote{The reader who is not interested in the more mathematical aspects can skip to the end of this section, read the remark \ref{r-EE_dyn} and go on. You are welcome!}.

\subsubsection{Preliminaries}
The following calculations simplify if, instead of the dyadic bit-like variables $a_{ij}=\{0,1\}$ we use the equivalent representation with spin-like variables $\sigma_{ij}=\{-1,1\}$. The two are related by the following\sidenote{Note that the analogy between graphs and spin systems is such that the dyads and not the nodes are equivalent to spins in classical statistical mechanics.}:
\begin{equation}\label{e-sp_dy}
    \sigma_{ij} = 2a_{ij}-1\ , \quad\quad a_{ij} = \frac{1+\sigma_{ij}}{2} \ .
\end{equation}
Analogous relations hold between the \emph{average graph density} $d\in[0,1]$ and the \emph{average magnetisation} $m\in[-1,1]$:
\begin{equation}\label{e-mag_den}
    m = \frac{1}{L}\sum_{i<j}\av{\sigma_{ij}} = 2d-1\ , \quad \quad d = \frac{1}{L} \sum_{i<j} \av{a_{ij}} = \frac{1+m}{2}\ .
\end{equation}

In this section, we will consider \eqref{e-EE} in the continuous time limit $\Delta t\rightarrow0$, implying that $\Delta t\varphi F\ll1$. The graph dynamics can be then described by the following master equation:
\begin{equation}\label{e-EE_c}
    \frac{d}{dt}P(G,t) \stackrel{(a)}{=} \mu\sum_{i<j} [P(M_{ij}G,t)-P(G,t)] + \varphi [F(G)-\av{F}_t]P(G,t) \ .
\end{equation}
\marginnote[-.5cm]{In $(a)$ we have used $e^{\pm x}\sim1\pm x$ for $x\sim 0$.}
The dynamic of the expected value (ensamble average) of any graph observable $O:\mathcal{G}\mapsto\mathbb{R}$ and its time-dependent probability distribution can be calculated by \eqref{e-NS_expectation} and \eqref{e-NS_dyn_obs}, respectively.

\subsection{No exploitation}\label{ss-EE_no_expl}
A trivial scenario is the one in which exploitation is turned off, $F=const$, let us discuss it briefly. Due to the influx of random dyadic inversions $M_{ij}\sigma_{ij}\rightarrow-\sigma_{ij}$, any initial graph structure is eventually corrupted and the system slides towards randomness\sidenote{On the meaning of \emph{randomness} for graphs, see also s.n. \ref{sn-ran_graph}}. The rapidity of this process is tuned by $\mu$. Formally,
\marginnote[2cm]{In $(a)$ we have used \eqref{e-EE_c}. In $(b)$ we took advantage of the symmetry of the spin-like representation: $\sum_G\sigma_{ij}P(M_{ij}G,t) = \sum_{G} -\sigma_{ij}P(G,t)$.}
\begin{equation}\label{e-EE_no_exp}
\begin{split}
    \frac{d}{dt}\av{\sigma_{ij}}_t &\stackrel{(a)}{=} \sum_{G} \sigma_{ij}\ \mu\sum_{k<l} [P(M_{kl}G,t)-P(G,t)] \\ 
    & = \mu\Big[ \sum_G\sigma_{ij}P(M_{ij}G,t)- \sum_G\sigma_{ij}P(G,t)\Big]\\
    &\stackrel{(b)}{=}  -2\mu\av{\sigma_{ij}}_t\ .
\end{split}
\end{equation}

The latter is the differential equation of an exponential decay with characteristic time $(2\mu)^{-1}$. The solution is straightforward, $\av{\sigma_{ij}}_t=e^{-2\mu t}\av{\sigma_{ij}}_{t_0}$. Since the dynamics of each $\sigma_{ij}$ are independent, the magnetization also exhibits the same exponential decay behavior,
\begin{equation}
    m_t=e^{-2\mu t}m_{t_0}\ .
\end{equation}
In terms of graph density, this implies that under the action of exploration alone, the average state of the system melts down in an Erd\H{o}s-Rényi random graph with connection probability $p=1/2$. 

\subsection{Energy-like biological function}\label{ss-EE_EL}
By \eqref{e-EE_F_def}, much of the complexity of the $F$ metric arises from that of the graph state space $\mathbb{R}^r$. A simple, non-trivial state space is the one that represents each graph by its number of edges, i.e., $x(G)=\sum_{i<j}a_{ij}\in\mathbb{N}$. Therefore, let us consider the following 
\begin{equation}\label{e-EE_EL}
    F(G)=-\frac{1}{L}\sum_{i<j}a_{ij}\ ,
\end{equation}
where each existing edge implies a fixed penalty\sidenote{The case of a fixed benefit is equivalent, modulo a minus sign in \eqref{e-EE_EL} and those that follow from it.}. The scenario considered is one in which the existence of any possible edge in the graph representation of the system is detrimental. In app. \ref{a-EE_math}, we plug \eqref{e-EE_EL} in \eqref{e-EE_c} to derive the following dynamic for the ensamble average of the $ij$ spin variable:
\begin{equation}\label{e-EL_av_spin}
    \frac{d}{dt}\av{\sigma_{ij}}_t = -2\mu\av{\sigma_{ij}}_t - \frac{\varphi}{2L}\sum_{k<l} \big[\av{\sigma_{ij}\sigma_{kl}}_t-\av{\sigma_{ij}}_t\av{\sigma_{kl}}_t\big]\ .
\end{equation}
The dynamics of the average spin variables are now coupled\sidenote{This may seem odd at first, since $F$ contains only single dyadic variables.\\ 
For instance, a Gibbs-Boltzmann distribution with an Hamiltonian of the form \eqref{e-EE_EL}, factorises in the spin variables and gives decoupled expected values. Accordingly, an MCMC dynamic based on such probability distribution does not introduce correlations between the spin variables. Where do the coupling come from?
While the $F$ metric  \eqref{e-EE_EL} is "energy-like", the dynamic \eqref{e-EE_c} is radically different from an MCMC dynamic in an energy landscape. It is the term $\av{F}$ in \eqref{e-EE_c} that is responsible for the coupling, since it contains an information about all dyadic variables, cf. \eqref{e-EE_EL}, app. \ref{a-EE_math}.
}. The coupling term is the sum of a row of the spin covariance matrix $C_t$ where $(C_{t})_{\sigma_{ij},\sigma_{kl}}= Cov_t(\sigma_{ij},\sigma_{kl}) = \av{\sigma_{ij}\sigma_{kl}}_t-\av{\sigma_{ij}}_t\av{\sigma_{kl}}_t$. 

\begin{figure}
    \centering
    \includegraphics{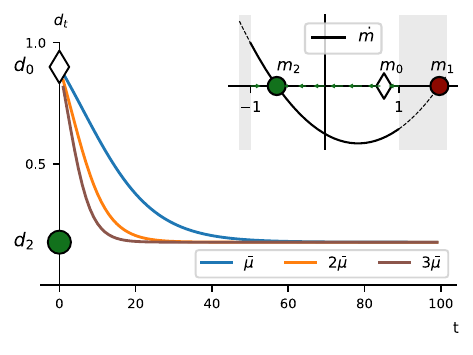}
    \caption{EE dynamic with an energy-like $F$ metric \eqref{e-EE_EL}, under decoupling approximation \eqref{e-EL_DA}. Here, $N=10$ ($L=45$), $\bar{\mu}=1/L, \bar{\rho}=2\times10^2$, $d_0=0.9$. Inset: plot of $\dot{m}=f(m)$, where $f$ is as in \eqref{e-EL_av_spin_DA}. There are two fixed points \eqref{e-EL_magn_fixed_points}, i.e., solutions of $f(m)=0$: $m_1$ (red, unstable) and $m_2$ (green, stable). Shaded area corresponds to the inaccessible regions, $m\in[-1,1]$. The initial condition $m_0$ (white diamond) and the one-dimensional vector field (green arrows) are indicated. Main: dynamic of the average graph density $d_t$, obtained from \eqref{e-EL_magn} and \eqref{e-mag_den}. The value of the asymptotic state $d_2$ solely depends on $L$ and $\bar{\rho}$ (held fixed) while the rapidity of the approach to it is tuned by the exploration rate $\mu$.
    }
    \label{f-EL}
\end{figure}

To proceed, we restrict ourselves to the case where the covariance matrix has an approximately diagonal form, i.e.,
\begin{equation}\label{e-EL_DA}
(C_{t})_{\sigma_{ij},\sigma_{kl}} \sim \mathcal{O}(\epsilon) \quad \text{for} \ \sigma_{ij}\ne\sigma_{kl}\ .
\end{equation}
Discarding all terms $\mathcal{O}(\epsilon)$ in \eqref{e-EL_av_spin}, we get:
\begin{equation}\label{e-EL_av_spin_DA}
    \frac{d}{dt}\av{\sigma_{ij}}_t \sim -2\mu\av{\sigma_{ij}}_t - \frac{\varphi}{2L}\big[1-\av{\sigma_{ij}}^2_t\big]\ .
\end{equation}
The latter is valid for $L\epsilon\ll1$, which means either small graph sizes (small $L$) or mild functional pressures $\rho$, for which the dynamics are close to those in sec. \ref{ss-EE_no_expl} (small $\epsilon$). By \eqref{e-EL_av_spin_DA}, the dynamics of the dyads are decoupled and this is the reason why we refer to \eqref{e-EL_DA} as the \emph{decoupling approximation}\sidenote{We stress that the independence of the dyad dynamics results here from restricting to a specific regime. In the absence of information about the structure of the covariance matrix, the decoupling approximation can be verified \emph{a posteriori}, as we will do in sec. \ref{ss-EE_sims_DL}.}. If we use the same initial conditions $\av{\sigma_{ij}}_{t_0}=\sigma_0\ \forall i,j$, the same differential equation \eqref{e-EL_av_spin_DA} holds for the magnetisation, since $m_t = \av{\sigma_{ij}}$. It can be explicitly integrated by partial fractions, the solution being: 
\begin{equation}\label{e-EL_magn}
    m_t = m_2 \Bigg[ 1+ \frac{m_1/m_2-1}{1+\frac{m_1-m_0}{m_0-m_2}\ e^{2\mu t\sqrt{1+(2L/\rho)^{-2}}}} \Bigg]\ ,
\end{equation}
where $m_0=m_{t_0}$ and 
\begin{equation}\label{e-EL_magn_fixed_points}
    m_{1}=2L/\rho +\sqrt{1+(2L/\rho)^{2}}\ ,\quad m_{2}=2L/\rho-\sqrt{1+(2L/\rho)^{2}}
\end{equation}
are the fixed points of the dynamic, fig. \ref{f-EL}. The one described by \eqref{e-EL_magn} is a relaxation dynamic to the stable fixed point $m_2$\sidenote[][-2cm]{\eqref{e-EL_magn} describes exhaustively the dynamics in the graph state space, this is what we call a \emph{solution} of the EE dynamic. The time course of any other graph feature, by construction, must be a function of \eqref{e-EL_magn}.}. Regarded as a function of $\rho$, we find consistently $m_2\rightarrow 0$ for mild functional pressures $(\rho\rightarrow0)$, which corresponds to the case of sec. \ref{ss-EE_no_expl}. In the opposite limit of perfect exploitation ($\rho\rightarrow\infty$), we have $m_2 \rightarrow-1$, corresponding to empty graphs\sidenote[][-1.5cm]{A straightforward generalisation of this discussion is the one for the case of a $F$ metric written as a general edge covariate \eqref{e-erg_ec}, i.e.,
\begin{equation}
    F(G)=\frac{1}{L}\sum_{i<j}\gamma_{ij}a_{ij}\ .
\end{equation}
Following the same steps, we get an expression which is analogous to \eqref{e-EL_av_spin_DA} but with the substitution $\varphi\rightarrow\varphi\gamma_{ij}$. In this case, however, even under decoupling approximation and the same initial conditions, the dyadic dynamics are different, since each is subject to a dyad-specific exploitation rate $\gamma_{ij}\varphi$. It is not possible to trade the dynamic equation of any $\av{\sigma_{ij}}$ for that of $m_t$.
}. 

As for the general characteristics of the dynamic \eqref{e-EL_magn}, we find that the value of the stable fixed point $m_2$ depends only on the ratio between the number of degrees of freedom (dyads) $L$ and the functional pressure $\rho$; the rapidity of the approach to the stable fixed point depends on the exploration rate $\mu$, with higher values yielding faster approaches; for any $\rho<\infty$ the asymptotic value is not attainable, as $m_2$ balances the strengths of the exploration and exploitation drivers. 

\subsection{Distance-like biological function}\label{ss-EE_DL}
The problem of determining the global maximum of an $F$ metric can be challenging\sidenote[][-1cm]{Worse, exponentially challenging. This is one of the many manifestations of the same, well-studied problem in the physics of disordered systems, namely the (NP-hard) problem of finding the ground state in a spin-glass landscape. \cite{mezard1987}.}. As theorists, however, we have the right to flip the script, with pen and paper. We can choose an optimal state and then shape our $F$ metric to suit it. A simple way to do this is to formulate $F$ as a distance function in the graph state space. If $\bm{x}:\mathcal{G}\mapsto\mathbb{R}^r$ defines the state space and $\bm{x}^*\in\mathbb{R}^r$ is (by definition) the optimal state, we can use, e.g., a squared distance\sidenote{Simple alternatives are the absolute value or the distance ($l^2$-norm). The choice of a squared distance, however, simplifies the math below.}
\begin{equation}
    F(G)\propto -[\bm{x}(G)-\bm{x}^*]^2 .
\end{equation}
Regardless of the complexity of the state space, the above $F$ metric has a simple structure: there is a single, global maximum corresponding to those graphs with the same statistics as $\bm{x}^*$. However, as in the previous section, we consider the simple, one-dimensional state space of the number of edges $x(G)=\sum_{i<j}a_{ij}\in\mathbb{N}$, which is amenable to analytical investigation. In particular, we consider 
\begin{equation}\label{e-EE_DL}
    F(G) = -\frac{1}{L^2}\Big( \sum_{i<j}a_{ij}-E^*\Big)^2,
\end{equation}
where $0<E^*<L$ is by construction the optimal number of edges in a graph. In app. \ref{a-EE_math}, we follow similar steps as in sec. \ref{ss-EE_EL} to derive an approximate dynamic equation for the magnetisation, under decoupling approximation\sidenote{
In particular, we generalise here the definition of decoupling approximation to the regime where 
\begin{equation}\label{e-DL_DA}
    \av{\sigma^{(1)}\dots\sigma^{(k)}}\sim\prod_{i=1}^k\av{\sigma^{(i)}} \ ,
\end{equation}
where the left-hand side contains no repeated spin variables. The approximate dynamic \eqref{e-DL_magn_DA} is expected to hold for $L\epsilon\ll1$ where $\epsilon$ is the order of magnitude of spin correlations. The validity of the hypothesis will be assessed \emph{a posteriori}, see sec. \ref{ss-EE_sims_DL}.
}. As a result, we get 
\begin{equation}\label{e-DL_magn_DA}
    \dot{m}_t = -2\mu m_t-\frac{\varphi}{L^2}(1-m_t^2)\Bigg[\frac{L-1}{2}m_t+\frac{L}{2}-E^*\Bigg] \ .
\end{equation}

\begin{figure}
    \centering
    \includegraphics{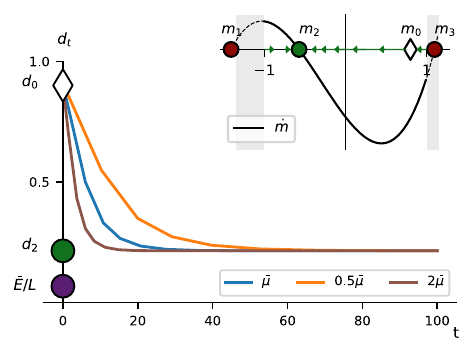}
    \caption{EE dynamic with an distance-like $F$ metric \eqref{e-EE_DL}, under decoupling approximation \eqref{e-DL_DA}. Here, $N=10$ ($L=45$), $\bar{\mu}=1/L, \bar{\rho}=5\times10^2$, $d_0=0.9$, $E^*=3$. Inset: plot of $\dot{m}=f(m)$, where $f$ is as in \eqref{e-EL_av_spin_DA}. There are three fixed points, i.e., solutions of $f(m)=0$: $m_1, m_3$ (red, unstable) and $m_2$ (green, stable). Shaded area corresponds to the inaccessible regions, $m\in[-1,1]$. The initial condition $m_0$ (white diamond) and the one-dimensional vector field (green arrows) are indicated. Main: dynamic of the average graph density $d_t$, obtained from $m_t$ using \eqref{e-mag_den}. The value of the asymptotic state $d_2$ solely depends on $L$ and $\bar{\rho}$ (held fixed) while the rapidity of the approach to it is tuned by the exploration rate $\mu$.
    }
    \label{f-DL}
\end{figure}

The latter can be integrated numerically, fixing the boundary conditions $m_{t_0}=\sigma_0$. Qualitatively, we find the same behaviour as in sec. \ref{ss-EE_EL}, fig. \ref{f-DL}. It is worth highlighting it:

\begin{remark}\label{r-EE_dyn}
\emph{\textbf{(Characteristics of the EE dynamics)}} The EE dynamics \eqref{e-EE_c} is such that (i) the asymptotic state ($t\rightarrow\infty$) depends on the functional pressure $\rho$ and approaches $\max F$ in the limit of perfect exploitation\sidenote{Note that, (a) for any nonzero exploration rate $\mu>0$, the $\max F$ cannot be reached exactly and (b) for $\mu=0$ there would be no dynamics at all. The limit of perfect exploitation should be intended as a the limit of $\varphi\rightarrow\infty$ for $\mu\ne0$, cf. \eqref{e-EE_c}.} $\rho\rightarrow\infty$; (ii) for fixed $\rho$, higher exploration rates $\mu$ correspond to faster approaches to the asymptotic state.
\end{remark}

While the above findings have the robustness of an analytical result, they were obtained under the rather limiting assumption of the decoupling approximation. Therefore, our conclusions need to be assessed numerically, which will be done in sec. \ref{ss-EE_sims_DL}.

\section{Population-based simulations}[Simulations]\label{s-EE_sims}
\emph{Where a simulation scheme for the EE dynamics -- inspired, once again, by evolution -- is deployed and its main features presented. It is to be used there where the pen cannot get.}\bigskip 

The computational problem of setting up simulations for equations of the form \eqref{e-EE} has recently been addressed, e.g., by \cite{zanini2012,mauri2021}, in the context of evolutionary models. We develop a similar computational framework, coded in \texttt{Python 3.9.7} and freely available on the GitHub folder \href{https://github.com/dichio/EE-graph-dyn}{\texttt{EE-graph-dyn}}\sidenote{Here, we discuss the general features of the simulations. A detailed description of the design of the code can be found as documentation in the GitHub folder.}. 

Once again, the core idea of the simulations is to mimic the evolutionary process by simultaneously tracking the dynamics of an entire population of individuals. Each individual is associated to a graph, i.e., to a binary strings with $L$ bit-like entries $a_{ij}=0/1$ (dyads). Our population-based simulations keep track of all the individuals \emph{existing} within the population, at each time $t$\sidenote[][-1.2cm]{In general, one has two possible strategies for simulating forward population dynamics: (i) tracking the number of individuals (or frequency) associated with each possible graph, or (ii) tracking the graphs associated with the individuals present in the population. We opt for the latter, since the former requires listing and tracking all $2^L$ possible graph configurations, which quickly becomes infeasible as $L$ increases.
}. To speed up the simulations, we group similar individuals into a \emph{clone}, that is a pair $(G,n)$, where $n$ is the number of individuals associated with the same graph $G$. At time $t$, the population is thus defined as the set of existing clones $\mathscr{P}(t) = (\bm{G}(t),\bm{n}(t))$. The population size (total number of individuals) $\sum_{\alpha}n_{\alpha}(t)=M$ is held fixed while the total number of clones $M_c(t)\le M$ fluctuates. 

\begin{figure*}[]
	\includegraphics[width=14cm]{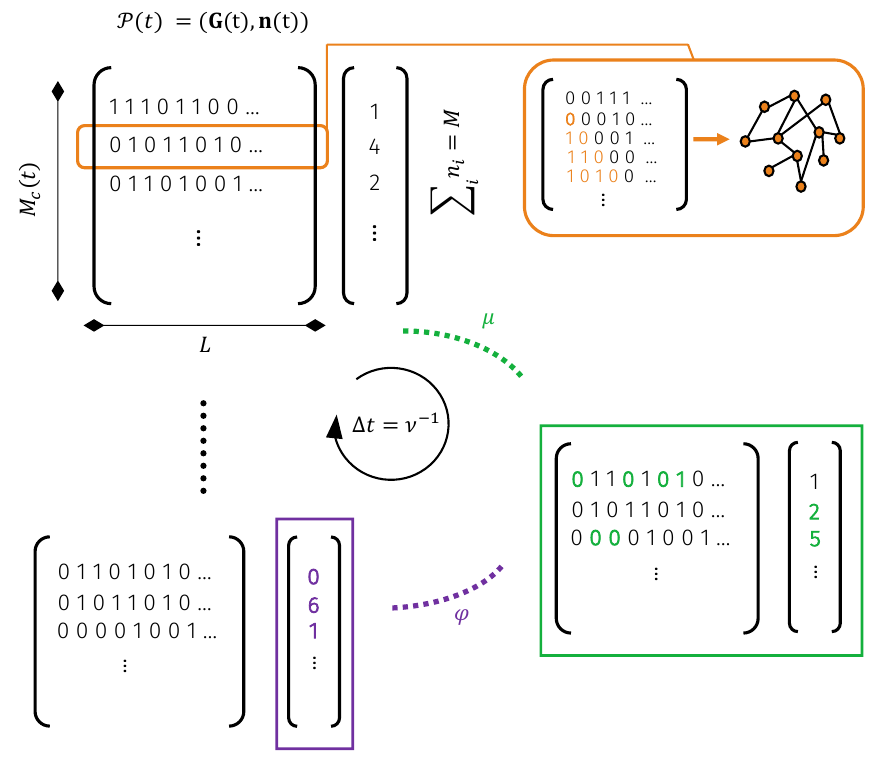}
	\caption{Population-based simulations for eq.(\ref{e-EE}), simulation step $\Delta t$. A population $\mathscr{P}(t)$ at time $t$ is made up of a matrix $\bm{G}(t)$ with dimensions $M_c(t)\times L$ along with a vector $\bm{n}(t)$ of length $M_{c}(t)$ (top left). Each row of $\bm{G}(t)$ corresponds to a unique graph configuration, specifically, to the vectorised lower triangular matrix of an adjacency matrix (top right). First, the configuration space is explored by introducing random dyadic mutations (edge toggles) at the rate $\mu$ in each individual of the population (green, bottom right). As a consequence, both the matrix $\bm{G}$ and the counts $\bm{n}$ change. As new configurations are found, the number of clones $M_c$ increases. Second, the more functional graphs proliferate within the population. The exploitation alters only the count vector $\bm{n}$ and is regulated by the parameter $\varphi$. Those clones that end up with zero-size are removed from $\bm{G}$ before the next time-step. The total number of individuals $M$ and time interval $\Delta t$ are free internal parameters of the simulations.}
    \label{f-sim_step}
\end{figure*}

At each simulation step, the population is updated, new clones are created by dyadic mutations (exploration), their size updated by functional selection (exploitation), fig. \ref{f-sim_step}. In particular:
\begin{itemize}
    \item \emph{Exploration}. Each dyad of each individual in the population mutates\sidenote{In the case of an edge toggle, one has $\sigma_{ij}\rightarrow-\sigma_{ij}$. Later in ch. \ref{c-Celegans}, we will use growth-only dyadic mutations, i.e., $\sigma_{ij}\rightarrow|\sigma_{ij}|$.} with probability $1-e^{-\Delta t\mu}\sim\Delta t\mu$. The exploration rate $\mu$ is uniform across dyads.
    \item \emph{Exploitation}. The $F$ values of the graphs associated to each clone are computed. The clone sizes are then updated by extracting $M$ independent samples from a multinomial distribution where each graph $G_{\alpha}$ is selected with probability
    \begin{equation}
        p_{\alpha} =  n_{\alpha}e^{\Delta t \varphi F(G_{\alpha})}/\sum_{\beta}n_{\beta}e^{\Delta t \varphi F(G_{\beta})}\ ,\quad \alpha\in 1,\dots,M_c(t) \ .
    \end{equation}
\end{itemize}

Our simulations have six parameters, summarised in tab. \ref{t-EE-params}. The structural parameters $N,T$ set the geometry of the simulations. The former is the (fixed) number of nodes of each graph; the latter is the size of the time window to be simulated. There are two internal degrees of freedom: the population size $M$ and the time step $\Delta t$ -- for technical convenience, it is often preferable to set the inverse time step $\nu=\Delta t^{-1}$. Finally, two parameters control the dynamics of the system, the exploration rate $\mu$ and the relative strength of exploitation $\rho$\sidenote[][-1cm]{In practice, the simulation step can always be defined as $\Delta t=1$ by rescaling accordingly:
\begin{equation*}
    T \rightarrow \nu T\qquad  \mu \rightarrow \mu/\nu \ .
\end{equation*}}.

{\LinesNumberedHidden
\begin{algorithm}[H]
    \DontPrintSemicolon
    \TitleOfAlgo{EE graph dynamics, forward simulations (pseudocode).}
    $\mathscr{P}(0) = (\bm{G}_0,\bm{n}_0)$\;
    $t=0$\;
    \While{t<T}{
        \emph{Exploration:} $\sigma_{ij}\rightarrow-\sigma_{ij}$ with probability $\Delta t\mu\ $ $\forall (i,j)$, $\forall G_{\alpha}$\;
        update $\mathscr{P}^* = (\bm{G}^*,\bm{n}^*)$\;
        compute $F(G_{\alpha}^*) \ \forall \ G_{\alpha}^*$\;
        \emph{Exploitation:} $M$ draws from a multinomial distribution with $p_{\alpha} = n_{\alpha}^*e^{\Delta t \varphi F(G_{\alpha}^*)}/\sum_{\beta} n_{\beta}^*e^{\Delta t \varphi F(G_{\beta}^*)}$ $\Rightarrow$ compute new counts $\bm{n}^{**}$\;
        set $\mathcal{P}(t)= (\bm{G}^*,\bm{n}^{**})$\;
        $t \mathrel{+}= \Delta t$
    } 
\end{algorithm}
}

\begin{table}[htbp]
  \centering
  \caption{Parameters of simulations for EE graph dynamics. Our computational framework has six degrees of freedom, which we group by color: structural parameters (top), internal degrees of freedom (middle) and parameters of the dynamics (bottom).}
  \begin{tabularx}{.7\textwidth}{>{\centering\arraybackslash\hsize=.5\hsize}X>{\raggedright\arraybackslash\hsize=1.5\hsize}X}
    \textbf{Parameter} & \textbf{Description} \\
    \toprule
    \rowcolor[HTML]{DFD7BF}
    $N$  &  number of nodes  \\
    \rowcolor[HTML]{DFD7BF}
    $T$  &  time window span  \\
    \addlinespace[3pt]
    \rowcolor[HTML]{F2EAD3}
    $M$  &  population size \\
    \rowcolor[HTML]{F2EAD3}
    $\nu$  &  inverse time step $\Delta t^{-1}$ \\
    \addlinespace[3pt]
    \rowcolor[HTML]{F5F5F5}
    $\mu$  & exploration rate \\
    \rowcolor[HTML]{F5F5F5}
    $\rho$  & functional pressure $\varphi/\mu$ \\
    \bottomrule
  \end{tabularx}
  \label{t-EE-params}
\end{table}

\begin{figure}
    \centering
    \includegraphics{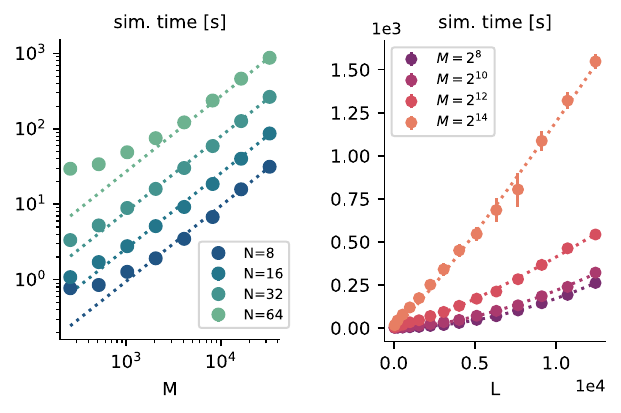}
    \caption{Simulation (running) time as a function of the population size $M$ and number of dyads $L$. Dots are average running times obtained from $10$ simulation runs, error bar are indicated when visible. Here $T=100, \nu=1, \mu=1.0\times10^{-4},\rho=5$, an $F$ metric \eqref{e-EE_EL}. (Left) The sim. time scales linearly with $M$ (log-log scale). Dotted lines are the a linear fit with the curve $y=aM$, where $a$ is a parameter. (Right) The sim. time has a mild exponential dependence on $L$. Dotted lines are a linear fit with the curve $y=aL^{b}$, where $a,b$ are parameters. We find $b=1.95, 1.71, 1.29, 1.12$ for increasing $M$. In the large $M$ limit, we expect $b\rightarrow1$, indicating an approximately linear dependence on $L$. The dependence on $L$ however, strongly depends on the complexity of the $F$ metric.
    }
    \label{f-sim_time}
\end{figure}

The running time of a single simulation has an obvious linear scaling with the inverse time step $\nu$, since the same operations described above are repeated a number $\nu T$ of times. A linear scaling is also observed with the population size $M$, which is reasonable since both the evaluation of dyadic mutations and of the $F$ metric have to be performed independently for each clone -- in the worst case, $M_c(t)\sim M$, fig. \ref{f-sim_time} (left). The dependence on the number of nodes $N$ (or equivalently, on $L$) is trickier, since it depends strongly on the complexity of the graph operations involved in the evaluation of $F$. For a $F$ metric as simple as \eqref{e-EE_EL}, and for large $M$, we observe an approximately linear scaling in $L$, fig. \ref{f-sim_time} (right).

\subsection{Gleaning dynamics from simulations}\label{ss-EE_sims_DL}
At each time $t$, the raw information provided by the simulations comes in the form of a snapshot of the population $\mathscr{P}(t) = (\bm{G}(t),\bm{n}(t))$. Assuming that it is a representative sample of the entire probability distribution, we can compute from it the distribution of any graph observable $O:\mathcal{G}\mapsto\mathbb{R}$ by adapting \eqref{e-NS_dyn_obs} to
\begin{equation}\label{e-emp_distr}
    P(O,t) = \frac{1}{M}\sum_{\alpha=1}^{M_c(t)} n_{\alpha}(t)\ \delta[O-O(G_{\alpha}(t)]\ ,
\end{equation}
where $\delta$ is the Dirac-delta ($\int dO \ \delta(O) = 1$). By consequence, the expected value of $O$ at time $t$ is 
\begin{equation}\label{e-emp_av}
    \av{O}_t\sim\frac{1}{M}\sum_{\alpha=1}^{M_c(t)} n_{\alpha}(t) \ O(G_{\alpha}(t))\ .
\end{equation}

\begin{figure*}[h!]
	\includegraphics[width=14cm]{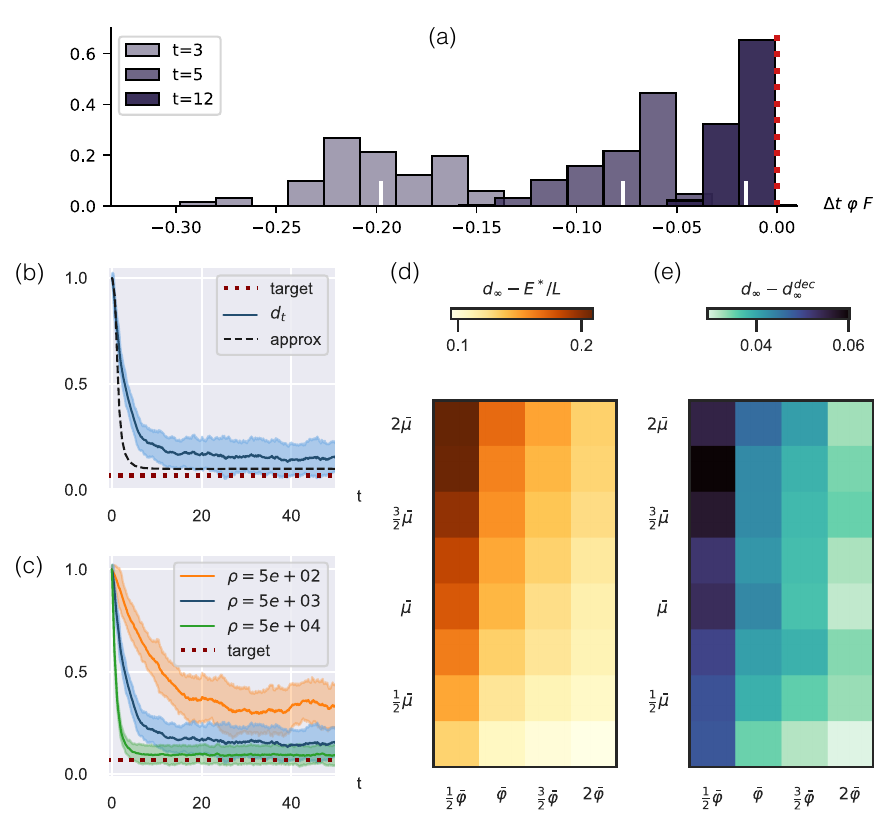}
	\caption{Exploration-exploitation (EE) dynamic \eqref{e-EE} with distance-like $F$ metric \eqref{e-EE_DL}. Parameters: $N=10$, $\bar{\mu}=1/L$, $\bar{\rho}=5\times10^{3}$, $T=50$, $\Delta t = 1/\nu = 10^{-2}$, $E^*=3$, $\Delta t =1$, $M=4096$. (a) Distribution of $F$ values at three different time points $t$, the corresponding $\av{F}_t$ are indicated (white bars), cf. fig. \reffigshort{f-F_dyn}. Exploitation amplifies the probability of those graphs exhibiting higher $F$ values than the ensemble average. Consequently, the distribution shifts towards higher $F$ values, until the max is reached (red dotted line). (b) The average graph density $d_t = \sum_{i<j}\av{a_{ij}}/L$ exhibits a relaxation dynamic towards the steady-state value $d_{\infty}$, in proximity to the target density $E^*/L$. Shaded area indicates the $95\%$ confidence interval, black dashed line represents the analytical solution from \eqref{e-EE_DL}. (c) The value $d_{\infty}$ depends on the functional pressure $\rho$, the target is approached for $\rho\rightarrow\infty$. The rapidity of the approach to the steady state value depends on $\rho$ and, for fixed $\rho$, on the exploration rate $\mu$ (not shown). (d) The distance from the target diminishes for increasing functional pressure $\rho$. Here, we illustrate this both as a function of the exploration rate $\mu$ and the exploitation rate $\varphi = \rho\mu$. (e) The value $d_{\infty}^{dec}$ is invariably nearer to the target than $d_{\infty}$, with the difference between the two tapering off for increasing $\varphi$ -- where both converge towards $E^*/L$.
    }
    \label{f-MRG}
\end{figure*}

To conclude, let us showcase an example of EE graph dynamics simulations. Let us reconsider the case discussed in sec. \ref{ss-EE_DL}, i.e., an EE dynamics with distance-like $F$ metric \eqref{e-EE_DL}, fig. \ref{f-MRG}. 

We use \eqref{e-emp_distr} to evaluate the instantaneous distribution of the $F$ values, at three illustrative time points, fig. \ref{f-MRG}(a). As a result of the exploitation, these distributions are skewed towards the high-$F$ direction, the more so the higher the functional pressure. The dynamics in the one-dimensional $F$ corroborate the pattern we have depicted in fig. \reffigshort{f-F_dyn}. We further use \eqref{e-emp_av} to evaluate the time dynamics of the average graph density $d_t$, as defined in \eqref{e-mag_den}. We find the same qualitative behaviour as for the analytical solution discussed in sec. \ref{ss-EE_DL}. Specifically, after an initial transient phase, a steady-state value is attained, striking a balance between the strengths of exploration and exploitation, fig. \ref{f-MRG}(b). This steady-state value is moderated by the functional pressure $\rho$ -- the higher the pressure, the closer the density approaches the target value, fig. \ref{f-MRG}(c-d). 

Finally, we perform a comparison between our simulations and the numerical solution we obtained from \eqref{e-DL_magn_DA}, under decoupling approximation. We present evidence that the asymptotic state of the latter consistently lies intermediate to the target density and the simulated dynamics, the difference between the three vanishing for increasing functional pressure, fig. \ref{f-MRG}(e). We conclude that the decoupling approximation universally exhibits qualitative agreement with simulations across the parameter space and quantitative agreement for large values of $\rho$.

This result, in turn, allows us to regard at the characteristics of the EE dynamics derived under rather specific conditions (Remark \ref{r-EE_dyn}) as universal attributes of the EE dynamics. The next natural step is to direct our attention towards more realistic systems with more complex functional landscapes.
\setchapterpreamble[u]{\margintoc}
\chapter{Weaving the mind of a worm}\label{c-Celegans}
\begin{quote}
\begin{flushright}
With four parameters I can fit an elephant,\\ and with five I can make him wiggle his trunk.

--- John von Neumann (attributed)
\end{flushright}
\end{quote}
\bigskip

The exploration-exploitation paradigm is general and theoretically sound for biological dynamics. Therefore it should apply elsewhere than in the evolutionary context. The aim of this chapter is to show that it does apply elsewhere. Between the \emph{should} and the \emph{does} there are a number of theoretical and methodological details to be worked out, approximations to be made along the way and, most importantly, plenty of biology to be learned. 

The beginning of wisdom -- as they say -- is the definition of terms\sidenote{This is often attributed to Socrates, but -- if I understand correctly -- there is no such quotation in the writings of Plato, Socrates' press office. To err on the side of caution, and to avoid offending any Greek philosophers who might pass through here, we refer to a generic "they".}. The process we analyse in this chapter is the development of a natural nervous system, from birth to adulthood -- or, as it has been called, the \emph{brain wiring dynamics}. To do so, we shall focus on a specific organism, for which a natural (and almost obliged) choice is a tiny, transparent worm, the nematode \emph{\Gls{caenorhabditis elegans}}, or simply \ce. 

We found it convenient to organise this chapter as a long-form scientific paper\sidenote{In \emph{Physical Review Letters} -- a flagship publication for physicists -- typical articles span just four or five pages. This concise format assumes that readers have a considerable amount of knowledge, as details are often distilled for brevity. Here, I do not. Our journey to the results might be a longer read, but it is supposed to be pedagogical. The main reference of this chapter \cite{dichio2023b}, on the other hand, is written in a short format, for the already expert, for the impatient, or simply for the lazy.}. We begin with a general discussion of the problem (sec. \ref{s-brain_wiring}) and a description of the essential features of the organisation of the worm nervous system, including the data we use (sec. \ref{s-mind_worm}). We then turn to describe in detail our EE model of the \ce\ brain maturation, including our main results (sec. \ref{s-EE_ce}). We provide a plausible interpretation of how the exploration-exploitation dynamics could be implemented at a fine-scale (sec. \ref{s-ce_interpr}) and conclude with an overview of the many possible generalisations of the model presented (sec. \ref{s-perspectives}).

\begin{kaopaper}[title=Main reference]
\faFile*[regular] \textbf{Vito Dichio} \& Fabrizio De Vico Fallani. \emph{The exploration-exploitation paradigm for networked biological systems}. In: arXiv e-prints 2306.17300 (2023) \cite{dichio2023b}.
\end{kaopaper}

\bigskip
\section{The brain wiring problem}\label{s-brain_wiring}
\emph{Where the (genetically encoded) growth of a brain -- furiously debated among neuroscientist -- is recognised as a specific instance of the EE dynamics. Where also the fundamental facts about the brain are: developmental variability, functional robustness. }\bigskip 

\begin{figure*}[htbp]
    \centering
    \includegraphics[width=16cm]{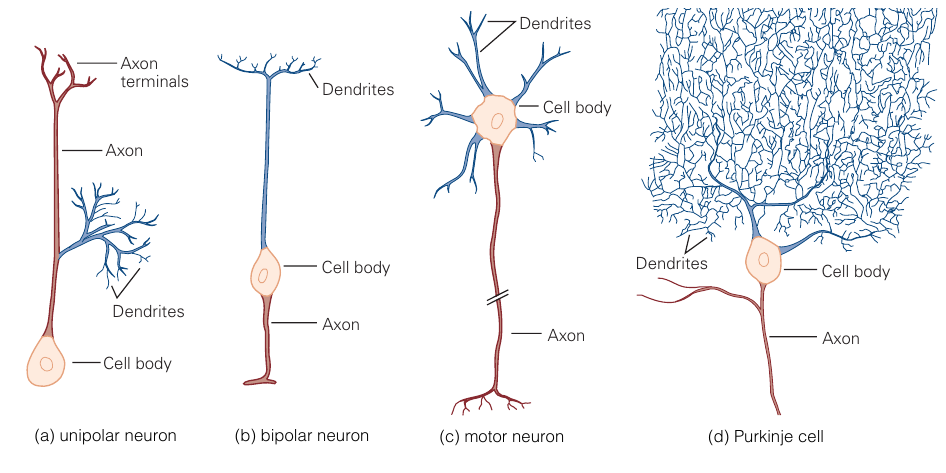}
    \caption{Four common type of neuron: (a) unipolar, (b) bipolar, (c-d) multipolar, in particular: motor neuron (c) and Purkinje cell (d). The \emph{cell body} (soma) is the neuron's core and houses the nucleus. This is where the neuron's fundamental metabolic activities occur. Extending from the cell body are \emph{dendrites}, which are tree-like projections that receive signals from other neurons and transmit them to the cell body. The \emph{axon}, on the contrary, is a long slender projection, that serves to transmit electrical impulses away from the cell body. This allows the neuron to communicate with other neurons, muscles or glands. The contact zone between dendrites and axons where two cells exchange chemical and electrical signals are called \emph{synapses}. Together, these components form the communication network within the nervous system. The examples here proposed showcase the great variety of neuronal geometries found in Nature. Adapted from Kandel et al., \emph{Principles of neural science} \cite{kandel2000}.
    }
    \label{f-neuro}
\end{figure*}

The first occurrence of the wording appears in a recent perspective article by \emph{Hassan and Hiensinger} \cite{hassan2015}. The first lines read:
\begin{quote}
    The brain, as we neuroscientists like to say, is \emph{really} complex. A good deal of our efforts are therefore dedicated to figuring out just how this apparent complexity is generated: where does the information to build a brain come from, and how is such information turned into synapse-specific wiring? We call this the ''brain wiring problem''.
\end{quote}
In the rest of this manuscript, we will work alongside our fellow neuroscientists and confront the same compelling challenge. The brain wiring\sidenote{\emph{Wires, circuits, junctions} (...) The language of neuroscience is steeped in the vocabulary of telecommunications. This is a long-standing metaphor that goes back to the days of the telegraph \cite{cobb2021}. As early as 1875, the German physicist Hermann von Helmoltz wrote: \emph{"Nerves have often and not unsuitably been compared to telegraph wires. Such a wire conducts one kind of electric current and no other; it may be stronger, it may be weaker, it may move in either direction; it has no other qualitative differences. Nevertheless (...) we can send telegraphic dispatches, ring bells, explode mines, decompose water, move magnets, magnetise iron, develop light, and so on. So with the nerves."} \cite{helmholtz2009}} dynamic is a developmental dynamic that unfolds during a lifespan and involves the formation, growth and establishment of an individual's \gls{nervous system} -- primarily, \glspl{neuron} and \glspl{synapse}, fig. \ref{f-neuro}.

The crucial empirical observation is that, although the functional outcomes are highly reproducible and almost invariable, the nervous system is not \emph{hardwired}. From worms to humans, neuroanatomical differences are observed between the nervous systems of any two individuals, even when they are genetically identical and even when environmental factors are controlled \cite{clarke2012}. For instance, the branching patterns of neuronal connections vary in lower isogenic animals such as worms, grasshoppers and locusts \cite{ward1975,goodman1978,steeves1983} but also in mammals, including monozygotic human twins \cite{schmitt2007}. This form of stochasticity, which (i) is not due to genetic differences, (ii) is not induced by the environment, and (iii) nevertheless leads to equally functional outcomes, has been referred to as \emph{genetically encoded} \cite{hassan2015} or \emph{intrinsic chance} \cite{finch2000}. Much of the solution to the brain wiring puzzle lies in answering the question: what is the origin of genetically encoded stochasticity?

One possibility is that the observed variability is noise of a molecular code. In a nutshell, genes encode molecules, molecular mechanisms drive the growth of the nervous system\sidenote{The main way this happens is through so-called \emph{guidance cues}, biochemical signals (molecules) that guide the growing axons of neurons.}, neuron by neuron, synapse by synapse. An appropriate spatiotemporal regulation of the latter, in turn, results in the synapse-specific wiring of the brain \cite{kolodkin2011,yogev2014,sanes2020}. This specification process is noisy and occasionally results in inaccurate outcomes. Whenever such misspecifications do not impair the system's functionality, they reveal themselves as variability in the observed systems.
A computer scientist faced with such a brain wiring algorithm would probably be appalled. Indeed, programming for each input/output is a highly inefficient coding strategy. Moreover, it seems implausible, at the very least, that a system as complex as the human brain ($\sim10^{15}$ synapses) can be exhaustively specified by a single genome down to the finest spatiotemporal scale\sidenote{This has been called the blueprint problem: a deterministic molecular code, accurate at every spatiotemporal scale, would be at least as complicated as the resulting wiring diagram \cite{hassan2015}.}. 

A contrasting view has recently emerged: it is not the precise result, but the wiring algorithm that is genetically encoded \cite{hassan2015,hiesinger2018,hiesinger2021}. Accordingly, neural circuits grow based on simple, genetically encoded, pattern formation rules\sidenote{These include, among others, spacing between axons, self-avoidance, lateral inhibition. For two recent examples in the \emph{Drosophila} brain, see \cite{langen2013,langen2015}.}. The variability of the outcomes is not due to misspecified molecular instructions, but rather is an intrinsic and essential feature of the dynamics. This because the brain wiring is a stochastic process that generates patterns, and patterns can be realised in a variety of different ways.
Our computer scientist would be relieved: encoding a finite set of (possibly simple) rules is certainly a less daunting programming task than fine-coding a nervous system. Indeed, from an algorithmic point of view, a stochastic process based on a few algorithmic constraints and otherwise random appears to be an efficient, flexible -- and maybe ideal -- way to explore an unknown environment.

It further follows from this view that the \emph{functionality} is an attribute of the algorithm, rather than of the outcome: a functional rule-set is the one that leads to functional configurations of the nervous system. If the set of such configurations is large enough, the whole brain wiring process turns out to be robust, since small configuration changes do not affect the system's functionality \cite{hiesinger2018}. In this sense, allowing for variability of the outcomes is an insurance policy against failure in the case of perturbations\sidenote{The amount of variability is regulated by the algorithm itself: strict and/or complex functional requirements will yield a narrow distribution of outcomes, while simple wiring rules will allow for a broad outcome variability. This \emph{degree of variability} is likely to be subject to evolutionary pressure and optimised by natural selection \cite{hiesinger2018}.}. 

Down to the neuronal scale, an experimental evidence consistent with this view is the fact that the synapse formation process is largely non-specific. The growth of each branch of a dendritic tree happens thorugh a series of stochastic local decisions in an unknown molecular environment\sidenote{One might think at a branch of a dendritic tree as a navigator in a maze, who ignores both the maze map and the position of the other navigators. It only has algorithmic information of the sort: "\emph{at each crossroad, choose the wider path}" or "\emph{if you see a lemon tree, turn around}" or \emph{"if possible, avoid passing by the owl's nest"} and so on. At each new point in the maze, the navigator makes decisions based on its rules and trying to accommodate constraints in the best possible way. 
}. This allows the process to cope with unforeseen environmental conditions. For example, neurons that innervate incorrect target regions will form synapses wherever they land, regardless of how inappropriate the targets may be \cite{kandel2000}. In the absence of other potential partners, they can even form perfectly functional synapses with themselves, known as autapses \cite{bekkers1991}.  

At this point, we cannot resist the temptation to draw a parallel with our exploration-exploitation framework. It is straightforward to rewrite the above paragraphs in the language of sec. \ref{s-EE_fun}: the brain wiring process is a self-referential (state-dependent) biological process, that unfolds as a random exploration of the configuration space under the action of a set of functional drivers and constraints. The dynamics \eqref{e-EE} endowed with a choice for the $F$ metric is precisely the way in which we specify a brain wiring algorithm, genetically encoded by assumption. The observed variability of the nervous system is nothing but the statistical uncertainty associated with the \emph{ensemble} probability distribution.

Below, we will follow this line of thought and tackle the problem of brain wiring with our EE dynamics. To work out the details, we need to focus on a specific context. The time has come to talk about worms.

\section{The mind of a worm}\label{s-mind_worm}
\emph{Where a biological minimum is provided of the neuroanatomy of a tiny little worm. Where also the unique and recently published experimental data are presented, colorful brain networks are flashed. }\bigskip 

\begin{marginfigure}[-4cm]
	\includegraphics[height=4cm]{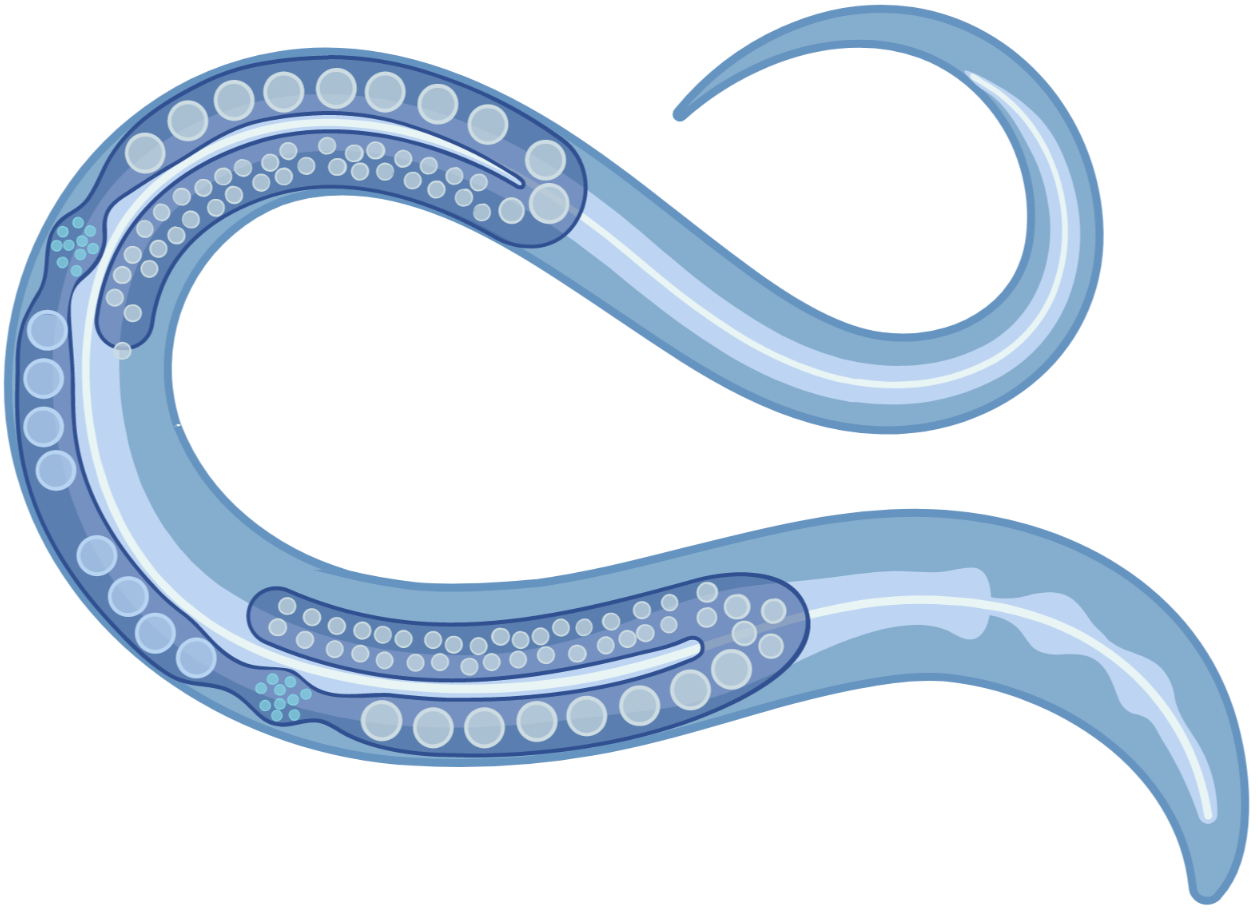}
\end{marginfigure}
It took thirteen years, the first time. But finally, in the 1986, Sydney Brenner and his collaborators published \emph{The structure of the nervous system of the nematode Caenorhabditis elegans}, the first complete anatomical reconstruction of a natural nervous system, a major milestone in the history of neuroscience \cite{white1986}. The report (over 300 pages) sported a rather solemn running title: \emph{the mind of a worm}\sidenote[][-2cm]{The running title comes from a bit of a lab inside joke, we all have some. The anecdote has been recently re-evoked: 
\emph{"We sent the text and figures to the publishers in a number of ring binders (no PDFs in those days). We had labelled these notebooks “the mind of a worm” in order to identify them on a shelf among all the other similar-looking notebooks. We were amused the printers had picked this up and used it as a running title, so we let the name stick."} \cite{white2018}
}.

The \emph{Caenorhabditis elegans}, or \ce, is a small ($\sim \SI{1}{\milli\meter}$ long), transparent nematode, or roundworm, that has been widely used as a model organism in biological research, from genetics to behaviour \cite{hillier2005,debono2005,kaletta2006,corsi2015,haag2018,ahamed2021}. It is especially valuable due to its relative simplicity. This organism has held a unique position in the field of neuroscience since, until very recently\sidenote{At the time of the writing of this manuscript -- summer 2023 -- a preprint has appeared on bioRxiv with the reconstruction of a new, complete nervous system -- that of the \emph{Drosophila melanogaster}, $\sim500\times$ larger than the one of the \ce\ \cite{dorkenwald2023}. Similar neural maps at the synapse level exist for other model systems at various stages of completion, including the mouse \cite{helmstaedter2013,lee2016,abbott2020}, the larval zebrafish \cite{hildebrand2017}, the tadpole larva \cite{ryan2016}. The experimental technique employed is for the reconstructions is the serial section electron microscopy \cite{helmstaedter2008,lichtman2014,mulcahy2018}.}, it has remained the only one with a fully reconstructed connectome -- i.e., a detailed charting of all its neural connections. Below, we provide a brief overview of the essential features of the \ce\ neural network.  
There are excellent resources available on the web for all aspects of the \ce\ biology, including the nervous system, in particular \href{https://www.wormatlas.org/handbookhome.htm}{Wormatlas} and \href{http://www.wormbook.org/}{WormBook}.

\subsection{\ce\ nervous system: a digest}
The nervous system of an adult hermaphrodite \ce\ consists of just 302 neurons, organised in two independent nervous system: a large somatic nervous system (282 neurons) and a small pharyngeal nervous system (20 neurons)\sidenote[][-1cm]{It is nothing short of impressive that such a tiny nervous system is able to support the wide behavioral array of a \ce. Beyond the basics of locomotion, foraging, and feeding, the worm can discern and navigate towards or away from various chemicals, odors, temperature gradients, and food sources. Furthermore, it demonstrates social awareness, detecting the presence, density, and even sex of neighboring nematodes \cite{altun2005,debono2005}.}, uniquely identifiable\sidenote[][+3.2cm]{Each neuron in the worm's nervous system is identified by a code. The nomenclature system consists in two or three letters (or, occasionally, numbers), followed by the position in worm's body D/V (dorsal/ventral), R/L (right/left) \cite{white1986}. For instance, the code AFDL stands for \textbf{A}mphid \textbf{F}inger-like Endings \textbf{D}orsal \textbf{L}eft.}. The vast majority of neurons have simple unipolar or bipolar morphology, fig. \ref{f-neuro} (a-b).

\begin{figure*}[t!]
	\includegraphics[width=15cm]{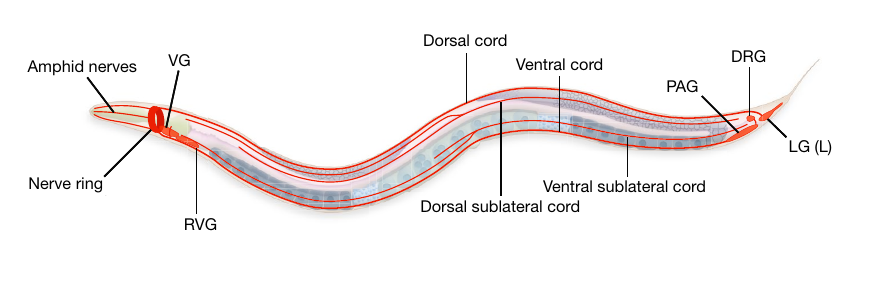}
	\caption{The majority of neuron cell bodies are found in clusters, named ganglia, located in the head -- VG, ventral ganglion; RVG, retrovesicular ganglion -- and in the tail -- PAG, pre-anal ganglion; DRG, dorsorectal ganglion; LG, umbar ganglion. Neuronal processes run in nerve bundles, the major is the nerve ring (head), the ventral and dorsal cord rung, that along the whole worm body. Adapted from Cook et al., \emph{Whole-animal connectomes of both Caenorhabditis elegans sexes} \cite{cook2019}.}
    \label{f-CE}
\end{figure*}

With few exceptions, neuron cell bodies are found in clusters, called \emph{ganglia}. Several head ganglia -- including the retrovesicular ganglion and ventral ganglion --, located around the nerve ring, host the largest collection of cell bodies. The second largest collection of cell bodies is found in the tail ganglia. Any projection from the cell bodies is generically called a \emph{process} - predominantly, axons and dendrites. Neuronal processes extend from the ganglia and travel in longitudinal nerve bundles to different regions of the nervous system. Out of the total 302 neurons in the adult hermaphrodite, 180 project axons/dendrites into the nerve ring. The most prominent nerve bundles are the nerve ring, ventral nerve cord and dorsal nerve cord, Fig. \ref{f-CE}. 

The neurons in the \ce\ nervous system are classified into different categories based on their morphology, function, and connectivity.
\begin{itemize}
    \item \emph{Sensory} neurons are the primary receptors of environmental stimuli, ranging from temperature changes to chemical signals. 
    \item \emph{Interneurons} are information processors, responsible for transmitting signals between other classes of neurons. 
    \item \emph{Motor} neurons control the contraction and relaxation of muscles, on which they primarily form synapses.
    \item \emph{Modulatory} neurons release neuromodulators, molecules that alter the activity of other neurons or influence the strength of the signals they send.
\end{itemize}

Neurons exchange information mainly through $\sim1500$ electrical and $\sim5000$ chemical synapses. The former, also called \emph{gap junctions}, are specialized channels that directly connect the cytoplasm of adjacent cells, allowing various molecules, ions, and electrical impulses to pass between the cells. The latter, \emph{chemical synapses}, function as specialised junctions that facilitate the one-way relay of chemical signals, or neurotransmitters, from a presynaptic to one or more postsynaptic cells. Chemical synapses are found between neighbouring processes. Therefore, it is the neighbourhood of the processes that predominantly determines the connectivity between neurons. The nerve ring hosts the highest density of these synapses, followed by the ventral and dorsal cord.

Recently, a number of studies have revised the original annotations of the hermaphrodite, adult \ce\ nervous system, updated its connectome,  and measured it in with the tools of network science \cite{varshney2011,towlson2013,bentley2016,cook2019}. However, it is important to understand that, strictly speaking, \emph{the C.elegans connectome does not exist}. This because (i) due to technical limitations, the whole animal connectome is constructed by patching together regions of the nervous system obtained from different animals. Some connections are not even observed, but inferred from the similarity patterns of certain regions \cite{cook2019}. Furthermore, (ii) although the overall structure is highly stereotyped, individual connectomes differ in detail due to natural developmental variability, sec. \ref{s-brain_wiring}.  

Given these limitations, and the availability of the adult connectome only, it is unsurprising that fewer studies have examined the growth of the \ce\ neural network throughout development (from embryo to adult)  \cite{varier2011,nicosia2013,alicea2018,pathak2020}. This circumstance took a sharp turn in 2021, as we are about to discuss.

\subsection{\emph{C.elegans} brain maturation}
The lifecycle of \ce\ unfolds in a series of distinct stages \cite{sulston1977}. The initial embryonic stage of \ce\ concludes with hatching (birth). The post-embryonic development consists of four larval stages $(L_1 - L_4)$ and adulthood. Along with other biological structures, the nervous system of the worm matures during development. Neurodevelopment in \ce\ is characterised by a stereotyped cell lineage, resulting in consistent neural placement across individuals\sidenote[][-2.5cm]{After the cellular differentiation, the neuron body cells migrate across the worm's body to take up their final positions in adulthood. This happens in a highly stereotyped way. Either simultaneously or after migration, neurons extend their processes to find partners for synaptic connections. Note, however, that the two processes (neuron establishment and synapse formation) are distinct \cite{rapti2020}.}. The formation of synapses, or synaptogenesis, and process outgrowth, underlie the specific connectivity patterns of the mature nervous system, which are enhanced through substantial remodeling during post-embryonic growth \cite{rapti2020,pathak2020}.

\begin{figure*}[h!]
    \centering
	\includegraphics[width=14cm]{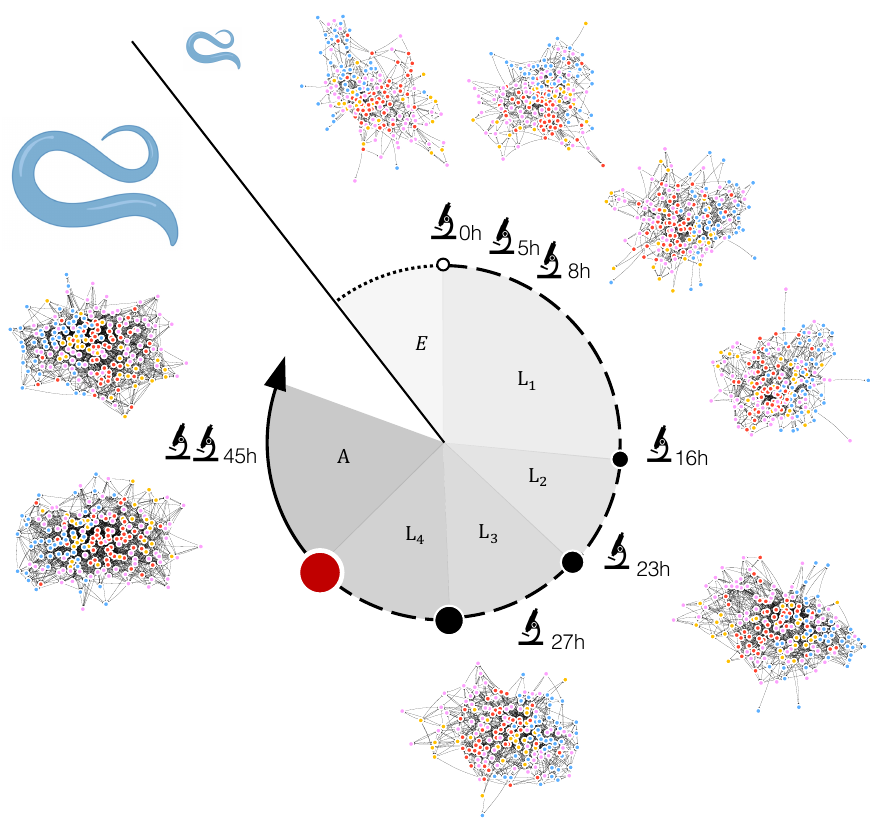}
	\caption{\ce\ brain networks at different developmental ages. The embryonic stage (E, dotted line) terminates with the hatching (birth, $\SI{0}{\hour}$, white circle). The post-embryonic stages include four larval ages ($L_1-L_4$, dashed line) and ends with the onset of adulthood (red circle). The adult stage (A, solid line) lasts about $2-3$ days. The dataset consists of eight snapshots (microscope icons), including one at birth  $\SI{0}{\hour}$, three $L_1$ $\sim\SIlist{5;8;16}{\hour}$, one $L_2$ $\sim\SI{23}{\hour}$, one $L_3$ $\sim\SI{27}{\hour}$ and two adults, both $\sim\SI{45}{\hour}$. The brain consists in the nerve ring and ventral ganglion, see fig. \ref{f-CE}. The nodes are neurons belonging to four different categories: inter- (red), modulatory (yellow), motor (blue) and sensory (pink) neurons, see tab. \ref{t-neuron_list}. Edges are here defined as synaptic connections, i.e., an edge exists between two nodes if at least one chemical synapse exists between them.
    } 
    \label{f-Celegans_witvliet}
\end{figure*}

\arrayrulecolor{white} 
\begin{table*}[h!]
  \centering
  \scriptsize
  \setlength{\arrayrulewidth}{2pt}
  \renewcommand{\arraystretch}{1.2}
  \caption{List of the $180$ neurons of the adult \emph{C.elegans} brain (hermaphrodite, N2), as reported in \cite{witvliet2021}. Interneurons in red, modulatory in yellow, motor in blue, sensory in pink. We have marked with an asterisk$^*$ those neurons that were not present at birth. Each neuron in the worm nervous system is uniquely identified by a code, which consists in two or three letters (or, occasionally, numbers), followed by the position in worm's body D/V (dorsal/ventral), R/L (right/left) \cite{white1986}. The left-right symmetry increases over time and reaches the $\sim90\%$ in the adult brain.}  \label{t-neuron_list}
  \begin{tabularx}{\columnwidth}{*{14}{|X}|}
    \hline
    \cellcolor{interneuron} ADAL & \cellcolor{interneuron} ADAR & \cellcolor{interneuron} AIAL & \cellcolor{interneuron} AIAR & \cellcolor{interneuron} AIBL & \cellcolor{interneuron} AIBR & \cellcolor{interneuron} AINL & \cellcolor{interneuron} AINR & \cellcolor{interneuron} AIYL & \cellcolor{interneuron} AIYR & \cellcolor{interneuron} AIZL & \cellcolor{interneuron} AIZR & \cellcolor{interneuron} AVAL & \cellcolor{interneuron} AVAR \\
    \hline
    \cellcolor{interneuron} AVBL & \cellcolor{interneuron} AVBR & \cellcolor{interneuron} AVDL & \cellcolor{interneuron} AVDR & \cellcolor{interneuron} AVEL & \cellcolor{interneuron} AVER & \cellcolor{interneuron} AVJL & \cellcolor{interneuron} AVJR & \cellcolor{interneuron} BDUL & \cellcolor{interneuron} BDUR & \cellcolor{interneuron} PVCL & \cellcolor{interneuron} PVCR & \cellcolor{interneuron} PVPL & \cellcolor{interneuron} PVPR \\
    \hline
    \cellcolor{interneuron} PVR & \cellcolor{interneuron} PVT & \cellcolor{interneuron} RIAL & \cellcolor{interneuron} RIAR & \cellcolor{interneuron} RIBL & \cellcolor{interneuron} RIBR & \cellcolor{interneuron} RIFL & \cellcolor{interneuron} RIFR & \cellcolor{interneuron} RIGL & \cellcolor{interneuron} RIGR & \cellcolor{interneuron} RIH & \cellcolor{interneuron} RIML & \cellcolor{interneuron} RIMR & \cellcolor{interneuron} RIPL \\
    \hline
    \cellcolor{interneuron} RIPR & \cellcolor{interneuron} RIR & \cellcolor{modulatory} ADEL & \cellcolor{modulatory} ADER & \cellcolor{modulatory} AIML & \cellcolor{modulatory} AIMR & \cellcolor{modulatory} ALA & \cellcolor{modulatory} AVFL$^*$ & \cellcolor{modulatory} AVFR$^*$ & \cellcolor{modulatory} AVHL & \cellcolor{modulatory} AVHR & \cellcolor{modulatory} AVKL & \cellcolor{modulatory} AVKR & \cellcolor{modulatory} AVL$^*$ \\
    \hline
    \cellcolor{modulatory} CEPDL & \cellcolor{modulatory} CEPDR & \cellcolor{modulatory} CEPVL & \cellcolor{modulatory} CEPVR & \cellcolor{modulatory} DVC & \cellcolor{modulatory} HSNL$^*$ & \cellcolor{modulatory} HSNR$^*$ & \cellcolor{modulatory} PVNL$^*$ & \cellcolor{modulatory} PVNR$^*$ & \cellcolor{modulatory} PVQL & \cellcolor{modulatory} PVQR & \cellcolor{modulatory} RICL & \cellcolor{modulatory} RICR & \cellcolor{modulatory} RID \\
    \hline
    \cellcolor{modulatory} RIS & \cellcolor{modulatory} RMGL & \cellcolor{modulatory} RMGR & \cellcolor{motor} IL1DL & \cellcolor{motor} IL1DR & \cellcolor{motor} IL1L & \cellcolor{motor} IL1R & \cellcolor{motor} IL1VL & \cellcolor{motor} IL1VR & \cellcolor{motor} RIVL & \cellcolor{motor} RIVR & \cellcolor{motor} RMDDL & \cellcolor{motor} RMDDR & \cellcolor{motor} RMDL \\
    \hline
    \cellcolor{motor} RMDR & \cellcolor{motor} RMDVL & \cellcolor{motor} RMDVR & \cellcolor{motor} RMED & \cellcolor{motor} RMEL & \cellcolor{motor} RMER & \cellcolor{motor} RMEV & \cellcolor{motor} RMFL$^*$ & \cellcolor{motor} RMFR$^*$ & \cellcolor{motor} RMHL$^*$ & \cellcolor{motor} RMHR$^*$ & \cellcolor{motor} SIADL & \cellcolor{motor} SIADR & \cellcolor{motor} SIAVL \\
    \hline
    \cellcolor{motor} SIAVR & \cellcolor{motor} SIBDL & \cellcolor{motor} SIBDR & \cellcolor{motor} SIBVL & \cellcolor{motor} SIBVR & \cellcolor{motor} SMBDL & \cellcolor{motor} SMBDR & \cellcolor{motor} SMBVL & \cellcolor{motor} SMBVR & \cellcolor{motor} SMDDL & \cellcolor{motor} SMDDR & \cellcolor{motor} SMDVL & \cellcolor{motor} SMDVR & \cellcolor{motor} URADL \\
    \hline
    \cellcolor{motor} URADR & \cellcolor{motor} URAVL & \cellcolor{motor} URAVR & \cellcolor{sensory} ADFL & \cellcolor{sensory} ADFR & \cellcolor{sensory} ADLL & \cellcolor{sensory} ADLR & \cellcolor{sensory} AFDL & \cellcolor{sensory} AFDR & \cellcolor{sensory} ALML & \cellcolor{sensory} ALMR & \cellcolor{sensory} ALNL$^*$ & \cellcolor{sensory} ALNR$^*$ & \cellcolor{sensory} AQR$^*$ \\
    \hline
    \cellcolor{sensory} ASEL & \cellcolor{sensory} ASER & \cellcolor{sensory} ASGL & \cellcolor{sensory} ASGR & \cellcolor{sensory} ASHL & \cellcolor{sensory} ASHR & \cellcolor{sensory} ASIL & \cellcolor{sensory} ASIR & \cellcolor{sensory} ASJL & \cellcolor{sensory} ASJR & \cellcolor{sensory} ASKL & \cellcolor{sensory} ASKR & \cellcolor{sensory} AUAL & \cellcolor{sensory} AUAR \\
    \hline
    \cellcolor{sensory} AVM$^*$ & \cellcolor{sensory} AWAL & \cellcolor{sensory} AWAR & \cellcolor{sensory} AWBL & \cellcolor{sensory} AWBR & \cellcolor{sensory} AWCL & \cellcolor{sensory} AWCR & \cellcolor{sensory} BAGL & \cellcolor{sensory} BAGR & \cellcolor{sensory} DVA & \cellcolor{sensory} FLPL & \cellcolor{sensory} FLPR & \cellcolor{sensory} IL2DL & \cellcolor{sensory} IL2DR \\
    \hline
    \cellcolor{sensory} IL2L & \cellcolor{sensory} IL2R & \cellcolor{sensory} IL2VL & \cellcolor{sensory} IL2VR & \cellcolor{sensory} OLLL & \cellcolor{sensory} OLLR & \cellcolor{sensory} OLQDL & \cellcolor{sensory} OLQDR & \cellcolor{sensory} OLQVL & \cellcolor{sensory} OLQVR & \cellcolor{sensory} PLNL$^*$ & \cellcolor{sensory} PLNR$^*$ & \cellcolor{sensory} SAADL & \cellcolor{sensory} SAADR \\
    \hline
    \cellcolor{sensory} SAAVL & \cellcolor{sensory} SAAVR & \cellcolor{sensory} SDQL$^*$ & \cellcolor{sensory} SDQR$^*$ & \cellcolor{sensory} URBL & \cellcolor{sensory} URBR & \cellcolor{sensory} URXL & \cellcolor{sensory} URXR & \cellcolor{sensory} URYDL & \cellcolor{sensory} URYDR & \cellcolor{sensory} URYVL & \cellcolor{sensory} URYVR \\
    \hline
  \end{tabularx}
\end{table*}

In 2021, the research into neural development has been boosted by the release of an unprecedented dataset. \emph{Witvliet et al.} have published the electron microscopy reconstruction of the \ce\ brain across different stages of the worm development \cite{witvliet2021}. 

More specifically, eight \ce\ -- wild-type N2, hermaphrodite, isogenic, reared in the same environment -- were selected for imaging at different post-embryonic stages\sidenote[][-1cm]{The developmental age of each specimen is estimated using the known and stereotypical cell division pattern \cite{sulston1977}. Thus, precise temporal annotation is not available.}. These comprise one at birth $\SI{0}{\hour}$, three $L_1$ $\sim\SIlist{5;8;16}{\hour}$, one $L_2$ $\sim\SI{23}{\hour}$, one $L_3$ $\sim\SI{27}{\hour}$ and two adults, both $\sim\SI{45}{\hour}$, fig. \ref{f-Celegans_witvliet}. The brain -- i.e., nerve ring and ventral ganglion\sidenote{As mentioned above, a large proportion of \ce\ neurons are located close to the brain or extend their processes and form synapses within it. However, it is worth stressing that, as in humans, the \emph{brain} \emph{not} the entire nervous system, fig. \ref{f-CE}.} -- of each specimen was entirely imaged by serial section electron microscopy. Each cell was identified based on its unique morphology and position, tab. \ref{t-neuron_list}. The totality of its chemical synapses and a subset of its gap junctions were manually annotated\sidenote{By focusing solely on the brain, one gains the ability to reconstruct it completely for an individual, without the need to stitch together segments from different specimens. This is important for the purposes of our investigation, because to capture individual variability, we prefer individual reconstructions to collages.}. From birth to adult, the number of nodes increased from $161$ to $180$, that of chemical synapses from $\sim1300$ to $\sim8000$. In contrast to mammals \cite{klintsova1999}, synaptic pruning does not occur, and the removal of synaptic connections is rarely observed.

\section{The EE development of a worm brain}[The EE brain maturation]\label{s-EE_ce}
\emph{Where the core results of this PhD are illustrated, a parsimonious white-box model of a worm brain development is formulated. Where some understanding is reached of a (seemingly complex, certainly fascinating) biological process.}\bigskip

Here, we will detail our exploration-exploitation model for the maturation of the \ce\ brain. Along the way, we mark any hypothesis and/or assumption with the icon \h$_i$ \sidenote{The ambition of this chapter is to speak out clearly every single assumption. This is why, every time we make one, we ring a bell \h$_i$ ($i$ counts the assumptions). We also try to resist the temptation to summon \emph{common practices} or \emph{established methods} to justify their use or, worse, to conceal them. 
}, and discuss them in sec. \ref{s-perspectives}.

\subsection{A minimal worm brain}
\subsubsection{Methods}
To start with, we exclude gap junctions from our analysis \h$_1$ because they were only partially annotated in \cite{witvliet2021}. 
As for the chemical synapses, a number of them can be found between each pair of neurons. The networks of chemical synapses are therefore weighted and directed. In this work, we transform them however into the form \eqref{e-graph_def}, i.e., we consider the unweighted \h$_2$ and undirected \h$_3$ networks of chemical synapses. 

\begin{figure}[h] 
    \centering
	\includegraphics[height=5cm]{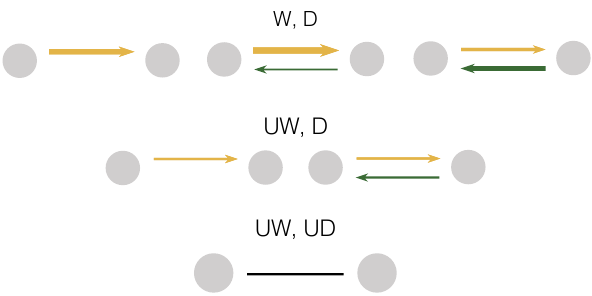}
    \caption{Multiple synapses in both direction can exist between two neurons. The network of chemical synapses is therefore weighted (W) (by the synapse number) and directed (D). A directed synaptic connection (UW,D) exists if at least one synapse is observed in the same direction. An undirected connection (UW, UD) exists if at least one directed connection is observed.}
    \label{f-connections}
\end{figure}

Casting weighted to unweighted networks entails casting the number of synapses between a pair of neurons in a binary state, which we call connection/non-connection. In particular, a (directed) connection exists from a presynaptic to a postsynaptic neuron if at least one synapse exist between the two, fig. \ref{f-connections}. The dynamics of synaptic connections is a low-dimensional projection of the dynamics of synapses. Importantly however, the formation of a new connection implies (by definition) that of a new synapse. Therefore, the biology of the connection formation process is, loosely speaking, the same as that of the synaptogenesis process\sidenote[][-3cm]{The contrary, however, is not true: a new synapse does not necessarily mean a new synaptic connection, in the case where it is added to and strengthens an existing one. This further implies that the biology of the connection removal process is not the same of that of the synapse elimination process. In our case, as we will see in sec. \ref{ss-wbm}, this is no cause for concern.\label{sn-syn_add}}. It has also been shown that both synapses and synaptic connections exhibit a qualitatively similar developmental dynamic, namely a near-linear rate of addition \cite{witvliet2021}.

\begin{figure}[htbp] 
    \centering
	\includegraphics{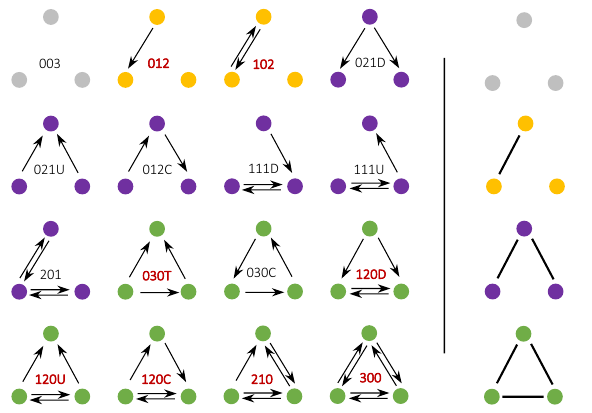}
    \caption{Triad census. Left: There exist sixteen directed graph patterns between three nodes. Each is denoted by a three-digit code, representing (i) the count of mutual links, (ii) of single links, (iii) and non-existent links, respectively. Additionally, (iv) a letter can be appended to indicate if the pattern has a cycle (C), a transitive (T), an upward (U), or a downward (D) connection structure. In bold red, the codes of those motifs that are over-represented in the adult \ce\ network of (directed) synaptic connections. With the exception of $030$C, all motifs involving connections between each pair of nodes (green) are over-represented. Those involving one empty dyad (violet) tend to be under-represented \cite{varshney2011,cook2019}. Right: There exist four unique connectivity patterns are possible among three nodes in an undirected graph. Color code highlights corresponding patterns.}
    \label{f-triad_census}
\end{figure}

The further reduction to an undirected network means that an undirected connection is placed between two neurons if there is at least one directed connection (thus, a synapse) between them, regardless of its direction. This is a more delicate assumption\sidenote[][-1cm]{In general, it can introduce spurious connection reciprocities and distort the network information flow -- e.g., by obscuring causal relationships between different connections.}, which we can motivate as follows. Previous analyses \cite{varshney2011,cook2019} of the (adult) \ce\ network of chemical synapses have examined its triad census, i.e., the counts of all possible directed connection patterns between triples of nodes -- there exist sixteen patterns. These studies have demonstrated that those connection patterns involving the same number of empty dyads exhibit consistent statistical characteristics, in the sense that they are all (with few exceptions) over- or under-represented relative to a randomised null model\sidenote[][-2.2cm]{The randomisation procedure used in both \cite{varshney2011,cook2019} preserves in-degree and out-degree and the numbers of bidirectional and unidirectional connections for each neuron. See Figure 7 in \cite{varshney2011} and an updated version of the same exact plot, Extended Data Figure 7 in \cite{cook2019}.}, fig. \ref{f-triad_census}. The simple patterns listed in the triad census are the building blocks of more intricate network motifs. In turn, this means that, when it comes to the analysis of graph patterns, the use of undirected connections results in a minor distortion of the original directed network.

As a coarse-graining procedure, the projection to an unweighted, undirected graph implies a loss of information\sidenote[][]{In fact, we can derive the edge dynamics of the unweighted undirected graph from that of the original weighted directed graph, but not vice versa. The arguments we have presented above are not intended to prove that our coarse graining is a mere rephrasing of the network's original information. Rather, they are intended to convince the reader that it is not meaningless, in the sense that it does not change the nature of the problem.}. This sacrifice, however, is not worthless. What we gain is the noteworthy possibility of writing a simple model for the \ce\ brain maturation in terms of a handful of graph motifs\sidenote{This is typically the point of a manuscript at which one summons the spirit of the 14th-century Franciscan friar, named William of Ockham, and its renowned razor. The relationship between the the \emph{Ockham's razor}  -- also, the \emph{principle of parsimony} -- and the scientific truth is riddled with nuances. A discussion of the use of the Ockham's razor as an abductive heuristic can be found in \cite{gauch2003,sober2015}.}, as we will explain in more detail in the following. 

\subsubsection{Results}
We obtain eight unweighted, undirected graphs\sidenote{The scripts for both the preprocessing step and preliminary data analyses have been written using \texttt{RStudio} with \texttt{R v4.0.4}. They can be found in the Github folder: \href{https://github.com/dichio/EE-graph-dyn}{EE-graph-dyn}.}. Unless otherwise specified, we will refer to the undirected synaptic connections simply as edges. A preliminary step in our modelling approach is to measure the network properties of interest and how they change during development. In tab. \ref{t-sum_stats}, we report the computation of a representative subset of standard graph metrics, app. \ref{a-measures}. 

The number of neurons (nodes) increases from $161$ at birth to $180$ in the adult stage, with a burst of neuronal births at the turn of the larval stages $L_1$ and $L_2$, consistent with what was previously reported\sidenote{The \ce\ neurons are born in two separate bursts of cell differentiation. The first, major one happens during the embryonic stage (before hatching) and lasts approximately four hours. The second, minor, happens over seventeen hours during the post-embryonic stage, as here observed \cite{varier2011}. \label{sn-neu_birth}} \cite{varier2011,nicosia2013}. In parallel, we observe a $2.7$-fold increase in neuronal connectivity, from $617$ edges at birth to $\sim 1650$ for the adult. This further results in an increase of both the number of two-stars (or, connected triples) and triangles. Also, the average geodesic distance between any two nodes decreases from $\sim3$ to $\sim2.2$. In summary, the adult \ce\ network develops to become more closely interconnected.

Arguably, this leads to enhanced functionality of the \ce\ brain network. We can gain insights into this by computing the average local efficiency and clustering coefficient, which both increase during development. The first suggests an deployment of a biological strategy for increasing the system's redundancy and robustness. The second is compatible with an increasingly modular organisation throughout development, and suggests an improvement in local information processing.

Crucially, all the network metrics here considered indicate a consistent and monotonous trend throughout the worm's brain maturation.

\begin{table*}[h]
\centering
\begin{tabular}{c|cccccc|c}
\multicolumn{1}{c}{\textbf{t[h]}} & \multicolumn{1}{c}{\textbf{nodes}} & \multicolumn{1}{c}{\textbf{edges}} & \multicolumn{1}{c}{\textbf{two-star}} & \multicolumn{1}{c}{\textbf{triangles}} & \multicolumn{1}{c}{\textbf{av.sh.path}} & \multicolumn{1}{c}{\textbf{glob.eff.}} & \multicolumn{1}{c}{\textbf{clust.coeff.}} \\ \midrule
\rowcolor[HTML]{F2EAD3}
$\bm{0}$ & $161$ & $617$ & $5976$ & $346$ & $2.993$ & $0.380$ & $0.208$ \\
$\bm{5}$ & $162$ & $782$ & $9273$ & $601$ & $2.712$ & $0.416$ & $0.232$ \\
$\bm{8}$ & $162$ & $788$ & $9299$ & $614$ & $2.712$ & $0.416$ & $0.245$ \\
$\bm{16}$ & $168$ & $907$ & $11838$ & $830$ & $2.617$ & $0.428$ & $0.246$ \\
$\bm{23}$ & $173$ & $1166$ & $18449$ & $1406$ & $2.430$ & $0.459$ & $0.262$ \\
$\bm{27}$ & $174$ & $1175$ & $18866$ & $1433$ & $2.429$ & $0.458$ & $0.274$ \\
\rowcolor[HTML]{DFD7BF}
$\bm{45}$ & $180$ & $1633$ & $34124$ & $2889$ & $2.217$ & $0.498$ & $0.286$ \\
\rowcolor[HTML]{DFD7BF}
$\bm{45}$ & $180$ & $1669$ & $35677$ & $3003$ & $2.206$ & $0.501$ & $0.292$ \\
\bottomrule
\end{tabular}
\caption{Properties of the \ce\ networks of undirected synaptic connections. Each row corresponds to a graph, the first (birth) and the last two (adulthood) are highlighted. We compute the number of \textbf{nodes}, \textbf{edges}, \textbf{two-stars} -- or connected triples --, \textbf{triangles}, the average shortest path (\textbf{av.sh.path}) -- or average geodesic distance -- , the average local efficiency (\textbf{loc.eff.}) and the average clustering coefficient (\textbf{clust.coeff.}). See app. \ref{a-measures} for the definitions.} 
\label{t-sum_stats}
\end{table*}

\subsection{Topography of the functional landscape}\label{s-fun_land}
The crucial methodological step in EE modelling is the specification of an $F$ metric and the inference of the topography of the resulting functional landscape. In this work, we do so by ERG inference, ch. \ref{c-ergm}.
In particular, without loss of generality, we can express the $F$ metric as a linear combination $F(G) = \bm{\theta}\cdot\bm{x}(G)$ of graph statistics $\bm{x}\in\mathbb{R}^r$ with linear coefficients $\bm{\theta}\in\mathbb{R}^r$. With loss of generality instead, we propose the following parsimonious, coarse-grained model for the \ce\ brain maturation \h$_4$:
\begin{equation}\label{e-FL_ce}
    F(G) = \theta_{gwd}\ x_{gwd}(G|\lambda_{gwd}) + \theta_{gwesp}\ x_{gwesp}(G|\lambda_{gwesp})\ ,
\end{equation}
where the model statistics 
\begin{equation}\label{e-FL_ce_sts}
    \bm{x}(G) = 
    \begin{bmatrix}
        x_{gwd}(G|\lambda_{gwd})\\
        x_{gwesp}(G|\lambda_{gwesp})
    \end{bmatrix}
    \in \mathbb{R}^2
\end{equation}
have been defined in \eqref{e-CS_GWD} and \eqref{e-CS_GWESP}, respectively\sidenote{Note that the two graph statistics used here are defined in the case of undirected, unweighted graphs. Simple representations allow for simple models. Furthermore, in defining this model, we have not included a term to represent the graph's edge count (or density). In fact, the latter is a degree of freedom which is controlled by the exploration rate in an EE dynamics, therefore it is not at the disposal of the $F$ metric specification. See later in sec. \ref{ss-wbm}.
}.

According to \eqref{e-FL_ce}, the biological function of a worm brain network can be characterised in terms of two complementary graph statistics $\bm{x}\in\mathbb{R}^2$. The first one, $x_{gwd}$, based on the graph degree distribution, highlights node connectivity. The second one, $x_{gwesp}$, based on the distribution of edgewise shared partners, captures relational patterns. Together, they provide a comprehensive view of both node attributes and network configurations.

\subsubsection{Methods}
Modulo a minus sign, the $F$ metric \eqref{e-FL_ce} corresponds to the graph Hamiltonian $\mathscr{H}$ obtained when constructing an ERG model with graph statistics $x_{gwd}, x_{gwesp}$.  Therefore, we can use the ERG inference to estimate the four parameters - two linear coefficients $\bm{\theta}$ and two decay parameters $\bm{\lambda}$ - of \eqref{e-FL_ce}. In other words, we employ the ERG methods to infer the topography of the functional landscape. 
To ensure that the correct (functional) balance of model statistics
can be achieved at the end of the developmental process \h$_5$, we use the two adult \ce\ brain snapshots $\bm{G}_{T}^*$ as input for the inference, $T=\SI{45}{\hour}$. 

\begin{lstlisting}[caption={ERG inference based on \eqref{e-FL_ce}, library \texttt{ergm v4.3.2} for \texttt{R v4.0.4}, code available in the Github folder \href{https://github.com/dichio/EE-graph-dyn}{EE-graph-dyn}. \texttt{G} represents the input graph for the inference, either $G_{T,1}^*$ or $G_{T,2}^*$ . The decay parameters of the curved statistics are estimated as well (\texttt{fixed=F}). The model is constrained to those graphs that have the same number of edges as the \texttt{G}. As initial guess of the four parameters, we use \texttt{(1,1,1,1)}.}, label={l-ergm_formula}, numbers=none]
# ERGM formula
ergm(formula = G ~ 
            gwdegree(fixed=F)+gwesp(fixed=F), 
            constraints = ~ edges,
            control=snctrl(init = c(1,1,1,1))
            )
\end{lstlisting}

As the ERG inference is defined for one single graph, an output procedure is therefore required. A simple choice is that to use the so-called \emph{mean-ERG} \h$_6$, originally proposed in \cite{simpson2012}. Accordingly, the inference is performed independently for each network, resulting in multiple estimates of each parameter. The final estimate is determined by averaging the corresponding values across all networks\sidenote{
This is an instance of the more general problem of constructing a \emph{group representative network} (GRN). There are several other methods, ranging from the cruder to the more sophisticated alternatives \cite{dichio2023b}. We consider the mean-ERG to be a lower limit of methodological complexity. We are essentially limited here by the availability of only two adult networks.
}. In doing so, we implicitly assume that each network is a different realisation of the same (bio)physical system. 

\subsubsection{Results}
The ERG inference based on \eqref{e-FL_ce} for the two \ce\ brain networks is performed as discussed in ch. \ref{c-ergm} and yields the estimates summarised in tab. \ref{t-FL_indiv}. 
\begin{table}[h]
\renewcommand{\arraystretch}{1.2}
\setlength{\arrayrulewidth}{1.5pt}
\centering
\begin{tabular}{c>{\columncolor[HTML]{F2EAD3}}cc>{\columncolor[HTML]{F2EAD3}}cc}
& $\theta^*_{gwd}$ & $\lambda^*_{gwd}$ & $\theta^*_{gwesp}$ & $\lambda^*_{gwesp}$\\
\hline
$G^*_{T,1}$ &  $0.45 \pm 0.20$ & $1.91 \pm 0.46$ &  $0.626 \pm 0.056$ & $1.432 \pm 0.067$ \\
\hline
$G^*_{T,2}$ &  $0.43 \pm 0.20$ & $1.97 \pm 0.48$ & $0.529 \pm 0.048$ &  $1.542 \pm 0.075$\\\bottomrule
\end{tabular}
\caption{ERG estimation based on \eqref{e-FL_ce} for the two adult worms $\bm{G}^*_{T}$. The maxent parameters $\bm{\theta}^*$ are both significant and positive for all networks. The parameters $\bm{\lambda}^*$ controlling for the geometric decays of the model statistics are significant -- and positive by construction.} 
\label{t-FL_indiv}
\end{table}

The emerging picture is of an adult \ce\ brain network characterised by a propensity for (i) the presence of highly connected nodes ($\theta_{gwd}>0$) and (ii) triadic closure ($\theta_{gwesp}>0$). The former is consistent with the presence of medium and large hub nodes. The latter can reflect an underlying graph modular structure and is compatible with a common neighbor rule -- i.e., neuron pairs with more shared neighbors have a higher likelihood of connection -- for the worm's neuronal wiring. These characteristics of the worm's adult brain network have been extensively documented in recent years \cite{towlson2013,arnatkeviciute2018,pathak2020,pan2010,azulay2016}.

The results of the mean-ERG construction for the parameters of the \ce\ functional landscape are shown in tab. \ref{t-FL_params}. 

\begin{table}[h]
\renewcommand{\arraystretch}{1.2}
\setlength{\arrayrulewidth}{1.5pt}
\centering
\begin{tabular}{c>{\columncolor[HTML]{F2EAD3}}cc>{\columncolor[HTML]{F2EAD3}}c}
$\theta^*_{gwd}$ & $\lambda^*_{gwd}$ & $\theta^*_{gwesp}$ & $\lambda^*_{gwesp}$\\
\hline
$0.44$ & $1.94$ &  $0.578$ & $1.487$ 
\end{tabular}
\caption{Parameters of the \ce\ functional landscape, as defined by the $F$ metric \eqref{e-FL_ce}. Mean-ERG based on the estimation obtained from two adult worm's brain, tab. \ref{t-FL_indiv}.} 
\label{t-FL_params}
\end{table}

We observe that an EE dynamics based on the $F$ metric \eqref{e-FL_ce} with the above parameters would favour the emergence of hub-like structures and the strengthening of triadic closure throughout development. This is consistent with two of the developmental principles highlighted in \cite{witvliet2021}, namely (i) that well-connected neurons receive more inputs and (ii) that network modularity increases with time. The latter, in turn, align with the trends described in tab. \ref{t-sum_stats}, i.e., with a developmental process that progressively weaves a more tightly connected, robust and efficient network topology.

Our minimal model of the \ce\ neurofunctional landscape is based on two graph statistics and has only four parameters that are amenable to biological interpretation and are inferred from the data. All that remains for us is to unleash an EE dynamic on it.

\subsection{Worm brain maturation tracked down}\label{ss-wbm}
Using an EE dynamics to model the \ce\ brain network development means representing it as a stochastic dynamics of a probability distribution on a functional landscape. The rationale for this has been discussed in sec. \ref{s-brain_wiring} in the more general context of the brain wiring problem and originates from the need to accommodate two fundamental observations: the existence of developmental variability\sidenote{In \cite{witvliet2021}, the $43\%$ of the (directed) synaptic connections were found not to be conserved between isogenic individuals, contrary to the common assumption that the \ce\ brain is hardwired. Interestingly, not all of these connections consist of only a few synapses.}
and the robustness of the functional outcome. 

In using this picture, we implicitly make an important assumption, which we might call the \emph{hypothesis of functional homogeneity} \h$_7$. This means that the same definition of biological function holds true throughout the whole developmental process\sidenote{Cf. with the discussion on the fitness landscape/seascape, sec. \ref{ss-fit_fun}.}. We will treat this hypothesis self-consistently and verify it a posteriori. 

\begin{figure*}[t!]
    \centering
	\includegraphics[width=15cm]{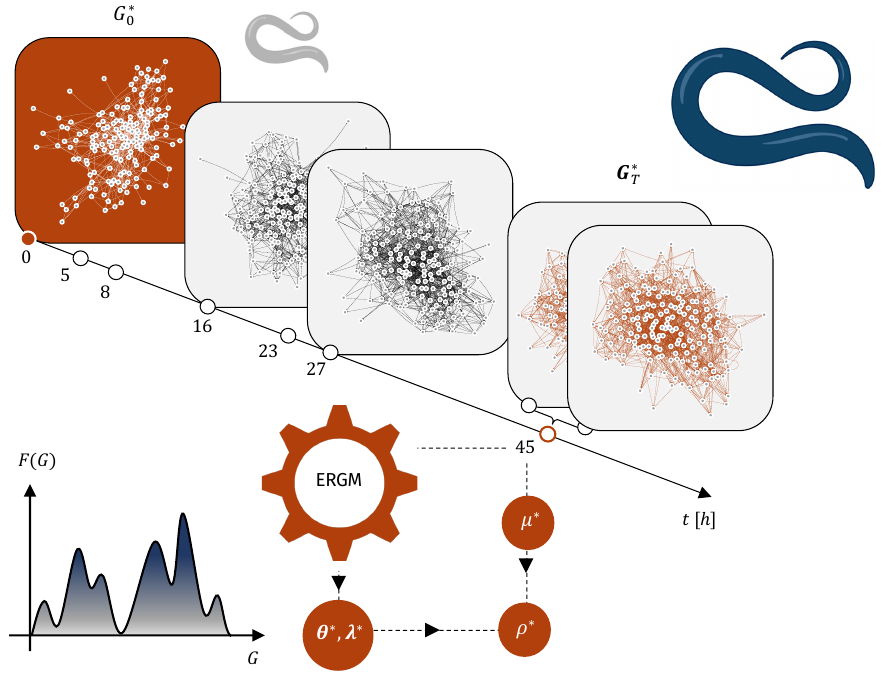}
	\caption{Exploration-exploitation dynamics for the \ce\ brain maturation, illustration. The process is depicted as the dynamics of a probability distribution on a functional landscape. Our starting point is the dataset presented by \emph{Witvliet et al.} \cite{witvliet2021}, which consists of eight reconstructions of the worm's brain (nerve ring and ventral ganglion), at different developmental ages $t=\SIlist{0;5;8;16;23;27;45}{\hour}\ (\times\ 2)$ after hatching (birth). In particular, we consider the unweighted, undirected networks of chemical synapses -- some are omitted for visual clarity. The configuration of the brain at birth $G_0^*$ (white graph, red background) is set as the starting point of the dynamics. The two adult brain networks $\bm{G}_T^*$ are used for the inference of the model parameters. In particular, we use the framework of exponential random graphs (ERG) to infer the the topography of the functional landscape $F(G)$ \eqref{e-FL_ce}, as encoded in the parameters $\bm{\theta}\in\mathbb{R}^2$ (Lagrange multipliers) and $\bm{\lambda}\in\mathbb{R}_{+}^2$ (decay parameters of the curved statistics). The final estimations are obtained by mean-ERG construction. As for the parameters of the dynamics, the exploration rate $\mu\in\mathbb{R}_{+}$ is assumed to be uniform across node pairs and constant in time. It is evaluated by computing the average increase of connections per dyad and per unit time, from birth to adulthood. Finally, the functional pressure $\rho\in\mathbb{R}_{+}$ is used to inform the EE dynamics about the age of the adult worm and is fixed by minimising the Mahalanobis distance between the experimental model statistics at adulthood (average) and their simulated distribution based on \eqref{e-EE}. Our model of the \ce\ brain maturation has six parameters, all of which lend themselves to biological interpretation. 
    } 
    \label{f-methods}
\end{figure*}

To our best effort, no analytical treatment is possible for an EE dynamic \eqref{e-EE} with the $F$ score in \eqref{e-FL_ce}, therefore, we resort to simulations, described in sec. \ref{s-EE_sims}. In the context of the EE dynamic, three elements need to be set: the boundary conditions and the two EE parameters, i.e., the exploration rate $\mu$ and the exploitation rate $\varphi$ -- or, equivalently, the functional pressure $\rho$. The collections of the methods we employ is summarised in fig. \ref{f-methods}.

\subsubsection{Methods}

The boundary conditions are fixed by setting the birth connectome as the starting point for the EE dynamic \h$_8$, i.e., 
\begin{equation}\label{e-ce_bc}
    P(G=G_0^*, t=0) = 1 \ .
\end{equation}
In fact, as reported by \emph{Nicosia et al.} \cite{nicosia2013} the embryonic and post-embryonic stages represent two distinct phases in the maturation of the \ce\ brain, with the hatching (birth) serving as a watershed. Qualitatively, these two phases are likely driven by the same developmental principles. Yet, in quantitative terms, they differ fundamentally \sidenote{For instance, there is a prominent difference is in the rate at which the connections appear during development, which is accelerated in the pre-embryonic phase ($\sim N^2$, where $N$ is the number of neurons) and linear after birth ($\sim N$).}.
All of the data we use were collected at different post-embryonic stages. Therefore, in line with the hypothesis of functional homogeneity, we restrict our modeling to the post-embryonic developmental phase, setting the birth connectome as the starting point.

The removal of existing synaptic connection is rarely observed during the worm's brain maturation. Therefore, we adopt an exploration scheme where only the formation of new connections is permitted\sidenote{Luckily! As discussed in s.n. \ref{sn-syn_add}, the biology of the process of synapse formation coincides with that of the process of connection formation, but the same is not true of synapse elimination. In practice, for the \ce\ the latter does not occur. Using the unweighted representation of synaptic connections is therefore less harmful in this case.} \h$_9$. As previously reported, the number of synapses increases approximately linearly with time \cite{witvliet2021}. For simplicity, we further assume that the formation of new connections occurs uniformly across all neuron pairs \h$_{10}$. The exploration rate $\mu$ -- i.e., the number of edges added to the graph per dyad and per unit time -- is then computed as 

\begin{equation}\label{e-ce_mu}
    \mu^*=\frac{1}{TL}\sum_{i<j}\big[\bar{a}_{ij}(\bm{G}_{T}^*)-a_{ij}(G_{0}^*)\big]\ ,
\end{equation}

where $T = \SI{45}{\hour}$, $\sum_{i<j}a_{ij}(G_{0}^*)$ is the number of edges of the birth connectome and $\sum_{i<j}\bar{a}_{ij}(\bm{G}_{T}^*)$ is the mean\sidenote{We use here the notation of the bar for the mean value, in this case: $$\bar{a}_{ij}(\bm{G}_{T}^*) = \frac{1}{2}[a_{ij}(G_{T,1}^*) + a_{ij}(G_{T,2}^*)]$$} number of edges between the two adult worms.

This leaves us with only one degree of freedom, the functional pressure $\rho$. The value of this parameter is subject to biological tuning and is peculiar to both the process and the system under consideration -- in our case, the brain wiring dynamics and the \ce\ nervous system. The regulation of the functional pressure ensures that, as development progresses, specialised functional circuits emerge and mature appropriately before the organism reaches its adult stage. For this reason, we use it to inform the EE dynamics about the age of the adult worms\sidenote[][-1cm]{It is indeed a matter of time. For every $\rho>0$, the distribution of $F$ values shifts towards higher values and, waiting long enough, it reaches any desired target. Thus, it is not a question of \emph{if} the EE dynamics reach a predetermined functional value, but \emph{when} they do so. High values of $\rho$ rapidly advance the $F$ distribution while low values instead imply a slow progress.}. Intuitively, we intend to harness the degree of freedom of the functional pressure to minimise the distance between our simulations and the experimental values, at the adult age $T = \SI{45}{\hour}$.

\begin{figure}[htbp]
  \centering
    \includegraphics[width=\textwidth]{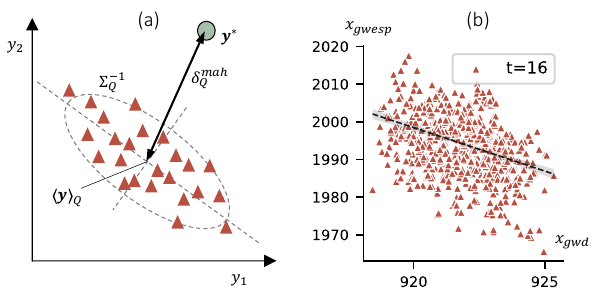}
    \caption{Mahalanobis distance in the space of model statistics \eqref{e-FL_ce_sts}. (a) Illustration of the Mahalanobis distance \eqref{e-mah_def}. It is a natural generalisation of the Euclidean distance to that between a point $\bm{y}^*\in\mathbb{R}^r$ (here, $r=2)$ and a multivariate distribution $Q$ in the same space, with mean $\av{\bm{y}}_Q$ and covariance matrix $\Sigma_Q^{-1}$. (b) Example of the correlation structure between the model statistics \eqref{e-FL_ce_sts} from an EE simulation of the \ce\ brain maturation. The two statistics show a clear anti-correlation. }
    \label{f-mah_dist}
\end{figure}

To do so, we resort to the Mahalanobis distance \h$_{11}$ \cite{mahalanobis1936}, fig. \ref{f-mah_dist}(a). Consider a a multivariate distribution $Q$ on $\mathbb{R}^r$ and point $\bm{y}^*\in\mathbb{R}^r$. The Mahalanobis distance  $\delta^{mah}_Q$ between the distribution $Q$ and the point $\bm{y}$ is defined as 
\begin{equation}\label{e-mah_def} 
    \delta^{mah}_Q = \sqrt{ (\av{\bm{y}}_Q-\bm{y^*})^{\top}\Sigma^{-1}_Q (\av{\bm{y}}_Q-\bm{y^*}) }
\end{equation}
where $\av{\bm{y}}_Q$ and $\Sigma_Q$ are the mean and covariance matrix of $Q$. It is a natural generalisation of the Euclidean distance and has two intriguing, i.e., (i) it accounts for the covariance structure of the distribution\sidenote[][-2cm]{It does so by computing the Euclidean distance of the \emph{whitened} (standardized) data. More precisely, consider the \emph{whitening} transformation \cite{kessy2018}
$$\bm{y}\rightarrow \Sigma_Q^{-\frac{1}{2}}\bm{y}\qquad   \forall\bm{y}\in\mathbb{R}^r\ , 
$$
where $\Sigma_Q^{-\frac{1}{2}}$ is the inverse principal square root of the covariance matrix $\Sigma_Q$. It is possible to show that (a) the transformed variables have unit diagonal (white) covariance matrix and (b) the distance \eqref{e-mah_def} corresponds to the Euclidean distance of the transformed variables.} and (ii) it is scale invariant\sidenote[][3cm]{This means that it is invariant under affine transformations of the form $$\bm{y}\rightarrow A \bm{y} + \bm{b} \ \ \forall \bm{y}\in\mathbb{R}^r\ ,$$ where $A$ is an $r\times r$ matrix and $\bm{b}\in\mathbb{R}^r$.
}.

Back to the EE dynamics, we compute \eqref{e-mah_def} where:
\begin{itemize}
    \item[$\circ$] $Q$ is the simulated two-dimensional distribution of model statistics $\bm{x} = (x_{gwd},x_{gwesp})$, computed as \eqref{e-emp_distr}, at the adult age $T = \SI{45}{h}$ -- we use $T$ to label the distribution.
    \item[$\circ$] $\av{\bm{y}}_Q = \av{\bm{x}}_T$ and $\Sigma_Q = \Sigma_T$ are the ensemble average and covariance matrix, at $T = \SI{45}{h}$, evaluated with \eqref{e-emp_av}.
    \item[$\circ$] $\bm{y}^*=\bar{\bm{x}}(\bm{G}_T^*)$ are the experimental values of the model statistics, averaged over the two adult connectomes, i.e., $y_i^* = [ x_i(G^*_{T,1}) + x_i(G^*_{T,2})]/2$.
\end{itemize}
The optimal functional pressure $\rho^*$ is then defined as: 
\begin{equation}
        \rho^* = \min_{\rho} \delta_T^{mah}\, \label{e-ce_rho}
\end{equation}
where
\begin{equation}
        \delta^{mah}_T = \sqrt{ (\av{\bm{x}}_T-\bar{\bm{x}}(\bm{G}_T^*))^{\top}\Sigma^{-1}_T (\av{\bm{x}}_T-\bar{\bm{x}}(\bm{G}_T^*)) }\ . \label{e-mah_T}
\end{equation}
The two properties of the Mahalanobis distance mentioned above turn out to be particularly useful in the present case. In fact (i) the two model statistics \eqref{e-FL_ce_sts} exhibit a pronounced anti-correlation, fig. \ref{f-mah_dist}(b). Moreover (ii) the distance \eqref{e-mah_T}, hence the estimation \eqref{e-ce_rho}, would not change if we scaled the statistics by their corresponding ERG parameter $\theta$. Therefore, \eqref{e-mah_T} offers a common ground for comparing models distinguished by varying sets of $\bm{\theta}$\sidenote{The same would not be true if instead we considered a distance function based on the $F$ metric.}.

Our model of the \ce\ brain maturation is exhaustively specified by the three expressions \eqref{e-ce_bc}, \eqref{e-ce_mu}, \eqref{e-ce_rho}. It has six parameters, all of which inferred by using the adult worms exclusively, all of which lend themselves to biological interpretation. 

If the theory is fully defined, our simulations have two more degrees of freedom that need to be set, i.e., the population size (number of samples) $M$ and the inverse time interval $\nu$, tab. \ref{t-EE-params}\sidenote{These are chosen in order to optimise the trade-off between computational time and reliability of the simulations. Small $M$ result in noisy empirical statistics \eqref{e-emp_av}, therefore $M$ should be as large as possible -- here, we have checked that the results do not change when using $M=2^{11},2^{12}$. As for $\nu$, setting it as described in the text helps in designing efficient simulations. The running time scales linearly with both $M,\nu$. With the settings described here, a single simulation run currently takes $\sim\SI{1.5}{\hour}$. Scripts available in the Github folder \href{https://github.com/dichio/EE-graph-dyn}{EE-graph-dyn}.
}. The first is set to $M=2^{10}$. The second is fixed by requiring that a single edge addition occurs in each graph during in a simulation step. This means choosing $\Delta t = (L\mu^*)^{-1}$.

As a final note, our simulations are currently not able to model the appearance of neurons (nodes)\sidenote{This affects $19$ of the $180$ adult brain neurons, see tab. \ref{t-neuron_list}. The majority of them appears during the post-embryonic burst of cell differentiation between the $L_1$ and $L_2$ stages, tab. \ref{t-sum_stats} and s.n. \ref{sn-neu_birth}.}. Consequently, we take the adult connectome as reference and embed the experimental networks at earlier developmental stages within larger networks that matched the adult node count of $180$ \h$_{12}$. 

\subsubsection{Results}
The exploration rate and the functional pressure of the EE dynamics for the \ce\ brain maturation -- evaluated by \eqref{e-ce_mu} and \eqref{e-ce_rho} -- turn out to be
\begin{align}
      \mu^* &= 1.426\times10^{-3}\  h^{-1}\ , \label{e-mu_est}\\
      \rho^* &= 9.017\times10^{2}\ . \label{e-rho_est}
\end{align}

\begin{figure}[h]
  \centering
    \includegraphics[width=7.37cm]{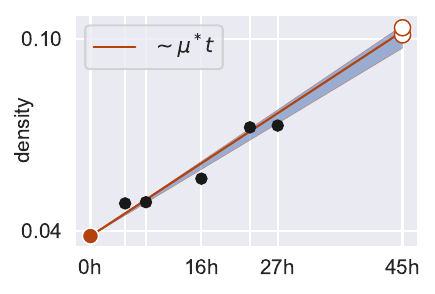}
    \caption{Exploration rate for the EE dynamics. 
    The number of connections at birth (red circle) and average of two adults (white circles) determine the constant exploration rate $\mu^*$ (red line) as per \eqref{e-ce_mu}. Thus, the number of connections grows linearly (red line). The shaded region represents a linear fit across all time points, including intermediate ages (black circles). We fit the line $y = \mu t + y_0$ and shade between lines with slopes $\mu^{**}\pm\delta\mu^{**}$ (same intercept $y_0^{**}$) -- double asterisks indicate the estimated values. The line $\sim\mu^* t$ falls within the shaded area. 
    }
    \label{f-ce_mu}
\end{figure}

\begin{figure}[htbp]
  \centering
    \includegraphics[width=8.36cm]{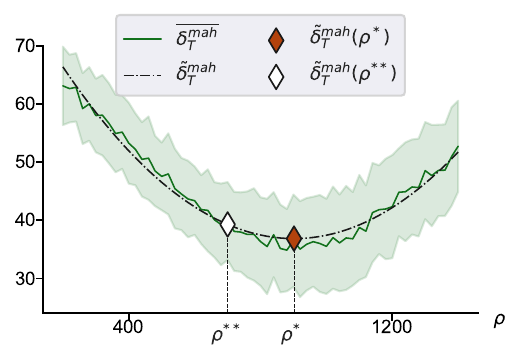}
    \caption{Functional pressure for the EE dynamics. We run $S=100$ simulations $\forall \rho\in \{200+20\ i, 0\le i\le60 \}$. For each $\rho$, we compute the mean and standard deviation of $\delta_T^{mah}$ \eqref{e-mah_T} over the $S$ simulations (green line and shaded area, respectively). We fit the mean values by a quadratic curve $\Tilde{\delta}_T^{mah}(\rho) = a\rho^2+b\rho+c$ (dash dotted line) and take its minimum $\rho^* = -b^*/2a^*$ as  estimate of the functional pressure for the EE dynamics. The same procedure, defining instead the Mahalanobis distance over the whole dataset \eqref{e-mah_allt}, gives $\rho^{**}$. The values $\Tilde{\delta}_T^{mah}(\rho^*)$ and $\Tilde{\delta}_T^{mah}(\rho^{**})$ are highlighted (red and white diamonds, respectively). The two overlap within one standard deviation.}
    \label{f-ce_rho}
\end{figure}

The exploration rate determines the frequency of structural changes, fig. \ref{f-ce_mu}. On average, $L\mu^*\sim 23$ new connections appear per hour. As for the exploitation strength, in fig. \ref{f-ce_rho} we show that the average Mahalanobis distance \eqref{e-mah_T} is a convex function of the functional pressure $\rho$. If $\rho$ is too low, the dynamics resemble a random formation of new connections, and are therefore unlikely to result in functional configurations. Less intuitively, if $\rho$ is too large, the dynamics also deviate from the experimental values, as they overestimated the strength of the functional selection. The \emph{true} biological process is therefore not an upper bound of the corresponding EE dynamics, in the $\rho\rightarrow\infty$ limit. Rather, it corresponds to a finite value of functional pressure that has been plausibly calibrated by its evolutionary history.

By design, the estimates \eqref{e-mu_est} and \eqref{e-rho_est} were obtained using only the adult stage of worm development. Nevertheless, we can legitimately ask how the estimates would change if instead we used the whole information available, including the configurations of the worm brain at intermediate developmental stages, i.e., those at $\SIlist{5;8;16;23;27}{\hour}$. For instance, (i) a linear fit of the rate at which new connections appear, using the entire time series, results in $\mu^{**}=(1.389\pm0.079)\times10^{-3}\ h^{-1}$, which encompasses the value in \eqref{e-mu_est}, fig. \ref{f-ce_mu}. Similarly, we can define for the functional pressure an equivalent minimisation problem to \eqref{e-ce_rho} over the whole dataset as:
\begin{equation}\label{e-mah_allt}
    \rho^{**} = \min_{\rho} \sum_{t \in \bm{t}^*, \ t>0 } \delta^{mah}_t \ ,
\end{equation}
where each $\delta^{mah}_t$ is defined as in \eqref{e-mah_T}, using the graph configuration(s) at time $t\in\bm{t^*}= (0,5,8,16,23,27,45)\ h$. This procedure results in $\rho^{**}= 7.001\times10^{2}$. The distance from the experimental values of the model statistics in the adult stage, using $\rho^{**}$, is compatible within one standard deviation with the one obtained using $\rho^*$, fig. \ref{f-ce_rho}. 

\begin{figure*}[h!]
  \centering
    \includegraphics[width=15cm]{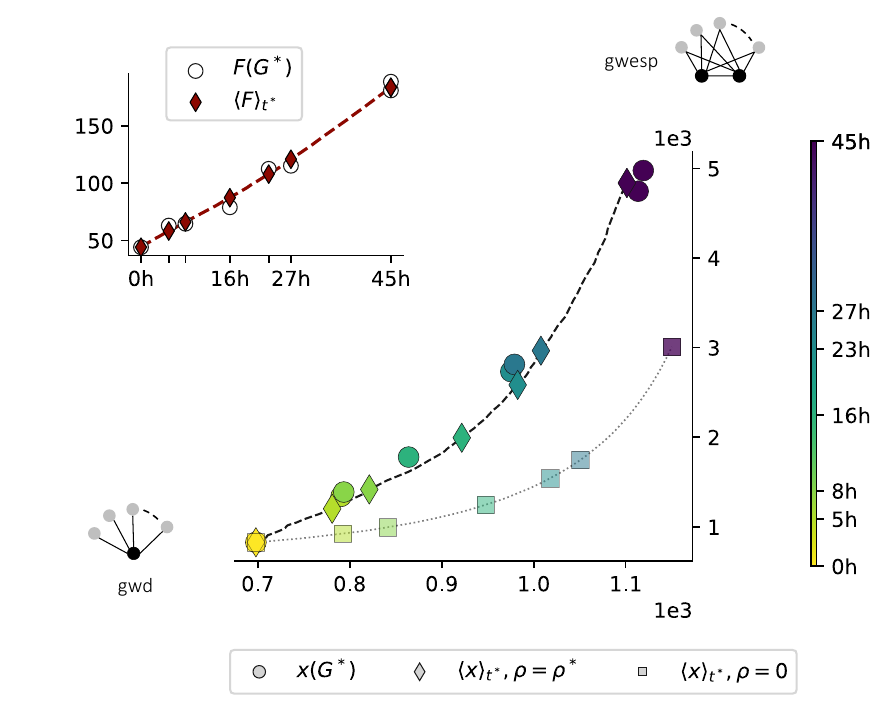}
    \caption{Optimal reconstruction of the \ce\ brain maturation. We fix $\mu^*, \rho^*$ as in \eqref{e-mu_est}, \eqref{e-rho_est} and run $500$ simulations. The one that minimises \eqref{e-mah_T} is here shown. Main: dynamics in the space of model statistics \eqref{e-FL_ce_sts}. Circles represent the eight experimental points for $t^*=\SIlist{0;5;8;16;23;27;45}{\hour}\ (\times\ 2)$. Diamonds indicate the expected values $\av{\bm{x}}_{t^*}$ based on the EE dynamics. Finally, squares indicate the expected values of the statistics using a null model in which the exploitation is turned off, $\rho=0$. Dashed and dotted lines shows the entire developmental trajectories, as obtained from the simulations. Inset: experimental values of the $F$ metric \eqref{e-FL_ce} (white circles), the corresponding expected values based on the EE dynamics (red diamonds and dashed line). Our optimal EE graph dynamic, informed by the birth and adult configurations of the worm brain network, also closely captures the developmental ages not included in the inference process. By construction however, our model cannot capture fine-scale details of the dynamics, e.g., the slow down of the experimental progression in the space of model statistics between $t=\SI{23}{\hour}$ and $t=\SI{27}{\hour}$.
    }.
    \label{f-ce_sim}
\end{figure*}

In essence, we find that using the full time series for inference, rather than just the adult stage, does not significantly alter the estimates of the parameters of the EE dynamics. This suggests that the adult stage captures most of the salient information about the brain wiring dynamics of the \ce, and other stages might have redundant or less significant impacts. 

To delve deeper into this observation, we can look at a single simulation of the \ce\ brain maturation process, obtained using the exploration rate and functional pressure $\mu^*,\rho^*$.  In fig. \ref{f-ce_sim}, we show the dynamic in the space of model statistics \eqref{e-FL_ce_sts}. For comparison, we also plot a null model of the same process in which $\rho=0$, i.e., a random formation of new connections at rate $\mu^*$. We find that the statistics at the adult stage are accurately reproduced by our simulation -- as it should be, in view of \eqref{e-ce_rho}. Remarkably, however, even though we did not use the other observed developmental ages for parameter inference, our simulations closely approximate them as well. This holds true for the statistics \eqref{e-FL_ce_sts} and, by consequence, for the $F$ metric \eqref{e-FL_ce}, which drives the dynamics.

This hints that our EE graph dynamics, informed about birth and adulthood, is able to capture the entire developmental trajectory\sidenote[][-2cm]{It is understood that this statement is contingent upon the representation chosen for both the system and the process. For example, by construction our model cannot reproduce any fine-scale detail of the worm brain networks related to node-specific effects -- since nodes are indistinguishable in \eqref{e-FL_ce} -- or to the directed nature of the connections -- since we use undirected graphs. Similarly, a constant mutation rate and functional pressure cannot reproduce transient dynamic patterns, which are apparent in fig. \ref{f-ce_sim}.}. Upon reflecting on the model's design, this provides \emph{a posteriori} validation for our hypotheses (i) that the model \eqref{e-FL_ce} encapsulates the fundamental drivers of the \ce\ brain's growth process, and that (ii) that the functional homogeneity discussed above holds true for the process considered. Importantly, our simulations are defined and open for inspection at all times $t\in[0,T]$, not just at the ages reconstructed in our source dataset. This opens up the possibility of using them to predict and generate reliable estimates for those stages of worm brain maturation for which there is currently no data.

\begin{figure*}[htbp]
  \centering
    \includegraphics[width=15cm]{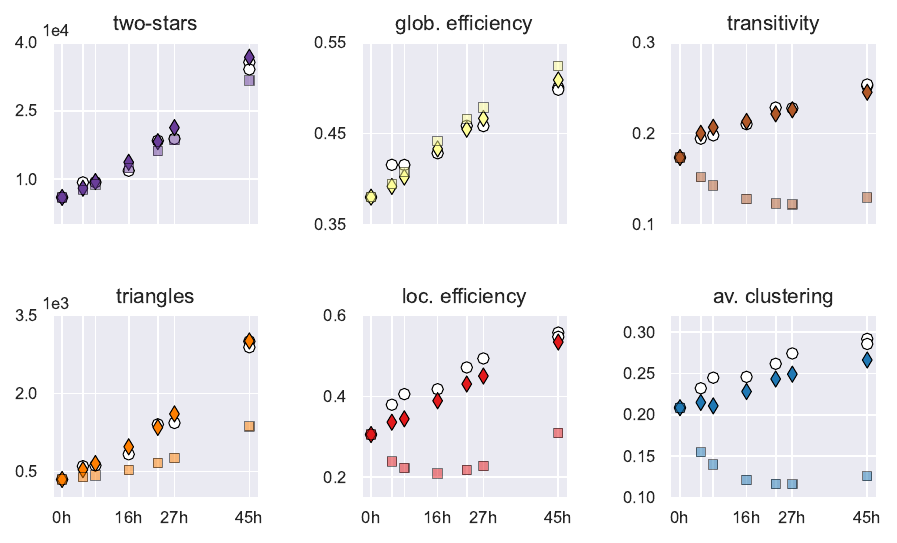}
    \caption{Feature generalisation for the EE model of the \ce\ brain network maturation. We compare the experimental values (circles) of a set of graph statistics with those obtained from simulations (diamonds) and those corresponding to a null model with no exploitation (squares). The same simulation illustrated in fig. \ref{f-ce_sim} is used here. The graph statistics include (i-ii) two motif counts (two-stars, triangles), which represent the fundamental building blocks of higher order graph patterns; (iii-iv) two efficiency measures (global and local efficiency), which quantify the efficiency of information transmission within the network at a global and local scale; (v-vi) two measures of clustering (transitivity and average clustering coefficient), which indicate the presence of tightly knit communities or groups. See app. \ref{a-measures} for a more detailed description of these (standard) graph measures. The null model (random formation of new connections) produces similar values to the experimental ones for the number of two stars and global efficiency, but fails elsewhere. On the contrary, our optimal EE model closely reproduces the experimental values for the entire set of graph statistics presented here.
    }
    \label{f-ce_gof}
\end{figure*}

\begin{figure}[h!]
  \centering
    \includegraphics[width=10cm]{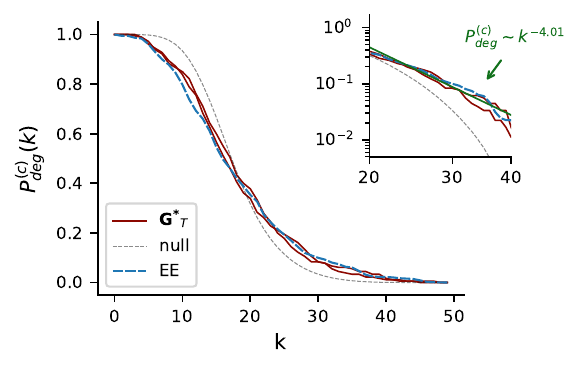}
    \caption{
    Complementary cumulative degree distribution at adulthood. The distributions associated to the two adult worms are shown as red solid lines. Blue dashed line for the curve obtained from simulations as $P_{deg}^{(c)}(k) = \frac{1}{N}\sum_{j\ge k}\av{x^{(j)}_d}$. Gray dashed line for the null model (no exploitation). Inset: zoom in the high-degree tail (loglog plot). Green dashed line corresponds to a power law behaviour. Our EE model is able to reproduce closely the cumulative degree distribution of the adult worms, including the high-degree tail behaviour. Similar plots can be obtained for each developmental age. 
    }
    \label{f-ce_gof_inst}
\end{figure}

A meaningful follow-up question to ask is whether our model is able to reproduce other features of the data -- not included in the formulation --, and which ones. We refer to this characteristic as \emph{feature generalisation}. In fig. \ref{f-ce_gof}, \ref{f-ce_gof_inst}, we demonstrate the ability of our model to generalise its predictions across different aspects of the data on which it was not explicitly trained. Once again, this is true not only at adulthood, but across the entire developmental process.

Collectively, these findings demonstrate that already a simple, low-dimensional, and coarse-grained model of the \ce\ brain maturation, properly informed about the final steady state of the process, can quantitatively replicate several features of the experimental data across development. Importantly, the model proposed has only six interpretable parameters -- four for the functional landscape, two for the dynamics\sidenote[][-2cm]{Back to the quote that opened this chapter: who knows what an elephant can do with six parameters! But for this thesis, this was our best effort. The whole anecdote related to that quote, as told by its main character, F. Dyson, can be found in  \cite{dyson2004}.}.
To the best of our knowledge, this is a first for neurodevelopmental biology studies\sidenote{Of course, this is mainly due to the fact that data such as that used here has only been available for a few years, but still.}. Returning to the general discussion of the brain wiring problem in sec. \ref{s-brain_wiring}, our EE dynamics provide an appropriate framework within which to specify a (genetically encoded) developmental algorithm. In the most conservative interpretation, the one described in this chapter is precisely an algorithm, consistent with the general features of the biological dynamics under investigation (in particular, the self-referentiality) and whose validity is demonstrated \emph{a posteriori} by the results discussed above. 

Nevertheless, we intend to put forward a bolder interpretation of the EE dynamics, at the local scale, consistent with our current understanding of the synapse formation process.

\section{Interpretation down to the synapse scale}[Interpretation]\label{s-ce_interpr}
\emph{Where a biological interpretation of EE dynamics is speculated. Where in particular individual neurons in a single developing system make local decisions based only on their limited knowledge.  }\bigskip 

An interpretation of our model requires the specification of how the EE dynamics is implemented by the biological process of synapse formation, which happens at the scale of the single neuron. Here, we speculate on a plausible biological interpretation of the EE dynamics, within a graph representation of the system.

\begin{figure*}[htbp]
  \centering
    \includegraphics[width=15cm]{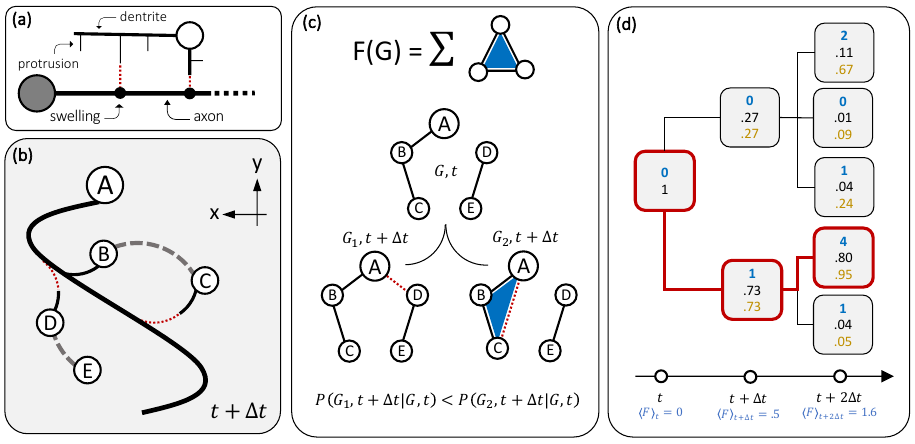}
    \caption{Interpretation of the EE dynamics. 
    (a) Synapse formation, schematics. Presynaptic sites, appear as swellings (black circles) on the axon shaft (thick black line) of a presynaptic neuron (gray circle). Postsynaptic neuronal processes - dendrites (black lines) and spine-like protrusions (thin black lines) - sprout from a postsynaptic neuron (white circle). Occasionally, they form synaptic connections with physically proximal presynaptic sites (red dashed lines). The presynaptic processes of the postsynaptic neuron and the postsynaptic processes of the presynaptic neuron are not shown here. (b) A simple scenario. We represent in a cartoonish physical space one presynaptic neuron $A$ and four postsynaptic neurons $B,C,D,E$. The axon shaft extending from $A$ is represented by the thick black line. At time $t$ there is a synaptic connection between the nodes $AB$ (black line connecting $B$ to the axon shaft). Additional connections exist between the neurons $BC$ and $DE$ (not shown in the physical space, indicated by the gray dashed line). After a time interval $\Delta t$, postsynaptic neuronal processes extend from the neurons $C,D$ towards the axon, potentially leading to new connections (red dotted lines). On the contrary, no such postsynaptic process exists for the neuron $E$. (c) Representation of the scenario in (b) in the corresponding graph space (undirected connections). We assume that the biological function ($F$ metric) simply consists in the count of triangles. The two potential connections between $AD$ and $AC$ at time $t+\Delta t$ can be represented as two different graph configurations, $G_1, G_2$, associated to different $F$ values. $G_2$, by virtue of its higher $F$, will be realised with higher probability. (d) Decision tree for two time steps of the EE dynamic (example). Here, $\Delta t =1$. Each square represents a graph. In blue, we indicate the $F$ values. In black, the unconditioned probabilities computed at each time as $\exp [F(G_i)]/\sum_j \exp [F(G_j)]$ where the sum runs over all graphs at that time (column). In brown, the probabilities conditioned on the previous time-point. They can be computed either as above, restraining the sum to those graphs that come from the same parent graph at the previous time, as in eq.(\ref{e-expl_ce}). Alternatively, they can be evaluated starting from the unconditioned EE probabilities and using $P(G_i,t+1|G_j,t)=P(G_i,t+1\cap G_j,t)/P(G_j,t)$, where $P(G_j,t)= \sum_k P(G_k,t+1\cap G_j,t)$. In bold-red we highlight the most likely developmental trajectory.
    }
    \label{f-ce_interpr}
\end{figure*}

Most \ce\ neurons have only one or two processes that extend in parallel bundles along the worm's body\sidenote{See sec. \ref{s-mind_worm} for an introduction to the \ce\ neurobiology.}. These processes grow during development, guided by molecular cues. Presynaptic sites appear as \emph{en passant} swellings on the shaft of the axon. The postsynaptic processes are dendrites or as spine-like protrusions \cite{witvliet2021}. Occasionally, these dendrites or protrusions form new synaptic connections, fig. \ref{f-ce_interpr}(a). 

To illustrate the essential idea, let us consider for simplicity the scenario illustrated in fig. \ref{f-ce_interpr}(b). At time $t$, there exist synaptic connections between neurons $AB$, $DE$, and $BC$. In the time interval $\Delta t$, postsynaptic processes from both neurons $C$ and $D$ grow sufficiently close to the axon shaft of $A$, and hold the potential to develop into new synaptic connections. Conversely, neuron $E$ has no such process, so it cannot form a connection with $A$. If $\Delta t$ is sufficiently small, only one of the two possible synaptic connections $AC$ or $AD$ is likely to be formed. Which of the potential connections materialises first is determined stochastically. However, the connection with a greater functional advantage will plausibly have a higher probability of forming. Let us suppose -- again, for simplicity -- that the notion of biological function for this system is simply represented by the number of triangles in the undirected graph representation of the system, fig. \ref{f-ce_interpr}(c). Consequently, we expect the $AC$ connection to form preferentially, given it leads to the formation of the $ABC$ triangle.

The example above illustrates how the EE dynamics could be implemented for a single developing system\sidenote[][-3.25cm]{The argument straightforwardly generalises to more complex notions of biological functions.}. From the standpoint of the individual neuron, the process of synaptogenesis consists of a series of stochastic decisions about which other neuron to connect with\sidenote[][-2.75cm]{It is worth stressing that this is a pictorial way of understanding the process and, clearly, only an effective description, which is based on (and assumes) a graph representation of the system. At the molecular level, the synaptogenesis is regulated by a complex forest of biochemical mechanisms. Strictly speaking, therefore, it is not necessary for neurons to "compute" any notion of biological function and "make decisions". Once again, the developmental rules are ultimately genetically encoded and biochemically implemented.}. These decisions are biased towards those connections that lead to higher functional gains, which in turn are evaluated based on the information available to the neuron at any given time.

According to this interpretation, the exploration consists in the formation (or extension) of neuronal processes that \emph{could} lead to a new synaptic connection and thus do not themselves correspond to the formation of physical connections\sidenote[][-1cm]{Note that this marks a sharp difference with the context of the evolutionary dynamics, where different exploration events are interpreted as a set of different genetic mutations, each associated with a distinct individual in a population. Here there is one and only one copy of the system, and the exploration events correspond to the formation of potential, not physical, connections. }. 

On the other hand, the exploitation consists in assigning higher probabilities of formation to those potential connections that would lead to higher functional gains. In particular, suppose that $G$ is the graph configuration at time $t$ and $\Tilde{\bm{G}}$ are the potential graph configurations at time $t+\Delta t$, then
\begin{equation}\label{e-expl_ce}
    P(G_i,t+\Delta t|G,t) = e^{\Delta t F(G_i)}/\sum_{G_j\in\bm{\tilde{G}}} e^{\Delta t F(G_j)} \ ,
\end{equation}
where we have taken $\varphi=1$ for simplicity. By definition, only one of them is eventually realised in $\Delta t$. The exploration-exploitation cycle is iterated throughout the process and defines the EE stochastic trajectory of a single, developing system in the configuration space.

We stress, however, that the information contained in \eqref{e-EE} is much more general, as it allows to compute the (unconditioned) probability of all possible configurations that could have appeared by the time $t$, including those that result from very unlikely developmental paths, fig. \ref{f-ce_interpr}(d). This provides the justification for using the EE dynamics to capture the intersubject variability of the brain wiring, that results from slightly different developmental trajectories.

\section{Paving ways}[Perspectives]\label{s-perspectives}
\emph{Where an agenda of the next steps can be found. Some easy to take, some more ambitious, some optimistic, none unfeasible. Enough for another PhD project, a bold postdoc or -- question by question, answer by answer -- an entire early career.}\bigskip

The model presented in this chapter for the brain wiring problem, like any other model, is shaped by the assumptions made during its formulation. Some relate to the choice of the representation, some to the model, some to the theory and some others to the biological system itself -- we list them all in tab. \ref{t-assumptions}. So far in this chapter, we have moved from broad themes to granular insights. In this final section, we do the opposite, rewinding the tape of our discussion from the specific to the general and pointing out a (small) subset of the (many) possible extensions and generalisations of the concepts and methods exposed above\sidenote[][-4.5cm]{Of all the sections typically found in a scientific paper, the Discussion is surely the most literary, the most replete with metaphors. 
Some papers claim to \emph{shed light} or \emph{open doors}. Others boldly \emph{navigate uncharted waters}, \emph{open horizons}, or \emph{break new ground}. Yet, some prefer the humility of merely \emph{scratching the surface}, being \emph{the tip of an iceberg}, or \emph{a drop in the ocean}. In our case, we do choose to \emph{pave ways}, a challenging task -- as most of the carpentry work --, but enduring and steadfast.
}.

\begin{table}[h]
  \arrayrulecolor{black}
  \centering
  \begin{tabularx}{.9\textwidth}{>{\centering\arraybackslash\hsize=.1\hsize}X>{\centering\arraybackslash\hsize=0.5\hsize}X>{\centering\arraybackslash\hsize=.3\hsize}X}
  \rowcolor[HTML]{F2EAD3}
    \h & \textbf{Description} & \textbf{Step beyond} \\
    \toprule
    1 & connections are synapses & $\bullet\bullet\bullet$ \\
    2 & graphs are unweighted & $\bullet\bullet$ \\
    3 & graphs are undirected & $\bullet$ \\
    4 & $F$ metric is ERG-like & $\bullet\bullet$ \\
    5 & inference from the adult stage & $\bullet\bullet$ \\
    6 & mean ERG ($F$ landscape) & $\bullet\bullet\bullet$\\
    7 & functional homogeneity & $\bullet\bullet$ \\
    8 & start from brain at birth & $\bullet\bullet$ \\
    9 & connections are only formed & $\bullet$ \\
    10 & uniform exploration rate & $\bullet$ \\
    11 & min. Mahalanobis distance  & $\bullet\bullet$ \\
    12 & no node dynamics & $\bullet$  \\
    \bottomrule
  \end{tabularx}
  \caption{A complete list of assumptions enforced in our \ce\ brain maturation model. They are listed in order of appearance in the text (first column). In the second column, we give a synthetic description. In the third column, we qualitatively indicate -- according to our  how difficult is the task of removing and/or generalising the assumptions: if easy ($\bullet$), challenging ($\bullet\bullet$) or arduous ($\bullet\bullet\bullet$). Those related to a lack of data are marked with three bullets, as there is not much we can do, as humble theoretical physicists.}
  \label{t-assumptions}
\end{table}

\subsubsection{The model}

First and foremost, our model of biological function \eqref{e-FL_ce} for the \ce\ brain. A number of factors (drivers and constraints) have the potential to play a role in molding the fine-scale details of an adult worm wiring, which have not been included in our coarse-grained model \cite{witvliet2021,pathak2020,brittin2021,moyle2021}. 

For instance, the graphs considered in this research are embedded in the physical space. One possible way of taking this into account would be to calculate the total cost of the \ce\ brain, i.e. the sum of the physical soma-soma distances between connected neurons. The latter is not strictly minimised \cite{perez2007,perez2009}, yet is likely to play a role in shaping the adult connectome\sidenote[][]{
However, this is a simplistic way of thinking at role of distance in the worm nervous system. As discussed, neuronal process mostly run in parallel bundles along the worm body. Therefore it is more the physical neighborhood of the neurons' processes that matters and in particular the contact area between adjacent processes. Such data is increasingly available \cite{brittin2021}. For instance, the contact area at birth between pairs of neurons correlates with the probability of a forming a new connections \cite{witvliet2021}.
} \cite{chen2006,nicosia2013,pathak2020}.

Homophily effects based on various cellular attributes influence the neuronal (synaptic) connectivity. Neurons in the adult worm brain are more likely to be connected if they differentiate close together in time \cite{varier2011} and if they belong to a bilaterally symmetric pair \cite{hobert2002}, tab. \ref{t-neuron_list}. In fact, homophily effects in the adult brain based on birth cohort (pre- or after-hatching) and symmetric pairing have been demonstrated \cite{pathak2020}. In addition, the symmetry of the wiring increases over the course of development \cite{witvliet2021}.

Synaptic connections are inherently directed, which is 
essential for detailing the information flow, fig. \ref{f-connections}. For example, hub neurons receive disproportionately more input connections while the number of outputs remains stable. In addition, synaptogenesis preferentially creates new connections in the direction from sensory to motor neurons, increasing the "feedforward bias", fig. \reffigshort{f-info_flow} \cite{witvliet2021}.

\begin{marginfigure}
	\includegraphics[height=4.5cm]{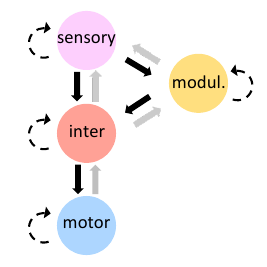}
    \caption{Information flow by cell type. Black, gray, and dashed black arrows indicate feedfoward, feedback and recurrent connections, as defined in \cite{witvliet2021}. The number (and strength) of feedforward connections during the \ce\ development.}
    \labfig{f-info_flow}
\end{marginfigure}

The ERG-like $F$ metric in our framework provides a straightforward means of incorporating all these effects into a model by formulating them as edge covariates \eqref{e-erg_ec}, with an appropriate choice of matrix $\gamma$, see sec. \ref{ss-mod_sel}. The ERG parameters estimation then has the role of weighting the relative importance of each factor.  

Finally, $F$ can be endowed with terms that can account for the presence of stable neuronal circuits in the wiring\sidenote{
A \emph{ functional circuit} is a  group of interconnected neurons that work together to perform a specific function or behavior \cite{hobert2003,metaxakis2018}. Examples are the navigation, touch sensitivity, chemosensation \cite{gray2005,chalfie1985,troemel1997}.
}. A more realistic picture of the wiring variability is that of a core circuit, that is conserved across individuals and development, embedded in a background of variable connectivity \cite{brittin2021,witvliet2021}. This (soft) constraint can be accounted for by a term in $F$ that penalises those graphs that do not contain a given functional circuit.

\subsubsection{The framework}

A different set of possible extensions of the model here presented concerns the EE framework illustrated in ch. \ref{c-EE}. 

A natural initial step is to integrate node formation dynamics alongside the EE edge dynamics. Indeed, 19 out of the 180 neurons in the adult worm brain differentiate post-hatching. The time of their emergence is documented in the literature \cite{varier2011}. 

It is more challenging to relax the assumption of time homogeneity of the process and to introduce a time dependence for the parameters of the dynamics, e.g., $\mu(t)$, $\rho(t)$.  The latter case is the one we would generally expect, since the worm brain maturation is influenced by several internal (and external) factors, that might manifest as a slow-down or speed-up of the developmental process\sidenote{The reader who has taken the trouble to read ch. \ref{c-Darwin} will recognise the parallel of this discussion (and others in the present section) with the one in sec. \ref{ss-fit_fun}. For instance, we are here proposing to move from a functional landscape to a \emph{functional seascape}, which might possibly co-develop with the system itself.
} -- e.g., fig. \ref{f-ce_sim}. Here, however, we are essentially limited by the data -- seven developmental ages --, which do not justify the use of more refined time dependencies. 

As we have discussed in sec. \ref{ss-wbm}, the developmental processes during the embryonic and post-embryonic stages operate in distinct regimes and the latter has been studied here. We argue that the general EE structure of the problem should not change, while the parameters and possibly the $F$ metric should. It would then be interesting to design an EE model of the embryonic phase of the \ce\ brain maturation and connected it to the post-embryonic model here described, at the moment of hatching.

Our EE simulation framework can accommodate (with varying degrees of effort) the ensemble of generalisations outlined here, mainly by adapting the mathematical structure of the parameters\sidenote[][-1.5cm]{For instance, a constant, uniform and scalar exploration rate would become a matrix to account for non-uniform rates, a vector to account for time variability, or a vector of matrices in the case of both.} and/or of the dynamic entities. 

\subsubsection{The system}
Time to get ambitious. If we insist on a model for the brain wiring problem grounded on the experimental evidence, data are vital. The kind of data relevant to the research question under consideration here - in particular, electron microscopic reconstructions of nervous systems - is becoming increasingly available for ever larger systems. Let us therefore imagine for a paragraph (hopefully near) future in which we have the data we are looking for. 

There is no fundamental reason to restrict our analysis to the connection defined as chemical synapses. A complete map of gap junctions (undirected connections) for the \ce\ across development would be easily incorporated into our analysis. 

There is also no fundamental reason to restrict to the nematode \ce\, as the general principles that drive the wiring development do not depend on the specific system by which they are implemented. An EE model of the brain wiring dynamics for the fruit fly, mouse, zebrafish, tadpole and more\sidenote[][-3cm]{We will resist the temptation to include the human brain in this list, since, to the present day, it would be ludicrous to even think of having such data. We leave here this side-note, in the hope of returning to it one day and being amazed at how quickly it has been disproved.} \cite{dorkenwald2023,helmstaedter2013,lee2016,abbott2020,hildebrand2017,ryan2016} might be within reach. It would be then interesting then to use the EE framework as a common ground to compare equivalent models across different natural nervous systems\sidenote[][-1cm]{For example, the dream plot we have in mind is a scatterplot, where on the two axes we have $\mu,\rho$ and each point represents an EE model of a different brain system.}.

Finally, it would be interesting to go beyond connectomics and look at functional connectivity\sidenote{Connectomics assumes that the anatomical map of connections, like synapses and gap junctions, is key to understanding neural functions. While essential, this overlooks aspects such as inhibitory or excitatory nature of synapses, extrasynaptic communication via signaling molecules \cite{bentley2016}, and the timescales of signals propagation.} through the same lens \cite{seung2011}. In fact, a more direct relation with the notion of biological function exists for the signal propagation atlas, as the one very recently reconstructed for the \ce\ by \emph{Randi et al.} \cite{randi2022}.
Here, individual neurons are excited by optogenetic stimulation and the activity induced in other neurons is recorded, thus defining a graph of directed, weighted functional connections.

\chapter{Conclusions} \label{c-conclusions}

\begin{quote}
\begin{flushright}
\emph{Addo' arriv' chiant' u zipp'} \sidenote{Tr.: Wherever you get, plant a stick.}.

-- Popular Lucanian wisdom
\end{flushright}
\end{quote}
\bigskip

\begin{figure*}[t!]
	\includegraphics[width=17cm]{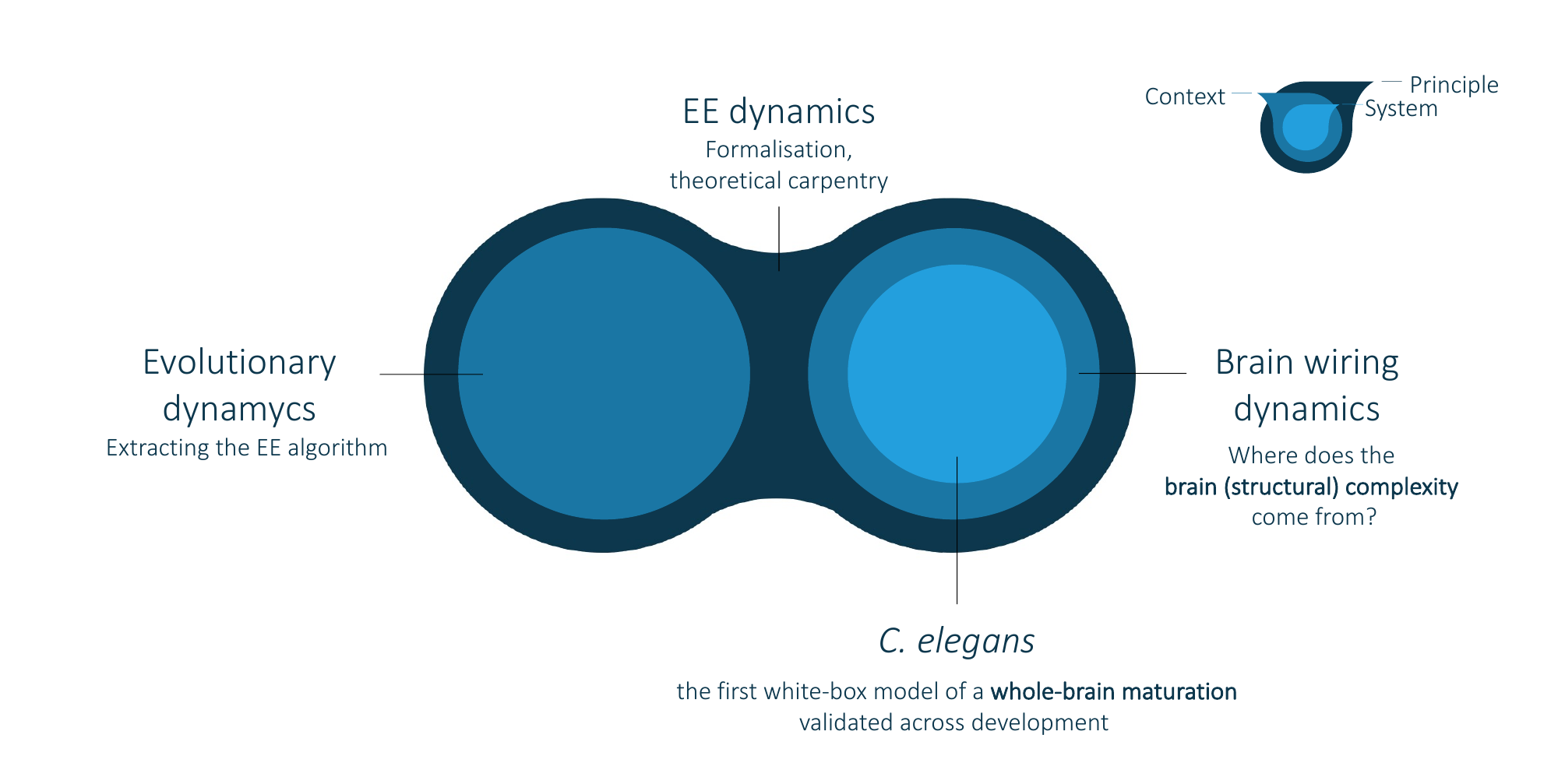}
	\caption{Visual summary of a PhD project. We started from the context of evolutionary dynamics to formulate a context-free exploration-exploitation (EE) problem, which is claimed to be general for a class of biological dynamics. We developed a theoretical framework, studied it analytically and developed simulations. We have then used it to tackle another type of biological dynamics, namely the developmental dynamics of a nervous system. In this context, we studied a specific system, the \ce\ brain. We have developed an EE white-box model of brain growth that has been validated throughout development.}
    \label{f-summary}
\end{figure*}

The discussion presented up to this point has chronicled the birth, development and implementation of what Schrodinger would have called a \emph{naive physicist's idea about organisms} \cite{schroedinger1944}. As we stand at the threshold of the conclusion of both this manuscript and this academic project, it is an opportune moment to step back, reflect upon the journey undertaken, and discern the patterns that have emerged from the collective body of work presented, fig. \ref{f-summary}.

At the heart of our scientific discourse was the exploration-exploitation (EE) paradigm, which was posited as a general dynamic principle for biological systems. It applies whenever the dynamics of a system arise from the interplay of (i) the variability introduced by stochastic state changes and (ii) a state-dependent optimisation of a biological function.

One obvious context in which the EE paradigm manifests itself is the evolutionary dynamics. In the simplest scenario, the latter results from the combined effect of random genetic mutations (exploration) and natural selection (exploitation). Crucially, this example showcases the self-referential nature of biological dynamics. For these reasons, we have devoted (ch. \ref{c-Darwin}) to discussing the core concepts of evolution, ranging from biological foundations to modelling efforts, to algorithms inspired by the evolutionary processes.

Our foray into the realm of evolutionary biology had an underlying purpose, which was made explicit in (ch. \ref{c-EE}). From the specific case of evolution, we learned the formal structure of a general exploration-exploitation dynamics \eqref{e-EE}\sidenote{This, holding fixed the mathematical representation of the system -- essentially a string of zeros and ones.}. We chiseled away the context-dependent details of the evolutionary dynamics to unveil and discuss the underlying context-free EE algorithm. The resulting theoretical picture is that of a stochastic evolution of a probability distribution on a functional landscape. The study of analytically tractable toy models allowed us to elucidate the main characteristics of the EE dynamics.

Then, we took the leap. We started (ch. \ref{c-Celegans}) by arguing that the brain wiring dynamics -- i.e., the development from birth to adulthood of a nervous system, here the graph of neurons and connections between them  -- is another manifestation of the EE paradigm. To work out the details, it is necessary to focus on a particular system, in our case it was the brain of the nematode \ce. Within the EE framework, we were able to specify a model of worm brain maturation with only six parameters, all inferred from data and all amenable to biological interpretation. We offered a putative, biologically realistic  interpretation of the EE dynamics in terms of the synapse formation process.

Our main result is that a parsimonious characterisation of the adult \ce\ brain combined with our EE dynamics is able to quantitatively reproduce the entire developmental trajectory, as reconstructed experimentally by serial section electron microscopy. To the best of our knowledge, this stands as the first theoretical model of system-wide neurodevelopmental dynamics for a living system, that is (i) firmly anchored in experimental data across development and (ii) wholly interpretable. More generally, our results support the recently proposed view of brain wiring dynamics as driven by a set of simple and genetically encoded wiring rules.

There is one thorny issue that is the main theoretical bottleneck of the EE approach. What is the \emph{biological function}? More specifically, what is the mathematical expression of the $F$ metric? How to identify the relevant features? How to learn the topography of an $F$ landscape from the data? These problems, of course, are much more general. These questions echo a foundational inquiry in physics: what is the energy function of a physical system? When it comes to complex systems -- which includes, but is not limited to biological systems --, there is no Delphi's oracle, the answer is nuanced, intricate\sidenote{The lexicon varies as a function of the context, so that energy f., cost f., utility f., fitness f., our biological f. (...) all essentially refer to the same mathematical entity.
}.

To build a functional landscape, we have proposed here the use of an inferential approach that is both principled and data-driven. In particular, we used the maxent inference scheme associated with the exponential random graph (ERG) models. This approach was apt for studying the brain wiring dynamics because (i) the data were naturally represented as a graph, and (ii) inference needed to be drawn from a singular realisation of the system. In (ch. \ref{c-ergm}), we dwelt on the theoretical underpinnings of the ERG models and offered a pedagogical guide to assist the interested users in their application.

A fruitful theory is a theory that leaves the theorist with more and more precise questions about the subject matter than at the beginning. Several have been sketched in sec. \ref{s-perspectives} for the brain wiring problem, and represent the agenda for the upcoming developments. Yet, I claimed generality\sidenote{Naturally, generality does not equate to explaining \emph{everything}. There are numerous dynamical processes in biology that cannot be adequately described or explained within the exploration-exploitation framework. Just as gravity offers little insight into the bizarre nature of strong nuclear interactions. It is trivial to say, but better to say it.} for the EE paradigm and for such a claim demonstrating its applicability in at least two distinct contexts -- evolution, brain wiring -- was only the bare minimum. This manuscript should therefore serve as a guide to unifying and approaching new problems in biology along the same lines -- some mentioned in the text, many more probably unforeseen by the writer himself. As is often the case at the beginnings, our progress is but a grain in the granary of what remains.

It is no job for the hasty, though. One profound lesson from physics resonates especially when it comes to the study of living systems. That is, general principles do not emerge unless we look for them. This is the central message that we hope will be a legacy of this work.

Here we stand. We shall resist the temptation to dismiss the phenomena of life as too messy for the physics-style of scientific inquiry. If we do so, then a theoretical physics of biological systems -- built on solid, compelling principles and grounded in experimental data -- becomes not only possible but also one of the most fascinating frontiers\sidenote{I may not be completely impartial.} of modern physics.

\appendix 

\pagelayout{wide} 
\addpart{Appendix}
\pagelayout{margin} 


\pagelayout{wide}
\setchapterstyle{kao}
\chapter{Pen and paper EE dynamics}\label{a-EE_math}
We here provide a step-by-step derivation of the results discussed in sec. \ref{ss-EE_EL}, \ref{ss-EE_DL}, for EE dynamics under simple scenarios. The three fundamental ingredients are (i) the EE dynamics in the continuous time limit \eqref{e-EE_c} (ii) a formal specification of the $F$ metric as in \eqref{e-EE_EL}, \eqref{e-EE_DL} and (iii) the dynamic of any graph observable $O:\mathcal{G}\mapsto\mathbb{R}$, i.e.,
\begin{equation}\label{e-EE_expectation}
    \frac{d}{dt}\av{O}_t = \frac{d}{dt}\sum_GO(G)P(g,t)=\sum_GO(G)\ \frac{d}{dt}P(G,t)\ .
\end{equation} 

\section*{Energy-like biological function}
Consider the $F$ metric \eqref{e-EE_EL}. The exploitation term in the EE dynamics \eqref{e-EE_c} can be written as
\begin{align}
    [F(G)-\av{F}_t]P(G,t) &\stackrel{}{=} -\frac{1}{L}\sum_{i<j}\big[a_{ij}-\av{a_{ij}}_t\big]P(G,t)\ , \\
    &\stackrel{(a)}{=} -\frac{1}{L}\sum_{i<j}\Bigg[\frac{1+\sigma_{ij}}{2}-\Big\langle\frac{1+\sigma_{ij}}{2}\Big\rangle_t\Bigg]P(G,t)\ , \\ 
    &\stackrel{}{=} -\frac{1}{2L}\sum_{i<j}\big[\sigma_{ij}-\av{\sigma_{ij}}_t\big]P(G,t)\ ,    
\end{align}
where in $(a)$ we have used \eqref{e-sp_dy}. Note that, when switching from the bit-wise to the spin-wise representation of the dyadic variables, it is also implied that:
\begin{equation}
    \sum_G = \sum_{a_{11}=0,1} \dots \sum_{a_{LL}=0,1} = \sum_{\sigma_{11}=\pm1} \dots \sum_{\sigma_{LL}=\pm1} \ .
\end{equation}
The dynamics of the expected value $\av{\sigma_{ij}}$ can be evaluated using \eqref{e-EE_expectation}.
\begin{align}
    \frac{d}{dt}\av{\sigma_{ij}}_t &\stackrel{}{=} \sum_G \sigma_{ij} \Bigg\{ \mu\sum_{k<l} [P(M_{kl}G,t)-P(G,t)]  -\frac{\varphi}{2L}\sum_{k<l}\big[\sigma_{kl}-\av{\sigma_{kl}}_t\big]P(G,t) \Bigg\}\notag \\
     &\stackrel{(a)}{=} -2\mu\av{\sigma_{ij}}_t-\frac{\varphi}{2L} \sum_{k<l} \sum_G \big[\sigma_{ij}\sigma_{kl}-\av{\sigma_{ij}}_t\av{\sigma_{kl}}_t\big]P(G,t) \notag \\
     &\stackrel{}{=} -2\mu\av{\sigma_{ij}}_t-\frac{\varphi}{2L} \sum_{k<l} \big[\av{\sigma_{ij}\sigma_{kl}}_t-\av{\sigma_{ij}}_t\av{\sigma_{kl}}_t\big] \ ,
\end{align}
where in $(a)$ we have used \eqref{e-EE_no_exp}. The latter corresponds to \eqref{e-EL_av_spin}. Under the hypothesis of decoupling approximation \eqref{e-EL_DA}, we discard all terms in the last sum except $\av{\sigma_{ij}\sigma_{ij}}_t-\av{\sigma_{ij}}_t\av{\sigma_{ij}}_t= 1-\av{\sigma_{ij}}_t^2$ and obtain \eqref{e-EL_av_spin_DA}. The same differential equation holds for the magnetisation,
\begin{equation}\label{e-EL_magn_DA}
    \dot{m_t} = -2\mu m_t - \frac{\varphi}{2L}\big[1-m^2_t\big]\ .
\end{equation}
Solving \eqref{e-EL_magn_DA} is a simple calculus exercise -- separable variables, partial fraction decomposition. The general solution is 
\begin{equation}
    m_t = m_2\Bigg[1+\frac{m_1/m_2-1}{1+ce^{2\mu t \sqrt{1+(\rho/2L)^2}}} \Bigg] \ .
\end{equation}
Fixing the constant $c$ by requiring $m_{t_0}=m_0$ results in \eqref{e-EL_magn}.

\section*{Distance-like biological function}
We follow the exact same steps as in the previous section. Consider the $F$ metric \eqref{e-EE_DL}, we can rewrite it as
\begin{align}\label{e-DL_F}
   F(G) &\stackrel{(a)}{=} -\frac{1}{L^2}\Big[\sum_{i<j}\frac{1+\sigma_{ij}}{2}-E^*\Big]^2\notag\\
    &\stackrel{}{=} -\frac{1}{L^2}\Big[\Big( \frac{L}{2}-E^* \Big)^2 + \Big( \frac{L}{2}-E^* \Big)\sum_{i<j}\sigma_{ij} + \frac{1}{4}\sum_{i<j,k<l}\sigma_{ij}\sigma_{kl}\Big]\notag\\
    &\stackrel{}{=} -\frac{1}{L^2}\Big[\Big( \frac{L}{2}-E^* \Big)^2 + \Big( \frac{L}{2}-E^* \Big)\sum_{i<j}\sigma_{ij} + \frac{L}{4}+\frac{1}{4}\sum_{\substack{i<j,k<l\\(i,j)\ne(k,l)}}\sigma_{ij}\sigma_{kl}\Big]
\end{align}

where in $(a)$ we used \eqref{e-sp_dy}. The exploitation term in the EE dynamics \eqref{e-EE_c} can be written as

\begin{equation}
    [F(G)-\av{F}_t]P(G,t) \stackrel{}{=} -\frac{1}{L^2}\Bigg[\Big( \frac{L}{2}-E^* \Big)\sum_{i<j}\big[\sigma_{ij}-\av{\sigma_{ij}}_t\big] + \frac{1}{4}\sum_{\substack{i<j,k<l\\(i,j)\ne(k,l)}}\big[\sigma_{ij}\sigma_{kl}-\av{\sigma_{ij}\sigma_{kl}}_t\big]\Bigg]P(G,t)\ ,
\end{equation}
where all the terms that are constant in \eqref{e-DL_F} cancel out. The dynamics of the expected value $\av{\sigma_{ij}}$ can be evaluated using \eqref{e-EE_expectation}.

\begin{alignat}{2}
    \frac{d}{dt}\av{\sigma_{ij}}_t &\stackrel{}{=} -2\mu\av{\sigma_{ij}}_t-\frac{\varphi}{L^2}\Bigg[&&\Big( \frac{L}{2}-E^* \Big)\sum_{k<l}\big[\av{\sigma_{ij}\sigma_{kl}}_t-\av{\sigma_{ij}}_t\av{\sigma_{kl}}_t\big] \ +\notag\\
    & &&+\frac{1}{4}\sum_{\substack{k<l,m<n\\(k,l)\ne(m,n)}}\big[\av{\sigma_{ij}\sigma_{kl}\sigma_{mn}}_t-\av{\sigma_{ij}}_t\av{\sigma_{kl}\sigma_{mn}}_t\big]\Bigg]P(G,t)\ ,\notag\\
    &\stackrel{}{=}-2\mu\av{\sigma_{ij}}_t-\frac{\varphi}{L^2}\Bigg[&&\Big( \frac{L}{2}-E^* \Big)\Big[1-\av{\sigma_{ij}}_t^2+\sum_{\substack{i<j,k<l\\(i,j)\ne(k,l)}}\big[\av{\sigma_{ij}\sigma_{kl}}_t-\av{\sigma_{ij}}_t\av{\sigma_{kl}}_t\big]\Big] \ +\notag\\
    & &&+\frac{1}{2}\sum_{\substack{k<l\\(i,j)\ne(k,l)}}\big[\av{\sigma_{kl}}_t-\av{\sigma_{ij}}_t\av{\sigma_{ij}\sigma_{kl}}_t\big]\ + \notag \\
    & &&+\frac{1}{4}\sum_{\substack{k<l,m<n\\(i,j)\ne(k,l)\ne(m,n)}}\big[\av{\sigma_{ij}\sigma_{kl}\sigma_{mn}}_t-\av{\sigma_{ij}}_t\av{\sigma_{kl}\sigma_{mn}}_t\big]\Bigg]P(G,t)\ .
\end{alignat}
Enforcing now the decoupling approximation \eqref{e-DL_DA} we get:
\begin{equation}
    \frac{d}{dt}\av{\sigma_{ij}}_t \stackrel{}{=} -2\mu\av{\sigma_{ij}}_t-\frac{\varphi}{L^2}\Bigg[ \Big( \frac{L}{2}-E^* \Big)\big[1-\av{\sigma_{ij}}_t^2\big] + \frac{1}{2}\big[1-\av{\sigma_{ij}}_t^2\big] \sum_{\substack{k<l\\(i,j)\ne(k,l)}}\av{\sigma_{kl}}_t\Bigg]
\end{equation}
Using the same initial conditions for all dyads, the last sum can be approximated as $\sim(L-1)\av{\sigma_{ij}}$. The same differential equation can then be written for the magnetisation $m_t$, the result is precisely \eqref{e-DL_magn_DA}.

\pagelayout{margin} 



\chapter{Network measures}\label{a-measures}

The general purpose of network measures is \emph{descriptive}. They do not add information, quite the opposite. Computing measures on a network entails excluding all information except that which relates to the specific attribute of the network we seek to illuminate. We here provide a synthetic summary\sidenote{Detailed discussions and long catalogues of other measures can be found in any monograph on network science, our main references are \cite{newman2018,latora2017}. We will not here enter into competition with the thousands of items in the academic literature that have covered these topics.} of the network measures used in sec. \ref{ss-wbm}. Popular libraries such as \texttt{NetworkX} for \texttt{Python} or \texttt{igraph} for \texttt{R/Python} can be used for computation. 

Let us consider, as usual, an undirected, unweighted graph (or network) $G$ \eqref{e-graph_def}. The degree of the node $i$, we recall, is defined as:
\begin{equation}\label{e-degree}
    k_i = \sum_{j}a_{ij}\ .
\end{equation}
We can categorise our measures into three groups, based on the specific network feature they examine: clustering, efficiency and degrees.

\subsubsection{Network motifs}
Much of the content of this manuscript is based on the enumeration of network \emph{motifs}. Essentially, these are identifiable patterns or subgraphs within a network that occur more frequently than would be statistically expected in a random network \cite{milo2002}. Identifying a motif count entails (i) counting the occurrences of a given subgraph and (ii) evaluating its statistical significance.

In a few fortunate cases\sidenote{These happen to span the totality of the cases discussed in ch. \ref{c-Celegans}. Our EE simulations gain considerable advantage -- in terms of computation time -- from explicit formulae for the computation of an $F$ metric building blocks.}, the count of network motifs can be expressed in terms of powers of the adjacency matrix. The simplest motif count is the number of edges, trivially. The number of connected triples $\#_{\wedge}$ and that of triangles $\#_{\triangle}$ can be expressed as:

\begin{marginfigure}[1cm]
    \centering
	\includegraphics[height=3cm]{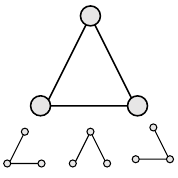}
    \caption{A triangle (top) is a a triple of nodes $i,j,q$ with $a_{ij}=a_{iq}=a_{ja}=1$. A connected triple (below) is a pair of edges $a_{ij}=a_{iq}=1$. Each triangle contains three connected triples.}
    \labfig{f-triangles}
\end{marginfigure}

\begin{align}
    \#_{\wedge}(G) &= \sum_{i<j,q}a_{iq}a_{jq} = \frac{1}{2}\sum_{i,j}(G^2)_{ij} - Tr(G^2)\ . \label{e-conn_triples}\\
    \#_{\triangle}(G) &= \sum_{i<j<q}a_{ij}a_{jq}a_{iq} = \frac{1}{6}\ Tr(G^3)\ ,\label{e-triangles}
\end{align}
where $Tr$ is the trace operator, fig. \reffigshort{f-triangles}. The number $x_d^{(k)}$ of nodes with degree $k$ is simply
\begin{equation}\label{e-deg_k}
    x_d^{(k)}(G) = \sum_i\delta_{k,k_i}\ ,
\end{equation}
where $\delta$ is the Kronecker delta, a $k_i$ as in \eqref{e-degree}. Finally, the number $x_{esp}^{(k)}$ of connected dyads whose extremal nodes share exactly $k$ partners is found as:
\begin{equation}
    x_{esp}^{(k)}(G) = \sum_{i<j} \delta_{k,B_{ij}} \ \text{with}\ B = G^2 \odot G
\end{equation}
where $\odot$ is the Hadamard (element-wise)
product, $B_{ij}=\sum_q a_{iq}a_{qj}a_{ij}$. In ch. \ref{c-Celegans}, the statistical significance of the motif counts was assessed either within the ERG framework (non-zero inferred parameter) or by direct comparison with a null model.

\subsubsection{Clustering}
A common property of a number of real-world networks -- in particular, social networks -- is the presence of tightly knit communities or groups \cite{watts1998,strogatz2001}. A straightforward manifestation of such a \emph{clustering behaviour} is a higher-than-random connection probability for two nodes that share a common partner. In other words, if nodes $i$ and $j$ are both connected to node $q$, they are more likely to be directly connected to each other as well. Such a behaviour can be quantified by the following two metrics:
    \begin{itemize}
    \item[$\circ$] \emph{Transitivity}, $T$. It is the ratio between the number of existing triangles $\#_{\triangle}$ \eqref{e-triangles} and the number of connected triples $\#_{\wedge}$ \eqref{e-conn_triples}. Formally,
        \begin{equation}\label{e-transitivity}
            T = 3\ \frac{\#_{\triangle}(G)}{\#_{\wedge}(G)}\ ,\quad  T \in [0,1]\ .
        \end{equation}
    \item[$\circ$] \emph{Average clustering coefficient}, $C$. The local clustering coefficient for a node $i$ is defined as
    \begin{marginfigure}[0cm]
    \centering
	\includegraphics[height=4cm]{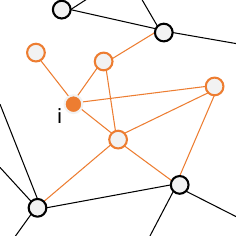}
    \caption{Subnetwork induced by the node $i$ and its neighbors (orange). Here, $i$ has four neighbors. There exist six possible pairs of neighbors, two of which are connected, therefore $C_i=1/3$.}
    \labfig{f-subnetwork}
    \end{marginfigure}
    \begin{equation}
        C_i = \frac{\text{\# connected pairs of neighbors of } i}{\text{\# pairs of neighbors of } i} = \frac{\sum_{j<l}a_{ij}a_{jl}a_{il}}{\frac{1}{2}k_i(k_i-1)}\ .
    \end{equation}
    It is computed by considering the subnetwork induced by the node $i$ and its first neighbors and quantifies the relative number of neighbors of $i$ that are also themselves neighbors, fig. \reffigshort{f-subnetwork}. 
    The average clustering coefficient then simply takes the average value over the node set:
    \begin{equation}\label{e-clustcoeff}
        C = \frac{1}{N} \sum_{i=1}^N C_i\ ,\quad  C \in [0,1]\ .
    \end{equation}
\end{itemize}
Although both $T$ and $C$ approach a value of $1$ in the limit of perfect transitivity, they do not convey identical information. The clustering coefficient is relatively more sensitive to nodes with low degrees because it averages across all nodes. On the contrary, transitivity is more affected by nodes with high degrees, as they are the ones that influence the number of triangles more. As a result, while transitivity provides a more comprehensive picture of the overall structure of the network, the clustering coefficient sheds more light on local structures or subnetworks embedded in the larger network.

\subsubsection{Efficiency}

Another class of metrics in network science is designed to quantify the \emph{efficiency} of information or resource transmission within the network \cite{latora2001}. This notion is based on the fundamental premise that the proximity of two nodes in a network graph strongly correlates with the efficiency of their information exchange. The distance $d_{ij}$ between two given nodes $i,j$ is defined as the length of the geodesic between them, i.e., the number of edges that form the shortest path from one to the other, fig. \reffigshort{f-subnetwork_2}. By definition, if $i,j$ are disconnected, $d_{ij}=\infty$.

\begin{marginfigure}[0cm]
    \centering
	\includegraphics[height=4cm]{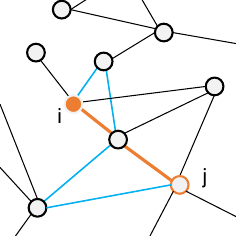}
    \caption{The geodesic between the nodes $i, j$ (orange) has length $d_{ij}=2$. For comparison, another, longer path connecting the same two nodes is highlighted (cyan).}
    \labfig{f-subnetwork_2}
\end{marginfigure}
    \begin{itemize}
        \item[$\circ$] \emph{Global efficiency}. It is defined as the harmonic mean of geodesic lengths. The global efficiency $E_{g}$ is defined as 
        \begin{equation}\label{e-efficiency}
            E_{g} = \frac{1}{L}\sum_{i < j}\frac{1}{d_{ij}}\ .
        \end{equation}
        A network exhibiting high global efficiency typically features brief paths connecting any two nodes, implying swift information distribution throughout the network. This is observed in random and 'small-world' networks. 
        \item[$\circ$] \emph{Local efficiency}. Once again, it is possible to measure the efficiency at a local level and then average this across all nodes. This metric, known as the local efficiency or $E_l$, is defined as:
        \begin{equation}
            E_l = \frac{1}{N} \sum_i E_g^{(i)}\ , 
        \end{equation}
        where $E_g^{(i)}$ is the global efficiency of the subgraph induced by the node $i$ and its neighbors. High local efficiency in a network means that removing a node would not significantly disrupt communication between its neighbours. For this reason, it is often considered a proxy for the robustness or resilience of the network to attacks. Much like global efficiency, local efficiency is a characteristic trait of small-world networks. However, unlike global efficiency, high local efficiency is not typically found in random networks.
    \end{itemize}    

\subsubsection{Degrees}
The degree distribution encapsulates fundamental information about the network's structure, robustness to failures, and information spreading dynamics \cite{albert2002}. The \emph{degree distribution} of a given network is simply given by
\begin{equation}
    P_{deg}(k) = x^{(k)}_d/N\ ,
\end{equation}
where $x^{(k)}_d$ is as in \eqref{e-deg_k}. Alternatively, one can look at the \emph{cumulative degree distribution} 
\begin{equation}
    P_{deg}^{(c)}(k) = P_{deg}(j\ge k) = \frac{1}{N}\sum_{j\ge k}x^{(j)}_d
\end{equation}
Cumulative distributions provide a more effective visualization of the degree structure, smoothing out fluctuations and making it easier to identify long-tail behaviors indicative of scale-free networks\sidenote{In particular, if the underlying distribution has a power-law behaviour $\sim k^{-\gamma}$, then the cumulative distribution goes as $\sim k^{-\gamma+1}$ \cite{caldarelli2007}}.



\backmatter 
\setchapterstyle{plain} 



\printglossary[title=Glossary: biology for physicists, toctitle=Glossary: biology for physicists] 


\defbibnote{bibnote}{Here are the references in citation order.\par\bigskip} 
\printbibliography[heading=bibintoc, title=Bibliography, prenote=bibnote] 

@book{sporns2016,
  title={Networks of the Brain},
  author={Sporns, Olaf},
  year={2016},
  publisher={MIT press}
}

@article{butts2009,
  title={Revisiting the foundations of network analysis},
  author={Butts, Carter T},
  journal={science},
  volume={325},
  number={5939},
  pages={414--416},
  year={2009},
  publisher={American Association for the Advancement of Science}
}

@article{peel2022,
  title={Statistical inference links data and theory in network science},
  author={Peel, Leto and Peixoto, Tiago P and De Domenico, Manlio},
  journal={Nature Communications},
  volume={13},
  number={1},
  pages={6794},
  year={2022},
  publisher={Nature Publishing Group UK London}
}

@book{cilliers2002,
  title={Complexity and postmodernism: Understanding complex systems},
  author={Cilliers, Paul},
  year={2002},
  publisher={routledge}
}

@incollection{wigner1990,
  title={The unreasonable effectiveness of mathematics in the natural sciences},
  author={Wigner, Eugene P},
  booktitle={Mathematics and science},
  pages={291--306},
  year={1990},
  publisher={World Scientific}
}

@article{lecun2015,
  title={Deep learning},
  author={LeCun, Yann and Bengio, Yoshua and Hinton, Geoffrey},
  journal={nature},
  volume={521},
  number={7553},
  pages={436--444},
  year={2015},
  publisher={Nature Publishing Group UK London}
}

@article{halevy2009,
  title={The unreasonable effectiveness of data},
  author={Halevy, Alon and Norvig, Peter and Pereira, Fernando},
  journal={IEEE intelligent systems},
  volume={24},
  number={2},
  pages={8--12},
  year={2009},
  publisher={IEEE}
}

@book{gitelman2013,
  title={Raw data is an oxymoron},
  author={Gitelman, Lisa},
  year={2013},
  publisher={MIT press}
}

@article{hosni2018,
  title={Data science and the art of modelling},
  author={Hosni, Hykel and Vulpiani, Angelo},
  journal={Lettera Matematica},
  volume={6},
  pages={121--129},
  year={2018},
  publisher={Springer}
}

@article{leonelli2014,
  title={What difference does quantity make? On the epistemology of Big Data in biology},
  author={Leonelli, Sabina},
  journal={Big data \& society},
  volume={1},
  number={1},
  pages={2053951714534395},
  year={2014},
  publisher={SAGE Publications Sage UK: London, England}
}

@article{anderson2008,
  title={The end of theory: The data deluge makes the scientific method obsolete},
  author={Anderson, Chris},
  journal={Wired magazine},
  volume={16},
  number={7},
  pages={16--07},
  year={2008}
}

@article{aebersold2016,
  title={Mass-spectrometric exploration of proteome structure and function},
  author={Aebersold, Ruedi and Mann, Matthias},
  journal={Nature},
  volume={537},
  number={7620},
  pages={347--355},
  year={2016},
  publisher={Nature Publishing Group UK London}
}

@article{briggman2012,
  title={Volume electron microscopy for neuronal circuit reconstruction},
  author={Briggman, Kevin L and Bock, Davi D},
  journal={Current opinion in neurobiology},
  volume={22},
  number={1},
  pages={154--161},
  year={2012},
  publisher={Elsevier}
}

@article{mardis2008,
  title={The impact of next-generation sequencing technology on genetics},
  author={Mardis, Elaine R},
  journal={Trends in genetics},
  volume={24},
  number={3},
  pages={133--141},
  year={2008},
  publisher={Elsevier}
}

@article{collins2003,
  title={The Human Genome Project: lessons from large-scale biology},
  author={Collins, Francis S and Morgan, Michael and Patrinos, Aristides},
  journal={Science},
  volume={300},
  number={5617},
  pages={286--290},
  year={2003},
  publisher={American Association for the Advancement of Science}
}

@article{koga2012,
  title={Thermal adaptation of the archaeal and bacterial lipid membranes},
  author={Koga, Yosuke},
  journal={Archaea},
  volume={2012},
  year={2012},
  publisher={Hindawi}
}

@article{girard2007,
  title={WormBook: the online review of Caenorhabditis elegans biology},
  author={Girard, Lisa R and Fiedler, Tristan J and Harris, Todd W and Carvalho, Felicia and Antoshechkin, Igor and Han, Michael and Sternberg, Paul W and Stein, Lincoln D and Chalfie, Martin},
  journal={Nucleic acids research},
  volume={35},
  number={suppl\_1},
  pages={D472--D475},
  year={2007},
  publisher={Oxford University Press}
}

@article{sanchez2020,
  title={On the origin of oxygenic photosynthesis and Cyanobacteria},
  author={S{\'a}nchez-Baracaldo, Patricia and Cardona, Tanai},
  journal={New Phytologist},
  volume={225},
  number={4},
  pages={1440--1446},
  year={2020},
  publisher={Wiley Online Library}
}

@article{cannon1939,
  title={The wisdom of the body},
  author={Cannon, Walter Bradford},
  year={1939},
  publisher={Norton \& Co.}
}

@article{ivanov2006,
  title={The development of the concepts of homeothermy and thermoregulation},
  author={Ivanov, KP},
  journal={Journal of Thermal Biology},
  volume={31},
  number={1-2},
  pages={24--29},
  year={2006},
  publisher={Elsevier}
}

@article{smith2003,
  title={Eutrophication of freshwater and coastal marine ecosystems a global problem},
  author={Smith, Val H},
  journal={Environmental Science and Pollution Research},
  volume={10},
  pages={126--139},
  year={2003},
  publisher={Springer}
}

@article{hessen2013,
  title={Ecological stoichiometry: an elementary approach using basic principles},
  author={Hessen, Dag O and Elser, James J and Sterner, Robert W and Urabe, Jotaro},
  journal={Limnology and Oceanography},
  volume={58},
  number={6},
  pages={2219--2236},
  year={2013},
  publisher={Wiley Online Library}
}

@article{frauenfelder2014,
  title={Ask not what physics can do for biology—ask what biology can do for physics},
  author={Frauenfelder, Hans},
  journal={Physical biology},
  volume={11},
  number={5},
  pages={053004},
  year={2014},
  publisher={IOP Publishing}
}

@article{logothetis2008,
  title={What we can do and what we cannot do with fMRI},
  author={Logothetis, Nikos K},
  journal={Nature},
  volume={453},
  number={7197},
  pages={869--878},
  year={2008},
  publisher={Nature Publishing Group UK London}
}

@article{barrangou2016,
  title={Applications of CRISPR technologies in research and beyond},
  author={Barrangou, Rodolphe and Doudna, Jennifer A},
  journal={Nature biotechnology},
  volume={34},
  number={9},
  pages={933--941},
  year={2016},
  publisher={Nature Publishing Group US New York}
}

@book{nelson2008,
  title={Biological physics: energy, information, life},
  author={Nelson, Philip Charles and Radosavljevi{\'c}, Marko and Bromberg, Sarina and Goodsell, David S},
  year={2008},
  publisher={WH Freeman New York}
}

@book{nas2022,
  title={Physics of life},
  author="National Academies of Sciences Engineering and Medicine",
  year={2022},
  publisher={The National Academies Press}
}

@article{frauenfelder1999,
  title={Biological physics},
  author={Frauenfelder, Hans and Wolynes, Peter G and Austin, Robert H},
  journal={Reviews of Modern Physics},
  volume={71},
  number={2},
  pages={S419},
  year={1999},
  publisher={APS}
}

@article{loofbourow1940,
  title={Borderland problems in biology and physics},
  author={Loofbourow, John R},
  journal={Reviews of modern physics},
  volume={12},
  number={4},
  pages={267},
  year={1940},
  publisher={APS}
}

@article{copenhagen2021,
  title={Topological defects promote layer formation in Myxococcus xanthus colonies},
  author={Copenhagen, Katherine and Alert, Ricard and Wingreen, Ned S and Shaevitz, Joshua W},
  journal={Nature Physics},
  volume={17},
  number={2},
  pages={211--215},
  year={2021},
  publisher={Nature Publishing Group UK London}
}

@article{meshulam2019,
  title={Coarse graining, fixed points, and scaling in a large population of neurons},
  author={Meshulam, Leenoy and Gauthier, Jeffrey L and Brody, Carlos D and Tank, David W and Bialek, William},
  journal={Physical review letters},
  volume={123},
  number={17},
  pages={178103},
  year={2019},
  publisher={APS}
}

@article{good2017,
  title={The dynamics of molecular evolution over 60,000 generations},
  author={Good, Benjamin H and McDonald, Michael J and Barrick, Jeffrey E and Lenski, Richard E and Desai, Michael M},
  journal={Nature},
  volume={551},
  number={7678},
  pages={45--50},
  year={2017},
  publisher={Nature Publishing Group UK London}
}

@article{cavagna2023,
  title={Natural swarms in 3.99 dimensions},
  author={Cavagna, Andrea and Di Carlo, Luca and Giardina, Irene and Grigera, Tomas S and Melillo, Stefania and Parisi, Leonardo and Pisegna, Giulia and Scandolo, Mattia},
  journal={Nature Physics},
  pages={1--7},
  year={2023},
  publisher={Nature Publishing Group UK London}
}

@book{schroedinger1944,
  author = {Schr{\"o}dinger, Erwin},
  publisher = {Cambridge University Press, Cambridge},
  title = {What is Life?},
  year = 1944
}

@article{randi2022,
       author = {{Randi}, Francesco and {Sharma}, Anuj K. and {Dvali}, Sophie and {Leifer}, Andrew M.},
        title = "{Neural signal propagation atlas of C. elegans}",
      journal = {arXiv e-prints},
         year = 2022,
        month = aug,
          eid = {arXiv:2208.04790},
        pages = {arXiv:2208.04790},
          doi = {10.48550/arXiv.2208.04790},
    }

@article{seung2011,
  title={Towards functional connectomics},
  author={Seung, H Sebastian},
  journal={Nature},
  volume={471},
  number={7337},
  pages={171--172},
  year={2011},
  publisher={Nature Publishing Group UK London}
}

@article{troemel1997,
  title={Reprogramming chemotaxis responses: sensory neurons define olfactory preferences in C. elegans},
  author={Troemel, Emily R and Kimmel, Bruce E and Bargmann, Cornelia I},
  journal={Cell},
  volume={91},
  number={2},
  pages={161--169},
  year={1997},
  publisher={Elsevier}
}

@article{chalfie1985,
  title={The neural circuit for touch sensitivity in Caenorhabditis elegans},
  author={Chalfie, Martin and Sulston, John E and White, John G and Southgate, Eileen and Thomson, J Nicol and Brenner, Sydney},
  journal={Journal of Neuroscience},
  volume={5},
  number={4},
  pages={956--964},
  year={1985},
  publisher={Soc Neuroscience}
}

@article{gray2005,
  title={A circuit for navigation in Caenorhabditis elegans},
  author={Gray, Jesse M and Hill, Joseph J and Bargmann, Cornelia I},
  journal={Proceedings of the National Academy of Sciences},
  volume={102},
  number={9},
  pages={3184--3191},
  year={2005},
  publisher={National Acad Sciences}
}

@article{hobert2002,
  title={Left--right asymmetry in the nervous system: the Caenorhabditis elegans model},
  author={Hobert, Oliver and Johnston Jr, Robert J and Chang, Sarah},
  journal={Nature Reviews Neuroscience},
  volume={3},
  number={8},
  pages={629--640},
  year={2002},
  publisher={Nature Publishing Group UK London}
}

@article{metaxakis2018,
  title={Multimodal sensory processing in Caenorhabditis elegans},
  author={Metaxakis, Athanasios and Petratou, Dionysia and Tavernarakis, Nektarios},
  journal={Open biology},
  volume={8},
  number={6},
  pages={180049},
  year={2018},
  publisher={The Royal Society}
}

@article{hobert2003,
  title={Behavioral plasticity in C. elegans: paradigms, circuits, genes},
  author={Hobert, Oliver},
  journal={Journal of neurobiology},
  volume={54},
  number={1},
  pages={203--223},
  year={2003},
  publisher={Wiley Online Library}
}

@article{moyle2021,
  title={Structural and developmental principles of neuropil assembly in C. elegans},
  author={Moyle, Mark W and Barnes, Kristopher M and Kuchroo, Manik and Gonopolskiy, Alex and Duncan, Leighton H and Sengupta, Titas and Shao, Lin and Guo, Min and Santella, Anthony and Christensen, Ryan and others},
  journal={Nature},
  volume={591},
  number={7848},
  pages={99--104},
  year={2021},
  publisher={Nature Publishing Group UK London}
}

@article{brittin2021,
  title={A multi-scale brain map derived from whole-brain volumetric reconstructions},
  author={Brittin, Christopher A and Cook, Steven J and Hall, David H and Emmons, Scott W and Cohen, Netta},
  journal={Nature},
  volume={591},
  number={7848},
  pages={105--110},
  year={2021},
  publisher={Nature Publishing Group UK London}
}

@article{chen2006,
  title={Wiring optimization can relate neuronal structure and function},
  author={Chen, Beth L and Hall, David H and Chklovskii, Dmitri B},
  journal={Proceedings of the National Academy of Sciences},
  volume={103},
  number={12},
  pages={4723--4728},
  year={2006},
  publisher={National Acad Sciences}
}

@article{perez2007,
  title={Optimally wired subnetwork determines neuroanatomy of Caenorhabditis elegans},
  author={P{\'e}rez-Escudero, Alfonso and de Polavieja, Gonzalo G},
  journal={Proceedings of the National Academy of Sciences},
  volume={104},
  number={43},
  pages={17180--17185},
  year={2007},
  publisher={National Acad Sciences}
}

@article{perez2009,
  title={Structure of deviations from optimality in biological systems},
  author={P{\'e}rez-Escudero, Alfonso and Rivera-Alba, Marta and de Polavieja, Gonzalo G},
  journal={Proceedings of the National Academy of Sciences},
  volume={106},
  number={48},
  pages={20544--20549},
  year={2009},
  publisher={National Acad Sciences}
}

@article{dyson2004,
  title={A meeting with Enrico Fermi},
  author={Dyson, Freeman},
  journal={Nature},
  volume={427},
  number={6972},
  pages={297--297},
  year={2004}
}

@article{kessy2018,
  title={Optimal whitening and decorrelation},
  author={Kessy, Agnan and Lewin, Alex and Strimmer, Korbinian},
  journal={The American Statistician},
  volume={72},
  number={4},
  pages={309--314},
  year={2018},
  publisher={Taylor \& Francis}
}

@article{mahalanobis1936,
  author = {Mahalanobis, Prasanta Chandra},
  journal = {Proceedings of the National Institute of Sciences (Calcutta)},
  pages = {49--55},
  title = {On the generalized distance in statistics},
  volume = 2,
  year = 1936
}

@article{klintsova1999,
  title={Synaptic plasticity in cortical systems},
  author={Klintsova, Anna Y and Greenough, William T},
  journal={Current opinion in neurobiology},
  volume={9},
  number={2},
  pages={203--208},
  year={1999},
  publisher={Elsevier}
}

@article{pan2010,
  title={Mesoscopic organization reveals the constraints governing Caenorhabditis elegans nervous system},
  author={Pan, Raj Kumar and Chatterjee, Nivedita and Sinha, Sitabhra},
  journal={PloS one},
  volume={5},
  number={2},
  pages={e9240},
  year={2010},
  publisher={Public Library of Science San Francisco, USA}
}

@article{azulay2016,
  title={The C. elegans connectome consists of homogenous circuits with defined functional roles},
  author={Azulay, Aharon and Itskovits, Eyal and Zaslaver, Alon},
  journal={PLoS computational biology},
  volume={12},
  number={9},
  pages={e1005021},
  year={2016},
  publisher={Public Library of Science San Francisco, CA USA}
}

@article{arnatkeviciute2018,
  title={Hub connectivity, neuronal diversity, and gene expression in the Caenorhabditis elegans connectome},
  author={Arnatkeviciute, Aurina and Fulcher, Ben D and Pocock, Roger and Fornito, Alex},
  journal={PLoS computational biology},
  volume={14},
  number={2},
  pages={e1005989},
  year={2018},
  publisher={Public Library of Science San Francisco, CA USA}
}

@article{towlson2013,
  title={The rich club of the C. elegans neuronal connectome},
  author={Towlson, Emma K and V{\'e}rtes, Petra E and Ahnert, Sebastian E and Schafer, William R and Bullmore, Edward T},
  journal={Journal of Neuroscience},
  volume={33},
  number={15},
  pages={6380--6387},
  year={2013},
  publisher={Soc Neuroscience}
}

@article{simpson2012,
  title={An exponential random graph modeling approach to creating group-based representative whole-brain connectivity networks},
  author={Simpson, Sean L and Moussa, Malaak N and Laurienti, Paul J},
  journal={Neuroimage},
  volume={60},
  number={2},
  pages={1117--1126},
  year={2012},
  publisher={Elsevier}
}

@book{gauch2003,
  title={Scientific method in practice},
  author={Gauch, Hugh G},
  year={2003},
  publisher={Cambridge University Press}
}

@book{sober2015,
  title={Ockham's razors},
  author={Sober, Elliott},
  year={2015},
  publisher={Cambridge University Press}
}

@article{witvliet2021,
  title={Connectomes across development reveal principles of brain maturation},
  author={Witvliet, Daniel and Mulcahy, Ben and Mitchell, James K and Meirovitch, Yaron and Berger, Daniel R and Wu, Yuelong and Liu, Yufang and Koh, Wan Xian and Parvathala, Rajeev and Holmyard, Douglas and others},
  journal={Nature},
  volume={596},
  number={7871},
  pages={257--261},
  year={2021},
  publisher={Nature Publishing Group UK London}
}

@article{rapti2020,
  title={A perspective on C. elegans neurodevelopment: from early visionaries to a booming neuroscience research},
  author={Rapti, Georgia},
  journal={Journal of Neurogenetics},
  volume={34},
  number={3-4},
  pages={259--272},
  year={2020},
  publisher={Taylor \& Francis}
}

@article{sulston1977,
  title={Post-embryonic cell lineages of the nematode, Caenorhabditis elegans},
  author={Sulston, John E and Horvitz, H Robert},
  journal={Developmental biology},
  volume={56},
  number={1},
  pages={110--156},
  year={1977},
  publisher={Elsevier}
}

@article{pathak2020,
  title={Developmental trajectory of Caenorhabditis elegans nervous system governs its structural organization},
  author={Pathak, Anand and Chatterjee, Nivedita and Sinha, Sitabhra},
  journal={PLoS computational biology},
  volume={16},
  number={1},
  pages={e1007602},
  year={2020},
  publisher={Public Library of Science San Francisco, CA USA}
}

@article{alicea2018,
  title={The emergent connectome in Caenorhabditis elegans embryogenesis},
  author={Alicea, Bradly},
  journal={BioSystems},
  volume={173},
  pages={247--255},
  year={2018},
  publisher={Elsevier}
}

@article{nicosia2013,
  title={Phase transition in the economically modeled growth of a cellular nervous system},
  author={Nicosia, Vincenzo and V{\'e}rtes, Petra E and Schafer, William R and Latora, Vito and Bullmore, Edward T},
  journal={Proceedings of the National Academy of Sciences},
  volume={110},
  number={19},
  pages={7880--7885},
  year={2013},
  publisher={National Acad Sciences}
}

@article{varier2011,
  title={Neural development features: spatio-temporal development of the Caenorhabditis elegans neuronal network},
  author={Varier, Sreedevi and Kaiser, Marcus},
  journal={PLoS computational biology},
  volume={7},
  number={1},
  pages={e1001044},
  year={2011},
  publisher={Public Library of Science San Francisco, USA}
}

@article{cook2019,
  title={Whole-animal connectomes of both Caenorhabditis elegans sexes},
  author={Cook, Steven J and Jarrell, Travis A and Brittin, Christopher A and Wang, Yi and Bloniarz, Adam E and Yakovlev, Maksim A and Nguyen, Ken CQ and Tang, Leo T-H and Bayer, Emily A and Duerr, Janet S and others},
  journal={Nature},
  volume={571},
  number={7763},
  pages={63--71},
  year={2019},
  publisher={Nature Publishing Group UK London}
}

@article{bentley2016,
  title={The multilayer connectome of Caenorhabditis elegans},
  author={Bentley, Barry and Branicky, Robyn and Barnes, Christopher L and Chew, Yee Lian and Yemini, Eviatar and Bullmore, Edward T and V{\'e}rtes, Petra E and Schafer, William R},
  journal={PLoS computational biology},
  volume={12},
  number={12},
  pages={e1005283},
  year={2016},
  publisher={Public Library of Science San Francisco, CA USA}
}

@article{varshney2011,
  title={Structural properties of the Caenorhabditis elegans neuronal network},
  author={Varshney, Lav R and Chen, Beth L and Paniagua, Eric and Hall, David H and Chklovskii, Dmitri B},
  journal={PLoS computational biology},
  volume={7},
  number={2},
  pages={e1001066},
  year={2011},
  publisher={Public Library of Science San Francisco, USA}
}

@inproceedings{altun2005,
  title={Nervous System - General Description},
  booktitle= {WormAtlas Hermaphrodite Handbook},
  author={Zeynep F. Altun and David H. Hall},
  year={2005},
}

@article{debono2005,
  title={Neuronal substrates of complex behaviors in C. elegans},
  author={de Bono, Mario and Villu Maricq, Andres},
  journal={Annu. Rev. Neurosci.},
  volume={28},
  pages={451--501},
  year={2005},
  publisher={Annual Reviews}
}

@article{mulcahy2018,
  title={A pipeline for volume electron microscopy of the Caenorhabditis elegans nervous system},
  author={Mulcahy, Ben and Witvliet, Daniel and Holmyard, Douglas and Mitchell, James and Chisholm, Andrew D and Samuel, Aravinthan DT and Zhen, Mei},
  journal={Frontiers in neural circuits},
  volume={12},
  pages={94},
  year={2018},
  publisher={Frontiers Media SA}
}

@article{helmstaedter2008,
  title={3D structural imaging of the brain with photons and electrons},
  author={Helmstaedter, Moritz and Briggman, Kevin L and Denk, Winfried},
  journal={Current opinion in neurobiology},
  volume={18},
  number={6},
  pages={633--641},
  year={2008},
  publisher={Elsevier}
}

@article{lichtman2014,
  title={The big data challenges of connectomics},
  author={Lichtman, Jeff W and Pfister, Hanspeter and Shavit, Nir},
  journal={Nature neuroscience},
  volume={17},
  number={11},
  pages={1448--1454},
  year={2014},
  publisher={Nature Publishing Group US New York}
}

@article{hildebrand2017,
  title={Whole-brain serial-section electron microscopy in larval zebrafish},
  author={Hildebrand, David Grant Colburn and Cicconet, Marcelo and Torres, Russel Miguel and Choi, Woohyuk and Quan, Tran Minh and Moon, Jungmin and Wetzel, Arthur Willis and Scott Champion, Andrew and Graham, Brett Jesse and Randlett, Owen and others},
  journal={Nature},
  volume={545},
  number={7654},
  pages={345--349},
  year={2017},
  publisher={Nature Publishing Group UK London}
}

@article{ryan2016,
  title={The CNS connectome of a tadpole larva of Ciona intestinalis (L.) highlights sidedness in the brain of a chordate sibling},
  author={Ryan, Kerrianne and Lu, Zhiyuan and Meinertzhagen, Ian A},
  journal={Elife},
  volume={5},
  pages={e16962},
  year={2016},
  publisher={eLife Sciences Publications, Ltd}
}

@article{abbott2020,
  title={The mind of a mouse},
  author={Abbott, Larry F and Bock, Davi D and Callaway, Edward M and Denk, Winfried and Dulac, Catherine and Fairhall, Adrienne L and Fiete, Ila and Harris, Kristen M and Helmstaedter, Moritz and Jain, Viren and others},
  journal={Cell},
  volume={182},
  number={6},
  pages={1372--1376},
  year={2020},
  publisher={Elsevier}
}

@article{lee2016,
  title={Anatomy and function of an excitatory network in the visual cortex},
  author={Lee, Wei-Chung Allen and Bonin, Vincent and Reed, Michael and Graham, Brett J and Hood, Greg and Glattfelder, Katie and Reid, R Clay},
  journal={Nature},
  volume={532},
  number={7599},
  pages={370--374},
  year={2016},
  publisher={Nature Publishing Group UK London}
}

@article{helmstaedter2013,
  title={Connectomic reconstruction of the inner plexiform layer in the mouse retina},
  author={Helmstaedter, Moritz and Briggman, Kevin L and Turaga, Srinivas C and Jain, Viren and Seung, H Sebastian and Denk, Winfried},
  journal={Nature},
  volume={500},
  number={7461},
  pages={168--174},
  year={2013},
  publisher={Nature Publishing Group UK London}
}

@article{dorkenwald2023,
  title={Neuronal wiring diagram of an adult brain},
  author={Dorkenwald, Sven and Matsliah, Arie and Sterling, Amy R and Schlegel, Philipp and Yu, Szi-chieh and McKellar, Claire E and Lin, Albert and Costa, Marta and Eichler, Katharina and Yin, Yijie and others},
  journal={bioRxiv},
  pages={2023--06},
  year={2023},
  publisher={Cold Spring Harbor Laboratory}
}

@article{kaletta2006,
  title={Finding function in novel targets: C. elegans as a model organism},
  author={Kaletta, Titus and Hengartner, Michael O},
  journal={Nature reviews Drug discovery},
  volume={5},
  number={5},
  pages={387--399},
  year={2006},
  publisher={Nature Publishing Group UK London}
}

@article{corsi2015,
  title={A transparent window into biology: a primer on Caenorhabditis elegans},
  author={Corsi, Ann K and Wightman, Bruce and Chalfie, Martin},
  journal={Genetics},
  volume={200},
  number={2},
  pages={387--407},
  year={2015},
  publisher={Oxford University Press}
}

@article{haag2018,
  title={From “the worm” to “the worms” and back again: the evolutionary developmental biology of nematodes},
  author={Haag, Eric S and Fitch, David HA and Delattre, Marie},
  journal={Genetics},
  volume={210},
  number={2},
  pages={397--433},
  year={2018},
  publisher={Oxford University Press}
}

@article{hillier2005,
  title={Genomics in C. elegans: so many genes, such a little worm},
  author={Hillier, LaDeana W and Coulson, Alan and Murray, John I and Bao, Zhirong and Sulston, John E and Waterston, Robert H},
  journal={Genome research},
  volume={15},
  number={12},
  pages={1651--1660},
  year={2005},
  publisher={Cold Spring Harbor Lab}
}

@article{ahamed2021,
  title={Capturing the continuous complexity of behaviour in Caenorhabditis elegans},
  author={Ahamed, Tosif and Costa, Antonio C and Stephens, Greg J},
  journal={Nature Physics},
  volume={17},
  number={2},
  pages={275--283},
  year={2021},
  publisher={Nature Publishing Group UK London}
}

@article{white1986,
  title={The structure of the nervous system of the nematode Caenorhabditis elegans},
  author={White, John G and Southgate, Eileen and Thomson, J Nichol and Brenner, Sydney and others},
  journal={Philos Trans R Soc Lond B Biol Sci},
  volume={314},
  number={1165},
  pages={1--340},
  year={1986},
  publisher={Citeseer}
}

@incollection{white2018,
  title={Getting into the mind of a worm—a personal view},
  author={White, John G},
  booktitle={WormBook: The Online Review of C. elegans Biology [Internet]},
  year={2018},
  publisher={WormBook}
}

@article{langen2015,
  title={The developmental rules of neural superposition in Drosophila},
  author={Langen, Marion and Agi, Egemen and Altschuler, Dylan J and Wu, Lani F and Altschuler, Steven J and Hiesinger, Peter Robin},
  journal={Cell},
  volume={162},
  number={1},
  pages={120--133},
  year={2015},
  publisher={Elsevier}
}

@article{langen2013,
  title={Mutual inhibition among postmitotic neurons regulates robustness of brain wiring in Drosophila},
  author={Langen, Marion and Koch, Marta and Yan, Jiekun and De Geest, Natalie and Erfurth, Maria-Luise and Pfeiffer, Barret D and Schmucker, Dietmar and Moreau, Yves and Hassan, Bassem A},
  journal={Elife},
  volume={2},
  pages={e00337},
  year={2013},
  publisher={eLife Sciences Publications, Ltd}
}

@book{kandel2000,
  title={Principles of neural science},
  author={Kandel, Eric R and Schwartz, James H and Jessell, Thomas M and Siegelbaum, Steven and Hudspeth, A James and Mack, Sarah and others},
  volume={4},
  year={2000},
  place={New York},
  publisher={McGraw-hill}
}

@article{bekkers1991,
  title={Excitatory and inhibitory autaptic currents in isolated hippocampal neurons maintained in cell culture.},
  author={Bekkers, JM and Stevens, CF},
  journal={Proceedings of the National Academy of Sciences},
  volume={88},
  number={17},
  pages={7834--7838},
  year={1991},
  publisher={National Acad Sciences}
}

@article{sanes2020,
  title={Synaptic specificity, recognition molecules, and assembly of neural circuits},
  author={Sanes, Joshua R and Zipursky, S Lawrence},
  journal={Cell},
  volume={181},
  number={3},
  pages={536--556},
  year={2020},
  publisher={Elsevier}
}

@article{yogev2014,
  title={Cellular and molecular mechanisms of synaptic specificity},
  author={Yogev, Shaul and Shen, Kang},
  journal={Annual review of cell and developmental biology},
  volume={30},
  pages={417--437},
  year={2014},
  publisher={Annual Reviews}
}

@article{kolodkin2011,
  title={Mechanisms and molecules of neuronal wiring: a primer},
  author={Kolodkin, Alex L and Tessier-Lavigne, Marc},
  journal={Cold Spring Harbor perspectives in biology},
  volume={3},
  number={6},
  pages={a001727},
  year={2011},
  publisher={Cold Spring Harbor Lab}
}

@book{finch2000,
  title={Chance, development, and aging},
  author={Finch, Caleb Ellicott and Kirkwood, Thomas BL},
  year={2000},
  publisher={Oxford University Press, USA}
}

@article{schmitt2007,
  title={Review of twin and family studies on neuroanatomic phenotypes and typical neurodevelopment},
  author={Schmitt, J Eric and Eyler, Lisa T and Giedd, Jay N and Kremen, William S and Kendler, Kenneth S and Neale, Michael C},
  journal={Twin Research and Human Genetics},
  volume={10},
  number={5},
  pages={683--694},
  year={2007},
  publisher={Cambridge University Press}
}

@article{steeves1983,
  title={Variability in the structure of an identified interneurone in isogenic clones of locusts},
  author={Steeves, John D and Pearson, Keir G},
  journal={Journal of Experimental Biology},
  volume={103},
  number={1},
  pages={47--54},
  year={1983},
  publisher={The Company of Biologists Ltd}
}

@article{goodman1978,
  title={Isogenic grasshoppers: genetic variability in the morphology of identified neurons},
  author={Goodman, Corey S},
  journal={Journal of Comparative Neurology},
  volume={182},
  number={4},
  pages={681--705},
  year={1978},
  publisher={Wiley Online Library}
}

@article{ward1975,
  title={Electron microscopical reconstruction of the anterior sensory anatomy of the nematode Caenorhabditis elegans},
  author={Ward, Samuel and Thomson, Nichol and White, John G and Brenner, Sydney},
  journal={Journal of Comparative Neurology},
  volume={160},
  number={3},
  pages={313--337},
  year={1975},
  publisher={Wiley Online Library}
}

@article{clarke2012,
  title={The limits of brain determinacy},
  author={Clarke, Peter GH},
  journal={Proceedings of the Royal Society B: Biological Sciences},
  volume={279},
  number={1734},
  pages={1665--1674},
  year={2012},
  publisher={The Royal Society}
}

@book{helmholtz2009, 
    place={Cambridge}, 
    edition={3}, 
    title={On the Sensations of Tone as a Physiological Basis for the Theory of Music}, 
    publisher={Cambridge University Press}, 
    author={Helmholtz, Hermann L. F.}, 
    editor={Ellis, Alexander J.Translator}, 
    year={2009}, 
    collection={Cambridge Library Collection - Music}}

@article{cobb2021,
  title={A brief history of wires in the brain},
  author={Cobb, Matthew},
  journal={Frontiers in Ecology and Evolution},
  volume={9},
  pages={760269},
  year={2021},
  publisher={Frontiers}
}

@article{hiesinger2021,
  title={Brain wiring with composite instructions},
  author={Hiesinger, P Robin},
  journal={BioEssays},
  volume={43},
  number={1},
  pages={2000166},
  year={2021},
  publisher={Wiley Online Library}
}

@article{hiesinger2018,
  title={The evolution of variability and robustness in neural development},
  author={Hiesinger, P Robin and Hassan, Bassem A},
  journal={Trends in Neurosciences},
  volume={41},
  number={9},
  pages={577--586},
  year={2018},
  publisher={Elsevier}
}

@article{hassan2015,
  title={Beyond molecular codes: simple rules to wire complex brains},
  author={Hassan, Bassem A and Hiesinger, P Robin},
  journal={Cell},
  volume={163},
  number={2},
  pages={285--291},
  year={2015},
  publisher={Elsevier}
}

@book{newman2018,
  title={Networks},
  author={Newman, Mark},
  year={2018},
  publisher={Oxford university press}
}

@book{latora2017,
  title={Complex networks: principles, methods and applications},
  author={Latora, Vito and Nicosia, Vincenzo and Russo, Giovanni},
  year={2017},
  publisher={Cambridge University Press}
}

@article{milo2002,
  title={Network motifs: simple building blocks of complex networks},
  author={Milo, Ron and Shen-Orr, Shai and Itzkovitz, Shalev and Kashtan, Nadav and Chklovskii, Dmitri and Alon, Uri},
  journal={Science},
  volume={298},
  number={5594},
  pages={824--827},
  year={2002},
  publisher={American Association for the Advancement of Science}
}

@article{latora2001,
  title={Efficient behavior of small-world networks},
  author={Latora, Vito and Marchiori, Massimo},
  journal={Physical review letters},
  volume={87},
  number={19},
  pages={198701},
  year={2001},
  publisher={APS}
}

@article{watts1998,
  title={Collective dynamics of ‘small-world’networks},
  author={Watts, Duncan J and Strogatz, Steven H},
  journal={nature},
  volume={393},
  number={6684},
  pages={440--442},
  year={1998},
  publisher={Nature Publishing Group}
}

@article{strogatz2001,
  title={Exploring complex networks},
  author={Strogatz, Steven H},
  journal={nature},
  volume={410},
  number={6825},
  pages={268--276},
  year={2001},
  publisher={Nature Publishing Group UK London}
}

@book{caldarelli2007,
  title={Scale-Free Networks - Complex Webs in Nature and Technology},
  author={Guido Caldarelli},
  year={2007},
}

@article{bialek2017,
  title={Perspectives on theory at the interface of physics and biology},
  author={Bialek, William},
  journal={Reports on Progress in Physics},
  volume={81},
  number={1},
  pages={012601},
  year={2017},
  publisher={IOP Publishing}
}

@book{bialek2012,
  title={Biophysics: searching for principles},
  author={Bialek, William},
  year={2012},
  publisher={Princeton University Press}
}

@article{heams2014,
  title={Randomness in biology},
  author={Heams, Thomas},
  journal={Mathematical Structures in Computer Science},
  volume={24},
  number={3},
  pages={e240308},
  year={2014},
  publisher={Cambridge University Press}
}

@book{garson2019,
  title={What biological functions are and why they matter},
  author={Garson, Justin},
  year={2019},
  publisher={Cambridge University Press}
}

@article{faisal2008,
  title={Noise in the nervous system},
  author={Faisal, A Aldo and Selen, Luc PJ and Wolpert, Daniel M},
  journal={Nature reviews neuroscience},
  volume={9},
  number={4},
  pages={292--303},
  year={2008},
  publisher={Nature Publishing Group UK London}
}

@article{guo2018,
  title={Functional importance of noise in neuronal information processing},
  author={Guo, Daqing and Perc, Matja{\v{z}} and Liu, Tiejun and Yao, Dezhong},
  journal={Europhysics Letters},
  volume={124},
  number={5},
  pages={50001},
  year={2018},
  publisher={IOP Publishing}
}

@article{balazsi2011,
  title={Cellular decision making and biological noise: from microbes to mammals},
  author={Bal{\'a}zsi, G{\'a}bor and Van Oudenaarden, Alexander and Collins, James J},
  journal={Cell},
  volume={144},
  number={6},
  pages={910--925},
  year={2011},
  publisher={Elsevier}
}

@article{raj2008,
  title={Nature, nurture, or chance: stochastic gene expression and its consequences},
  author={Raj, Arjun and Van Oudenaarden, Alexander},
  journal={Cell},
  volume={135},
  number={2},
  pages={216--226},
  year={2008},
  publisher={Elsevier}
}

@article{kaern2005,
  title={Stochasticity in gene expression: from theories to phenotypes},
  author={Kaern, Mads and Elston, Timothy C and Blake, William J and Collins, James J},
  journal={Nature Reviews Genetics},
  volume={6},
  number={6},
  pages={451--464},
  year={2005},
  publisher={Nature Publishing Group UK London}
}

@book{kaneko2006,
  title={Life: an introduction to complex systems biology},
  author={Kaneko, Kunihiko},
  year={2006},
  publisher={Springer}
}

@article{tsimring2014,
  title={Noise in biology},
  author={Tsimring, Lev S},
  journal={Reports on Progress in Physics},
  volume={77},
  number={2},
  pages={026601},
  year={2014},
  publisher={IOP Publishing}
}

@article{altan2020,
  title={Quantitative immunology for physicists},
  author={Altan-Bonnet, Gr{\'e}goire and Mora, Thierry and Walczak, Aleksandra M},
  journal={Physics Reports},
  volume={849},
  pages={1--83},
  year={2020},
  publisher={Elsevier}
}

@article{wilson2021,
  title={Balancing exploration and exploitation with information and randomization},
  author={Wilson, Robert C and Bonawitz, Elizabeth and Costa, Vincent D and Ebitz, R Becket},
  journal={Current opinion in behavioral sciences},
  volume={38},
  pages={49--56},
  year={2021},
  publisher={Elsevier}
}

@article{hills2015,
  title={Exploration versus exploitation in space, mind, and society},
  author={Hills, Thomas T and Todd, Peter M and Lazer, David and Redish, A David and Couzin, Iain D},
  journal={Trends in cognitive sciences},
  volume={19},
  number={1},
  pages={46--54},
  year={2015},
  publisher={Elsevier}
}

@article{sims2008,
  title={Scaling laws of marine predator search behaviour},
  author={Sims, David W and Southall, Emily J and Humphries, Nicolas E and Hays, Graeme C and Bradshaw, Corey JA and Pitchford, Jonathan W and James, Alex and Ahmed, Mohammed Z and Brierley, Andrew S and Hindell, Mark A and others},
  journal={Nature},
  volume={451},
  number={7182},
  pages={1098--1102},
  year={2008},
  publisher={Nature Publishing Group UK London}
}

@article{bengough2006,
  title={Root responses to soil physical conditions; growth dynamics from field to cell},
  author={Bengough, A Glyn and Bransby, M Fraser and Hans, Joachim and McKenna, Stephen J and Roberts, Tim J and Valentine, Tracy A},
  journal={Journal of experimental botany},
  volume={57},
  number={2},
  pages={437--447},
  year={2006},
  publisher={Oxford University Press}
}

@article{czaczkes2015,
  title={Trail pheromones: an integrative view of their role in social insect colony organization},
  author={Czaczkes, Tomer J and Gr{\"u}ter, Christoph and Ratnieks, Francis LW},
  journal={Annual review of entomology},
  volume={60},
  pages={581--599},
  year={2015},
  publisher={Annual Reviews}
}

@article{dickson2008,
  title={Wired for sex: the neurobiology of Drosophila mating decisions},
  author={Dickson, Barry J},
  journal={Science},
  volume={322},
  number={5903},
  pages={904--909},
  year={2008},
  publisher={American Association for the Advancement of Science}
}

@article{dukas2015,
  title={Fruit fly courtship: The female perspective},
  author={Dukas, Reuven and Scott, Andrew},
  journal={Current Zoology},
  volume={61},
  number={6},
  pages={1008--1014},
  year={2015},
  publisher={Oxford University Press Oxford, UK}
}

@article{coen2014,
  title={Dynamic sensory cues shape song structure in Drosophila},
  author={Coen, Philip and Clemens, Jan and Weinstein, Andrew J and Pacheco, Diego A and Deng, Yi and Murthy, Mala},
  journal={Nature},
  volume={507},
  number={7491},
  pages={233--237},
  year={2014},
  publisher={Nature Publishing Group UK London}
}

@article{reid2012,
  title={Slime mold uses an externalized spatial “memory” to navigate in complex environments},
  author={Reid, Chris R and Latty, Tanya and Dussutour, Audrey and Beekman, Madeleine},
  journal={Proceedings of the National Academy of Sciences},
  volume={109},
  number={43},
  pages={17490--17494},
  year={2012},
  publisher={National Acad Sciences}
}

@article{dyer2002,
  title={The biology of the dance language},
  author={Dyer, Fred C},
  journal={Annual review of entomology},
  volume={47},
  number={1},
  pages={917--949},
  year={2002},
  publisher={Annual Reviews 4139 El Camino Way, PO Box 10139, Palo Alto, CA 94303-0139, USA}
}

@article{seeley1991,
  title={Collective decision-making in honey bees: how colonies choose among nectar sources},
  author={Seeley, Thomas D and Camazine, Scott and Sneyd, James},
  journal={Behavioral Ecology and Sociobiology},
  volume={28},
  pages={277--290},
  year={1991},
  publisher={Springer}
}

@article{gruter2009,
  title={The honeybee waggle dance: can we follow the steps?},
  author={Gr{\"u}ter, Christoph and Farina, Walter M},
  journal={Trends in ecology \& evolution},
  volume={24},
  number={5},
  pages={242--247},
  year={2009},
  publisher={Elsevier}
}

@article{sourjik2012,
  title={Responding to chemical gradients: bacterial chemotaxis},
  author={Sourjik, Victor and Wingreen, Ned S},
  journal={Current opinion in cell biology},
  volume={24},
  number={2},
  pages={262--268},
  year={2012},
  publisher={Elsevier}
}

@article{webre2003,
  title={Bacterial chemotaxis},
  author={Webre, Daniel J and Wolanin, Peter M and Stock, Jeffry B},
  journal={Current Biology},
  volume={13},
  number={2},
  pages={R47--R49},
  year={2003},
  publisher={Elsevier}
}

@book{cox2006,
  title={Principles of statistical inference},
  author={Cox, David Roxbee},
  year={2006},
  publisher={Cambridge university press}
}

@article{bianconi2014,
  title={Triadic closure as a basic generating mechanism of communities in complex networks},
  author={Bianconi, Ginestra and Darst, Richard K and Iacovacci, Jacopo and Fortunato, Santo},
  journal={Physical Review E},
  volume={90},
  number={4},
  pages={042806},
  year={2014},
  publisher={APS}
}

@article{hunter2007,
  title={Curved exponential family models for social networks},
  author={Hunter, David R},
  journal={Social networks},
  volume={29},
  number={2},
  pages={216--230},
  year={2007},
  publisher={Elsevier}
}

@article{efron1975,
  title={Defining the curvature of a statistical problem (with applications to second order efficiency)},
  author={Efron, Bradley},
  journal={The Annals of Statistics},
  pages={1189--1242},
  year={1975},
  publisher={JSTOR}
}

@article{efron1978,
  title={The geometry of exponential families},
  author={Efron, Bradley},
  journal={The Annals of Statistics},
  pages={362--376},
  year={1978},
  publisher={JSTOR}
}

@article{hunter2006,
  title={Inference in curved exponential family models for networks},
  author={Hunter, David R and Handcock, Mark S},
  journal={Journal of Computational and Graphical Statistics},
  volume={15},
  number={3},
  pages={565--583},
  year={2006},
  publisher={Taylor \& Francis}
}

@article{snijders2006,
  title={New specifications for exponential random graph models},
  author={Snijders, Tom AB and Pattison, Philippa E and Robins, Garry L and Handcock, Mark S},
  journal={Sociological methodology},
  volume={36},
  number={1},
  pages={99--153},
  year={2006},
  publisher={Wiley Online Library}
}

@article{schweinberger2020,
  title={Exponential-Family Models of Random Graphs: Inference in Finite, Super and Infinite Population Scenarios},
  author={Schweinberger, Michael and Krivitsky, Pavel N and Butts, Carter T and Stewart, Jonathan R},
  journal={Statistical Science},
  volume={35},
  number={4},
  pages={627--662},
  year={2020},
  publisher={Institute of Mathematical Statistics}
}

@article{schweinberger2011,
  title={Instability, sensitivity, and degeneracy of discrete exponential families},
  author={Schweinberger, Michael},
  journal={Journal of the American Statistical Association},
  volume={106},
  number={496},
  pages={1361--1370},
  year={2011},
  publisher={Taylor \& Francis}
}

@article{rinaldo2008,
  title={On the geometry of discrete exponential families with application to exponential random graph models},
  author={Rinaldo, Alessandro and Fienberg, Stephen E and Zhou, Yi},
  journal={Electronic Journal of Statistics},
  year={2008},
  volume={3},
  pages={446-484}
}

@article{snijders2002,
  title={Markov chain Monte Carlo estimation of exponential random graph models},
  author={Snijders, Tom A. B.},
  journal={Journal of Social Structure},
  volume={3},
  number={2},
  pages={1--40},
  year={2002}
}

@inproceedings{kira1992,
  title={The feature selection problem: Traditional methods and a new algorithm},
  author={Kira, Kenji and Rendell, Larry A},
  booktitle={Proceedings of the tenth national conference on Artificial intelligence},
  pages={129--134},
  year={1992}
}

@article{guyon2003,
  title={An introduction to variable and feature selection},
  author={Guyon, Isabelle and Elisseeff, Andr{\'e}},
  journal={Journal of machine learning research},
  volume={3},
  number={Mar},
  pages={1157--1182},
  year={2003}
}

@book{burnham1998,
  title={Practical use of the information-theoretic approach},
  author={Burnham, Kenneth P and Anderson, David R and Burnham, Kenneth P and Anderson, David R},
  year={1998},
  publisher={Springer}
}

@software{krivitsky2022,
  title = {Statnet:  Tools for the Statistical Modeling of Network Data},
  author = {Krivitsky, Pavel N. and Handcock, Mark S. and Hunter, David R. and Butts, Carter T. and Klumb, Chad and Goodreau, Steven M. and Morris, Martina},
  date = {2003/2022},
  howpublished = {\url{https://statnet.org}},
  organization = {Statnet Development Team}
}

@article{krivitsky2023,
  title={ergm 4: New features for analyzing exponential-family random graph models},
  author={Krivitsky, Pavel N. and Hunter, David R. and Morris, Martina and Klumb, Chad},
  journal={Journal of Statistical Software},
  volume={105},
  pages={1--44},
  year={2023}
}

@article{hunter2008,
  title={ergm: A package to fit, simulate and diagnose exponential-family models for networks},
  author={Hunter, David R and Handcock, Mark S and Butts, Carter T and Goodreau, Steven M and Morris, Martina},
  journal={Journal of statistical software},
  volume={24},
  number={3},
  pages={nihpa54860},
  year={2008},
  publisher={NIH Public Access}
}

@article{geyer1991,
  title={Markov chain Monte Carlo maximum likelihood},
  author={Geyer, Charles J},
  year={1991},
  publisher={Interface Foundation of North America}
}

@book{parisi1998,
  title={Statistical Field Theory},
  author={Parisi, Giorgio},
  year={1998},
  publisher={Avalon Publishing}
}

@article{park2004b,
  title={Solution of the two-star model of a network},
  author={Park, Juyong and Newman, Mark EJ},
  journal={Physical Review E},
  volume={70},
  number={6},
  pages={066146},
  year={2004},
  publisher={APS}
}

@article{park2005,
  title={Solution for the properties of a clustered network},
  author={Park, Juyong and Newman, Mark EJ},
  journal={Physical Review E},
  volume={72},
  number={2},
  pages={026136},
  year={2005},
  publisher={APS}
}

@article{castellani2005,
  title={Spin-glass theory for pedestrians},
  author={Castellani, Tommaso and Cavagna, Andrea},
  journal={Journal of Statistical Mechanics: Theory and Experiment},
  volume={2005},
  number={05},
  pages={P05012},
  year={2005},
  publisher={IOP Publishing}
}

@book{rossi2018,
  title={Mathematical statistics: an introduction to likelihood based inference},
  author={Rossi, Richard J},
  year={2018},
  publisher={John Wiley \& Sons}
}

@book{mackay2003,
  title={Information theory, inference and learning algorithms},
  author={MacKay, David JC},
  year={2003},
  publisher={Cambridge university press}
}

@article{erdhos1960,
  title={On the evolution of random graphs},
  author={Erd{\H{o}}s, Paul and R{\'e}nyi, Alfr{\'e}d and others},
  journal={Publ. math. inst. hung. acad. sci},
  volume={5},
  number={1},
  pages={17--60},
  year={1960}
}

@article{albert2002,
  title={Statistical mechanics of complex networks},
  author={Albert, R{\'e}ka and Barab{\'a}si, Albert-L{\'a}szl{\'o}},
  journal={Reviews of modern physics},
  volume={74},
  number={1},
  pages={47},
  year={2002},
  publisher={APS}
}

@article{nguyen2017,
  title={Inverse statistical problems: from the inverse Ising problem to data science},
  author={Nguyen, H Chau and Zecchina, Riccardo and Berg, Johannes},
  journal={Advances in Physics},
  volume={66},
  number={3},
  pages={197--261},
  year={2017},
  publisher={Taylor \& Francis}
}

@article{chen2019,
  title={Searching for collective behavior in a small brain},
  author={Chen, Xiaowen and Randi, Francesco and Leifer, Andrew M and Bialek, William},
  journal={Physical Review E},
  volume={99},
  number={5},
  pages={052418},
  year={2019},
  publisher={APS}
}

@article{mora2010,
  title={Maximum entropy models for antibody diversity},
  author={Mora, Thierry and Walczak, Aleksandra M and Bialek, William and Callan Jr, Curtis G},
  journal={Proceedings of the National Academy of Sciences},
  volume={107},
  number={12},
  pages={5405--5410},
  year={2010},
  publisher={National Acad Sciences}
}

@book{salmon2006,
  title={Four decades of scientific explanation},
  author={Salmon, Wesley C},
  year={2006},
  publisher={University of Pittsburgh Press}
}

@article{hempel1948,
  title={Studies in the Logic of Explanation},
  author={Hempel, Carl G and Oppenheim, Paul},
  journal={Philosophy of science},
  volume={15},
  number={2},
  pages={135--175},
  year={1948},
  publisher={Williams and Wilkins Co.}
}

@article{besag1974,
  title = {Spatial {{Interaction}} and the {{Statistical Analysis}} of {{Lattice Systems}}},
  author = {Besag, Julian},
  year = {1974},
  journal = {Journal of the Royal Statistical Society. Series B (Methodological)},
  volume = {36},
  number = {2},
  eprint = {2984812},
  eprinttype = {jstor},
  pages = {192--236},
  publisher = {{[Royal Statistical Society, Wiley]}},
  issn = {0035-9246}
}

@article{bianconi2009,
  title={Entropy of network ensembles},
  author={Bianconi, Ginestra},
  journal={Physical Review E},
  volume={79},
  number={3},
  pages={036114},
  year={2009},
  publisher={APS}
}

@book{huang1987,
  author = {Huang, Kerson},
  edition = 2,
  publisher = {John Wiley \& Sons},
  title = {Statistical Mechanics},
  year = 1987
}

@article{dienes2011,
  title={Bayesian versus orthodox statistics: Which side are you on?},
  author={Dienes, Zoltan},
  journal={Perspectives on Psychological Science},
  volume={6},
  number={3},
  pages={274--290},
  year={2011},
  publisher={Sage Publications Sage CA: Los Angeles, CA}
}

@article{aurell2016,
  title={The maximum entropy fallacy redux?},
  author={Aurell, Erik},
  journal={PLoS computational biology},
  volume={12},
  number={5},
  pages={e1004777},
  year={2016},
  publisher={Public Library of Science San Francisco, CA USA}
}

@article{auletta2017,
  title={On the relevance of the maximum entropy principle in non-equilibrium statistical mechanics},
  author={Auletta, Gennaro and Rondoni, Lamberto and Vulpiani, Angelo},
  journal={The European Physical Journal Special Topics},
  volume={226},
  number={10},
  pages={2327--2343},
  year={2017},
  publisher={Springer}
}

@book{jaynes2003,
  title={Probability theory: The logic of science},
  author={Jaynes, Edwin T},
  year={2003},
  publisher={Cambridge university press}
}

@article{shannon1948,
  title={A mathematical theory of communication},
  author={Shannon, Claude E.},
  journal={The Bell system technical journal},
  volume={27},
  number={3},
  pages={379--423},
  year={1948},
  publisher={Nokia Bell Labs}
}

@book{cover1999,
  title={Elements of information theory},
  author={Cover, Thomas M and Thomas, Joy A.},
  year={1999},
  publisher={John Wiley \& Sons}
}

@article{jaynes1982,
  title={On the rationale of maximum-entropy methods},
  author={Jaynes, Edwin T.},
  journal={Proceedings of the IEEE},
  volume={70},
  number={9},
  pages={939--952},
  year={1982},
  publisher={IEEE}
}

@article{jaynes1957,
  title={Information theory and statistical mechanics},
  author={Jaynes, Edwin T.},
  journal={Physical review},
  volume={106},
  number={4},
  pages={620},
  year={1957},
  publisher={APS}
}

@article{jaynes1957b,
  title={Information theory and statistical mechanics. II},
  author={Jaynes, Edwin T.},
  journal={Physical review},
  volume={108},
  number={2},
  pages={171},
  year={1957},
  publisher={APS}
}

@article{fisher1922,
  title={On the mathematical foundations of theoretical statistics},
  author={Fisher, Ronald A},
  journal={Philosophical Transactions of the Royal Society of London. Series A, Containing Papers of a Mathematical or Physical Character},
  volume={222},
  number={594-604},
  pages={309--368},
  year={1922},
  publisher={The Royal Society London}
}

@book{lusher2013,
  title={Exponential random graph models for social networks: Theory, methods, and applications},
  author={Lusher, Dean and Koskinen, Johan and Robins, Garry},
  year={2013},
  publisher={Cambridge University Press}
}

@article{ghafouri2020,
  title={A survey on exponential random graph models: an application perspective},
  author={Ghafouri, Saeid and Khasteh, Seyed Hossein},
  journal={PeerJ Computer Science},
  volume={6},
  pages={e269},
  year={2020},
  publisher={PeerJ Inc.}
}

@article{cimini2019,
  title={The statistical physics of real-world networks},
  author={Cimini, Giulio and Squartini, Tiziano and Saracco, Fabio and Garlaschelli, Diego and Gabrielli, Andrea and Caldarelli, Guido},
  journal={Nature Reviews Physics},
  volume={1},
  number={1},
  pages={58--71},
  year={2019},
  publisher={Nature Publishing Group UK London}
}

@article{park2004,
  title={Statistical mechanics of networks},
  author={Park, Juyong and Newman, Mark EJ},
  journal={Physical Review E},
  volume={70},
  number={6},
  pages={066117},
  year={2004},
  publisher={APS}
}

@article{holland1981,
  title={An exponential family of probability distributions for directed graphs},
  author={Holland, Paul W and Leinhardt, Samuel},
  journal={Journal of the american Statistical association},
  volume={76},
  number={373},
  pages={33--50},
  year={1981},
  publisher={Taylor \& Francis}
}

@article{ove1986,
  title = {Markov {{Graphs}}},
  author = {Frank, Ove and Strauss, David},
  year = {1986},
  journal = {Journal of the American Statistical Association},
  volume = {81},
  number = {395},
  eprint = {2289017},
  eprinttype = {jstor},
  pages = {832--842},
  publisher = {{[American Statistical Association, Taylor \& Francis, Ltd.]}},
  issn = {0162-1459}
}

@article{strauss1986,
  title={On a general class of models for interaction},
  author={Strauss, David},
  journal={SIAM review},
  volume={28},
  number={4},
  pages={513--527},
  year={1986},
  publisher={SIAM}
}

@inproceedings{pitman1936,
  title={Sufficient statistics and intrinsic accuracy},
  author={Pitman, Edwin James George},
  booktitle={Mathematical Proceedings of the cambridge Philosophical society},
  volume={32},
  pages={567--579},
  year={1936},
  organization={Cambridge University Press}
}

@article{koopman1936,
  title={On distributions admitting a sufficient statistic},
  author={Koopman, Bernard O},
  journal={Transactions of the American Mathematical society},
  volume={39},
  number={3},
  pages={399--409},
  year={1936},
  publisher={JSTOR}
}

@article{darmois1935,
  title={Sur les lois de probabilit{\'e}a estimation exhaustive},
  author={Darmois, Georges},
  journal={CR Acad. Sci. Paris},
  volume={260},
  number={1265},
  pages={85},
  year={1935}
}

@book{brown1986,
  title={Fundamentals of statistical exponential families: with applications in statistical decision theory},
  author={Brown, Lawrence D},
  year={1978},
  publisher={Ims}
}

@book{barndorff2014,
  title={Information and exponential families: in statistical theory},
  author={Barndorff-Nielsen, Ole},
  year={2014},
  publisher={John Wiley \& Sons}
}

@book{sundberg2019,
  title={Statistical modelling by exponential families},
  author={Sundberg, Rolf},
  volume={12},
  year={2019},
  publisher={Cambridge University Press}
}

@article{nielsen2009,
       author = {{Nielsen}, Frank and {Garcia}, Vincent},
        title = "{Statistical exponential families: A digest with flash cards}",
      journal = {arXiv e-prints},
         year = 2009,
        month = nov,
        pages = {arXiv:0911.4863},
}

@book{darwin1859,
  author = {Darwin, Charles R.},
  publisher = {Murray},
  title = {On the Origin of Species by Means of Natural Selection},
  year = 1859
}

@book{rose2000,
  title={Darwin's spectre: evolutionary biology in the modern world},
  author={Rose, Michael R},
  year={2000},
  publisher={Princeton University Press}
}

@article{mayr2001,
  title={The philosophical foundations of Darwinism},
  author={Mayr, Ernst},
  journal={Proceedings of the American Philosophical Society},
  volume={145},
  number={4},
  pages={488--495},
  year={2001},
  publisher={JSTOR}
}

@article{huxley1942,
  title={Evolution. The modern synthesis.},
  author={Huxley, Julian},
  journal={Evolution. The Modern Synthesis.},
  year={1942},
  publisher={London: George Alien \& Unwin Ltd.}
}

@article{adami2002,
  title={What is complexity?},
  author={Adami, Christoph},
  journal={BioEssays},
  volume={24},
  number={12},
  pages={1085--1094},
  year={2002},
  publisher={Wiley Online Library}
}

@article{adami2012,
  title={The use of information theory in evolutionary biology},
  author={Adami, Christoph},
  journal={Annals of the New York Academy of Sciences},
  volume={1256},
  number={1},
  pages={49--65},
  year={2012},
  publisher={Wiley Online Library}
}

@article{wolf2018,
  title={Physical foundations of biological complexity},
  author={Wolf, Yuri I and Katsnelson, Mikhail I and Koonin, Eugene V},
  journal={Proceedings of the National Academy of Sciences},
  volume={115},
  number={37},
  pages={E8678--E8687},
  year={2018},
  publisher={National Acad Sciences}
}

@article{goldenfeld2011,
  title={Life is physics: evolution as a collective phenomenon far from equilibrium},
  author={Goldenfeld, Nigel and Woese, Carl},
  journal={Annu. Rev. Condens. Matter Phys.},
  volume={2},
  number={1},
  pages={375--399},
  year={2011},
  publisher={Annual Reviews}
}

@article{fisher1953,
  title={Croonian Lecture-Population genetics},
  author={Fisher, Ronald A.},
  journal={Proceedings of the Royal Society of London. Series B-Biological Sciences},
  volume={141},
  number={905},
  pages={510--523},
  year={1953},
  publisher={The Royal Society London}
}

@book{hamilton2021,
    author    = {Hamilton, Matthew B.},
    title     = {Population Genetics},
    year      = {2021},
    edition   = {2},
    publisher = {John Wiley \& Sons},
    isbn   = {9781444362459},
}

@article{manrubia2021,
  title={From genotypes to organisms: State-of-the-art and perspectives of a cornerstone in evolutionary dynamics},
  author={Manrubia, Susanna and Cuesta, Jos{\'e} A and Aguirre, Jacobo and Ahnert, Sebastian E and Altenberg, Lee and Cano, Alejandro V and Catal{\'a}n, Pablo and Diaz-Uriarte, Ramon and Elena, Santiago F and Garc{\'\i}a-Mart{\'\i}n, Juan Antonio and others},
  journal={Physics of Life Reviews},
  volume={38},
  pages={55--106},
  year={2021},
  publisher={Elsevier}
}

@article{rao1992,
  title={Ronald A. Fisher: The founder of modern statistics},
  author={Rao, C. Radhakrishna},
  journal={Statistical science},
  pages={34--48},
  year={1992},
  publisher={JSTOR}
}

@book{hartl1997,
  title={Principles of population genetics},
  author={Hartl, Daniel L. and Clark, Andrew G.},
  volume={116},
  year={1997},
  edition = {4},
  publisher={Sinauer associates}
}

@book{klug2019,
  title = {Concepts of Genetics},
  author = {Klug, William S. and Cummings, Michael R. and Spencer, C. and Palladino, M. and Killian, D.},
  date = {2019},
  series = {Always Learning},
  publisher = {Pearson},
  edition = {12},
  isbn = {978-1-292-26532-2}
}

@article{charlesworth2017,
  title={Population genetics from 1966 to 2016},
  author={Charlesworth, Brian and Charlesworth, Deborah},
  journal={Heredity},
  volume={118},
  number={1},
  pages={2--9},
  year={2017},
  publisher={Nature Publishing Group}
}

@book{crow2017,
  title={An introduction to population genetics theory},
  author={Crow, James F.},
  year={2017},
  publisher={Scientific Publishers}
}

@article{pence2018,
  title={Sir John FW Herschel and Charles Darwin: Nineteenth-century science and its methodology},
  author={Pence, Charles H.},
  journal={Hopos: the Journal of the International Society for the History of Philosophy of Science},
  volume={8},
  number={1},
  pages={108--140},
  year={2018},
  publisher={University of Chicago Press Chicago, IL}
}

@article{neher2011,
  title={Statistical genetics and evolution of quantitative traits},
  author={Neher, Richard A and Shraiman, Boris I},
  journal={Reviews of Modern Physics},
  volume={83},
  number={4},
  pages={1283},
  year={2011},
  publisher={APS}
}

@book{fisher1930,
  title={The genetical theory of natural selection},
  author={Fisher, Ronald A.},
  year={1930},
  publisher={Clarendon Press }
}

@article{sella2005,
  title={The application of statistical physics to evolutionary biology},
  author={Sella, Guy and Hirsh, Aaron E},
  journal={Proceedings of the National Academy of Sciences},
  volume={102},
  number={27},
  pages={9541--9546},
  year={2005},
  publisher={National Acad Sciences}
}

@article{rao2022,
  title={Evolutionary dynamics, evolutionary forces, and robustness: A nonequilibrium statistical mechanics perspective},
  author={Rao, Riccardo and Leibler, Stanislas},
  journal={Proceedings of the National Academy of Sciences},
  volume={119},
  number={13},
  pages={e2112083119},
  year={2022},
  publisher={National Acad Sciences}
}

@article{peliti1997,
    author = {{Peliti}, Luca},
    title = "{Introduction to the statistical theory of Darwinian evolution}",
    journal = {arXiv e-prints},
    year = 1997,
    month = dec,
    pages = {cond-mat/9712027},
    archivePrefix = {arXiv}
}

@book{gardiner1985,
  author = {Gardiner, Crispin W.},
  publisher = {Springer},
  title = {Handbook of stochastic methods for physics, chemistry, and the natural sciences},
  year = 1985
}

@article{mauri2021,
  title={Gaussian closure scheme in the quasi-linkage equilibrium regime of evolving genome populations},
  author={Mauri, Eugenio and Cocco, Simona and Monasson, R{\'e}mi},
  journal={Europhysics Letters},
  volume={132},
  number={5},
  pages={56001},
  year={2021},
  publisher={IOP Publishing}
}

@article{neher2009,
  title={Competition between recombination and epistasis can cause a transition from allele to genotype selection},
  author={Neher, Richard A and Shraiman, Boris I},
  journal={Proceedings of the National Academy of Sciences},
  volume={106},
  number={16},
  pages={6866--6871},
  year={2009},
  publisher={National Acad Sciences}
}

@article{zeng2020b,
  title={Inferring genetic fitness from genomic data},
  author={Zeng, Hong-Li and Aurell, Erik},
  journal={Physical Review E},
  volume={101},
  number={5},
  pages={052409},
  year={2020},
  publisher={APS}
}

@article{zanini2012,
  title={FFPopSim: an efficient forward simulation package for the evolution of large populations},
  author={Zanini, Fabio and Neher, Richard A},
  journal={Bioinformatics},
  volume={28},
  number={24},
  pages={3332--3333},
  year={2012},
  publisher={Oxford University Press}
}

@inproceedings{wright1932,
  title={The roles of mutation, inbreeding, crossbreedingand selection in evolution.},
  author={Wright, Sewall},
  booktitle={Proceedings  of  the  Sixth  InternationalCongress  on  Genetics},
  pages={356–366},
  year={1932}
}

@article{neher2018,
    author = {{Neher}, Richard A. and {Walczak}, Aleksandra M.},
    title = "{Progress and open problems in evolutionary dynamics}",
    journal = {arXiv e-prints},
    year = 2018,
    month = apr,
    pages = {arXiv:1804.07720},
    archivePrefix = {arXiv},
}

@article{moran2014,
  title={The tiniest tiny genomes},
  author={Moran, Nancy A and Bennett, Gordon M},
  journal={Annual review of microbiology},
  volume={68},
  pages={195--215},
  year={2014},
  publisher={Annual Reviews}
}

@book{mezard1987,
  title={Spin glass theory and beyond: An Introduction to the Replica Method and Its Applications},
  author={M{\'e}zard, Marc and Parisi, Giorgio and Virasoro, Miguel Angel},
  volume={9},
  year={1987},
  publisher={World Scientific Publishing Company}
}

@article{sherrington1975,
  title={Solvable model of a spin-glass},
  author={Sherrington, David and Kirkpatrick, Scott},
  journal={Physical review letters},
  volume={35},
  number={26},
  pages={1792},
  year={1975},
  publisher={APS}
}

@article{devisser2014,
  title={Empirical fitness landscapes and the predictability of evolution},
  author={De Visser, J. Arjan G.M. and Krug, Joachim},
  journal={Nature Reviews Genetics},
  volume={15},
  number={7},
  pages={480--490},
  year={2014},
  publisher={Nature Publishing Group UK London}
}

@article{fowler2014,
  title={Deep mutational scanning: a new style of protein science},
  author={Fowler, Douglas M. and Fields, Stanley},
  journal={Nature methods},
  volume={11},
  number={8},
  pages={801--807},
  year={2014},
  publisher={Nature Publishing Group US New York}
}

@article{lassig2017,
  title={Predicting evolution},
  author={L{\"a}ssig, Michael and Mustonen, Ville and Walczak, Aleksandra M},
  journal={Nature ecology \& evolution},
  volume={1},
  number={3},
  pages={0077},
  year={2017},
  publisher={Nature Publishing Group UK London}
}

@article{mustonen2009,
  title={From fitness landscapes to seascapes: non-equilibrium dynamics of selection and adaptation},
  author={Mustonen, Ville and L{\"a}ssig, Michael},
  journal={Trends in genetics},
  volume={25},
  number={3},
  pages={111--119},
  year={2009},
  publisher={Elsevier}
}

@book{holland1975,
  author = {Holland, John H.},
  publisher = {MIT Press},
  title = {Adaptation in Natural and Artificial Systems},
  year = {1992}
}

@article{holland1992,
  title={Genetic algorithms},
  author={Holland, John H},
  journal={Scientific american},
  volume={267},
  number={1},
  pages={66--73},
  year={1992},
  publisher={JSTOR}
}

@Book{bierlaire2018,
author = {Michel Bierlaire},
title = {Optimization: Principles and Algorithms},
publisher = {EPFL Press},
address = {Lausanne},
edition = {2nd},
ISBN = {9782940222780},
year = {2018}}

@book{bellman1961,
  author = {Bellman, Richard E},
  title = {Adaptive Control Processes: A Guided Tour},
  publisher = {Princeton Univ. Press},
  year = 1961
}

@article{kirkpatrick1983,
  title={Optimization by simulated annealing},
  author={Kirkpatrick, Scott and Gelatt Jr, C Daniel and Vecchi, Mario P},
  journal={science},
  volume={220},
  number={4598},
  pages={671--680},
  year={1983},
  publisher={American association for the advancement of science}
}

@book{gendreau2010,
  title={Handbook of metaheuristics},
  author={Gendreau, Michel and Potvin, Jean-Yves},
  volume={2},
  year={2010},
  publisher={Springer}
}

@book{goldberg1989,
  author = {Goldberg, David E.},
  publisher = {Addison-Wesley},
  title = {Genetic Algorithms in Search, Optimization, and Machine Learning},
  year = 1989
}

@incollection{dejong1993,
  title={Genetic algorithms are NOT function optimizers},
  author={De Jong, Kenneth A.},
  booktitle={Foundations of genetic algorithms},
  volume={2},
  pages={5--17},
  year={1993},
  publisher={Elsevier}
}

@inproceedings{kennedy1995,
  title={Particle swarm optimization},
  author={Kennedy, James and Eberhart, Russell},
  booktitle={Proceedings of ICNN'95-international conference on neural networks},
  volume={4},
  pages={1942--1948},
  year={1995},
  organization={IEEE}
}

@article{dorigo2006,
  title={Ant colony optimization},
  author={Dorigo, Marco and Birattari, Mauro and Stutzle, Thomas},
  journal={IEEE computational intelligence magazine},
  volume={1},
  number={4},
  pages={28--39},
  year={2006},
  publisher={IEEE}
}

@article{reeves2010,
  title={Genetic algorithms},
  author={Reeves, Colin R.},
  journal={Handbook of metaheuristics},
  pages={109--139},
  year={2010},
  publisher={Springer}
}

@book{reeves2002,
  title={Genetic algorithms: principles and perspectives: a guide to GA theory},
  author={Reeves, Colin and Rowe, Jonathan E},
  volume={20},
  year={2002},
  publisher={Springer Science \& Business Media}
}

@article{mccandlish2011,
  title={Visualizing fitness landscapes},
  author={McCandlish, David M},
  journal={Evolution},
  volume={65},
  number={6},
  pages={1544--1558},
  year={2011},
  publisher={Blackwell Publishing Inc Malden, USA}
}

@article{kaplan2008,
  title={The end of the adaptive landscape metaphor?},
  author={Kaplan, Jonathan},
  journal={Biology \& Philosophy},
  volume={23},
  pages={625--638},
  year={2008},
  publisher={Springer}
}

@article{zeng2020,
  title={Global analysis of more than 50,000 SARS-CoV-2 genomes reveals epistasis between eight viral genes},
  author={Zeng, Hong-Li and Dichio, Vito and Rodr{\'\i}guez Horta, Erwin and Thorell, Kaisa and Aurell, Erik},
  journal={Proceedings of the National Academy of Sciences},
  volume={117},
  number={49},
  pages={31519--31526},
  year={2020},
  publisher={National Acad Sciences}
}

@article{zeng2021,
  title={Inferring epistasis from genomic data with comparable mutation and outcrossing rate},
  author={Zeng, Hong-Li and Mauri, Eugenio and Dichio, Vito and Cocco, Simona and Monasson, Remi and Aurell, Erik},
  journal={Journal of Statistical Mechanics: Theory and Experiment},
  volume={2021},
  number={8},
  pages={083501},
  month=august,
  year={2021},
  publisher={IOP Publishing}
}

@article{zeng2022,
  title={Temporal epistasis inference from more than 3 500 000 SARS-CoV-2 genomic sequences},
  author={Zeng, Hong-Li and Liu, Yue and Dichio, Vito and Aurell, Erik},
  journal={Physical Review E},
  volume={106},
  number={4},
  pages={044409},
  month=october,
  year={2022},
  publisher={APS}
}

@article{dichio2023,
  title={Statistical genetics in and out of quasi-linkage equilibrium},
  author={Dichio, Vito and Zeng, Hong-Li and Aurell, Erik},
  journal={Reports on Progress in Physics},
  year = 2023,
  month = april,
  volume = {86},
  number = {5},
  pages = {052601},
  publisher={IOP Publishing}
}

@article{dichio2023b,
       author = {{Dichio}, Vito and {De Vico Fallani}, Fabrizio},
        title = "{The exploration-exploitation paradigm for networked biological systems}",
      journal = {arXiv e-prints},
         year = 2023,
        month = jun,
archivePrefix = {arXiv},
       eprint = {2306.17300}
}

@article{dichio2023c,
  title={Statistical models of complex brain networks},
  author={Dichio, Vito and Fallani, Fabrizio De Vico},
  journal={Reports on Progress in Physics},
  year = 2023,
  month = august,
  volume = {86},
  number = {10},
  pages = {102601},
  publisher={IOP Publishing}
}








\end{document}